\documentclass[a4paper,12pt]{report}

\usepackage[english,brazilian,french]{babel}
\usepackage[utf8]{inputenc}

\usepackage{cmap}

\usepackage[T1]{fontenc}

\usepackage{lmodern}

\usepackage{microtype}
\microtypesetup{activate={true,nocompatibility}}
\microtypesetup{factor=1100, stretch=10, shrink=10}
\microtypesetup{tracking=true, spacing=true, kerning=true}
\SetTracking{encoding={T1}, shape=sc}{0}
\DisableLigatures[f]{encoding={T1}}
\microtypesetup{final, babel}

\usepackage{amsmath,amssymb,amsfonts,textcomp}

\usepackage[tmargin=3.0cm, bmargin=2.5cm, rmargin=2.5cm, lmargin=3.0cm, a4paper]{geometry}

\usepackage{setspace}

\usepackage{graphicx}
\usepackage{grffile}

\usepackage{color}
\usepackage[usenames,dvipsnames,svgnames]{xcolor}

\usepackage{subfig}
\usepackage[font=footnotesize]{caption}

\usepackage{multicol}

\usepackage{array}

\usepackage{booktabs}

\usepackage{longtable}

\usepackage{lscape}

\usepackage[notintoc,portuguese]{nomencl}
\makenomenclature

\usepackage{footnote}

\makesavenoteenv{tabular}

\usepackage{lastpage}

\usepackage{natbib}

\usepackage[nottoc,notlof,notlot]{tocbibind}

\usepackage{icomma}

\usepackage[tight]{units}

\usepackage{fancyhdr} 
\newcommand{\nomedoaluno}{Elisson Saldanha da Gama de Almeida}
\newcommand{\titulo}{Disques et vents des étoiles chaudes : apport de la spectroscopie et de l'interférométrie multi-bandes}

\usepackage{hyperref}
\hypersetup{colorlinks=true, linkcolor=black, citecolor=black, filecolor=black, urlcolor=black,
            pdfauthor={\nomedoaluno},
            pdftitle={\titulo},
            pdfsubject={Assunto do Projeto},
            pdfkeywords={palavra-chave, palavra-chave, palavra-chave},
            pdfproducer={LaTeX},
            pdfcreator={pdfTeX}}

\DeclareRobustCommand{\ion}[2]{\textup{#1\,\textsc{\lowercase{#2}}}}

\usepackage{adjustbox}

\usepackage{enumitem}
\setlist{parsep=0pt,listparindent=\parindent}

\usepackage[figuresright]{rotating}

\usepackage{tablefootnote}

\usepackage{footmisc}

\providecommand{\e}[1]{\ensuremath{\times 10^{#1}}}

\usepackage[flushleft]{threeparttable}

\usepackage{float}

\usepackage{lipsum}
\usepackage{pifont}

\makeatletter
\newcommand*{\centerfloat}{%
  \parindent \z@
  \leftskip \z@ \@plus 1fil \@minus \textwidth
  \rightskip\leftskip
  \parfillskip \z@skip}
\makeatother

\usepackage{indentfirst}

\usepackage{siunitx}

\newcommand\omicron{o}

\usepackage{pdfpages}

\usepackage{relsize}

\usepackage[thinc]{esdiff}

\usepackage{epstopdf}
\AppendGraphicsExtensions{.gif}

\usepackage{esvect}

\usepackage{commath}

\usepackage{makecell}

\newcommand\blankpage{
    \clearpage
    \begingroup
      \null
      \thispagestyle{empty}%
      \addtocounter{page}{-1}%
      \hypersetup{pageanchor=false}%
      \clearpage
    \endgroup
}

\usepackage{afterpage}

\usepackage{dirtytalk}

\usepackage{listings}

\definecolor{codegreen}{rgb}{0,0.6,0}
\definecolor{codegray}{rgb}{0.5,0.5,0.5}
\definecolor{codepurple}{rgb}{0.58,0,0.82}
\definecolor{backcolour}{rgb}{0.95,0.95,0.92}

\lstdefinestyle{mystyle}{
    backgroundcolor=\color{backcolour},   
    commentstyle=\color{codegreen},
    keywordstyle=\color{magenta},
    numberstyle=\tiny\color{codegray},
    stringstyle=\color{codepurple},
    basicstyle=\ttfamily\footnotesize,
    breakatwhitespace=false,         
    breaklines=true,                 
    captionpos=b,                    
    keepspaces=true,                 
    numbers=left,                    
    numbersep=5pt,                  
    showspaces=false,                
    showstringspaces=false,
    showtabs=false,                  
    tabsize=2
}

\usepackage{listofitems}

\usepackage{etoc}
\usepackage{minitoc}

\begin{document}


\pagestyle{empty}


\includepdf[pages=-]{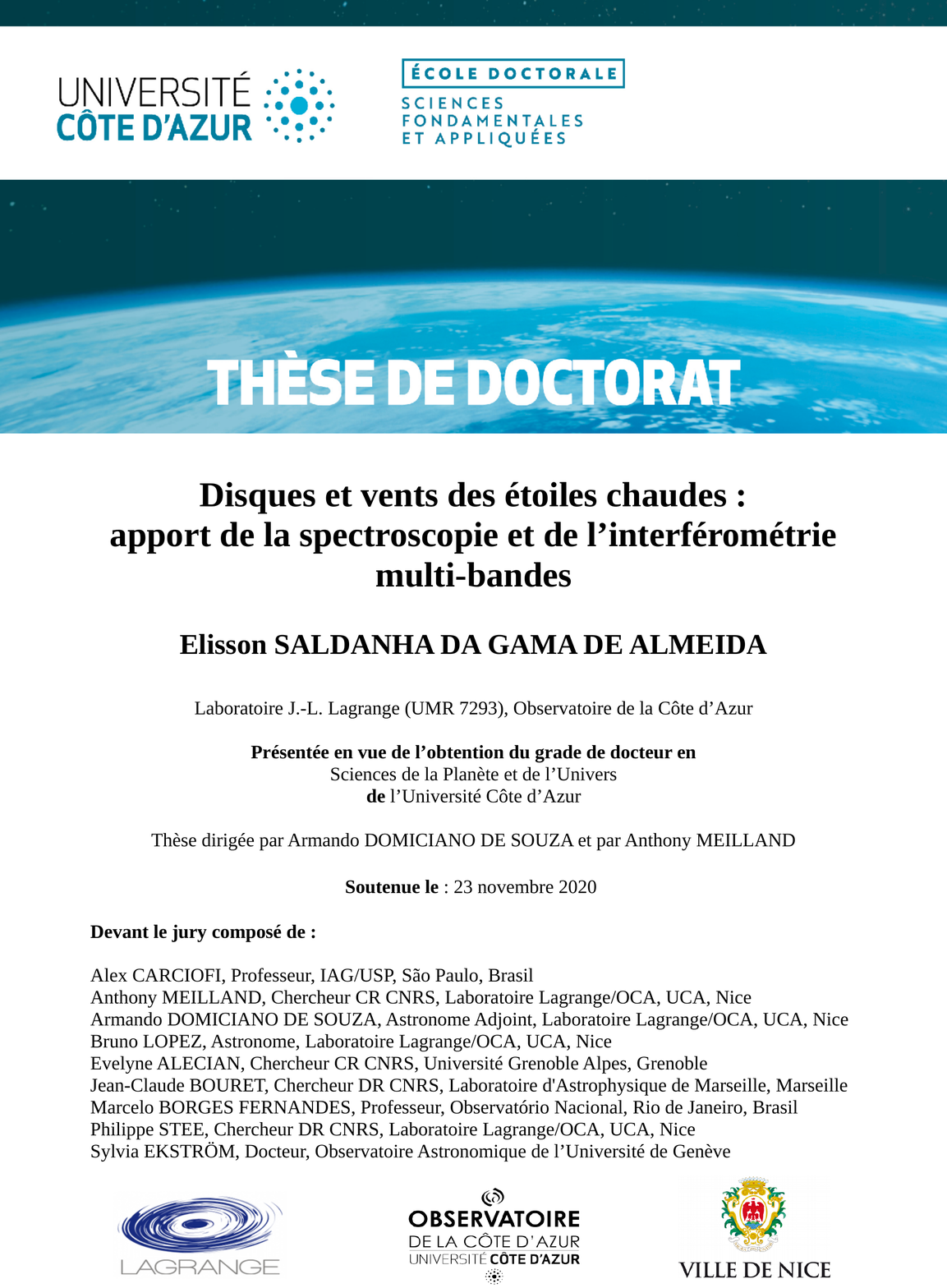}

\afterpage{\blankpage}

\cleardoublepage

\pagestyle{fancy}

\pagenumbering{roman}

\def\blankpage{%
      \clearpage%
      \thispagestyle{empty}%
      \addtocounter{page}{-1}%
      \null%
      \clearpage}



\includepdf[pagecommand={\thispagestyle{plain}}, pages=-]{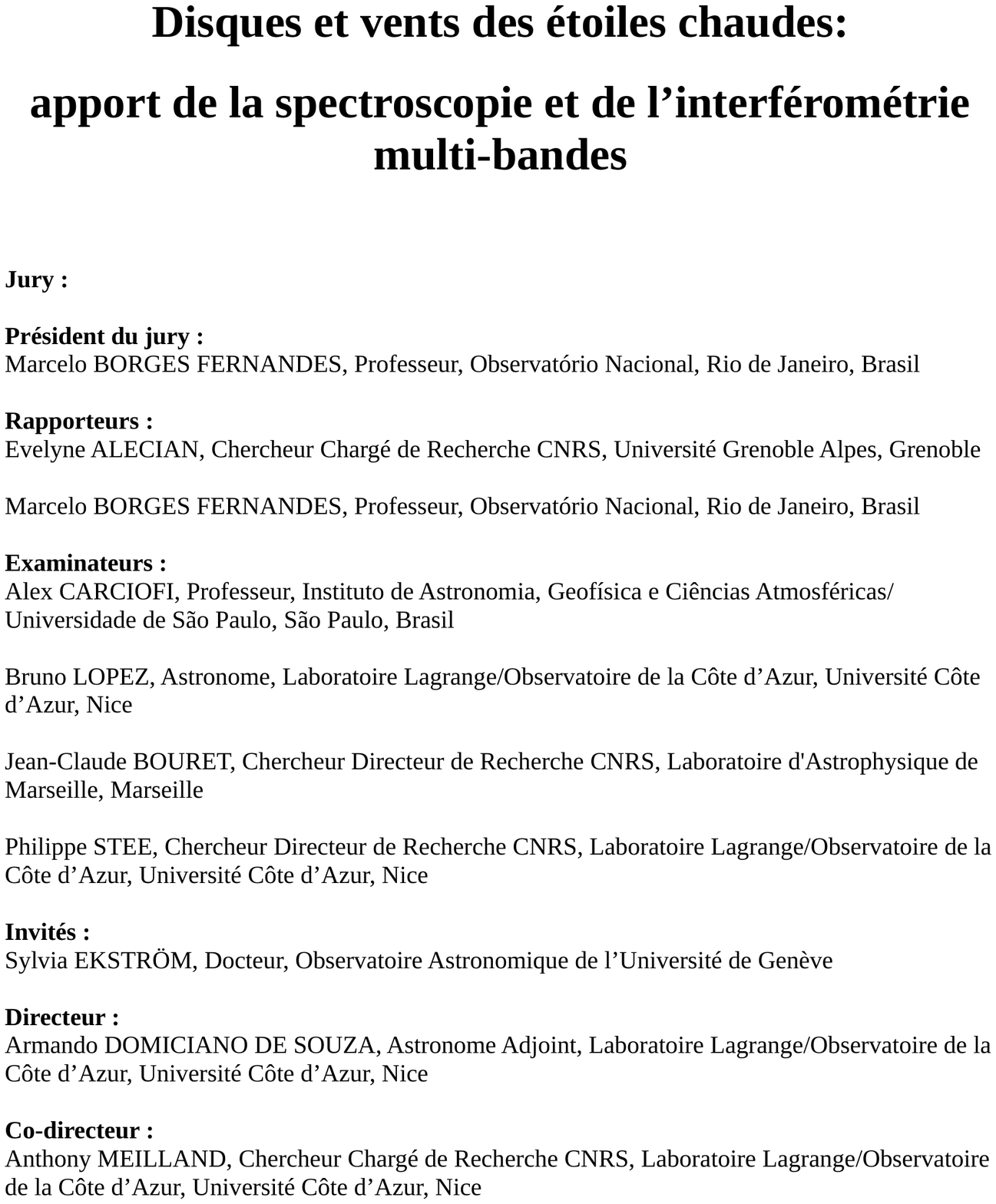}

\vspace*{0.75\textheight}
\begin{flushright}
  \emph{Aos meus pais, Elionai e Sônia, por todo o apoio e carinho.}
\end{flushright}
\newpage



\doublespacing

\noindent{\LARGE\textbf{Acknowledgments/Agradecimentos}}\\[0.6in]
\indent
I would like to thank my advisors, Armando Domiciano de Souza and Anthony Meilland, for all their support, commitment with my work, and for introducing me to the interferometric journey of hot stars. I also thank you for helping me with the practical stuff of life during my stay in Nice. Thank you a lot for helping me with the heavy baggage when I arrived in Nice!\par

I thank Yves Rabbia for sharing his time to teach me the theory of stellar interferometry, and also Denis Mourard and Frédéric Morand for giving me the opportunity to learn about interferometric observations with the CHARA/VEGA instrument at Plateau de Calern. I thank Alain Spang for working on the reduction of new VLTI/AMBER data of Rigel (and also for your occasional visits to my office and nice handshakes), Farrokh Vakili for inviting me to study P Cygni using intensity interferometry, and Philippe Stee for participating of my PhD meetings with Armando and Anthony (also for your constructive comments on my paper about $\omicron$ Aquarii). Philippe and Farrokh, thank you very much for reading and commenting on my thesis manuscript! I also thank Christine Delobelle, Delphine Saissi, and Sophie Rousset for helping me with the french bureaucracy.\par

I thank Julieta Sanchez, Fei Hua, and Mircea Moscu for sharing with me good moments during my stay in Nice. Julieta, thank you very much for receiving me at your home in La Plata. Also many thanks for Alban Ceau, Marc-Antoine Martinod, Pierre Janin-Potiron, Romain Laugie, and Vicent Hocdé: thank you for being patient with my French. I hope to meet all you again to have a glass of beer or a cup of coffee. Massinissa Hadjara, thank you for being a very nice neighbor and for the good green grapes!\par

I would like to thank Alex Carciofi, Bruno Lopez, Evelyne Alecian, Jean-Claude Bouret, Marcelo Borges, Philippe Stee, and Sylvia Ekström for being part of the jury of my thesis and for their constructive comments on my work. Em português, agradeço ao Alex e Marcelo pelos seus comentários detalhados para a versão final da minha tese. Deixo meu agradecimento ao Marcelo por aceitar ser presidente da minha banca e por me ajudar com os preparativos da defesa online no Observatório Nacional.\par

Agradeço aos meus pais, Elionai e Sônia, e minha tia Lila, por todo amor e apoio desde sempre. Também aos melhores amigos do mundo: David Campos, Icaro Rossignoli, Marcio Monteiro, Morgana Romão e Poema Portela. Muito obrigado por estarem sempre ao meu lado em todos os momentos, seja online ou de forma presencial. Obrigado também por suportarem minhas reclamações. Poema e Marcio, obrigado por me visitarem em Nice! Pai, mãe e tia Lila, a visita de vocês também está guardada para sempre comigo com muito carinho. Também deixo meu agradecimento à Alinges Lenz pela escuta recente e pelo incentivo para seguir em frente.\par

\newpage


\begin{center}
  \begin{large}Abstract\end{large}\label{abstract}
\vspace{2pt}
\end{center}
\noindent

{
\small
{\setstretch{1.1}

Hot stars are the main source of ionization of the interstellar medium and its enrichment due to heavy elements. Constraining the physical conditions of their environments is crucial to understand how these stars evolve and their impact on the evolution of galaxies.\par

Spectroscopy allows to access the physics, the chemistry, and the dynamics of these objects, but not the spatial distribution of these objects. Only long-baseline interferometry can resolve photospheres and close environments, and, combining spectroscopy and interferometry, spectro-interferometry allows to draw an even more detailed picture of hot stars.\par 

The objective of my thesis was to investigate the physical properties of the photosphere and circumstellar environment of massive hot stars confronting multi-band spectroscopic or spectro-interferometric observations and sophisticated non-LTE radiative transfer codes.\par

My work was focused on two main lines of research. The first concerns radiative line-driven winds. Using UV and visible spectroscopic data and the radiative transfer code CMFGEN, I investigated the weak wind phenomenon on a sample of nine Galactic O stars. This study shows for the first time that the weak wind phenomenon, originally found for O dwarfs, also exists on more evolved O stars and that future studies must evaluate its impact on the evolution of massive stars.\par

My other line of research concerns the study of classical Be stars, the fastest rotators among the non-degenerated stars, and which are surrounded by rotating equatorial disks. I studied the Be star $\omicron$ Aquarii using H$\alpha$ (CHARA/VEGA) and Br$\gamma$ (VLTI/AMBER) spectro-interferometric observations, the radiative transfer code HDUST, and developing new automatic procedures to better constrain the kinematics of the disk. This multi-band study allowed to draw the most detailed picture of this object and its environment, to test the limits of the current generation of radiative transfer models, and paved the way to my future work on a large samples of Be stars observed with VEGA, AMBER, and the newly available VLTI mid-infrared combiner MATISSE.

} 

\par
\vspace{1em}

}

\vspace*{10pt}
\newpage
\begin{center}
  \begin{large}Résumé\end{large}\label{abstract}
\vspace{2pt}
\end{center}

\selectlanguage{french}
\noindent

{
\small 
{\setstretch{1.1}

Les étoiles chaudes sont la principale source d'ionisation du milieu interstellaire et de son enrichissement en éléments lourds. Contraindre les conditions physiques de leur environnement est crucial pour comprendre comment ces étoiles évoluent et leur impact sur l'évolution des galaxies.\par

La spectroscopie permet d'accéder à la physique, la chimie et la dynamique de ces objets, mais pas à la distribution spatiale de ces objets. L’interférométrie à longue base est la seule technique permettant de résoudre la photosphère et les environnements, et, en combinant spectroscopie et interférométrie, la spectro-interférométrie permet de dresser une image encore plus détaillée des étoiles chaudes.\par

L'objectif de ma thèse était d'étudier les propriétés physiques de la photosphère et de l'environnement circumstellaire d'étoiles chaudes massives, en confrontant des observations spectroscopiques et spectro-interférométriques sur différents domaines de longueur d'onde à des modèles sophistiqués de transfert radiatif hors-ETL.\par

Mon travail s’est focalisé sur deux axes. La première concerne les vents radiatifs. En utilisant des données spectroscopiques UV et visible et le code CMFGEN, j'ai étudié le phénomène des vents faibles sur un échantillon de neuf géantes O galactiques. Cette étude montre pour la première fois que le phénomène des vents faibles, trouvé à l'origine pour les naines O, existe également pour des étoiles O plus évoluées et que des prochaines études doivent évaluer leur effet sur l'évolution des étoiles massives.\par

Mon autre axe de recherche concerne l'étude des étoiles Be classiques, les rotateurs les plus rapides parmi les étoiles non dégénérées et qui sont entourées par des disques équatoriaux en rotation. J'ai étudié l'étoile Be $\omicron$ Aquarii en utilisant des données spectro-interférométriques obtenues en H$\alpha$ (CHARA/VEGA) et Br$\gamma$ (VLTI/AMBER), le code de transfert radiatif HDUST, et en développant de nouvelles procédures automatiques pour mieux contraindre la cinématique des disques. Cette étude multi-bande a permis d’obtenir la vue la plus complète de cet objet et de son environnement, de tester les limites de la génération actuelle de modèles de transfert radiatif, et d’ouvrir la voie à des travaux futurs sur un échantillon large d’étoiles Be observées avec VEGA, AMBER et MATISSE, le nouvel instrument infrarouge thermique du VLTI.\par

} 

\par
\vspace{1em}
\noindent\textbf{Mots-clés:} étoiles: massive, raie d’émission, atmosphères, vents, matière circumstellaire; techniques: spectroscopique, interférométrique.

}

\selectlanguage{english}


\microtypesetup{protrusion=false}

\cleardoublepage

{\setstretch{1.25}
\listoffigures

\listoftables
}

\printnomenclature

\dominitoc

\tableofcontents

\microtypesetup{protrusion=true}


{\setstretch{1.25}

\cleardoublepage

\pagestyle{fancy}

\pagenumbering{arabic}

\chapter{Introduction}
\label{chapter_introduction}
\minitoc

\begin{center}
\makebox[\textwidth][c]{\includegraphics[width=0.45\paperwidth]{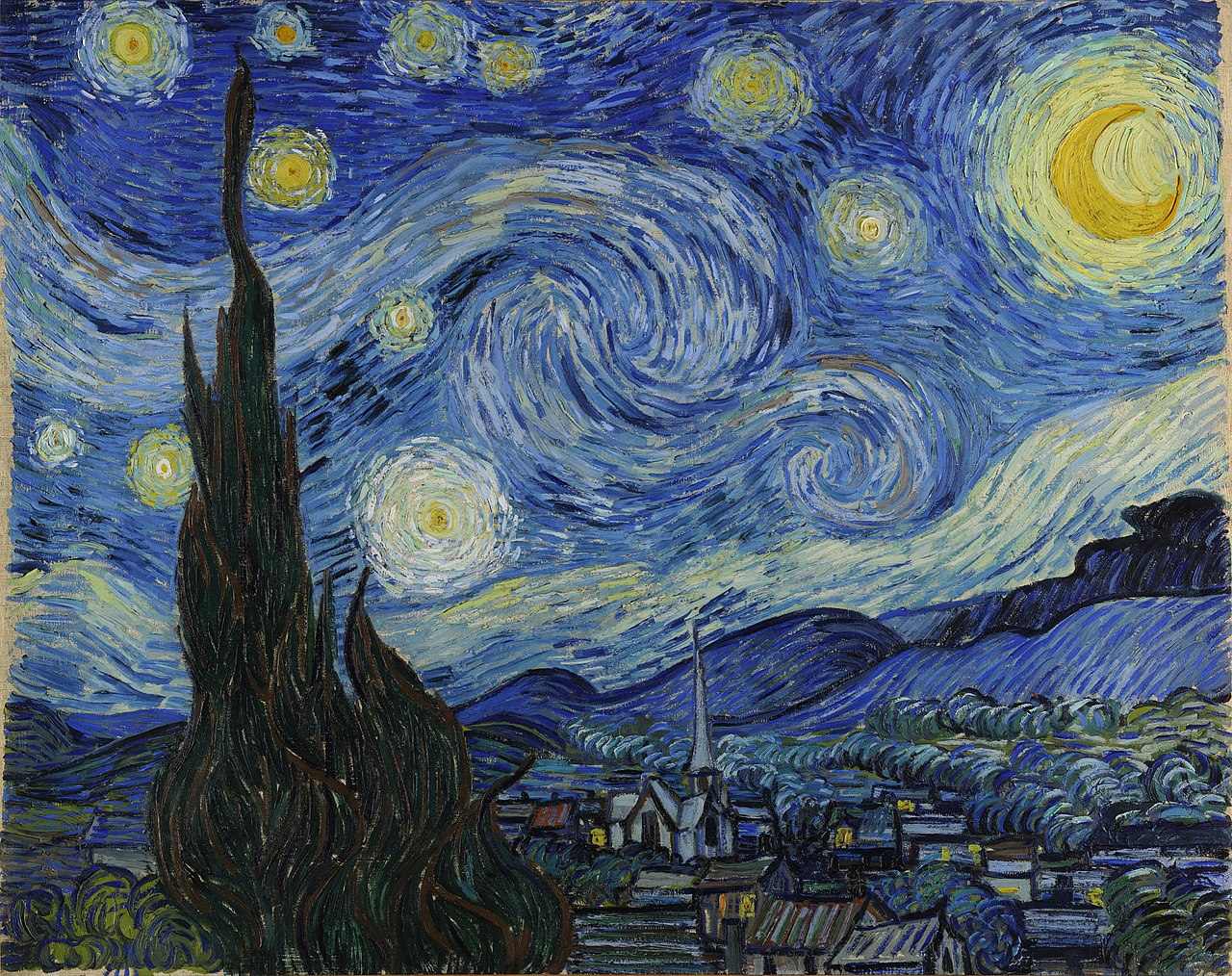}}
\end{center}

\vspace*{0.75cm}

Stars are unique astrophysical objects to test our understanding about the basic structure of nature. For example, observations of solar neutrinos provided the first clear evidence of the quantum phenomenon called neutrino oscillation, implying that at least one of the neutrino flavors has non-zero mass~\citep[e.g., see the Nobel Lecture of][]{mcdonald16}. This disagrees with the so-called Standard Model of elementary particles, and thus opens an entire branch of theoretical research of physics beyond the Standard Model. Conversely, apart from the current understanding of star formation, the study of stars is very well-founded in terms of fundamental physics: combining the background from hydrodynamics, thermodynamics and statistical physics, atomic and molecular physics, and electromagnetism, among others.\par

The aim of this thesis is to analyse the physical properties and mass loss processes of massive hot stars and study their close environments, ranging from radiatively driven winds (with velocities up to a thousand km s\textsuperscript{-1}) to quasi-stable thin equatorial disks in rotation. For this purpose, different observational techniques and also modeling methods are employed in this work. In this chapter, I overview in details the current understanding about OB-type stars.\par

\section{OB-type stars}
\label{sec_intro_hot_massive_stars}

\subsection{Physical properties and stellar evolution}
\label{sec_intro_evol_stages}

O-type stars ($\mathrm{M_{ZAMS}}$\footnote{Zero-age main sequence mass, the initial stellar mass (onset of the H-burning phase).} $\gtrsim$ 17 $\mathrm{M_\odot}$) are subject to extreme physical conditions, showing the highest value of luminosity and effective temperature ($T_{\mathrm{eff}}$) among the canonical Morgan-Keenan system of stellar classification~\citep{morgan43}. Based on sophisticated spectral type calibrations~\citep{martins05_calibration}, a typical late-type O dwarf (O9.5V) has $T_{\mathrm{eff}}$ $\sim$ 30500 K and bolometric luminosity $L_{\star}$ $\sim$ 41700 $\mathrm{L_{\mathrm{\odot}}}$. On the other hand, an earlier and also (supposedly) more evolved O star is expected to have a quite higher effective temperature and luminosity: $T_{\mathrm{eff}}$ $\sim$ 42600 K and $L_{\star}$ $\sim$ 1000000 $L_{\mathrm{\odot}}$ (e.g., O3I).\par

\begin{table}
\caption{\label{table_parameters_OB_dwarfs} Summary of the stellar parameters of OB-type dwarfs (luminosity class V) from spectral type calibrations in the literature. Parameters for B stars are from~\citet{townsend04} and O stars from~\citet{martins05_calibration}.}
\centering
\begin{adjustbox}{width=1.05\textwidth}
\begin{tabular}{lllll|lllll}
\toprule
\toprule
\multicolumn{1}{l}{ST} & \multicolumn{1}{c}{$M_{\star}$ ($\mathrm{M_{\odot}}$)} & \multicolumn{1}{c}{$R_{\star}$ ($\mathrm{R_{\odot}}$)} & \multicolumn{1}{c}{$T_{\mathrm{eff}}$ (K)}  & \multicolumn{1}{c}{$\log L_{\star}/L_{\odot}$} & \multicolumn{1}{|l}{ST} & \multicolumn{1}{c}{$M_{\star}$ ($\mathrm{M_{\odot}}$)} & \multicolumn{1}{c}{$R_{\star}$ ($\mathrm{R_{\odot}}$)} & \multicolumn{1}{c}{$T_{\mathrm{eff}}$ (K)} & \multicolumn{1}{c}{$\log L_{\star}/L_{\odot}$}\\
\midrule 
B9  & 3.4  & 2.8 & 12294 & 2.20 & O9.5 & 16.5 & 7.4 & 30488 & 4.62 \\
B8  & 3.8  & 3.0 & 13250 & 2.39 & O9   & 18.0 & 7.7 & 31524 & 4.72 \\
B7  & 4.2  & 3.2 & 14148 & 2.56 & O8.5 & 19.8 & 8.1 & 32522 & 4.82 \\
B6  & 4.8  & 3.5 & 15355 & 2.78 & O8   & 22.0 & 8.5 & 33383 & 4.90 \\
B5  & 5.5  & 3.8 & 16726 & 3.00 & O7.5 & 24.2 & 8.9 & 34419 & 5.00 \\
B4  & 6.4  & 4.2 & 18267 & 3.24 & 07   & 26.5 & 9.4 & 35531 & 5.10 \\
B3  & 7.7  & 4.7 & 20287 & 3.52 & O6.5 & 29.0 & 9.8 & 36826 & 5.20 \\
B2.5& 8.6  & 5.0 & 21566 & 3.68 & O6   & 31.7 & 10.2 & 38151 & 5.30 \\
B2  & 9.6  & 5.4 & 22887 & 3.85 & O5.5 & 34.2 & 10.6 & 40062 & 5.41 \\
B1.5& 10.8 & 5.7 & 24424 & 4.01 & O5   & 37.3 & 11.1 & 41540 & 5.51 \\
B1  & 12.5 & 6.3 & 26068 & 4.21 & O4   & 46.2 & 12.3 & 43419 & 5.68 \\
B0.5& 14.6 & 6.9 & 27948 & 4.41 & O3   & 58.2 & 13.8 & 44616 & 5.83 \\
B0  & 17.5 & 7.7 & 30201 & 4.64 &  &  &  &  &  \\
\bottomrule
\end{tabular}
\end{adjustbox}
\end{table}

With lower effective temperature, B-type dwarfs encompass both the range of intermediate- and high-mass stars, having $\mathrm{M_{ZAMS}}$ between $\sim$3 $\mathrm{M_\odot}$ (B9V, $T_{\mathrm{eff}}$ $\sim$ 12000 K) and $\sim$18 $\mathrm{M_\odot}$ (B0V, $T_{\mathrm{eff}}$ $\sim$ 30000 K). The physical parameters of OB-type dwarf stars are summarized in Table \ref{table_parameters_OB_dwarfs}\footnote{In Table~\ref{table_parameters_OB_dwarfs}, the quoted value for the initial mass of a B0 dwarf is somewhat larger than for an O9.5 dwarf. We emphasize that this fact is not physically reliable and it happens since the results complied in this table come from different studies in the literature.}. Thus, mainly depending on their initial mass, B dwarfs are progenitors of either planetary nebulae ($\mathrm{M_{ZAMS}}$ $\lesssim$ 8-9 $\mathrm{M_\odot}$), resulting in white dwarfs as stellar remnant, or core-collapse supernovae that result in neutron stars from the more massive progenitors~\citep[e.g., see Fig.~1 of][]{heger03}.\par

Despite being much more abundant than massive stars, based on a typical Salpeter initial mass function~\citep{salpeter55}, low- and intermediate-mass stars ($\mathrm{M_{ZAMS}}$ $\lesssim$ 8 $\mathrm{M_\odot}$) are only able to carry the stellar nucleosynthesis up to the core helium-burning phase, resulting in the production of carbon and oxygen during their more evolved phases as asymptotic giant branch (AGB) stars. From Table \ref{table_parameters_OB_dwarfs}, one sees that this applies to B stars among the types B9 and B3. However, massive stars ($\mathrm{M_{ZAMS}}$ $\gtrsim$ 8 $\mathrm{M_\odot}$) are able to continue the nucleosynthesis in their cores beyond the helium-burning phase up to the production of iron group elements during the silicon-burning phase, resulting in core-collapse supernovae (type II, Ib, or Ic).\par 

Fig.~\ref{sec_intro_schematic_massive_agb} presents a basic scheme of the chemical stratification in the structure of evolved massive stars, as well as of evolved low-intermediate-mass stars (AGB phase). One sees that concentric layers are successively composed by lighter elements (as H and He) toward the stellar surface. We stress that this description for the fate of stars according to the initial mass is a simplified picture. For instance, it is still largely unknown the pathways of stars with initial masses between $\sim$8-10 $\mathrm{M_\odot}$. In this case, depending on the mass loss, they can either evolve to the AGB phase, ending their evolution as planetary nebulae, or evolve up to core-collapse supernovae~\citep[e.g.,][]{nomoto84, nomoto87, jones13}.\par

The evolution of massive stars results in different types of supernovae, depending mainly on the initial stellar mass and metallicity~\citep[e.g., see Fig.~2 of][]{heger03}. In Fig.~\ref{sec_intro_ekstrom13_table1_conti_scenario}, we show the modified Conti scenario for the evolution of single non-rotating stars with solar-metallicity, as a function only of their initial masses~\citep{ekstrom13}. This is inspired on the scenario originally proposed by~\citet{conti75}, the first study to propose the evolutionary connection between O-type and Wolf-Rayet stars. Nevertheless, this scheme, presented in Fig.~\ref{sec_intro_ekstrom13_table1_conti_scenario}, is a summary of modern results found by state-of-the-art evolutionary models for massive stars calculated using the Geneva stellar evolution code~\citep{ekstrom12}.\par

\begin{figure}
  \begin{adjustbox}{minipage=\textwidth,scale=1.00}
  \centering
  \includegraphics[width=0.49\columnwidth]{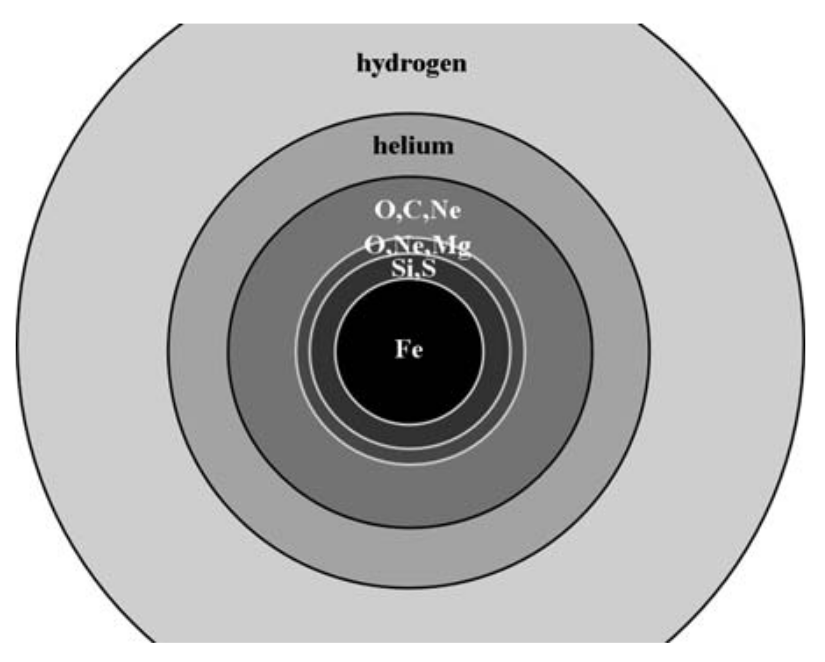}
  \medskip
  \includegraphics[width=0.49\columnwidth]{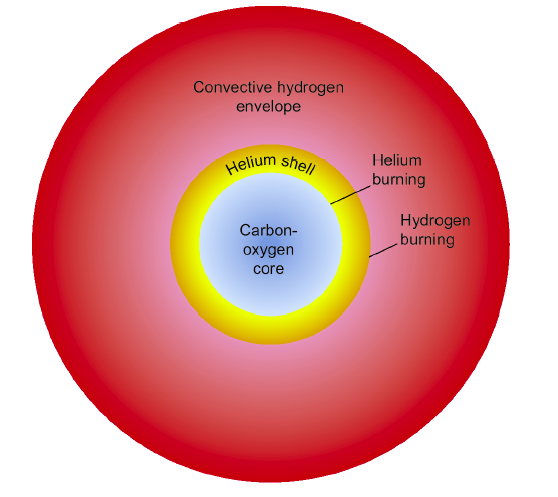}
  \end{adjustbox}
  \caption[Schemes of chemical stratification in evolved massive stars (left) and evolved low-intermediate-mass (right) stars.]
  {Schemes of chemical stratification in evolved massive stars (left) and evolved low-intermediate-mass (right) stars. Left: reproduced from~\citet{maeder09}. Right: reproduced from~\citet{herwig05}.}
  \label{sec_intro_schematic_massive_agb}
\end{figure}

From Fig.~\ref{sec_intro_ekstrom13_table1_conti_scenario}, we see that stars with $\mathrm{M_{ZAMS}}$ lower than about 30 $\mathrm{M_\odot}$ will end their evolution as red supergiants~\citep[$T_{\mathrm{eff}}$ up to $\sim$ 4000 K,][]{levesque05}, before exploding in type-II supernovae (hydrogen-rich core-collapse supernovae). On the other hand, massive stars with $\mathrm{M_{ZAMS}}$ larger than 30 $\mathrm{M_\odot}$ will finish their evolution as Wolf-Rayet stars~\citep[$T_{\mathrm{eff}}$ up to $\sim$ 200000 K,][]{tramper15}, before exploding in type-Ib or type-Ic supernovae (hydrogen-deficient core-collapse supernovae). It is interesting to note that stars with initial masses between 30 and 40 $\mathrm{M_\odot}$ are expected to loop between the two extreme regions of the HR diagram, passing from the blue to the red supergiant phases and then returning to the blue region of the HR diagram as Wolf-Rayet stars. As pointed out by~\citet{ekstrom13}, for these more massive stars that can evolve to red supergiants, stellar winds during this phase must be intense enough, in comparison with stars with 10 $\mathrm{M_\odot}$ < $\mathrm{M_{ZAMS}}$ < 30 $\mathrm{M_\odot}$, to remove the hydrogen from their atmospheres and thus resulting in Wolf-Rayet stars, which show very weak or absent hydrogen lines in their spectra.\par

\begin{figure}
\centerfloat
\centerline{\resizebox{1.00\textwidth}{!}{\includegraphics{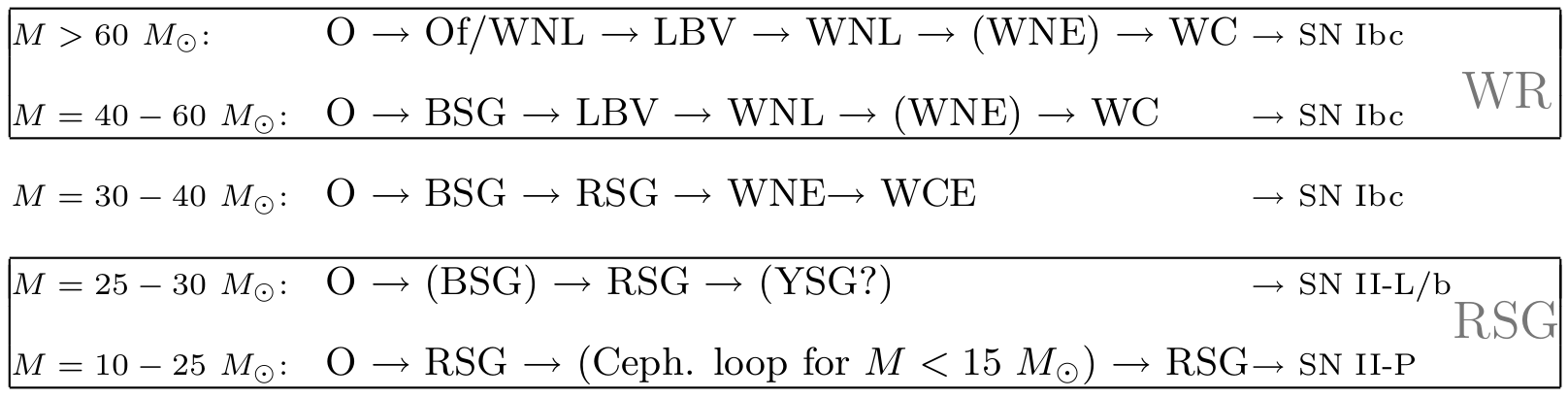}}}
\caption[Modified Conti scenario for the evolutionary scheme of massive stars as a function of zero-age main sequence mass.]
{Modified Conti scenario for the evolutionary scheme of massive stars as a function of zero-age main sequence mass. Reproduced from~\citet{ekstrom13}.}
\label{sec_intro_ekstrom13_table1_conti_scenario}
\end{figure}

We stress that the scheme shown above is based on non-rotating models for single stars with solar-metallicity. Other relevant physical processes, which are not taken into account here, must affect these results: metallicity, rotation, magnetic fields, tidal interaction and mass transfer in binary systems, and more accurate values for the wind mass-loss rate at different evolutionary stages. For instance, it is known that a large fraction (up to $\sim$70\%) of massive stars are born in multiple system and a significant fraction (up to $\sim$30\%) of the current single massive stars are in fact merger products from binary system formed in the past \cite[e.g.,][]{sana12, sana13b, demink14}. A more detailed discussion on the effects of magnetic field, rotation, and multiplicity, on the evolution of massive stars can be found in~\citet{meynet11},~\citet{meynet15}, and~\citet{sana13a}, respectively.\par

In short, our discussion above evidences the importance of mass loss processes on the properties and evolution of massive stars. In particular, Sect.~\ref{sec_intro_radiative_line_driven_winds} is devoted to discuss in details the phenomenon of stellar winds.\par

As a quantitative example, we show, in Fig.~\ref{sec_intro_groh14_fig3_e_evol_massloss_mass}, how the mass-loss rate of the stellar wind is expected to change as a function of time for a non-rotating single star with $\mathrm{M_{ZAMS}}$ = 60 $\mathrm{M_\odot}$. These values for the mass-loss rate are taken in account in the Geneva models analysed in~\citet{groh14}. Based on synthetic spectra (calculated with the radiative transfer code CMFGEN) at each evolutionary step, they studied the spectroscopic properties of a non-rotating single 60 $\mathrm{M_\odot}$ star, evaluating how the spectra type changes as the modeled star evolves from the H-burning phase to the pre-supernova phase.\par

Interestingly,~\citet{groh14} showed that a single non-rotating star with $\mathrm{M_{ZAMS}}$ = 60 $\mathrm{M_\odot}$ must appear in the zero-age main sequence as a O3-4 supergiant (luminosity class I), not as a dwarf star (class V). After that, the star evolves to the LBV and Wolf-Rayet phases before ending its evolution. From Fig.~\ref{sec_intro_groh14_fig3_e_evol_massloss_mass} one sees that the mass-loss rate changes highly through the evolution, from $\sim$$10^{-6}$ $\mathrm{M_\odot}$ yr\textsuperscript{-1} at the beginning of the H-core burning phase, reaching a maximum of $\sim$$10^{-3}$ $\mathrm{M_\odot}$ yr\textsuperscript{-1} by the end of the cool LBV phase. Due to the mass loss, this star with an initial mass of 60 $\mathrm{M_\odot}$ will end its evolution with about 13 $\mathrm{M_\odot}$, as a WO1 star until the supernova explosion~\citep{groh14}.\par

\begin{figure}
\centerfloat
\centerline{\resizebox{0.80\textwidth}{!}{\includegraphics{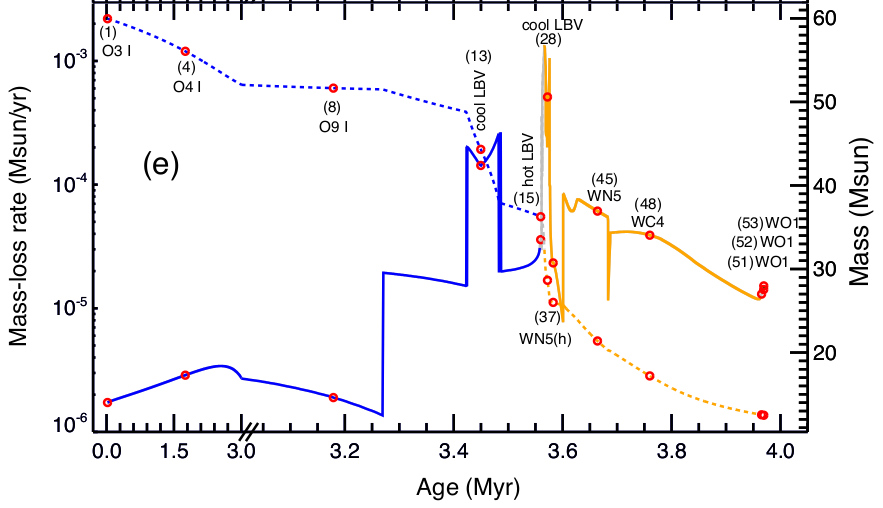}}}
\caption[Mass-loss rate (left axis, solid line) and stellar mass (right axis, dashed line) as function of age in the Geneva evolutionary model for a non-rotating single star with $\mathrm{M_{ZAMS}}$ = 60 $\mathrm{M_\odot}$.]
{Mass-loss rate (left axis, solid line) and stellar mass (right axis, dashed line) as function of age in the Geneva evolutionary model for a non-rotating single star with $\mathrm{M_{ZAMS}}$ = 60 $\mathrm{M_\odot}$. Colors encode different evolutionary phases: H-core (blue), H-shell and H+He-shell (gray), and He-core (orange). The evolution of the spectral type predicted by~\citet{groh14} is also indicated here. Due to the large amount of mass loss since the beginning of the main sequence, the star ends its evolution with about 13 $\mathrm{M_\odot}$. Adapted from~\citet{groh14}.}
\label{sec_intro_groh14_fig3_e_evol_massloss_mass}
\end{figure}

\subsection{Enrichment of the interstellar medium}
\label{sec_intro_effects_interstellar}

Stars transfer mechanical and radiative energy to the interstellar medium (ISM) by different ways: stellar winds (in addition to episodic mass loss), radiation, and supernovae. In Fig.~\ref{sec_intro_30doradus}, we show the central part of the \ion{H}{II} region 30 Doradus (Tarantula Nebula), called R136, in the Large Magellanic Cloud. This region has been widely studied regarding very massive stars ($\mathrm{M_{ZAMS}}$ $\approx$ 100-300 $\mathrm{M_\odot}$) and multiplicity properties of massive stars~\citep[see, e.g.,][]{crowther10, chene11, sana13}. From that, we note that R136 is a very crowded stellar environment. In fact, it is the region of highest stellar density in Tarantula Nebula, being populated by a large number of OB-type stars. The spectroscopic analysis of~\citet{doran13} confirmed about 500 early-type stars in this region. Massive hot stars are the main source of ultraviolet (UV) radiation, ionizing the nebular gas, remnant of the primordial molecular cloud, and then originating \ion{H}{II} regions. Moreover, we see that the structure of 30 Doradus is highly shaped due to the interaction of the intense winds and radiation fields from early-type stars with the ISM gas.\par

Therefore, it is conspicuous the importance of massive stars in the enrichment of the ISM, both from a physical (transfer of kinematic energy) and chemical (production and transfer of metals) views.~\citet{abbott82} was one of the first quantitative studies to investigate the effects of the winds of massive stars on the ISM. This author evaluated the mechanical and radiative energy contribution from O-type, BA supergiants, and  Wolf-Rayet stars to the enrichment of the ISM within a distance of $\sim$3 kpc. He found that these stars transfer mass (by means of winds) and radiation to the ISM at a rate of $9\e{-5}$ $\mathrm{M_\odot}$ yr\textsuperscript{-1} kpc\textsuperscript{-2} and $2\e{38}$ erg s\textsuperscript{-1} kpc\textsuperscript{-1}, respectively.\par

\begin{figure}[t]
\centerfloat
\centerline{\resizebox{0.60\textwidth}{!}{\includegraphics{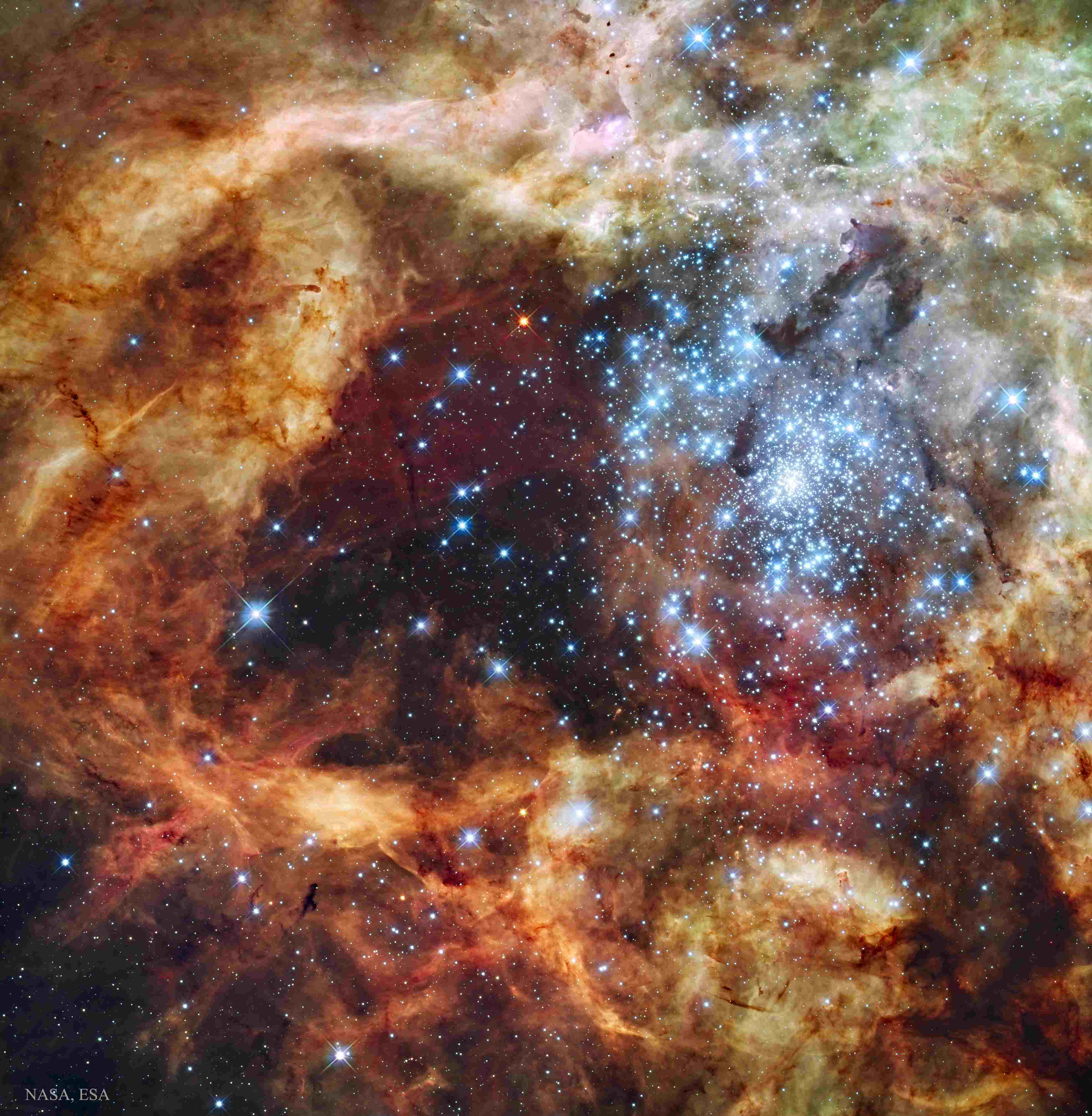}}}
\caption[The R136 open cluster located in the \ion{H}{II} region 30 Doradus (Tarantula Nebula) in the Large Magellanic Cloud.]
{The R136 open cluster located in the \ion{H}{II} region 30 Doradus (Tarantula Nebula) in the Large Magellanic Cloud. Imaging from the Hubble Space Telescope/Wide Field Camera instrument in the visible region (photometry in the U-, B-, V-, I-, and H$\alpha$-bands). This image has field of view of about 46 pc. Source: \url{https://apod.nasa.gov/apod/ap160124.html}.}
\label{sec_intro_30doradus}
\end{figure}

\begin{figure}[t]
\centerfloat
\centerline{\resizebox{0.65\textwidth}{!}{\includegraphics{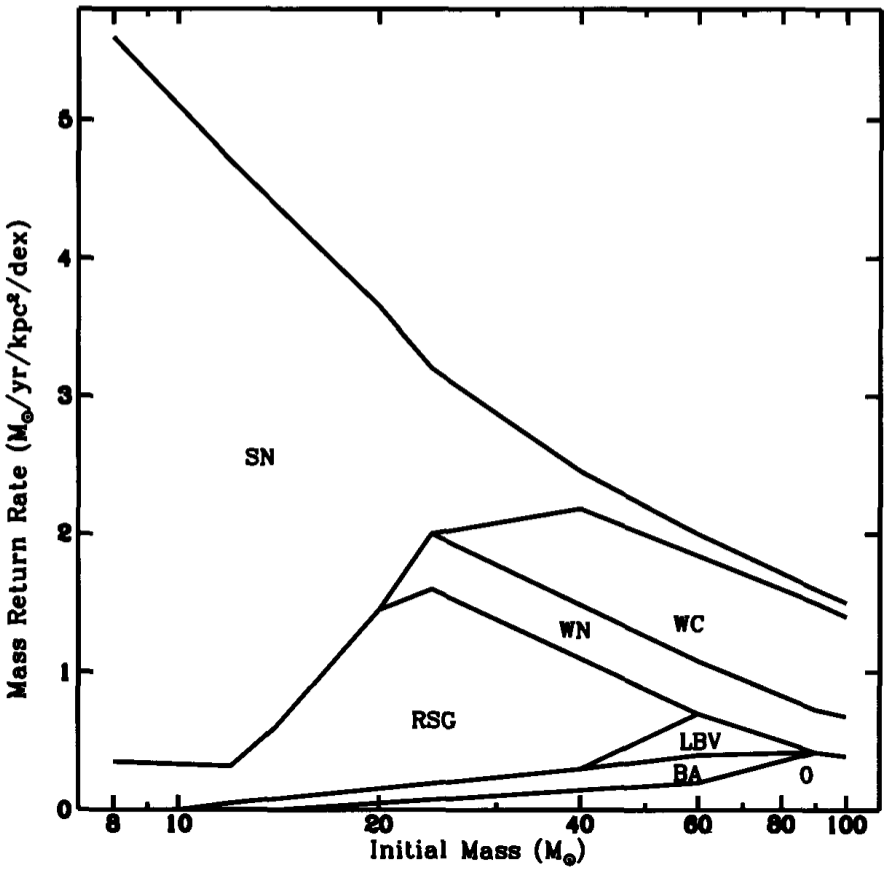}}}
\caption[Mass return rate (logarithmic scale) of massive stars as a function of stellar mass.]
{Mass return rate (logarithmic scale) of massive stars as a function of stellar mass. The evolutionary phases are indicated at the different regions here: O-type, BA supergiant, luminous blue variable (LBV), red supergiant (RSG), carbon Wolf-Rayet (WC), and nitrogen Wolf-Rayet (WN) stars. The region of core-collapse supernovae is indicated by SN. The stellar mass is shown in logarithmic scale. Apart from the stellar mass that returns to the interstellar medium due to supernovae, the peak of mass return comes from stars with initial mass of about 40 $\mathrm{M_\odot}$ during the Wolf-Rayet phase. Reproduced from~\citet{lamers_cassinelli99}.}
\label{lamers99_fig12.10_mass_return_ism}
\end{figure}

Fig.~\ref{lamers99_fig12.10_mass_return_ism} shows the results found by~\citet{castor93} for the rate of mass that is injected into the interstellar medium as a function of initial stellar mass. This rate (in units of $\mathrm{M_\odot}$ yr\textsuperscript{-1} kpc\textsuperscript{-2}) is weighted by the initial mass function from~\citet{garmany82}. In short, it expresses the contribution to the enrichment of the ISM given the initial mass through the different evolutionary phases of massive stars. First, notice that stars with $\mathrm{M_{ZAMS}}$ $\sim$ 8-10 $\mathrm{M_\odot}$ will transfer mass to the ISM mainly at the end of their evolutionary paths, when exploding in core-collapse supernovae of type-II (Fig.~\ref{sec_intro_ekstrom13_table1_conti_scenario}). Despite having less intense winds than the more massive stars, these stars with $\mathrm{M_{ZAMS}}$ $\sim$ 8-10 $\mathrm{M_\odot}$ also significantly contribute to the mass return to the ISM, taking the SN contribution into account, since they are more numerous in comparison with the more massive objects. Apart from the mass injection by means of supernovae, the largest individual contribution to the enrichment of the ISM comes from stars with initial masses of $\sim$40 $\mathrm{M_\odot}$ during their evolutionary stages as Wolf-Rayet stars.\par

\begin{figure}[t]
\centerfloat
\centerline{\resizebox{0.55\textwidth}{!}{\includegraphics{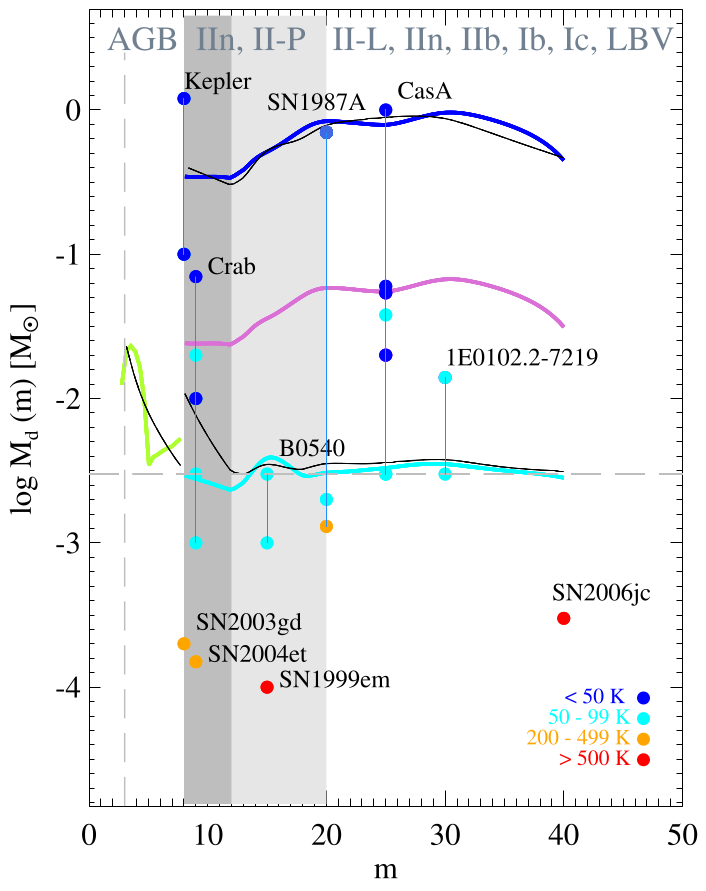}}}
\caption[Amount of dust mass produced by AGB stars and supernovae as a function of progenitor initial mass from~\citet{gall11b}.]
{Amount of dust mass produced by AGB stars and supernovae as a function of progenitor initial mass from~\citet{gall11b}. The theoretical curve for AGB stars is shown in green, while the curves for supernovae are shown in blue, pink, and cyan according to the dust production efficiency that is used in the models. Measured values for supernovae remnants are shown in colored points according to different values of dust temperature. Further details on these theoretical curves and measured values are found in Sect.~7.2 of~\citet{gall11b}. See text for discussion. Reproduced from~\citet{gall11b}.}
\label{sec_intro_gall11_fig6}
\end{figure}

However, as pointed out in Sect.~\ref{sec_intro_evol_stages}, B dwarfs encompasses both intermediate- mass ($\mathrm{M_{ZAMS}}$ of $\sim$ 3  $\mathrm{M_\odot}$, type B9) and high-mass stars ($\mathrm{M_{ZAMS}}$ of $\sim$ 18 $\mathrm{M_\odot}$, type B0). These lower-mass B stars will then evolve to the AGB phase. Our discussion above was focused on the contribution of massive stars to the enrichment of the ISM, but AGB stars are also important contributors of gas and dust to the ISM by their intense winds. Due to winds, mainly during the AGB phase, a star with $\mathrm{M_{ZAMS}}$ of $\sim$ 4 $\mathrm{M_\odot}$ is expected to loose about 80\% of its initial mass~\citep[e.g.,][]{cummings16}.\par

\begin{figure}
\centerfloat
\centerline{\resizebox{0.75\textwidth}{!}{\includegraphics{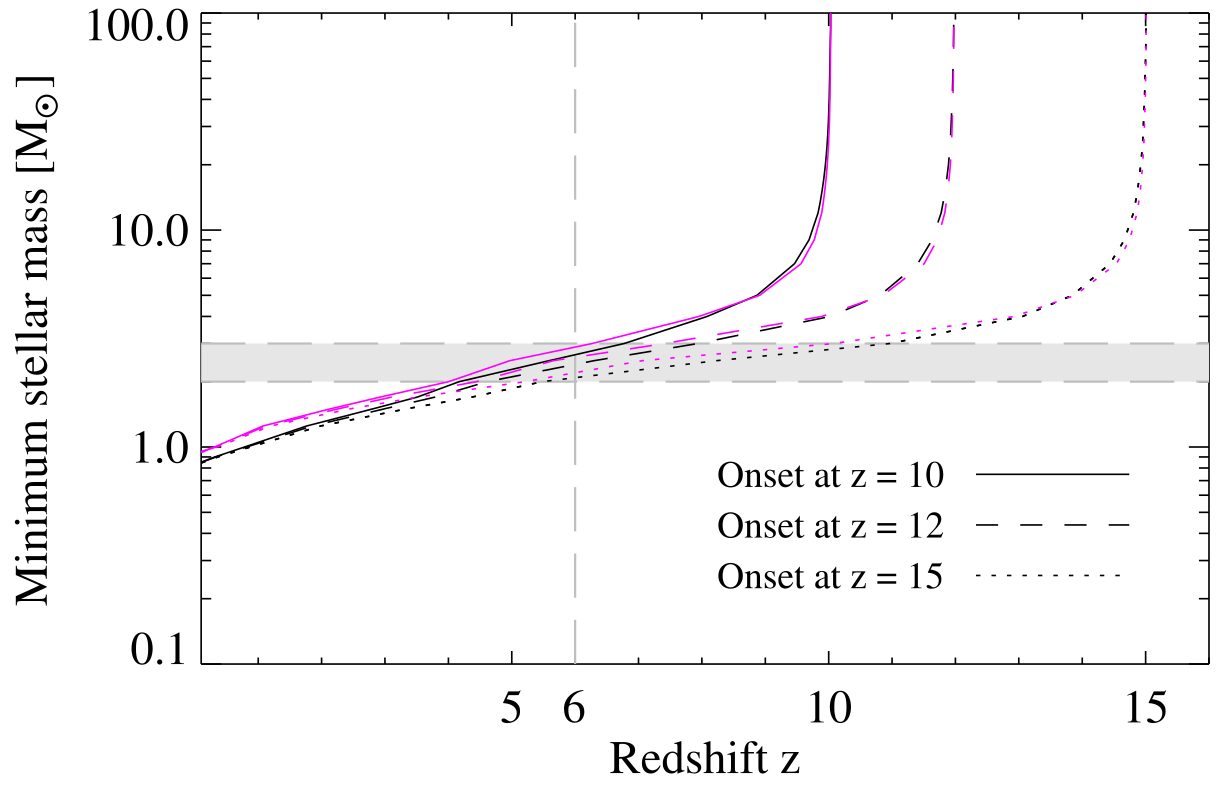}}}
\caption[Minimum $\mathrm{M_{ZAMS}}$ that allows a complete stellar evolution at a certain value of cosmological redshift ($z$).]
{Minimum $\mathrm{M_{ZAMS}}$ that allows a complete stellar evolution at a certain value of cosmological redshift ($z$). Stellar formation starts on three different value of redshift: $z$ = 10 (solid line), $z$ = 12 (dashed line), and $z$ = 15 (dotted line). Black lines corresponds to stellar metallicity $Z$ = 0.001 and pink lines to  $Z$ = 0.040. The gray region indicates the minimum initial mass ($\sim$2-3 $\mathrm{M_\odot}$) for stars contributing (in this case, during the AGB phase) to the injection of dust into the ISM in the early Universe ($z$ = 6). Reproduced from~\citet{gall11b}.}
\label{sec_intro_gall11_fig1}
\end{figure}

AGB stars are responsible to about the half of the recycled gas that is used in the star formation process~\citep[e.g.,][]{maeder92}, and they are important sources of dust in the Universe~\citep[see, e.g.,][and references therein]{valiante09, gall11b}. Fig.~\ref{sec_intro_gall11_fig6} shows the predictions from~\citet{gall11b} for the amount of dust yield both by AGB stars and supernovae. These theoretical curves are compared with measured values for supernovae remnants from the literature. Here, the theoretical curves for supernovae are calculated considering different regimes of dust production efficiency, which is related to the total amount of dust injected into the ISM~\citep[see Eq.~21 of][]{gall11a}. From that, one sees that the amount of dust produced by AGB stars is somewhat comparable to the one by supernovae regardless of the different theoretical scenarios that are analysed by these authors.\par

In addition, Fig.~\ref{sec_intro_gall11_fig1} shows the minimum initial stellar mass for a star dying at a certain epoch (expressed in terms of cosmological redshift $z$). Considering the onset of star formation at $z$ = 10, just very massive stars with mass up to $\sim$100 $\mathrm{M_\odot}$ are able to contribute to the enrichment of the ISM at this epoch~\citep{gall11b}.\par 

Based on Fig.~\ref{sec_intro_gall11_fig1}, we see that only stars with $\mathrm{M_{ZAMS}}$ larger than $\sim$3 $\mathrm{M_\odot}$ are able to pollute the ISM at the epoch $z$ = 6 (Universe with age of $\sim$ $10^{9}$ yr). Stars with initial mass lower than this threshold evolve in time-scales longer than the age of the Universe at $z$ = 6. In short, OB-type dwarfs are the progenitors of the sources of the ISM gas and dust at the early Universe.\par

In conclusion, massive star are rare due to the initial mass function and also to their shorter lifetimes of $\sim$$10^{6}$-$10^{8}$ Myr. As a rough estimate, there are only $\sim$14000-18000 O-type stars in the Galaxy~\citep{maiz13}. Despite their rarity, these stars are important because they enrich physically and chemically the ISM with their strong outflows, since the beginning of the main sequence until evolved stages as BSG, RSG, LBV, and Wolf-Rayet stars. Moreover, they are progenitors of more exotic astrophysical objects, neutron stars and black holes, that are connected to high-energy astrophysical phenomena, as gamma-ray bursts and gravitational waves~\citep[e.g.,][]{gehrels13, abbott16}.\par

\section{Radiative line-driven winds}
\label{sec_intro_radiative_line_driven_winds}

\subsection{The phenomenon of stellar winds}
\label{sec_intro_stellar_winds}

Stellar winds are characterized by a continuous process of mass loss from the stellar atmosphere (i.e., the photosphere\footnote{Throughout this thesis, the terms ``(stellar) atmosphere'' and ``extended atmosphere'' are occasionally used to design the photosphere and the circumstellar environment, respectively.}). The reader should pay attention here to the property of continuity because massive stars lose mass by means of other mechanisms during their evolution, such as supernovae or episodic mass eruptions that eject a large amount of matter in a limited amount of time (time-scale of years), as the giant mass loss eruption that occurred at P Cygni (LBV star, spectral type B1-2Ia) in the 17th century~\citep[see, e.g.,][]{israelian99, smith14}.\par

The mass-loss rate ($\dot{M}$) and the terminal velocity ($v_\infty$) are the main fundamental parameters to describe the hydrodynamics of the wind. With respect to a larger astrophysical context, it is important to constrain the real values for these two parameters in massive stars because they directly provide the kinematic energy injected into the interstellar medium through winds by the wind power $P_{W}$~\citep{abbott82}:

\begin{equation}
P_{W} = \frac{1}{2}\dot{M}{v_\infty}^{2}.
\end{equation}

The mass-loss rate is defined as the quantity of mass that the star loses via winds per unit time:

\begin{equation}
\dot{M} = \displaystyle -\frac{dM_{\star}}{dt},
\end{equation}
where $M_{\star}$ is the stellar mass (expressed as a function of time).\par

The terminal velocity is the wind velocity that is reached at a sufficiently large distance from the stellar surface ($r \to \infty$) for enabling null-acceleration (as a result of null-force) on the wind.\par

These two basic parameters of the wind, $\dot{M}$ and $v_\infty$, are related to each one by the equation of mass continuity:

\begin{equation}
\dot{M} = 4\pi r^{2}\rho(r)v(r),
\label{eq:mass_continuity}
\end{equation}
where $\rho(r)$ is the wind density structure and $v(r)$ is the wind velocity structure for a certain distance $r$ from the center of the star ($r \geq R_\star$).\par 

Eq.~\ref{eq:mass_continuity} stands in the case of a stationary (i.e., time-independent, $\diffp{v(r,t)}{t} = 0$), smooth, and spherically symmetrical wind. It expresses the conservation of mass through the wind, that is, the same quantity of mass (gas) flows, per unit time, through a sphere with area given by $4\pi r^{2}$ (at any value of $r$).\par 

Based on solutions for the momentum equation of the wind from~\citet{castor75}, the wind velocity structure, $v(r)$, is usually parameterized in the literature by the so-called $\beta$-law approximation:

\begin{equation}
v(r) = v_{0} + (v_\infty - v_{0}) \left(1 - \frac{R_\star}{r}\right)^{\beta},
\label{eq:beta_law_geral}
\end{equation}
where $R_\star$ is the stellar radius, $v_{0}$ is the velocity at the photosphere (i.e., the base of the wind is set at the photosphere), and $r$ is given as a function of $R_\star$. Considering that $v_\infty >> v_{0}$, Eq.~\ref{eq:beta_law_geral} can be approximated as follows: 

\begin{equation}
v(r) = v_\infty\left(1 - \frac{R_\star}{r}\right)^{\beta}.
\label{eq:beta_law}
\end{equation}

Thus, given the terminal velocity $v_\infty$, the wind velocity is fully parameterized by the exponent $\beta$. Based on spectroscopic and hydrodynamical studies, $\beta$ is typically found around 0.8-1.0 for O dwarfs (luminosity class V) stars~\citep[e.g.,][]{bouret13, muijres12}. On the other hand, $\beta$ can reach larger values up to $\sim$ 2.0-3.0 in OB supergiants~\citep[e.g.,][]{crowther06, martins_marcolino15}. From Eq.~\ref{eq:beta_law}, we note that higher values of $\beta$ implies that the wind accelerates slower.~\citet{cure04} investigated the effect of rotation on the wind acceleration of massive stars, and they verified that solutions for slow velocity winds are acceptable in the case of fast rotators, having rotational velocities higher than about 75\% the critical value (Sect.~\ref{sec_intro_stellar_rot_oblateness}). This could be in part one possibility to explain such large values of $\beta$ that are typically observed in more evolved massive stars.\par

We point out that winds are common in different types of stars: both in young and evolved low-mass stars (e.g., T-Tauri, solar-type, and AGB stars) and in young and evolved massive stars. Despite being created by different physical mechanisms, this means that winds happens virtually in all types of stars, having different effects on stellar evolution depending on the intensity of the mass-loss rate.\par

For instance, the Sun shows a stable outflow that is well characterized by a quiescent mass-loss rate of $\sim$$10^{-14}$ $\mathrm{M_\odot}$ yr\textsuperscript{-1} and a terminal velocity of $\approx$ 400 km s\textsuperscript{-1}~\citep[e.g.,][]{sturrock86}. The solar wind is driven by gas pressure due to the high temperature of the solar corona~\citep[e.g.,][]{noble63}\@.~\citet{parker58} was the first to show quantitatively that a static solar corona is impossible, introducing the term ``stellar wind''.~Further details on coronal winds can be found in Chapter 5 of~\citet{lamers_cassinelli99}. For a better comparison with these values mentioned for the Sun, O-type stars show mass-loss rates up to $\sim$$10^{-6}$ $\mathrm{M_\odot}$ yr\textsuperscript{-1} and terminal velocities up to $\approx$3000 km s\textsuperscript{-1} for the earlier spectral types (e.g., O3-4): their mass-loss rates are $10^{8}$ times larger than in winds of solar-type stars during the main sequence.\par

More generally, intermediate- and low-mass stars develop winds with higher mass-loss rates just during evolved (more luminous) evolutionary phases, such as the AGB phase. In this case, mass-loss rates vary from $\sim$$10^{-7}$ $\mathrm{M_\odot}$ yr\textsuperscript{-1} at the initial AGB phase up to $\sim$$10^{-4}$ $\mathrm{M_\odot}$ yr\textsuperscript{-1} at the end of the AGB phase. This later and more intense wind phase is called as the AGB super-wind phase~\citep{renzini81, bowen91}. It is thought that such very slow winds ($v_\infty$ up to $\sim$30 km s\textsuperscript{-1}) showing a large amount of mass loss are driven during the AGB phase as follows~\citep[e.g.,][]{hofner07}: pulsations transfer gas from the stellar surface to the outer atmospheric layers, where the temperature is low enough to enable the formation of dust grains. From that, the radiative pressure on the coupled system of gas and dust grains (carbonaceous or silicates) drives a steady outflow. The reader interested on further details regarding the mass loss of AGB stars is refereed to the review of~\citet{hofner18}.\par

In short, intermediate- and low-mass stars are important to the enrichment of the interstellar medium, as discussed in Sect.~\ref{sec_intro_effects_interstellar}, but this contribution is limited to later evolutionary phases. Their contribution is also limited in terms of heavy elements: they will mostly enrich the ISM with the injection of carbon, nitrogen, and oxygen. On the other hand, winds of massive stars are relevant during all their evolutionary phases, since the main sequence phase up to their later stages. Due to their high effective temperatures and luminosities, up to $\sim$$10^6$ times the solar luminosity, OB-type stars are able to develop radiative line-driven winds, as discussed below.\par

\subsection{Elementary concepts of radiative line-driven winds}
\label{sec_intro_theory_radiative_line_winds}

Radiative line-driven winds means that the wind acceleration arises from the scattering of the stellar continuum flux by line transitions of (mainly) elements heavier than hydrogen and helium, creating spectral absorption and emission lines. Such photon-matter interaction transfers linear momentum from the stellar radiation field to the gas that composes the photosphere and the wind, and thus dropping the assumption of hydrostatic equilibrium for the stellar atmosphere.\par

This mechanism for driving winds is not limited to massive hot stars, such as OB dwarfs, OBA supergiants, LBVs, and WR stars. After the post-AGB phase, low- and intermediate-mass stars will develop radiative-line driven winds during the phase as central stars of planetary nebula (CSPN) ~\citep[e.g.,][]{cerruti85, prinja90}. Despite having quite low luminosities, when compared with massive stars, CSPNs show very high effective temperature, reaching extreme values as high as $\sim$150000 K~\citep[e.g.,][]{herald11, keller11}. Moreover, the line-driven mechanism is promising to explain the origin of winds in the accretion disks of quasars~\citep[e.g.,][]{shlosman85, proga00}. Concerning the massive stars with low $T_{\mathrm{eff}}$, as the RSGs, the mechanisms for the wind acceleration are thought to be quite similar to the ones described for AGBs stars in Sect.~\ref{sec_intro_stellar_winds}: a combination of stellar pulsations and radiative pressure on the coupled system of gas and dust grains (that are formed in outer atmospheric layers).\par

Considering the simplest case of stationary winds, and that the only forces exerted are due the gravity, the gas pressure gradient, and the radiation, the momentum equation of a radiation line-driven wind is expressed as follows:

\begin{equation}
v \diff{v}{r} = -\frac{GM_{\star}}{r^2} + \frac{1}{\rho} \diff{p(r)}{r} + g_{\mathrm{rad}(r)},
\label{eq:wind_momentum}
\end{equation}
where $G$ is the gravitational constant, $M_{\star}$ is the stellar mass, and $p$ is the gas pressure. Assuming the case of a ideal gas with isothermal temperature $T$, the state equation is described by:

\begin{equation}
p(r) = \frac{RT}{\mu} \rho(r),
\label{eq:state_gas}
\end{equation}
where $\sqrt{\frac{RT}{\mu}}$ is the isothermal sound speed in the wind, $R$ is the ideal gas constant and $\mu$ is the mean molecular weight of the gas ($\mu$ = 0.602 for an atmosphere with solar-metallicity).

On the right hand of Eq.~\ref{eq:wind_momentum}, the first two terms are common when describing the hydrodynamics of different types of stellar winds. However, the third-term  in the equation of motion represents the total radiative acceleration acting on the wind and is expressed as follows:

\begin{equation}
g_{\mathrm{rad}}(r) = g_{e}(r) + g_{\mathrm{line}}(r),
\label{eq:acceleration_rad}
\end{equation}
since this results from the momentum transfer by two different ways of photon-atom interaction: 

\begin{enumerate}[label=(\roman*)]
\setlength\itemsep{1em}

\item $g_{e}$ is the radiative acceleration due to Thomson scattering (i.e., elastic scattering of photons by electrons, the contribution from the continuum opacity).

\item $g_{\mathrm{line}}$ is the contribution due to line opacity (bound-bound transitions).

\end{enumerate}

In addition, free-free and bound-free transitions also contribute to the radiative force due to continuum opacity~\citep[see, e.g., Eq.~26 of][]{sander15}.\par

Note that both terms are explicitly written here as a function of $r$. The first term in Eq.~\ref{eq:acceleration_rad} is given by:

\begin{equation}
g_{e}(r) = \frac{\kappa_{e} L_{\star}}{4\pi r^2 c},
\label{eq:thomson_acceleration}
\end{equation}
where $\kappa_{e}$ is the opacity for electron scattering, $L_{\star}$ is the stellar luminosity, and $c$ is the light speed in the vacuum. The term $\kappa_{e}$ is dependent on the gas metallicity and also the degree of gas ionization that typically ranges from $\sim$0.28 to 0.35 $\mathrm{cm^2 g^{-1}}$ in early-type stars.\par 

From Eq.~\ref{eq:thomson_acceleration}, the Eddington parameter $\Gamma_{e}$ is defined as the ratio between the radiative acceleration contribution from Eq.~\ref{eq:thomson_acceleration} and the gravitational acceleration $g(r)$, which is given by $G M_{\star}/r^{2}$:

\begin{equation}
\Gamma_{e} = \frac{g_{e}(r)}{g(r)} = \frac{ \kappa_{e} L_{\star} }{4\pi G M_{\star} c},
\label{eq:eddington_parameter}
\end{equation}
where $M_{\star}$ is the stellar mass.\par 

Physically, this parameter expresses how close a star is to the gravitationally unbound limit ($\Gamma_{e} \to 1$, the so-called classical Eddington limit), just considering the radiative force due to electron scattering. From that, most massive hot stars have $\Gamma_{e}$ up to a factor of 2 lower than the Eddington limit. In advance of discussion, one sees, from Fig.~\ref{sec_intro_zeta_puppis_uv_spectrum}, that electron scattering contributes much more to the radiative force at the base of the wind (photosphere) than at large distances through the wind. As these stars display strong winds, this evidences the large contribution from line opacity ($g_{\mathrm{line}}$) to the total radiative force in Eq.~\ref{eq:acceleration_rad}.\par 

The density stratification $\rho(r)$ in the momentum equation is related to the velocity $v(r)$ by the equation of mass continuity (Eq.~\ref{eq:mass_continuity}). Given a certain value for the mass-loss rate (constant term in Eq.~\ref{eq:mass_continuity}), the momentum equation can be re-written replacing the term $\rho(r)$ by $\dot{M}/(4 \pi r^2 v(r))$. Therefore, the hydrodynamics of the wind is fully described by Eqs. \ref{eq:mass_continuity} and \ref{eq:state_gas} together with Eq.~\ref{eq:wind_momentum}\footnote{Note that the state equation (Eq.~\ref{eq:state_gas}) was assumed for an isothermal wind . However, it is not a realistic approximation since the temperature and the mean molecular weight must be radially dependent~\citep[see., e.g.,][]{sander17}.}.

\begin{figure}[t]
\centerfloat
\centerline{\resizebox{0.50\textwidth}{!}{\includegraphics{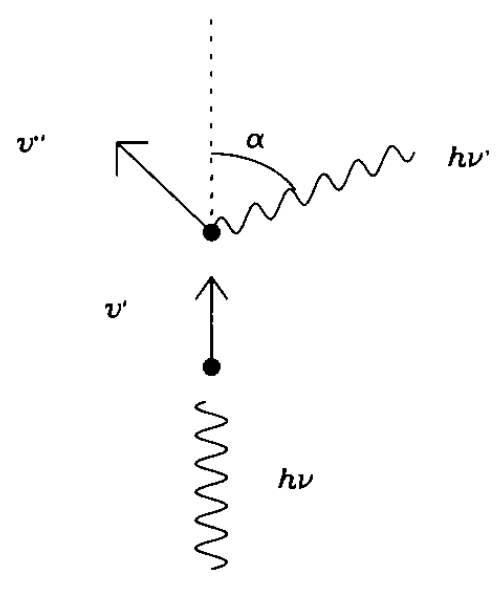}}}
\caption[Schematic representation of iteration between a photon with linear momentum $h \nu/c$ (energy given by $h \nu$) and a particle of mass $m$.]
{Schematic representation of iteration between a photon with linear momentum $h \nu/c$ (energy given by $h \nu$) and a particle of mass $m$. The particle has velocity $v'$ due to the gain of momentum by absorbing the photon. The velocity of the particle will change to $v''$ after re-emitting a photon with linear momentum $h \nu'/c$, which forms an angle $\alpha$ with respect to the initial photon's path (linear momentum $h \nu/c$). See text for discussion. Reproduced from~\citet{lamers_cassinelli99}.}
\label{sec_intro_lamers99_fig8_1}
\end{figure}

Clearly, the evaluation of the radiative acceleration due to line transitions imposes the biggest challenge to solve the equation of motion of the wind. Consider the absorption of a photon of frequency $\nu_{0}$ by an atom with mass $m$ of the stellar atmosphere, and the subsequently re-emission of a photon by this same atom. The variation of the atom's radial linear momentum $p_{\mathrm{r}}$ (in the same direction of the initially absorbed photon path) is given by: 

\begin{equation}
\Delta  p_{\mathrm{r}} = m \Delta  v_{\mathrm{r}} =  \frac{h \nu_{0}}{c} (1 - \cos \alpha),
\label{eq:delta_linear_momentum}
\end{equation}
where $\alpha$ is the angle formed by the initially absorbed photon path and the subsequent re-emitted photon path (as represented in Fig.~\ref{sec_intro_lamers99_fig8_1}). Thus, in the particular case of $\alpha$ = 0, that is, the re-emitted photon has the same direction of the initially absorbed photon, the atom does not present any net gain of radial linear momentum. On the other hand, the maximum gain of momentum is achieved when the photon is re-emitted in the particular case of opposite direction ($\alpha$ = \ang{180}): $\Delta  p_{\mathrm{r}} = \frac{2 h \nu_{0}}{c}$.\par

One of the basic idea behind the transfer of momentum in radiative line-driven winds is that, statistically, the photon re-emission must follow approximately an isotropic distribution. Thus, after a large number of iterations with photons with linear momentum $h \nu/c$ coming from the radial direction (as indicated in Fig.~\ref{sec_intro_lamers99_fig8_1}), we can evaluate the mean variation of linear momentum of this atom of mass $m$, by the integration of Eq.~\ref{eq:delta_linear_momentum} over $4\pi$ radian (sphere) as follows\footnote{We point out a small mistake in the lower ($-\pi/2$) and upper ($\pi/2$) limits of the integration shown in Eq.~8.5 of~\citet{lamers_cassinelli99}}:

\begin{equation}
\langle \Delta p_{\mathrm{r}} \rangle = \frac{h\nu_{0}}{c} \frac{1}{4\pi} \int_{0}^{\pi} 2\pi (1 - \cos \alpha) \sin \alpha  d \alpha.
\end{equation}

Here, we found that the mean variation of the linear momentum, due to the interaction with photons with $\nu_{0}$ coming from a certain direction, is given by the following expression:

\begin{equation}
\langle \Delta p_{\mathrm{r}} \rangle = \frac{h\nu_{0}}{c}.
\end{equation}

Thus, the radiative force due to lines that is exerted on a volume element of the wind with mass $\Delta m$, during a interval $\Delta t$, can be written as follows:

\begin{equation}
g_{\mathrm{line}} = \mathlarger{\sum}_{i=1}^{N} \frac{{\langle \Delta p_{\mathrm{r}} \rangle}_{i}}{\Delta m \Delta t},
\end{equation}
where $i$ denotes different line transitions participating in the transfer of linear momentum to the gas.\par

Despite being didactic, the formulation of the line-radiative acceleration as shown above is not useful to evaluate the wind momentum equation.~\citet{lucy70} was one of the first works to attempt this task and to determine the mass-loss rate of massive hot stars from first principles (i.e., solving the momentum equation). These authors identified the absorption resonance lines\footnote{Lines formed due to the electron transition between the ground energy level and the first excited state.} in the ultraviolet region of ion metals, such as \ion{Si}{IV}, \ion{C}{IV}, and \ion{N}{V}, as the mechanism to break the hydrostatic equilibrium in the atmosphere of these stars. By setting regularity condition at the sonic point of the wind \footnote{Considering an isothermal wind, the sonic point is defined by the distance $r_{s}$ in the wind where the velocity is equal to the isothermal sound speed, i.e.,  $v(r=r_{s})$ = $\sqrt{RT/\mu}$.}, they introduced the so-called reversing moving layer theory, which was recently updated considering more sophisticated non-LTE (Local Thermodynamics Equilibrium) radiative transfer calculations~\citep[see][]{lucy07, lucy10, lucy10_grid}. One of the main findings of this pioneer work was to estimate an upper limit on the mass-loss rate for line-driven winds due to the contribution from just one line:

\begin{equation}
\dot{M} \lesssim \frac{L_{\star}}{c^2}.
\end{equation}

As one could expect, the mass-loss rate of radiative line-driven winds is dependent on the stellar luminosity: higher values of mass-loss rate must be achieved for the more luminous stars. This result found by~\citet{lucy70} can be explained using a simple physical argument as follows.\par 

Consider that just one absorption line in certain rest-frame $\nu_{0}$ contributes to the radiative acceleration and the wind is optically very thick in such line. The latter hypothesis, i.e., optically very thick line, means that all the photons with $\nu_{0}$ are absorbed in the wind. Thus, we will have the following quantity of linear momentum that is transferred from the radiation field to the wind:

\begin{equation}
\dot{M} v_{\infty}  =  \frac{1}{c} \int_{\nu_{0}}^{\nu_{0} (1 + v_{\infty}/c)} 4 \pi R^{2}_{\star} F_{\nu}^{*} d\nu,
\label{eq:momentum_gas_radiation}
\end{equation}
where $F_{\nu}^{*}$ is the stellar flux in the frequency $\nu$ at the stellar radius.\par 

Due to the Doppler effect, Eq.~\ref{eq:momentum_gas_radiation} states that photons with frequency between $\nu_{0}$ and $\nu_{0}(1 + v_\infty/c)$ contribute to the formation of the spectral line with frequency rest-frame frequency $\nu_{0}$. Thus, we integrate Eq.~\ref{eq:momentum_gas_radiation} through the whole wind extension, i.e., from the base of the wind with $v(r)$ = 0 up to the outermost wind region where $v(r)$ = $v_{\infty}$. This equation expresses the gain of linear momentum of the wind, $\dot{M}v_{\infty}$, due to the absorbed radiation (on the right hand of Eq.~\ref{eq:momentum_gas_radiation}).\par 

Note that the velocity stratification in wind (see, again, Eq \ref{eq:beta_law}) is important regarding the wind acceleration of massive hot stars, which have high values of terminal velocities up to $\sim$3000 km s\textsuperscript{-1}. As a result of the Doppler effect, photons with frequencies higher than $\nu_{0}$ will be absorbed through the whole wind extension, also contributing to form such a line: in the outermost part of the wind, photons launched from the stellar surface with rest-frame $\nu_{0}(1 + v_{\infty}/c)$.

Therefore, the Doppler effect enables to keep the wind driving due to lines even up to large distances from the stellar surface.\par

Within a good approximation, we can assume $F_{\nu}^{*}$ as constant inside the integration interval in frequency, writing Eq.~\ref{eq:momentum_gas_radiation} as follows:

\begin{equation}
\dot{M} v_{\infty} \sim  \frac{4 \pi R^{2}_{\star} F_{\nu_{0}}^{*} \nu_{0} v_{\infty}}{c^2},
\label{eq:momentum_gas_radiation_approx}
\end{equation}
where $F_{\nu_{0}}^{*}$ is the stellar flux in the rest-frame frequency $\nu_{0}$.\par

Assuming a black body radiation for the stellar flux and that our adopted optically very thick lines happens in the intensity maximum peak\footnote{This is a reasonable assumption since hot stars emit the most part of their energy in the UV and the relevant lines to drive the wind are formed in this spectral region.}, $F_{\nu_{0}}^{*} \nu_{0} \sim 0.62 \sigma {T_{\mathrm{eff}}}^4$, where $\sigma$ is the Stefan-Boltzmann constant. Thus, from Eq.~\ref{eq:momentum_gas_radiation_approx} and the Stefan-Boltzmann equation:

\begin{equation}
L_{\star} = 4 \pi \sigma {R_{\star}}^2 {T_{\mathrm{eff}}}^4,
\label{eq:stefan_boltzmann}
\end{equation}
we have the following estimation for the mass-loss rate due to such very thick line:

\begin{equation}
\dot{M} \lesssim 0.62\frac{L_\star}{c^2} \sim \frac{L_\star}{c^2}.
\label{eq:estim_mass_loss}
\end{equation}

For sure, Eq.~\ref{eq:estim_mass_loss} must be seen just as rough estimation for the maximum value of $\dot{M}$, assuming here that all the stellar luminosity is used to accelerate the wind. Massive hot stars present winds that are driven by a large number of lines. Thus, the resulting mass loss of a wind driven by $N_{\mathrm{thick}}$ very thick lines can be expressed as:

\begin{equation}
\dot{M} \sim N_{\mathrm{thick}}\frac{L_\star}{c^2},
\label{eq:estim_mass_loss_n_lines}
\end{equation}
since these transitions are independent among themselves.\par

\begin{figure}
\centerfloat
\centerline{\resizebox{1.00\textwidth}{!}{\includegraphics{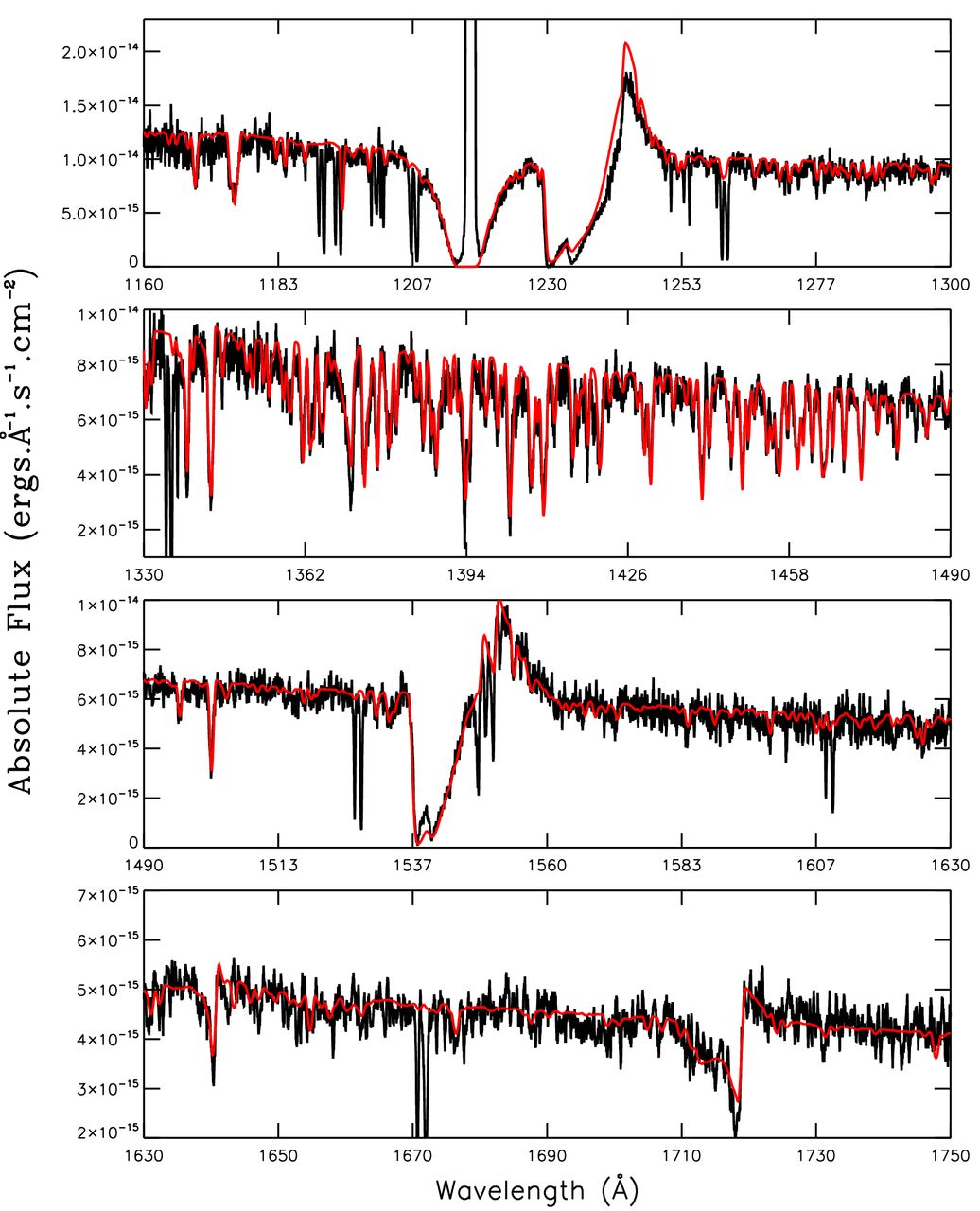}}}
\caption[Ultraviolet spectrum of the O supergiant (type O9.5I) IC 1613-B11 (located in the dwarf galaxy IC 1613) is shown in black line.]
{Ultraviolet spectrum of the O supergiant (type O9.5I) IC 1613-B11 (located in the dwarf galaxy IC 1613) is shown in black line. The best-fit CMFGEN found by~\citet{bouret15} is shown in red line. See text for discussion. Reproduced from~\citet{bouret15}.}
\label{sec_intro_uv_spectrum}
\end{figure}

In Fig.~\ref{sec_intro_uv_spectrum}, we show spectroscopic data from the Copernicus and the International Ultraviolet Explorer (IUE) satellites, covering the region of $\sim$1100-1760 {\AA}, for the early O supergiant IC 1613-B11 (O9.5I). The best-fit radiative transfer model from~\citet{bouret15}, calculated with the code CMFGEN, is also shown here. Later, we discuss in details about this code in Sect.~\ref{sec_radiative_transfer_modeling_cmfgen}. From fitting the observed spectrum and the spectral energy distribution using CMFGEN, these authors derived the following parameters for this star $T_{\mathrm{eff}}$ = 30000 K, $\log g$ = 3.25, $\log L_{\star}/L_{\odot}$ = 5.45, $\dot{M}$ = $1.5\e{-8}$ $\mathrm{M_\odot}$ yr\textsuperscript{-1}, and $v_{\infty}$ = 1300 km s\textsuperscript{-1}, among other stellar and wind parameters.\par

Most spectral lines seen in Fig.~\ref{sec_intro_uv_spectrum} are formed in the photosphere, being useful to probe the stellar parameters such as the effective temperature. The usage of spectroscopy to derive the stellar and wind parameters will be better discussed in Sect.~\ref{chapter_spectroscopy}. These transitions are mainly due to heavy elements from the iron group (e.g., iron, nickel, and cobalt), forming the so-called iron forest in the ultraviolet region.\par 

Furthermore, one sees that other lines are formed showing both absorption and emission components, such as \ion{N}{V} $\lambda$1240, \ion{C}{IV} $\lambda$$\lambda$ 1548,1551, and \ion{N}{IV} $\lambda$1718. These lines have a significant fraction of formation in the wind and thus displaying the so-called P Cygni profiles (Sect.~\ref{sec_pcygni_profiles}). As discussed above,~\citet{lucy70} calculated the theoretical mass-loss rate for a O-type star based only on resonant transitions (as \ion{C}{IV} $\lambda$$\lambda$ 1548,1551). From Eq.~\ref{eq:estim_mass_loss}, one sees that the limit mass-loss rate due to a very thick line is $\dot{M}$ = $2.0\e{-8}$ $\mathrm{M_\odot}$ yr\textsuperscript{-1}, being somewhat larger than the value derived from~\citet{bouret15} of $\dot{M}$ = $1.5\e{-8}$ $\mathrm{M_\odot}$ yr\textsuperscript{-1} for IC 1613-B11 ($\log L_{\star}/L_{\odot}$ = 5.45).\par

Indeed, from Fig.~\ref{sec_intro_uv_spectrum}, we see that a large amount of lines of metals contribute to the transfer of linear momentum to the stellar atmosphere, and then to the wind driving in a O-type star. In this work,~\citet{bouret15} analysed three O-type stars in low-metallicity galaxies. First, since the momentum transfer is mainly due to the iteration of the radiation with metal ions, as one may expect the mass-loss rate is scaled to the stellar metallicity: lower metallicity, and thus smaller opacity in the important lines to accelerate the wind, means lower values of mass-loss rate and also terminal velocity. This dependence between the wind basic parameters and the stellar metallicity is very well supported both from theoretical predictions~\citep[e.g.,][]{vink01, lucy12_metallicity} and spectroscopic modeling~\citep[][]{mokiem07b, bouret03}. For example,~\citet{mokiem07b} derived the relation of $\dot{M} \propto Z^{0.83 \pm 0.16}$ based on the analysis of O stars and early-B supergiants in the Galaxy and in the Magellanic Clouds.\par

A quantitative example of the contribution by each chemical element to the total radiative acceleration is presented in Fig.~\ref{sec_intro_zeta_puppis_uv_spectrum}. These calculations were performed by~\citet{sander17} to analyse the Galactic O supergiant $\zeta$ Puppis (O4I) using the radiative transfer code Potsdam Wolf-Rayet~\citep[PoWR, ][]{hamann85, hamann86, grafener02, hamann03, sander17}. From Eq.~\ref{eq:acceleration_rad}, in addition to the line contribution, we need to take into account the radiative acceleration due to Thomson scattering. This is shown, in Fig.~\ref{sec_intro_zeta_puppis_uv_spectrum}, in pink line, and one sees how clearly the acceleration due to the electron scattering becomes much less important, in comparison with the contribution due to lines, toward larger distances from the photosphere (e.g., $r$ > 2 $R_{\star}$).\par

\begin{figure}[t]
\centerfloat
\centerline{\resizebox{0.75\textwidth}{!}{\includegraphics{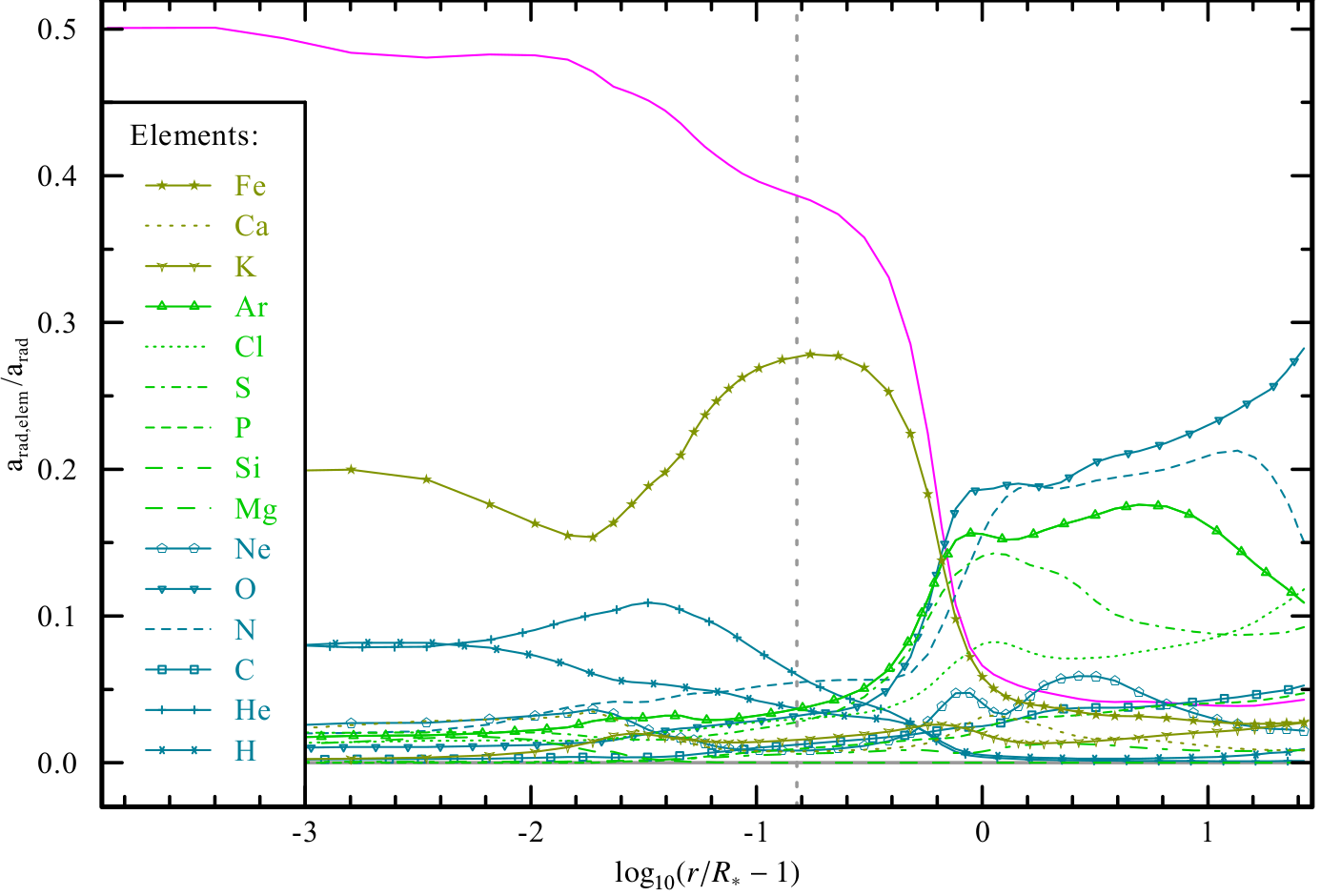}}}
\caption[Fraction of the radiative acceleration due to each chemical element, compared with the total radiative force ($a_{\mathrm{rad}}$), as a function of distance from the star, in the PoWR model of~\citet{sander17} to the analysis of $\zeta$ Puppis.]
{Fraction of the radiative acceleration due to each chemical element, compared with the total radiative force ($a_{\mathrm{rad}}$), as a function of distance from the star, in the PoWR model of~\citet{sander17} to the analysis of $\zeta$ Puppis. The gray dashed line indicates the critical point of the wind ($r$ $\sim$ 1.16 $R_{\star}$) for the solution of the hydrodynamic equation~\citep[Eq.~13 of][]{sander17}, corresponding, in this case, to the sonic point. The contribution from electron scattering (continuum) is shown in pink solid color. Note how the continuum contribution to the total radiative force decreases toward larger distances in the wind region. Together with the line acceleration due to iron, the continuum mainly contributes to transfer momentum in the innermost part of the wind. Carbon, nitrogen, and oxygen have a more important role to sustain the wind acceleration at larger distances from the stellar surface. Adapted from~\citet{sander17}.}
\label{sec_intro_zeta_puppis_uv_spectrum}
\end{figure}

Moreover, the acceleration contribution due to hydrogen and helium is also clearly less important than the ones from heavy elements, as iron, carbon, nitrogen, and oxygen. Interestingly, considering a solar-like chemical abundance, one may note that all these relevant elements (metals) to the wind driving compose about just about 1\% of the stellar atmosphere (or even less in low metallicity environments such as for IC 1613-B11). This means that the bulk of the material (typically, hydrogen and helium), composing the photosphere and the wind, are not the principal source of momentum transfer to drive a steady outflow.\par 

We can only to explain the wind driving based on  effect of Coulomb coupling. In short, atoms of hydrogen and helium are not mainly accelerated by lines, but by the 
electric field originated from the heavy ions. This process results in a slowing down of the metals and creates a steady outflow of the wind. Since this implies in reducing the acceleration of the heavy ions, the efficiency of the Coulomb coupling is measured in terms of two characteristic time-scales: $t_{s}$ is the time-scale to slow-down the heavy ions due to the Coulomb coupling and $t_{d}$ is the time-scale to increase the velocity of the metal ions due to gain of momentum by line transitions. As shown by~\citet{lucy70}, it is necessary to satisfy the condition $t_{s} < t_{d}$ to have an efficient process of Coulomb coupling, and then driving a steady outflow. In terms of basic stellar and wind parameters, this condition is satisfied if:

\begin{equation}
\frac{L_{\star} v_{\infty}}{\dot{M}} < 5.9\e{16}.
\label{eq:condition_coulomb}
\end{equation}

For more details, we refer the reader to the Section 8.1.2 of~\citet{lamers_cassinelli99}. As pointed out by these authors, despite being satisfied for O-type stars, this condition stated in Eq.~\ref{eq:condition_coulomb} for the phenomenon of Coulomb coupling may not be satisfied in the case of late B and A dwarfs~\citep[see, e.g.,][]{babel95, babel96}. As we will discuss in Sect.~\ref{sec_intro_radiative_winds_hr_diagram}, radiative line-driven winds are not expected on late B-type stars.\par

Despite being a ground-breaking work to explain the mechanism of mass-loss on massive hot stars,~\citet{lucy70} could not provide an accurate quantitative prediction for the mass-loss rate. As commented above, these authors computed the radiative acceleration in the wind due to a few resonance lines. From our previous discussion, a large number of lines (both optically thin and thick) indeed are contributing to the net momentum transfer in the wind of these stars.\par 

\citet{castor75} was the first to develop a formalism to take into account the acceleration contribution from a large ensemble of lines. This is the so-called CAK-theory for radiative line-driven winds and it has been updated and extended by a series of works~\citep[e.g.,][]{abbott82_line, pauldrach86, shimada94, kudritzki02, cure11}. For essentially the same stellar parameters, the CAK-theory predicts a mass-loss rate of about $10^2$ times larger than the one predicted by the work of~\citet{lucy70}.\par

In the CAK-theory, the distribution function of lines $N$ contributing to the radiative acceleration, is approximated by a power-law as follows:

\begin{equation}
\diff{N}{\kappa} \sim \left(\frac{\kappa}{\kappa_{0}}\right)^{\alpha -2},
\label{eq:line_distribution_cak}
\end{equation}
where $\kappa$ is the line opacity and $\kappa_{0}$ is a normalization constant such as $\kappa_{0} \diff{N}{\kappa}$ = 1. Here, the constant $\alpha$ is given in the interval between 0 and 1. Physically, $\alpha$ = 0 would mean purely a contribution from optically thin lines and $\alpha$ = 1 from optically thick lines.\par

In addition, based on the Sobolev approximation\footnote{Introduced by~\citet{sobolev60}, this method simplifies the radiative transfer treatment in stellar winds. For a given value of frequency and direction, the Sobolev approximation treats the photon-particle interaction as happening at a certain value of optical depth in the wind (Sobolev optical depth point). Likewise, the Sobolev approximation works in terms of an extended region in the wind to happen the photon-particle interaction (defined by the Sobolev length). Further details about the Sobolev approximation can be found in Sect.~8.4 of~\citet{lamers_cassinelli99}},~\citet{castor75} showed the following relation for the radiative force due to lines $g_{\mathrm{line}}$:

\begin{equation}
g_{\mathrm{line}} \sim \left(\frac{1}{\rho} \diff{v}{r}\right)^{\alpha}.
\label{eq:gl_cak_1}
\end{equation}

From Eq.~\ref{eq:line_distribution_cak} and \ref{eq:gl_cak_1}, the CAK-theory proposes to express $g_{\mathrm{line}}$ as a function of radiative acceleration due to electron scattering:

\begin{equation}
g_{\mathrm{line}} = g_{e} M(t),
\label{eq:gl_cak_2}
\end{equation}
where the variable $M(t)$ introduced in the formalism of~\citet{castor75} is the so-called the force multiplier, and it gives the ratio between the line radiative force and the radiative force due to the continuum, reaching up to $\sim$$10^{4}$~\citep[e.g., see. Fig.~8.8 of][]{lamers_cassinelli99}. $M(t)$ is defined as follows:

\begin{equation}
M(t) = kt^{-\alpha}s^{\delta} \sim \left(\frac{1}{\rho} \diff{v}{r}\right)^{\alpha} ,
\label{eq:force_multiplier}
\end{equation}
where the term $s$, later introduced by~\citet{abbott82_line} to taken into account ionization effects, is dependent on the electron density and the geometrical dilution factor, and thus being dependent on the distance from the star~\citep[see Eq.~8.86 of][]{lamers_cassinelli99}. The constants $k$, $\alpha$, and $\delta$ are the so-called force multiplier parameters and the variable $t$ is the optical depth parameter:

\begin{equation}
t \sim \rho \diff{r}{v}.
\label{eq:optical_depth_parameter}
\end{equation}

One may note that the definition of the optical depth parameter is related to the distance at the wind, where lower values means larger distances from the photosphere, i.e., $M(t)$ changes at each value of distance in the wind. Typical values of $\alpha$ lies between 0.45 and 0.65, depending on the stellar parameters, such as the effective temperature. Physically, this means that both optically thin ($\alpha$ = 0) and thick lines ($\alpha$ = 1) contribute to the wind acceleration. For a literature compilation of values for the force multiplier parameters, we refer the reader to Table 8.2 in Sect.~8.6.1 of~\citet{lamers_cassinelli99}.\par

Lastly, from the CAK-theory, we have the following scaling relations for the mass-loss rate~\citep[see, e.g.,][]{puls08}:

\begin{equation}
\dot{M} \sim (kL_{\star})^{1/\alpha'} (M(1 - \Gamma_{e}))^{1 - 1/\alpha'},
\label{eq:mass_loss_cak_theory }
\end{equation}
where $\alpha'$ = $\alpha$ - $\delta$ and $\Gamma_{e}$ is the Eddington factor (Eq.~\ref{eq:eddington_parameter}).\par

The terminal velocity is simply given by Eq.~\ref{eq:beta_law} with $\beta$ = 0.5 in the standard CAK-theory. As discussed in Sect.~\ref{sec_intro_stellar_winds}, more recent theoretical works, such as~\citet{cure04} that investigate the effects of rotation on the solution for the wind acceleration, are able to derive quite larger values of $\beta$, in this case being compatible to the spectroscopic results found for OB-supergiants.\par

\begin{figure}[t]
\centerfloat
\centerline{\resizebox{0.75\textwidth}{!}{\includegraphics{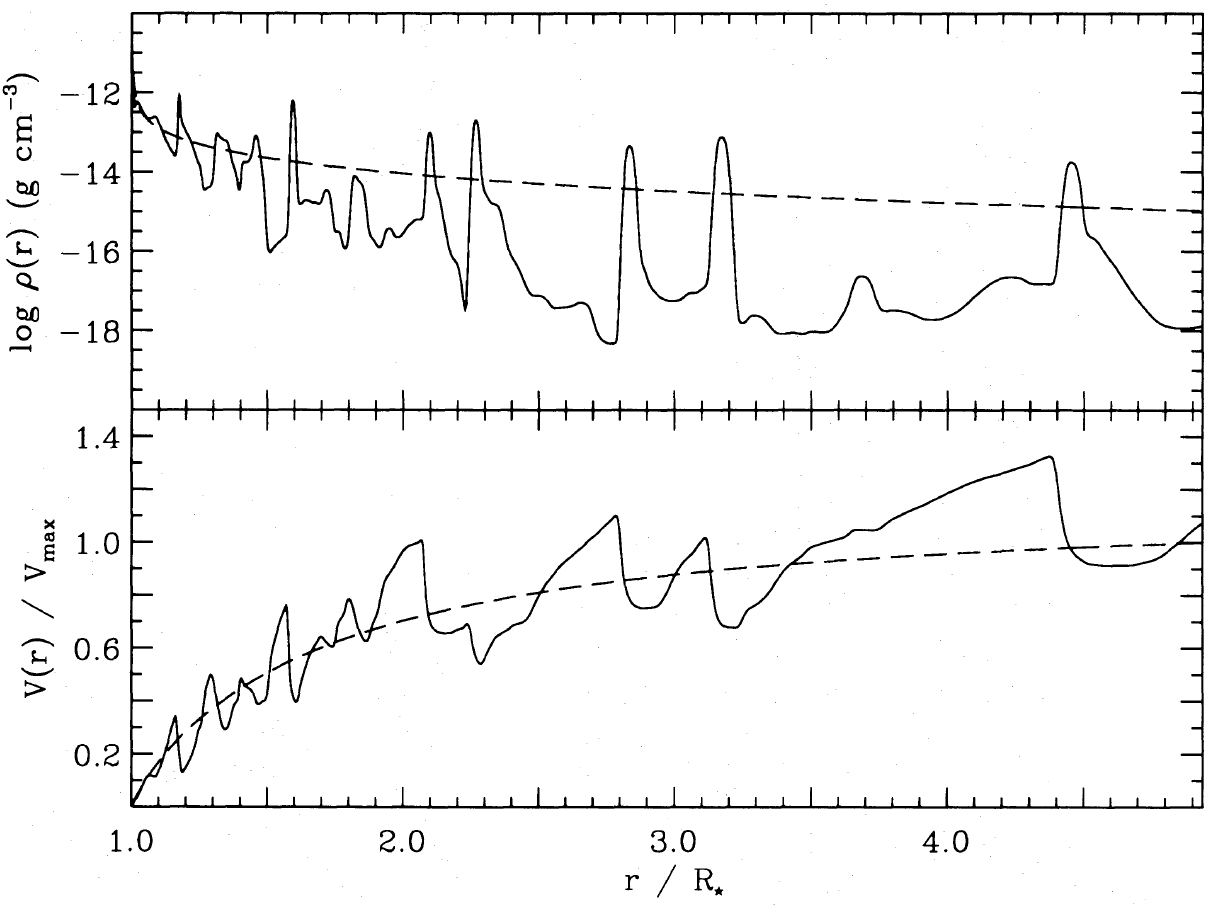}}}
\caption[Comparison between a initial smooth wind structure (dashed line, modified CAK-theory) and one predicted from time-dependent hydrodynamical simulation from~\citet{puls93} after 60000 s (solid line)]{Comparison between a initial smooth wind structure (dashed-line, modified CAK-theory) and one predicted from time-dependent hydrodynamical simulation from~\citet{puls93} after 60000 s (solid line). See text for discussion. Reproduced from~\citet{puls93}.}
\label{sec_intro_puls93_fig1}
\end{figure}

Finally, we point out that massive stars does not have smooth stationary winds, as simplified in the CAK-theory described above. It is well-established that massive stars, such as O-type, Wolf-Rayet, and LBV stars, have inhomogeneous and variable winds, that is, their wind density structures show local fluctuations, deviating from a smooth distribution, due to the agglomeration of matter, forming clumps of matter~\citep[see, e.g.,][]{eversberg98, bouret05, markova05, lepine08}.\par 

For instance, the amount of X-ray emission and the appearance of discrete absorption component in the P-Cygni profiles are understood as arising from wind inhomogeneity. In Fig.~\ref{sec_intro_puls93_fig1}, we show the comparison between the velocity and density structures from an modified version of the CAK-theory (homogeneous winds) and the results found by~\citet{puls93} for non-homogeneous winds (time-dependent hydrodynamical simulations). From the theoretical point-of-view, the formation of local agglomeration of material in the wind is understood to arise from intrinsic instabilities in the line force~\citep[e.g.,][and reference therein]{owocki88, owocki92}.\par

\subsection{Radiative winds in the HR diagram}
\label{sec_intro_radiative_winds_hr_diagram}

Fig.~\ref{sec_intro_smith14_fig3} provides an overview of the mass loss in massive stars as a function of luminosity. This diagram clarifies how the mass-loss rate of stellar winds changes with respect to their evolution. Here, both theoretical and measured mass-loss rates are given for OB-type stars, red/yellow supergiants, luminous blue variable, and Wolf-Rayet stars. As expected for radiative line-driven winds, we see that the mass-loss rate increases toward higher luminosity. Late O dwarfs (O9V) show $\dot{M}$ $\sim$ $10^{-10}$-$10^{-9}$ $\mathrm{M_\odot}$ yr\textsuperscript{-1}, while the wind is drastically stronger in Wolf-Rayet and luminous blue variable stars with $\dot{M}$ up to $10^{-5}$-$10^{-3}$ $\mathrm{M_\odot}$ yr\textsuperscript{-1}, reaching the mass loss limit for line-drive winds (gray line).\par

\begin{figure}[t]
\centerfloat
\centerline{\resizebox{0.70\textwidth}{!}{\includegraphics{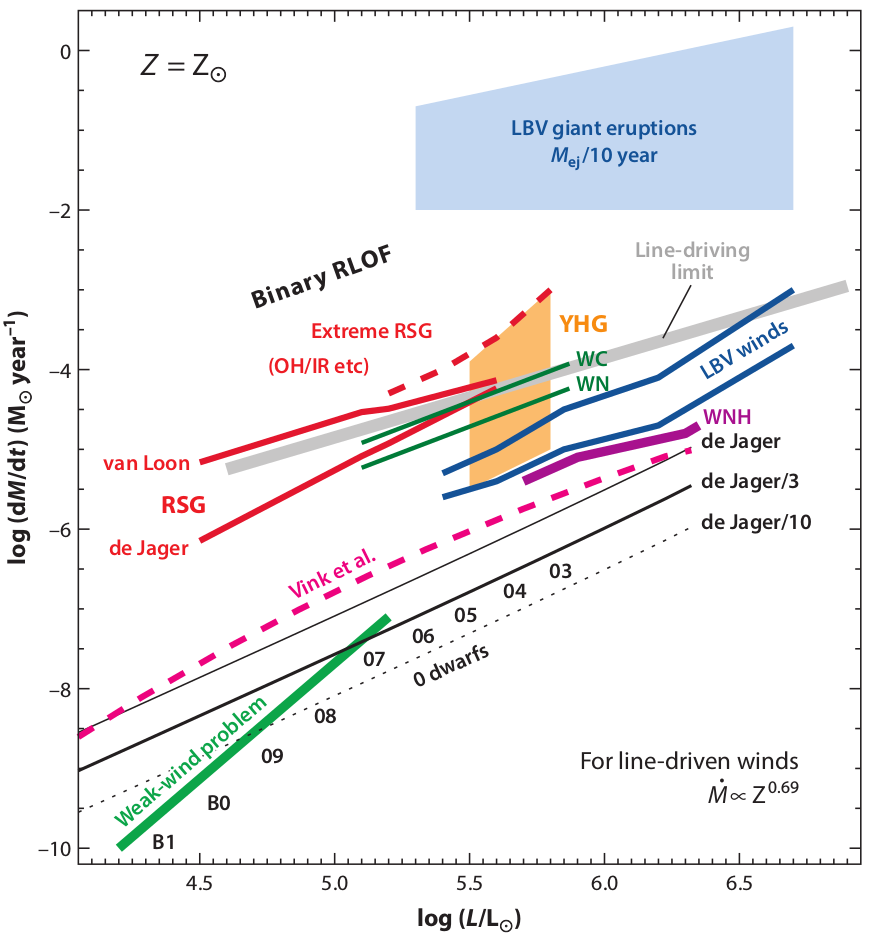}}}
\caption[Mass-loss rate as a function of stellar luminosity at different evolutionary stages of massive stars.]
{Mass-loss rate as a function of stellar luminosity at different evolutionary stages of massive stars. This diagram compiles both theoretical and spectroscopic (measured) mass-loss rates for massive stars in the Galaxy. The gray line indicates the estimated maximum mass-loss rate of a line-driven wind given by $\sim L_{\star}/c^2$ (Eq.~\ref{eq:estim_mass_loss}). The green line indicates the measured mass-loss rates using atmosphere models of O-type stars for which are found weak winds. See text for discussion. Reproduced from~\citet{smith14}.}
\label{sec_intro_smith14_fig3}
\end{figure}

In addition, the largest rate of mass transfer that are expected in Roche-lobe overflow in binary systems is shown too. It is also indicated the quantity of mass that is ejected by episodic eruption in luminous blue variable stars. As previously commented, due to lower values of effective temperature, winds from cool evolved massive stars are driven differently to the ones in massive hot stars (radiative line-driven winds). Nevertheless, note that the mass-loss rate also scales with the stellar luminosity since these cool evolved massive stars have radiative-dust driven winds~\citep[empirical prescriptions from][]{dejager88, vanloon05}. Extreme RSGs are the most luminous objects among RSGs, having the stronger winds (comparable to the ones found in LBVs). The stellar wind is so intense in extreme RSGs that the central star is obscured in the visible region, resulting in strong maser emission from OH, silicates, and $\mathrm{H_{2}}$O (OH/IR star), and infrared excess due to dust grains~\citep[e.g.,][]{smith01, decin06}.\par

For OB-type dwarf stars, both predicted and measured (derived from spectroscopic analysis) mass-loss rates are given in Fig.~\ref{sec_intro_smith14_fig3}. The theoretical prediction for the mass-loss rate from~\citet{vink00, vink01} is shown in pink dashed color. For comparison, the empirical prescription from~\citet{dejager88} is also indicated with different scales in order to account the effects due to wind clumping in the models used to measure the values shown in green line. Here, it is highlighted the expected region of the so-called weak wind problem.\par 

The phenomenon of weak winds is characterized by measured mass-loss rates that are much lower, by up to two orders of magnitude, than the theoretical values from~\citet{vink00}. This astrophysical problem is one of the main topics of this thesis and will be discussed in details in Sect.~\ref{sec_results_ostars}, in the context of my work about giant O-type stars.

\begin{figure}[t]
\centerfloat
\centerline{\resizebox{0.80\textwidth}{!}{\includegraphics{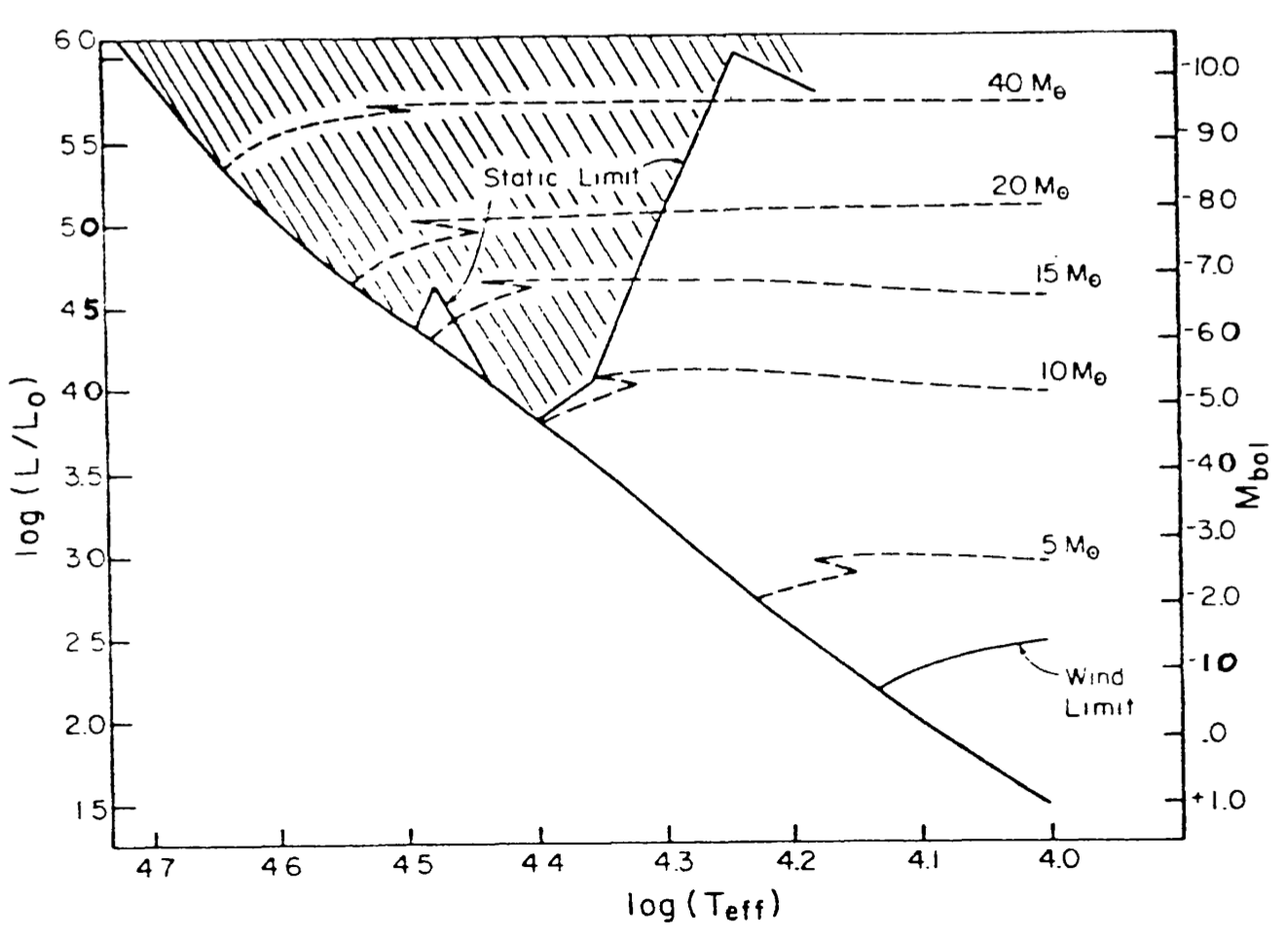}}}
\caption[Limits in the HR diagram for acceleration winds due to lines.]
{Limits in the HR diagram for acceleration winds due to lines. Evolutionary tracks for stars with different initial masses are shown in dashed lines with the ZAMS region (solid line across the HR diagram). Stars with high enough luminosity and effective temperature are able to ignite and sustain winds driven by lines (dashed region). Note that stars with initial mass higher than $\sim$10 $\mathrm{M_\odot}$ (B2V) must develop radiative line-driven winds. Reproduced from~\citet{abbott79}.}
\label{sec_intro_abbott79}
\end{figure}

We see that just the earliest B-type stars are indicated in Fig.~\ref{sec_intro_smith14_fig3} (B0 and B1). This corresponds to the most massive B stars with initial masses between $\sim$13 and 18 $\mathrm{M_\odot}$. In later B stars, much lower mass-loss rates are expected due to lower values of stellar luminosity. For a better visualisation, we show, in Fig.~\ref{sec_intro_abbott79}, the results from~\citet{abbott79} for the wind dynamics across the HR diagram. There are three region to be highlighted here:

\begin{enumerate}[label=(\roman*)]
\setlength\itemsep{1em}

\item The shaded region, which encompasses the most luminous and hotter stars, indicates the case of radiative-line driven winds that are self-initiated due to the instabilities in the stellar atmosphere. For these stars, hydrostatic atmospheres are not reliable physical solutions for the equation of motion. Note that these results from~\citet{abbott79} indicate that stars with initial masses higher than $\sim$ 10 $\mathrm{M_\odot}$ necessarily have stellar winds driven by lines.

\item For stars lying between the shaded region and the ``wind limit'' line, both hydrostatic atmospheres and radiative winds are reliable solutions. In this case, radiative line-driven winds can be sustained, once ignited, but these stars have too low luminosities to self-initiate line-driven winds.

\item Lastly, it is impossible to self-initiate or sustain a radiative line-driven winds in stars below the ``wind limit'' line.

\end{enumerate}

\begin{figure}[t]
\centerfloat
\centerline{\resizebox{0.65\textwidth}{!}{\includegraphics{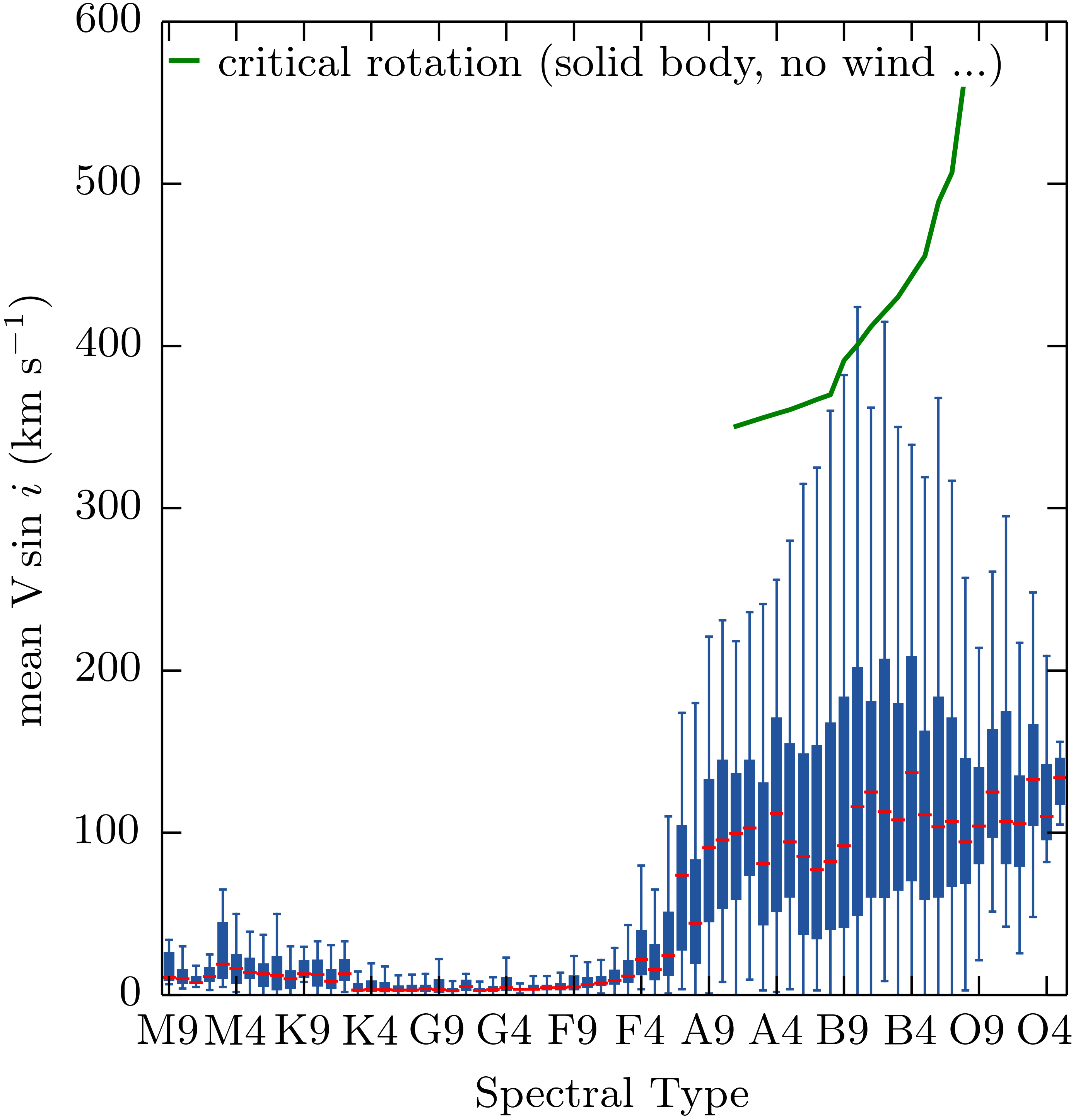}}}
\caption[Boxplot distribution of the projected stellar rotation velocity ($v\sin i$) as a function of spectral type.]
{Boxplot distribution of the projected stellar rotation velocity ($v\sin i$) as a function of spectral type. The mean value of $v\sin i$ is indicated in red and the interquartile range in blue boxes. The predicted value of critical rotation for the earlier types is shown in green line. Note that late-type stars (M-F types) rotates much more slower than the early-type stars (A-O types). Also note that B stars rotates closer to the critical value than O stars. Source: \url{http://aa.oma.be/stellar_rotation}}
\label{sec_intro_vsini_spectral_type}
\end{figure}

As indicated in Fig.~\ref{sec_intro_abbott79}, the lowest initial mass between the regimes of ``static limit'' and ``wind limit'' is somewhat lower than 5 $\mathrm{M_\odot}$, thus corresponding to B6-B9 dwarfs with $L_{\star}$ $\sim$ 600-160 $\mathrm{L_{\mathrm{\odot}}}$. We point out that these results from~\citet{abbott79} have been updated by a series of more recent studies. Using more sophisticated atmosphere models (e.g, non-LTE),~\citet{krticka14} showed that B-type stars with $T_{\mathrm{eff}}$ lower than $\sim$15000 K cannot self-initiate or sustain radiative line-driven winds. This is in fair agreement with the results from~\citet{abbott79} that indicate $T_{\mathrm{eff}}$ $\sim$ 17000 K. As found by~\citet{krticka14}, considering solar-like abundances, the hotter B stars ($T_{\mathrm{eff}}$ = 30000 K) present homogeneous radiative wind with $\dot{M}$ of $\sim$$2.0\e{-9}$ $\mathrm{M_\odot}$ yr\textsuperscript{-1}, while this value is decreased to $\sim$$7.0\e{-13}$ $\mathrm{M_\odot}$ yr\textsuperscript{-1} for B stars with $T_{\mathrm{eff}}$ = 16000 K.\par

This means that physical mechanisms other than line-driving, such as rotation, are needed in order remove material from the atmospheres of the less massive B stars. In Fig.~\ref{sec_intro_vsini_spectral_type}, we plot the distribution of the measured $v\sin i$ as a function of spectral type for main sequence stars ranging from M9 to O4. For sure, this quantity is dependent on the stellar inclination angle ($i$). Nevertheless, since there is no correlation between $i$ and the spectral type, we must consider this distribution as a real trend for the equatorial stellar velocity as a function of spectral type.\par

First, we clearly see that low-mass stars rotate much slower than the massive ones of type B and O. This happens due to a very efficient process of magnetic braking that removes a large quantity of angular momentum from the star~\citep{schatzman62, weber67}. Low-mass stars have convective dynamo in their outer layers that create a dipolar magnetic field that is extended beyond the stellar surface. The stellar wind is coupled to the magnetic field lines, forcing the wind to co-rotate with the star, and thus decreasing the stellar rotation due to the conservation of angular momentum.\par

As one can see from Fig.~\ref{sec_intro_vsini_spectral_type}, the mean value of $v\sin i$ starts to increase near the spectral type F4 up to values higher than $\sim$100 km s\textsuperscript{-1}. This can be understood since the presence of a convection layer is expected for low-mass stars up to this spectral type, but not for the more massive ones. Nevertheless, strong magnetic field have been detected for about 7-10\% of OBA-type stars, with intensity ranging from $\sim$$10^2$ to $10^4$ G. Due to their rarity, long-term stability (up to decades), and simple dipolar topology, it is understood that massive stars can have fossil magnetic field, that is, originated from the primordial molecular cloud that formed the star~\citep[see, e.g.,][and references therein]{neiner15, schultz18}. For further details about magnetic field in massive stars, see the series of paper from the Magnetic in Massive Stars (MiMeS) survey~\citep{wade16a} and the B fields in OB stars (BOB) survey~\citep{morel15}.\par

Fig.~\ref{sec_intro_vsini_spectral_type} also evidences that the highest values of stellar rotation are found for late B stars. The value of the outliers in the boxplot distribution decreases from the B9 stars to the more massive of type O4. As discussed above, the stellar wind intensity increases with the spectral type for OB stars: the more massive O stars lose more efficiently angular momentum through stellar winds than the B stars, and then resulting in lower rotational velocities. Indeed, some late-type B stars (around B9-B7) rotates so fast that they are close to the critical value of rotation.\par

Such fast rotation modifies the physical properties of the stellar interiors and surfaces~\citep[e.g.,][]{maeder05, brott11}. For massive stars, it can lead to a break of spherical symmetry of radiative winds~\citep[e.g.,][]{muller14}. In the most extreme cases, such as for classical Be stars (Sect.~\ref{sec_intro_Be_stars}), it can lead to the formation of dense circumstellar disks at the equatorial plane (Sect.~\ref{sec_intro_Be_disks}). In the following section, we introduce the effects of rotation on the stellar surface and their consequences in the case of fast rotation.\par

\section{Stellar rotation}

\subsection{Effects on the stellar shape}

Considering a non-rotating star under the Roche model~\citep[e.g.,][]{domiciano02, ekstrom08}, the equipotential surface at a distance $r$ from the center of the star is due to the gravitational potential energy:

\begin{equation}
\Phi(r) = -\frac{GM_{r}}{r},   
\end{equation}
where G is the universal gravitational constant and $M_{r}$ is the stellar mass inside the distance $r$.

Thus, the gravity acceleration $g$ is given by the gradient of the potential energy: 

\begin{equation}
\vv{g}(r) = -\vec{\nabla}{\Phi(r)} = \frac{GM_{r}}{r^2} \hat{r}.
\label{eq:g}
\end{equation}

This quantity $\Phi$ is not affected by rotation. Thus, taking into account the potential $V$ due to the centrifugal force in a rotating star, the total equipotential surface $\Psi$ is simply modified as follows:

\begin{equation}
\Psi(r,\theta) = \Phi(r) + V(r,\theta).
\label{eq:equi_potential_rot_1}
\end{equation}

Here, the centrifugal potential $V$ introduced by rotation is:

\begin{equation}
V(r,\theta) = -\frac{\Omega^2 {\varpi}^2(r,\theta)}{2},
\end{equation}
where $\Omega$ is the angular velocity that is constant along the distance $r$ and $\overline{\omega} = r \sin \theta$, being $\theta$ the co-latitudinal angle formed between the stellar rotational axis and position $\vec{r}$. Hence, the quantity $\varpi$ gives the distance from the position $\vec{r}$ to the rotational axis.

Eq.~\ref{eq:equi_potential_rot_1} is then expressed as follows:

\begin{equation}
\Psi(r,\theta) = -\frac{GM_{r}}{r} -\frac{\Omega^2 r^2 \sin^2 \theta}{2}. 
\label{eq:equi_potential_rot_2}
\end{equation}

$\Psi$ defines the shape of the stellar surface for a rotating star with angular velocity $\Omega$, and is constant at certain point defined by $r$ and $\theta$ (equipotential surface). Since the centrifugal force is null over the polar axis ($\theta$ = 0, so V = 0), it is useful to express Eq.~\ref{eq:equi_potential_rot_2} as function of polar radius $R_p$:

\begin{equation}
\Psi(r=R_{p},\theta=0) = \mathrm{constant} = \frac{GM_{\star}}{R_{p}} = \frac{GM_{\star}}{R(\theta)} + \frac{\Omega^2 R(\theta)^2 \sin^2 \theta}{2},  
\label{eq:equi_potential_rot_4}
\end{equation}
where $M_{\star}$ is the stellar mass and R($\theta$) is the stellar radius expressed as a function of co-latitude. For simplicity, the dependence of $R$ on $\theta$ is omitted in the following.\par

The acceleration at the stellar surface is then modified by the inclusion of rotation as follows:

\begin{equation}
\vv{g}_{\mathrm{eff}}(R,\theta) = -\vec{\nabla}{\Psi(R,\theta)},
\label{eq:geff_phi}
\end{equation}

\begin{equation}
\vv{g}_{\mathrm{eff}}(R,\theta) = \left(-\frac{GM_{\star}}{R^2} +  \Omega^2 R \sin^2 \theta \right)\hat{r} + \left(\Omega^2 R \sin \theta \cos \theta \right)\hat{\theta}.
\end{equation}

$\vv{g}_{\mathrm{eff}}$ is called the effective surface acceleration and its modulus ${g}_{\mathrm{eff}}$ = $\abs{\vv{g}_{\mathrm{eff}}}$ is then given by:

\begin{equation}
{g}_{\mathrm{eff}}(R,\theta) =  \sqrt{ \left(-\frac{GM_{\star}}{R^2} +  \Omega^2 R \sin^2 \theta \right)^2 +  \left(\Omega^2 R \sin \theta \cos \theta \right)^2 }.
\label{eq:geff}
\end{equation}

In the case of a non-rotation star ($\Omega$ = 0, thus $R(\theta)$ = $R_{p}$ = $R_{\star}$), Eq.~\ref{eq:geff} is simply reduced to the case of the acceleration due just to the gravitational field  (Eq.~\ref{eq:g}, $r$ = $R_{\star}$):

\begin{equation}
g = \frac{GM_{\star}}{R_{\star}^2}.
\end{equation}

\subsection{Stellar rotational rate and oblateness}
\label{sec_intro_stellar_rot_oblateness}

First, we define the critical velocity, also called the break-up velocity. This happens when the modulus of gravitational force is equal to the modulus of centrifugal force at the stellar equator. In this case, any radial contribution (in the same direction of the centrifugal force) to the equation of motion, as due to the radiative force, implies the break of the hydrostatic equilibrium of the stellar atmosphere, and thus mass loss. This means that the effective surface acceleration is null at the equator if the star rotates at critical velocity. Thus, from Eq.~\ref{eq:geff}, we have:

\begin{equation}
\Omega_{\mathrm{crit}} = \sqrt{ \frac{GM_{\star}}{R^3_{\mathrm{eq,crit}}} },
\label{eq:omega_critical_velocity}
\end{equation}
where $R_{\mathrm{eq,crit}}$ is the stellar equatorial radius in the case of critical rotation $\Omega = \Omega_{\mathrm{crit}}$.\par

Using this value of angular velocity in Eq.~\ref{eq:equi_potential_rot_4}, we see that the equatorial radius is 3/2 times larger than the polar one if $\Omega = \Omega_{\mathrm{crit}}$:

\begin{equation}
\frac{R_{\mathrm{eq,crit}}}{R_{\mathrm{p,crit}}} = \frac{3}{2},
\label{eq:req_rpol}
\end{equation}
where $R_{\mathrm{p,crit}}$ is the polar radius in critical rotation.\par

One useful quantity that is commonly used in the literature is the (angular) rotational rate $w$ defined as follows:

\begin{equation}
\omega = \frac{\Omega}{\Omega_{\mathrm{crit}}},
\label{eq:angular_rotational_rate}
\end{equation}
in order to express how close a certain star with angular velocity $\Omega$ is to the break-up limit. Conversely, this also can be expressed in terms of linear rotational velocity $v_{\mathrm{rot}}$ and $v_{\mathrm{crit}}$:

\begin{equation}
v_{\mathrm{crit}}= \Omega_{\mathrm{crit}} R_{\mathrm{eq,crit}},
\end{equation}
where $v_{\mathrm{crit}}$ is the linear equatorial velocity in the case of critical rotation. Thus, the angular and linear rotational rates are related to each other as follows:

\begin{equation}
\frac{\Omega}{\Omega_{\mathrm{crit}}} = \frac{v_{\mathrm{rot}}}{v_{\mathrm{crit}}} \, \frac{R_{\mathrm{eq,crit}}}{R_\mathrm{eq}},
\end{equation}
where $R_\mathrm{eq}$ is the actual equatorial radius and $v_{\mathrm{rot}}$ the actual linear equatorial rotational velocity.\par

Another useful quantity, which is related to the rotational rate, is the stellar oblateness $f$~\citep[e.g.,][]{domiciano02, ekstrom08} that is defined as the ratio between the equatorial and polar stellar radii:

\begin{equation}
f = \frac{R_{\mathrm{eq}}}{R_{\mathrm{p}}}.
\end{equation}

Finally, another common quantity used in literature~\citep[e.g.,][]{espinosa11, domiciano18} to express the stellar oblateness is by the flattening parameter $\epsilon$:

\begin{equation}
\epsilon = 1 - \frac{R_{\mathrm{p}}}{R_{\mathrm{eq}}}.
\end{equation}

From Eq.~\ref{eq:equi_potential_rot_4}, considering $\theta = \pi/2$ (stellar equator):

\begin{equation}
\frac{GM_{\star}}{R_{\mathrm{p}}} = \frac{GM_{\star}}{R_{\mathrm{eq}}} + \frac{\Omega^2 R_{\mathrm{eq}}^2}{2}.  
\label{eq:equi_potential_rot_5}
\end{equation}

Thus, we find the following relation between the stellar oblateness $f$ and the polar radius $R_{\mathrm{p}}$:

\begin{equation}
R_{\mathrm{p}} =  \left(\frac{GM_{\star}}{\Omega} \right)^{1/3} \left(\frac{2(f -1)}{f^3}\right)^{1/3}.
\label{eq:equi_potential_rot_6}
\end{equation}

From Eqs. \ref{eq:omega_critical_velocity} and \ref{eq:req_rpol}, the angular critical velocity $\Omega_{\mathrm{crit}}$ can also be expressed in terms of the polar radius:

\begin{equation}
\Omega_{\mathrm{crit}} = \sqrt{ \left(\frac{2}{3}\right)^3 \frac{GM_{\star}}{R_{\mathrm{p, crit}}^{3}}       }.
\label{eq:omega_critical_velocity_polar}
\end{equation}

Therefore, from Eqs. \ref{eq:equi_potential_rot_6} and \ref{eq:omega_critical_velocity_polar}, the angular rotational ratio and the stellar oblateness are related as follows:

\begin{equation}
\frac{\Omega}{\Omega_{\mathrm{crit}}} = \sqrt{ \left(\frac{3 R_{\mathrm{p,crit}}}{2 R_{\mathrm{p}}}\right)^{3} \frac{2(f-1)}{f^3}}.
\label{eq:omega_critical_velocity_oblateness}
\end{equation}

At the critical case, with $\Omega$ = $\Omega_{\mathrm{crit}}$ and $R_{\mathrm{p}} = R_{\mathrm{p,crit}}$, we found $f$ = 3/2, in agreement with Eq.~\ref{eq:req_rpol}.\par

Note, from Eq.~\ref{eq:omega_critical_velocity_oblateness}, that is necessary to know the value of the polar radius at the critical value in order to relate the rotational rate and the stellar oblateness. However, in good approximation, $R_{\mathrm{p,crit}} \sim R_{\mathrm{p}}$, being a better assumption as the stellar rotational velocity approaches the break-up case~\citep[e.g., see Fig.~2 of][]{ekstrom08}. Therefore, Eq.~\ref{eq:omega_critical_velocity_oblateness}, can be expressed as the following approximation:

\begin{equation}
\frac{\Omega}{\Omega_{\mathrm{crit}}} \sim \sqrt{ \left(\frac{3}{2}\right)^{3} \frac{2(f-1)}{f^3}.       }
\label{eq:omega_critical_velocity_oblateness_2}
\end{equation}

Alternatively, another definition, presented by~\citet{rivinius13}, for the rotational rate $W$ is given with respect to the Keplerian circular orbital velocity $v_{\mathrm{orb}}$ at the stellar equator:

\begin{equation}
v_{\mathrm{orb}} = \sqrt{\frac{GM_{\star}}{R_{\mathrm{eq}}}},
\end{equation}
being $W$ defined as follows:

\begin{equation}
W = \frac{v_{\mathrm{rot}}}{v_{\mathrm{orb}}}.
\label{eq:rotational_rate_w}
\end{equation}

As pointed out by~\citet{rivinius13}: ``it defines what velocity boost is required for a given star to launch material into the closest possible orbit, i.e., just above the photosphere at the equator''. Despite not being widely used in the literature, one advantage of this definition relies on its independence on the assumed stellar rotating model. Notice that the previous definition $\omega$, given by Eqs. \ref{eq:omega_critical_velocity} and  \ref{eq:angular_rotational_rate}, is based on the Roche model.\par

\subsection{The von Zeipel effect}
\label{sec_intro_stellar_rotation_von_zeipel_effect}

Under the assumption of LTE\footnote{Good approximation for stellar interiors as the temperature gradient is typically small over a mean free path~\citep[see, e.g.,][]{maeder09}. Further details on the hypothesis of local thermodynamics equilibrium will be provided in Sect.~\ref{sec_radiative_transfer_modeling_elementary_concepts}.}, and radiative equilibrium (i.e., all the energy is transported by radiation in the star), the flux $F(r)$ at the certain distance $r$ of the stellar interior is given by:

\begin{equation}
F(r) = \frac{L_{r}}{4\pi r^2} = -C_{\mathrm{rad}} \diff{T}{r},
\label{eq:flux_radiative_equilibrium}
\end{equation}
where $L_{r}$ is the stellar luminosity at $r$, and the coefficient of radiative conductivity $C_{\mathrm{rad}}$ is defined as follows:

\begin{equation}
C_{\mathrm{rad}} = \frac{4 a c T^3}{3 \kappa \rho},
\end{equation}
where $\kappa$ and $\rho$ are the local opacity and mass volume density, $a = 4\sigma/c$, $T$ is the local temperature following a black body distribution.\par

In the case of a star rotating with constant angular velocity $\Omega$, the transport of radiative flux, given by the temperature gradient, is then modified as follows:

\begin{equation}
\vec{F}(\Omega, \theta) = -C_{\mathrm{rad}} \vec{\nabla}{T(\Omega,\theta)},
\label{eq:flux}
\end{equation}
being analogous to Eq.~\ref{eq:flux_radiative_equilibrium}, but as a function of $\Omega$ and the co-latitude.\par

The equipotentials and isobars (i.e., lines of constant pressure in the star) coincide in the case of star rotating like a solid body. This means that the temperature and density in the star are expressed as a function only of the equipotential, thus we can rewrite Eq.~\ref{eq:flux} using Eq.~\ref{eq:geff_phi}:

\begin{equation}
\vec{F} = -C_{\mathrm{rad}} \diff{T}{\Psi} \vec{\nabla}{\Psi} =  C_{\mathrm{rad}} \diff{T}{\Psi} \vec{g}_{\mathrm{eff}},
\;
\vec{F} = -C_{\mathrm{rad}} \diff{T}{P} \vec{\nabla}{P} = -\rho C_{\mathrm{rad}} \diff{T}{P} \vec{g}_{\mathrm{eff}},
\label{eq:flux_geff}
\end{equation}
where the term $\rho C_{\mathrm{rad}} \diff{T}{P}$ is constant at a certain equipotential surface $\Psi$ and the relation between the pressure gradient ($\vec{\nabla}{P}$) and the effective acceleration, $\vec{\nabla}{P}/\rho = -\vec{\nabla}{\Psi} = \vec{g}_{\mathrm{eff}} $, states the hydrostatic equilibrium in the star.\par

Note that Eq.~\ref{eq:flux_geff} is valid at a certain distance $r$. At the stellar surface, the so-called von Zeipel theorem~\citep{vonzeipel24} states the following relation between the radiative flux and the effective surface acceleration:

\begin{equation}
\vec{F}(\Omega, \theta) = - \frac{L_{\star}}{4 \pi G M_{\star}} \vec{g}_{\mathrm{eff}}(\Omega, \theta)  \propto \vec{g}_{\mathrm{eff}}(\Omega, \theta),
\label{eq:von_zeipel1}
\end{equation}
where $\frac{L_{\star}}{4 \pi G M_{\star}}$ comes from the evaluation of the constant (for a certain distance $r$) $\rho C_{\mathrm{rad}} \diff{T}{P}$ at the stellar surface~\citep[see, e.g., Sect.~4.2.2 of][]{maeder09}.\par

Hence, the von Zeipel theorem states that the radiative flux is proportional to the effective surface acceleration. This effect is also called in the literature as gravity darkening, or von Zeipel effect.\par

One very important consequence from the von Zeipel theorem is the relation between the effective temperature and surface acceleration at a certain co-latitudinal angle. From the relation between the radiative flux and luminosity (Eq.~\ref{eq:flux_radiative_equilibrium}), and the Stefan-Boltzmann theorem (Eq.~\ref{eq:stefan_boltzmann}), the von Zeipel theorem says that:

\begin{equation}
T_{\mathrm{eff}}(\Omega, \theta) =  \sqrt[4]{ \frac{L_{\star}}{4 \pi \sigma G M_{\star}} } \sqrt[4]{g_{\mathrm{eff}(\Omega, \theta)} } \propto \sqrt[4]{g_{\mathrm{eff}}(\Omega,\theta)},
\label{eq:von_zeipel3}
\end{equation}
being the effective temperature dependent on the co-latitude.\par 

We stress that this relation $T_{\mathrm{eff}} \propto g_{\mathrm{eff}}^{0.25}$, found by~\citet{vonzeipel24}, is valid under the hypothesis of a uniformly rotating star in radiative equilibrium. However, barotropicity and radiative equilibrium are incompatible hypotheses at the same time in order to describe a rotating star~\citep[see, ][and references therein]{eddington25, rieutord09}. For instance,~\citet{lucy67} found $T_{\mathrm{eff}} \propto g_{\mathrm{eff}}^{0.08}$ for the gravity darkening effect, when considering a convective envelope (i.e., dropping the assumption of radiative equilibrium). More generally, this effect follows the relation $T_{\mathrm{eff}} \propto g_{\mathrm{eff}}^{\beta}$, where the exponent $\beta$ is lower than 0.25~\citep[e.g.,][]{vanbelle06, monnier07, hadjara18, domiciano18}. This means that the von Zeipel theorem, as originally stated, is likely to overestimate the difference of temperature between the pole and the equator. A more recent version of the gravity darkening effect is given by~\citet{espinosa11}\par

\subsection{Fast rotation}
\label{sec_intro_fast_rotation}

From Eq.~\ref{eq:geff}, we note that the effective acceleration is stratified with respect to the co-latitudinal angle $\theta$. At the stellar surface, the maximum value of effective acceleration is reached at the pole, while it decreases toward higher values of co-latitudinal angle down to the equator due to rotation. From the von Zeipel theorem, the effective temperature is then larger at the pole than at the equator, considering a rotating star model (Eq.~\ref{eq:von_zeipel3}). Also note that this effect increases with $\Omega$ (up to $\Omega_{\mathrm{crit}}$).\par

\begin{figure}[t]
\centerfloat
\centerline{\resizebox{0.75\textwidth}{!}{\includegraphics{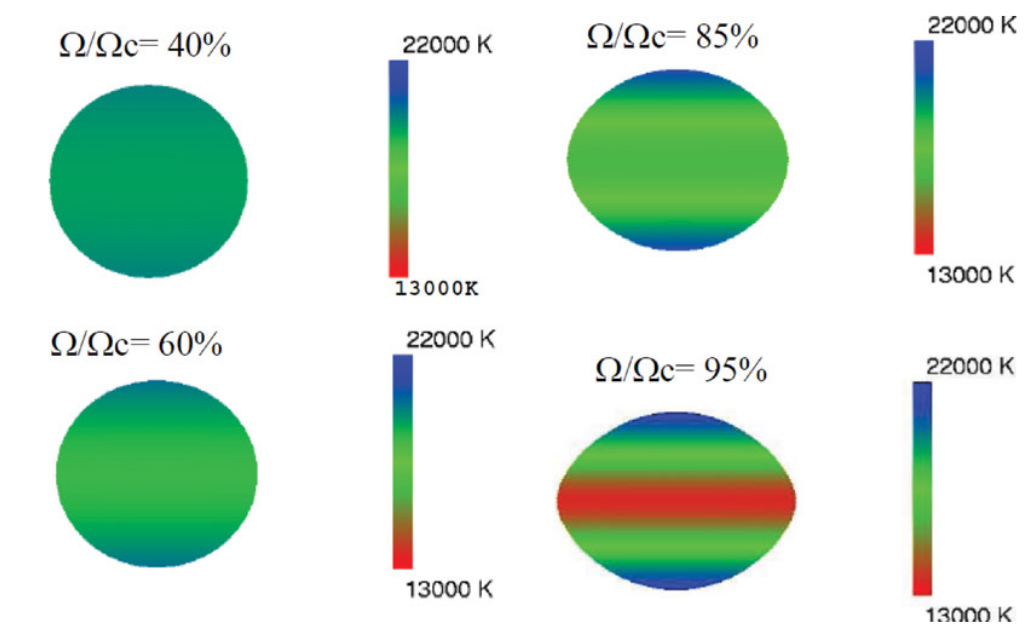}}}
\caption[Gravity darkening and geometrical oblateness effects due to rotation.]{Gravity darkening and geometrical oblateness effects due to rotation. Roche models for a solid body computed for different values of equatorial stellar velocity from $\Omega/\Omega_{\mathrm{crit}}$ = 0.40 up to $\Omega/\Omega_{\mathrm{crit}}$ = 0.95. The effective temperature is indicated by color bars. For the faster rotating star model, the effective temperature stratification reaches a difference of $\sim$9000 K with respect to the stellar pole and equator. Reproduced from~\citet{martayan11}.}
\label{sec_intro_martayan11_fig4}
\end{figure}

For example, in Fig.~\ref{sec_intro_martayan11_fig4}, we show results calculated under the Roche approximation of a solid body for different values of rotational rate $\Omega/\Omega_{\mathrm{crit}}$ ranging from 0.40 up to 0.95. First, the effect of geometrical oblateness is clear as $\Omega$ approaches its critical value. Moreover, we also clearly see how the von Zeipel effect increases toward higher values of $\Omega/\Omega_{\mathrm{crit}}$. While a star with lower rotational velocity ($\Omega/\Omega_{\mathrm{crit}}$ = 0.40) shows a fairly homogeneous effective temperature, the effective temperature is highly stratified, with respected to $\theta$, for stellar rotation close to the critical value. For the highest value of equatorial rotational velocity ($\Omega/\Omega_{\mathrm{crit}}$ = 0.95), the effective temperature reaches $\sim$22000 K at the stellar pole and $\sim$13000 K at the equator.\par

As discussed in Sect.~\ref{sec_intro_radiative_winds_hr_diagram}, from an observational point-of-view, B-type stars are much more likely to be fast rotators than O-type stars. In addition, Fig.~\ref{sec_intro_meynet2000_fig10_fig11} shows the variation of the linear equatorial velocity, and the respective angular rotational rate, according to the stellar lifetime in Geneva evolutionary models. These results are presented for massive stars with different initial masses from 12 $\mathrm{M_{\odot}}$ (around the type B1V) up to 60 $\mathrm{M_{\odot}}$ (O3V).\par 

\begin{figure}[t]
\centerfloat
\centerline{\resizebox{1.00\textwidth}{!}{\includegraphics{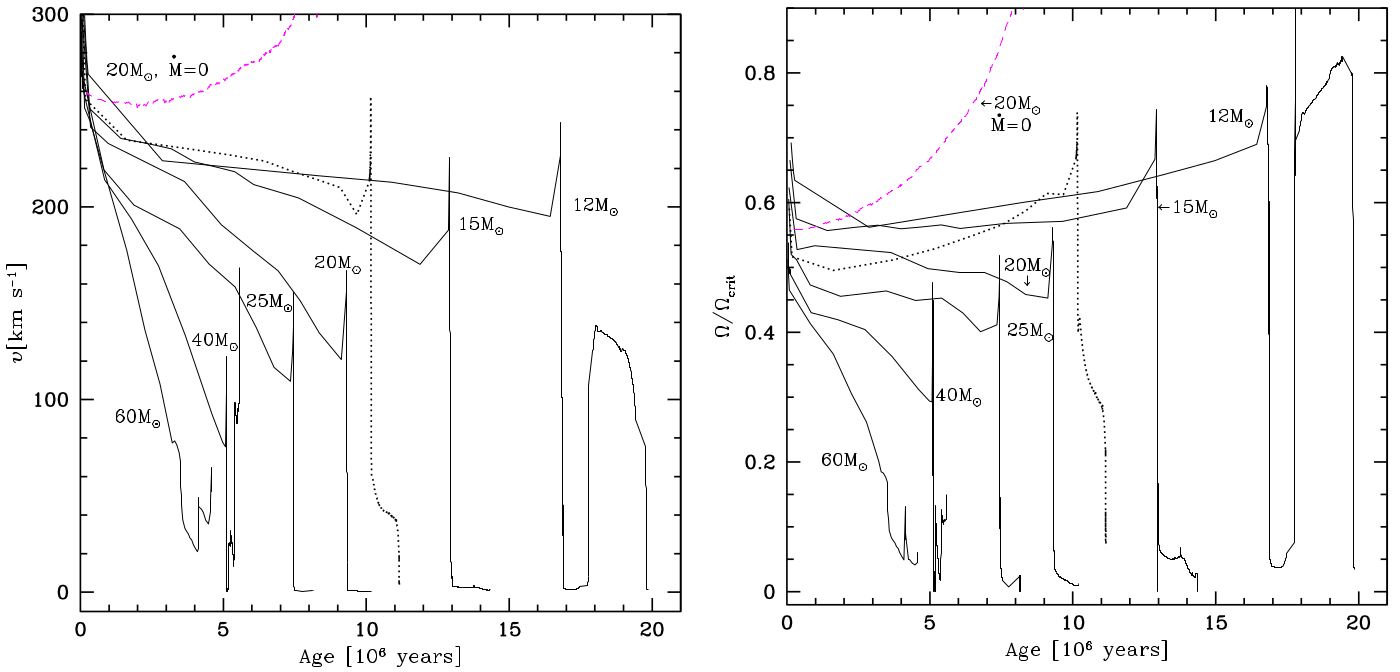}}}
\caption[Temporal evolution of the linear stellar equatorial velocity (left panel) from Geneva evolutionary models for massive stars with different initial masses. The corresponding angular rotational rate is shown in the right panel.]
{Temporal evolution of the linear stellar equatorial velocity (left panel) from Geneva evolutionary models for massive stars with different initial masses. The corresponding angular rotational rate is shown in the right panel. Stellar winds are taken in account in the models shown in solid black line, but not in pink line (20 $\mathrm{M_\odot}$, $\dot{M}$ = 0). A star with initial mass of 12 $\mathrm{M_\odot}$ (B1V) ends the main sequence phase with a high rotational rate ($\Omega/\Omega_{\mathrm{crit}}$ $\sim$ 0.8), while O-type stars will have quite lower values of rotational rate due to stronger stellar winds. Reproduced from~\citet{meynet00}.}
\label{sec_intro_meynet2000_fig10_fig11}
\end{figure}

First, we note how $\Omega/\Omega_{\mathrm{crit}}$ drastically evolves for a 20 $\mathrm{M_{\odot}}$ model, when not taking into account the loss of angular momentum due to the stellar winds ($\dot{M}$ = 0, pink dashed line). In this case, $\Omega/\Omega_{\mathrm{crit}}$ increases up to the critical limit around the end of the main sequence phase (around the age of 8 Myr). We stress that such result is valid for a solid rotating model, meaning instantaneous transport of angular momentum from the center of star to the surface (constant $\Omega$ as a function of $r$, as discussed in Sect.~\ref{sec_intro_stellar_rot_oblateness}). When taking diffusive processes in account in the stellar interior, the increase of the rotational velocity must be smaller in this case. On the other hand, considering the same initial mass of 20 $\mathrm{M_{\odot}}$, but taking into account the removal of angular momentum by winds, $\Omega/\Omega_{\mathrm{crit}}$ gradually decreases up to the end of the main sequence phase. This effect is stronger as the initial mass is larger due to higher values of mass-loss rate, even during the main sequence phase.\par

\begin{figure}
\centerfloat
\centerline{\resizebox{0.70\textwidth}{!}{\includegraphics{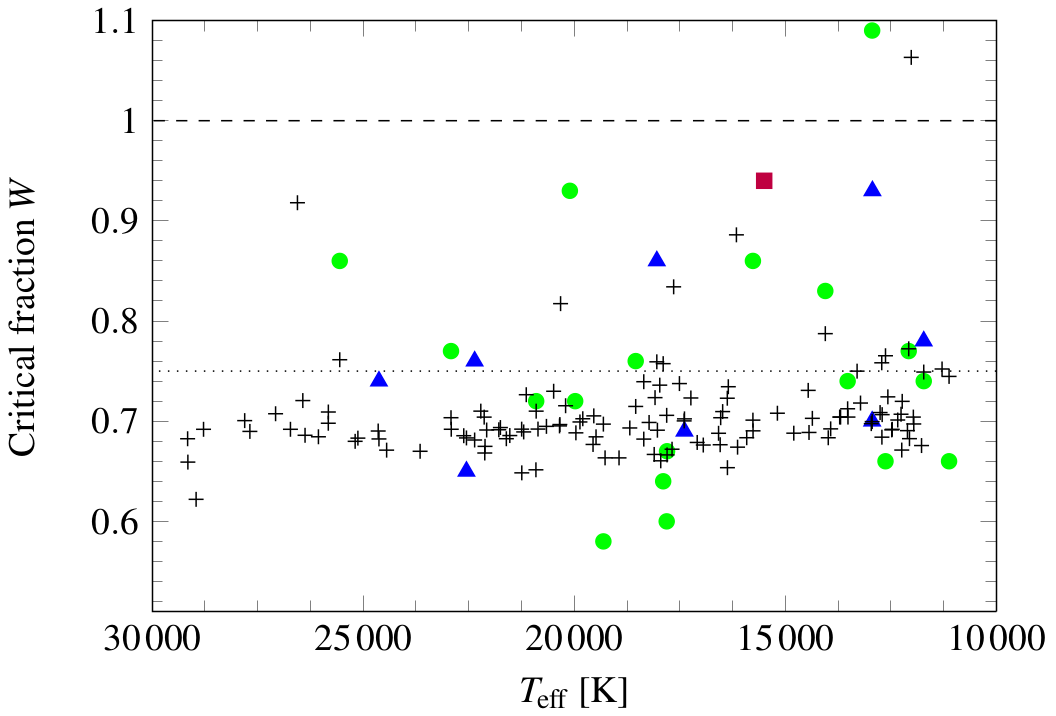}}}
\caption[Rotational rate $W$ of Be-type stars as a function of effective temperature from different studies in the literature.]
{Rotational rate $W$ of Be-type stars as a function of effective temperature from different studies in the literature. Black crosses:~\citet{fremat05}. Green points:~\citet{rivinius06}. Blue triangles:~\citet{meilland12}. Red square (Achernar, B6V):~\citet{domiciano12}. The mean value of $W$ is shown in dotted line and the critical case is indicated in dashed line. Despite the lack of clear correlation between $W$ and the effective temperature, late-type Be stars tend to have higher values of rotational rate than the earlier Be stars. Reproduced from~\citet{rivinius13}}
\label{sec_intro_rivinius_carciofi13_fig9}
\end{figure}

For the lowest mass star with $\mathrm{M_{ZAMS}}$ = 12 $\mathrm{M_\odot}$, we see that the rotational rate increases during the main sequence evolution up to about $\Omega/\Omega_{\mathrm{crit}}$ = 0.8. This happens due to the very low mass-loss rates by stellar winds in comparison with the other models for more massive stars shown here. Moreover, one can see the high decrease of $\Omega/\Omega_{\mathrm{crit}}$ after the H-burning phase due to the growth of the stellar radius. Further details on these models can be found in~\citet{meynet00}.\par

The discussion above clarifies how the stellar rotation is important to remove angular momentum, in particular for B stars. Interesting, a 12 $\mathrm{M_\odot}$ model, having a weak stellar wind (low $\dot{M}$), are more likely to reach the break-up limit and then to lead to a large amount of mass loss toward the end of the main sequence due to fast rotation.\par

Among the early-type stars, Be stars are well-known to present the highest rotational velocities~\citep[e.g., see][and references therein]{slettebak66, slettebak70}, and then enhanced gravity darkening and geometrical oblateness effects. This means that the stellar rotation is important to break the hydrostatic equilibrium in the atmosphere of Be stars. Indeed, Be stars also present larger (by up to $\sim$10 times) values of mass-loss rates than ``normal'' B-type stars, that is, which have never shown the Be phenomenon~\citep[e.g.,][]{prinja89}.\par

In Fig.~\ref{sec_intro_rivinius_carciofi13_fig9}, we provide a compilation of literature results for the rotational rate ($W$, Eq.~\ref{eq:rotational_rate_w}) of Be stars as a function of effective temperature. These results are obtained by different types of methodologies, based both on spectroscopic~\citep{fremat05, rivinius06} and interferometric analyses~\citep{domiciano12, meilland12}. One sees that Be stars are fast rotators (mean value of $W$ $\sim$ 0.75), but they are not necessarily rotating close to the critical case ($W$ = 1). Thus, in addition to the centrifugal force due to rotation, other mechanisms, as the radiative force and pulsations, are needed to explain the mass loss in Be stars that are rotating well below the break-up case.\par 

Moreover, we also note the lack of clear correlation between the rotational rate and the effective temperature. However, still from Fig.~\ref{sec_intro_rivinius_carciofi13_fig9}, cooler Be stars seems to be more likely close to the critical value than the hotter objects. This could be understood since the stellar wind are stronger in the hotter Be stars than in the cooler ones, then removing a large amount of angular momentum during the evolution in the main sequence. 

\begin{figure}
  \begin{adjustbox}{minipage=\textwidth,scale=1.00}
  \centering
  \includegraphics[width=0.49\columnwidth]{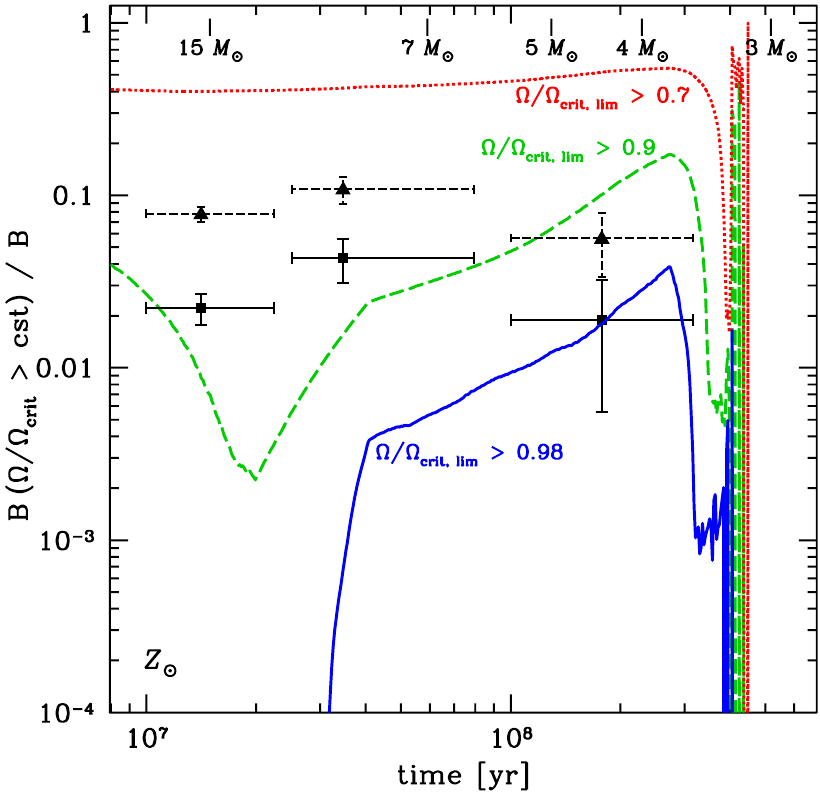}
  \medskip
  \includegraphics[width=0.49\columnwidth]{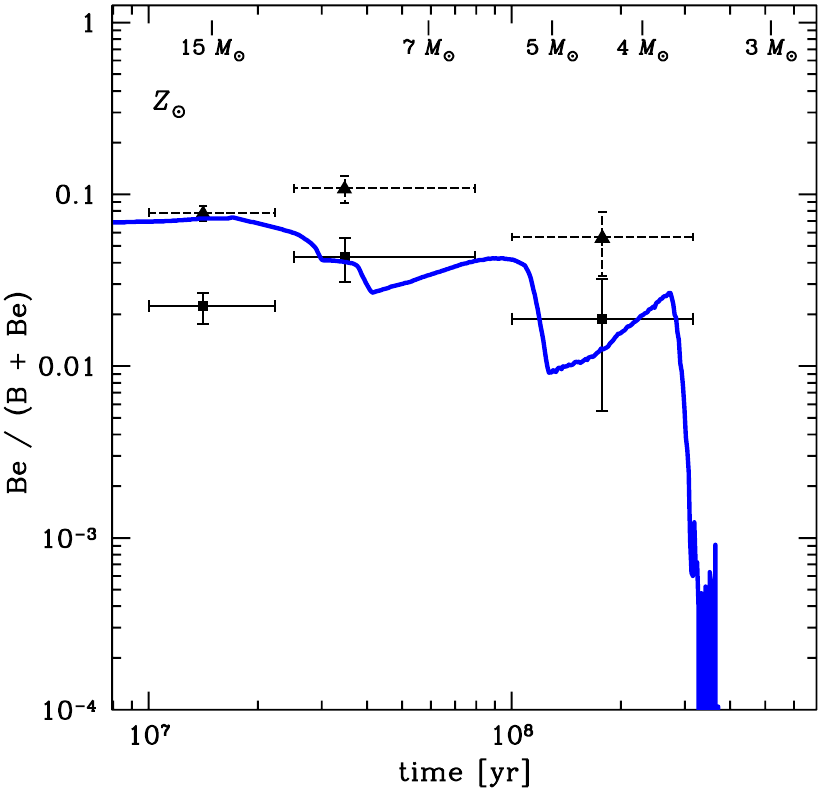}
  \end{adjustbox}
  \caption[Comparison between the fraction of stars rotating faster than a minimum value (theoretical fraction in color lines) and the observed fraction of Be stars in Galatic clusters (left panel).]
  {Comparison between the fraction of stars rotating faster than a minimum value (theoretical fraction in color lines) and the observed fraction of Be stars in Galatic clusters (left panel). The age of the cluster is indicated in the bottom horizontal axis, corresponding to the lifetime of stars with different initial masses (top horizontal axis). In the right panel, the theoretical fraction, calculated assuming a dependence between $\Omega/\Omega_{\mathrm{crit}}$ and the effective temperature, is also compared with data. See text for discussion. Reproduced from~\citet{granada13}.}
\label{sec_intro_granada13_fig2_fig9}
\end{figure}

Indeed, the dependence of the rotational rate of Be stars and stellar parameters (such as the effective temperature) is still an open issue in the literature, affecting our understating about the Be phenomenon itself. Interferometric surveys of Be stars  could not provide a clear correlation between the stellar rotation and the effective temperature~\citep[e.g.,][]{meilland12, cochetti19}. Nonetheless, as stressed by these authors, more precise determinations of the rotational rate of Be stars are needed to better clarify this issue. On the other hand, from the statistical analysis of~\citet{cranmer05}, late-type Be stars (with $T_{\mathrm{eff}}$ $\lesssim$  21000 K) are more likely to rotate closer to the critical value than the earlier objects.\par

These results from~\citet{cranmer05} are supported by other more recent studies as~\citet{huang10} and~\citet{granada13}. In Fig.~\ref{sec_intro_granada13_fig2_fig9}, we show the results from~\citet{granada13} for the temporal evolution of the fraction of stars rotating faster than a certain threshold. Their prediction from Geneva evolutionary models are compared with the observed fraction of Be stars, that is, Be/(Be + B), in Galactic clusters. Despite the few observational data, one sees that their theoretical fraction of Be stars can fairly reproduce the observed trend for Be stars (right panel), when considering a dependence of the rotational rate and the effective temperature, assuming the results of~\citet{cranmer05}.\par

In short, rotation is likely to be a major source of mass loss in late-type Be stars, when compared with the earlier ones. One could expect this fact since earlier Be stars have higher luminosities, and thus a larger contribution from the radiative force to break the hydrostatic equilibrium of the stellar atmosphere.\par

Lastly,~\citet{martayan06a} and~\citet{martayan07b} found that the rotational rates of Be stars are higher in the Large Magellanic Cloud (LMC) and SMC, in comparison with the Galaxy: up to $\sim$75-100\% of the critical value in the SMC (lower metallicity than the Galaxy and LMC). From Fig.~\ref{sec_intro_rivinius_carciofi13_fig9}, one sees a lower limit of about 60\%. Again, this effect could be expected since the mass-loss rate of the stellar wind is smaller as the stellar metallicity is reduced, and thus the process of removing angular momentum from star by winds is less efficient. This effect on stellar rotation explains the different fractions of Be stars that are observed in the Galaxy and in lower metallicity environments, as the LMC and SMC. From the few observational data shown in Fig.~\ref{sec_intro_granada13_fig2_fig9} for the fraction of Be stars in Galactic clusters, one sees that the highest value is about 10\%. Indeed, larger studies found that the fraction of Be stars is about 17\%, considering both late and early-type B stars~\citep{zorec97}. On the other hand, the fraction of Be stars is quite enhanced up to 40\% in the SMC~\citep[e.g.,][]{martayan07a,bodensteiner20}. In short, these observational facts emphasize the relation between fast rotation and the occurrence of the Be phenomenon.\par

In conclusion, as they are the faster rotators among all the non-degenerate stars, Be stars are unique objects to better understand the effects of near-critical rotation on the stellar structure and evolution, as well as on the mass-loss mechanism in early-type stars. In the following section, we discuss in details the so-called Be phenomenon.\par

\section{The Be phenomenon}
\label{sec_intro_Be_stars}

\subsection{Characterization and physical origin}
\label{sec_intro_be_phenomenon}

Classical Be stars are B-type stars in the main sequence phase (luminosity class from III-V) that show, or showed at some time, Balmer lines in emission~\citep[e.g.,][]{rivinius13}. Other spectral lines of \ion{He}{I} and low ionized metal such as \ion{Fe}{II} are also found in emission (see Fig.~\ref{sec_intro_rivinius_carciofi_fig1}). The Be phenomenon is found in the entire range of spectral type of B stars: $\mathrm{M_{ZAMS}}$ from $\sim$3 $\mathrm{M_\odot}$ for B9 stars ($T_{\mathrm{eff}}$ $\sim$ 12000 K) up to $\sim$18 $\mathrm{M_\odot}$ for B0 stars with $T_{\mathrm{eff}}$ $\sim$ 30000 K~\citep{townsend04}.\par

As pointed out by~\citet{rivinius13}, this definition for classical Be stars is made in order to exclude other early-type stars with Balmer lines in emission, as the Oe and Ae-type stars, since it is still an open question in literature whether these objects present such spectral features due to the same physical origin of classical Be stars. This question will be discussed in details in Sect.~\ref{sec_intro_Oe_Ae_stars}. Furthermore, another group of peculiar B stars, the so-called B[e] stars, also show Balmer lines in emission. Here, the qualifier ``[ ]'' indicates that these stars show in the visible region permitted and forbidden emission lines due to metals, such as \ion{Fe}{II} and \ion{O}{I}, in addition to strong Balmer emission lines~\citep[e.g.,][]{zickgraf86}. Stars showing the B[e] phenomenon form a highly inhomogeneous group of objects and such characteristics are found in supergiant stars, young stellar objects, symbiotic binaries, and compact planetary nebulae~\citep[see, e.g.,][]{lamers98}.\par

In addition to this peculiar spectral feature of emission-lines, Be stars also have other important observational characteristics that distinguish them to the ``normal'' B-type stars. They present a large excess in the spectral energy distribution from the infrared up to the radio region, that is, that cannot be explained as being originated from a pure-photospheric contribution~\citep[e.g.,][]{gehrz74, cote87, waters91}. Moreover, the majority of Be stars present non-null intrinsic polarization in emission-lines and in the continuum region~\citep[see, e.g.,][]{coyne76, wood97}.\par

All these observational characteristics of Be stars are well explained as arising from a dust-free gaseous thin equatorial disk that is supported by rotation with a very small radial velocity of a few km s\textsuperscript{-1}~\citep[e.g.,][]{rivinius99, stee11}. In order to explain the double-peaked Balmer emission in Be stars,~\citet{struve31} was the first to propose that these stars are surrounded by an equatorial gaseous disk due to their fast rotation.\par 

\begin{figure}[t]
\centerfloat
\centerline{\resizebox{0.70\textwidth}{!}{\includegraphics{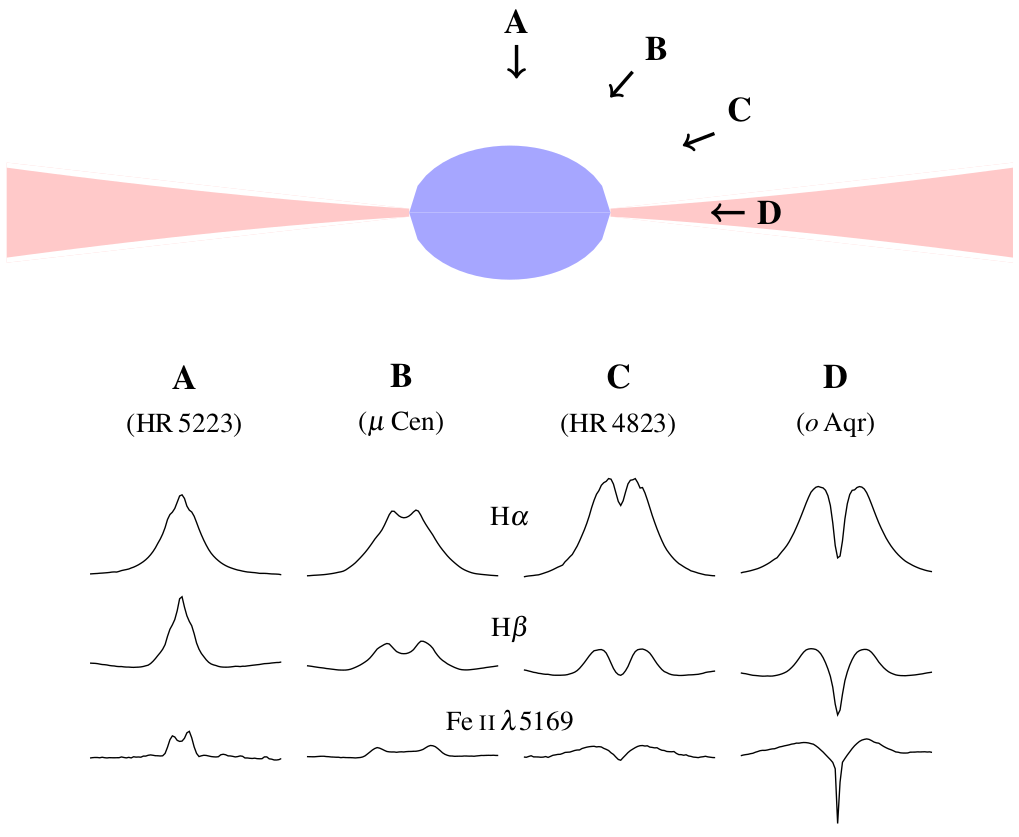}}}
\caption[Struve's picture to explain the emission line profiles of Be stars.]
{Struve's picture to explain the emission line profiles of Be stars. Observed spectral profiles are shown for Be stars seen under different values of inclination angle, from the pole-on case (HD 5223) up to the edge-on case ($\omicron$ Aquarii). Note how the H$\alpha$ profile of $\omicron$ Aquarii shows highly broadened peaks due to the circumstellar disk velocity field. See text for discussion. Reproduced from~\citet{rivinius13}.}
\label{sec_intro_rivinius_carciofi_fig1}
\end{figure}

A basic scheme of the Struve's picture is presented in Fig.~\ref{sec_intro_rivinius_carciofi_fig1}. Here, we present the observed spectra of Be stars seen under different inclination angles: from close to the pole-on case (HD 5223) to close to the edge-on case ($\omicron$ Aquarii). We see how the morphology of the double-peaked emission-lines changes as a function of inclination angle. While HR 5223 shows a single-peaked H$\alpha$ profile, the Be-shell star $\omicron$ Aquarii shows a broadened double-peaked H$\alpha$ profile due to the disk rotation. For the case of $\omicron$ Aquarii, the very deep absorption in the core of the H$\alpha$ line arises from the large amount of column density in the observer's line of sight through a large extension of disk.\par

As stressed in the first paragraph of this section, the part ``or showed at some time'' is important for the description of the Be stars. Such phenomenon is intrinsically transient as the circumstellar disks pass by different phases of dissipation and rebuilt due to mass injection from the central star. Once classified as Be star, a star will remain so in the future, even with the absence of emission-lines in its spectrum in observations a posteriori.\par

For example, we show, in Fig.~\ref{sec_intro_clark03_fig7}, the temporal variation of the  H$\alpha$ line profile of the Be-shell star $\omicron$ Andromedae (B6IIIpe) from~\citet{clark03}. These authors analysed H$\alpha$ spectroscopic data taken over a long period of about 17 years (from 1985 to 2002) in order to better constrain the disk variability for this star. To illustrate the time-scale variability of the H$\alpha$ line profile due to changes on the disk, they present observations taken over a period of 2 years (between 1986 and 1988).\par

\begin{figure}[t]
\centerfloat
\centerline{\resizebox{0.65\textwidth}{!}{\includegraphics{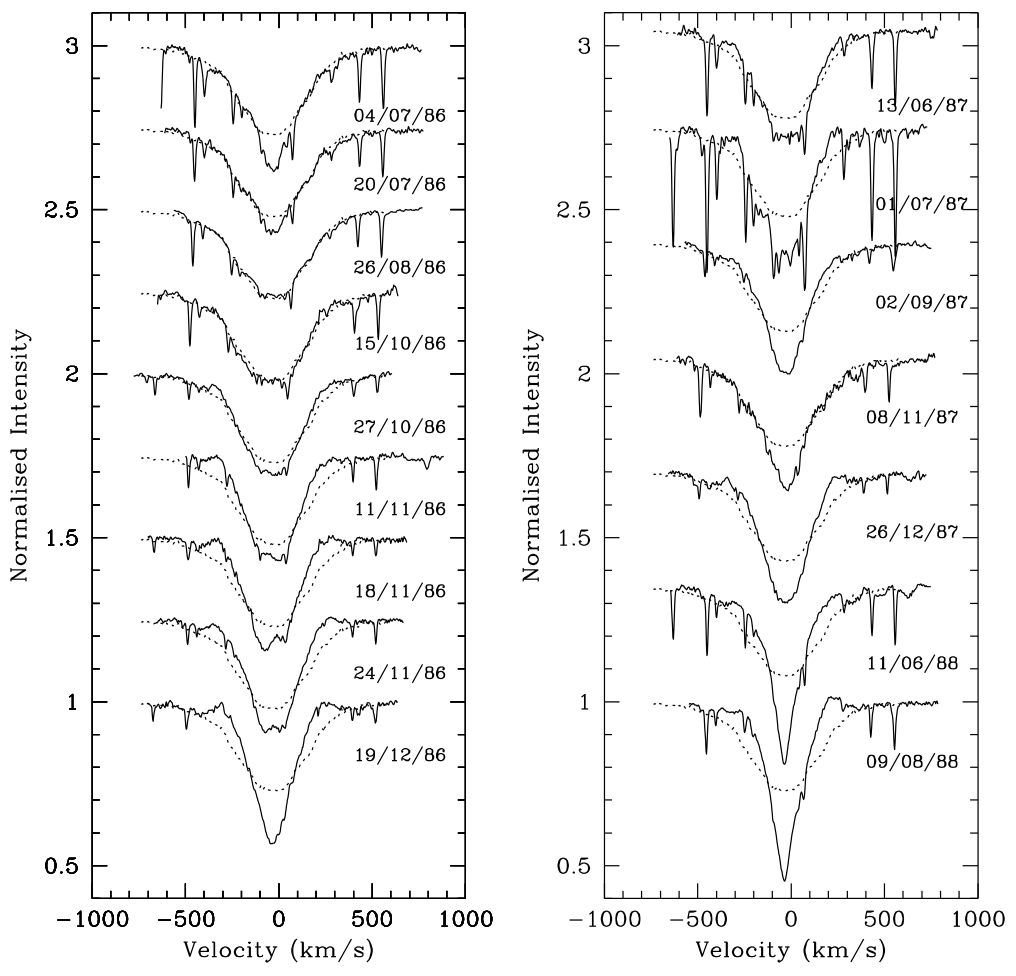}}}
\caption[Temporal variation from 1986 to 1988 of the observed H$\alpha$ profile (solid black lines) of $\omicron$ Andromedae (B6IIIpe).]
{Temporal variation from 1986 to 1988 of the observed H$\alpha$ profile (solid black lines) of $\omicron$ Andromedae (B6IIIpe). The mean H$\alpha$ profile from observations taken in 1985 is shown in dotted black line, being well reproduced by a pure photospheric model. Note the transition from the B-shell type to the Be-shell type between 1987 and 1988 due to a new accretion phase of the circumstellar disk. See text for discussion. Reproduced from~\citet{clark03}.}
\label{sec_intro_clark03_fig7}
\end{figure}

Their mean spectra taken from 1985 is shown in dashed line as a reference for a H$\alpha$ line profile formed (almost) purely in the stellar atmosphere, that is, none or very weak contribution due to the circumstellar material. These authors verified that the H$\alpha$ line profile is very stable during the year of 1985 and can be well-modeled as a pure photospheric line. It is clear from this modeling that $\omicron$ Andromedae was passing by a normal B-type phase during 1985. As pointed out by~\citet{clark03}, the transition from the B phase to the B-shell phase occurs during 1986, and finally, in 1988, the Be(-shell) phase is onset, characterized by a distinguishable emission observed in the H$\alpha$ line wings (in comparison with the pure photospheric model).\par

\subsection{Variability in Be stars}
\label{sec_intro_variability_be_stars}

In the case of $\omicron$ Andromedae, we see that the H$\alpha$ profile is significantly variable within a time-scale of about 2 years, due to mass injection from the central star to rebuild the circumstellar disk~\citep{clark03}. More generally, Be stars are known to present variability in a wide range of time-scales, from hours up to several decades, based on different types of observables (photometry, polarimetry, spectroscopy). Such large range of time-scales is well understood as arising from different physical phenomena from both the stellar atmosphere and the circumstellar material. For a more comprehensive discussion, we refer the reader to the review of~\citet{rivinius13}.\par 

The shortest variations in Be stars with a time-scale of hours to days concern H and He lines and photometric data are due to non-radial stellar pulsations~\citep[see, e.g.,][]{rivinius03}. Such short-term variability is more likely to be verified in the earlier Be stars than in the later spectral types since the amplitude of pulsations is larger for earlier-type B stars than for the later ones~\citep[e.g.,][]{percy04, baade16}.\par 

\begin{figure}[!ht]
\centerfloat
\centerline{\resizebox{0.50\textwidth}{!}{\includegraphics{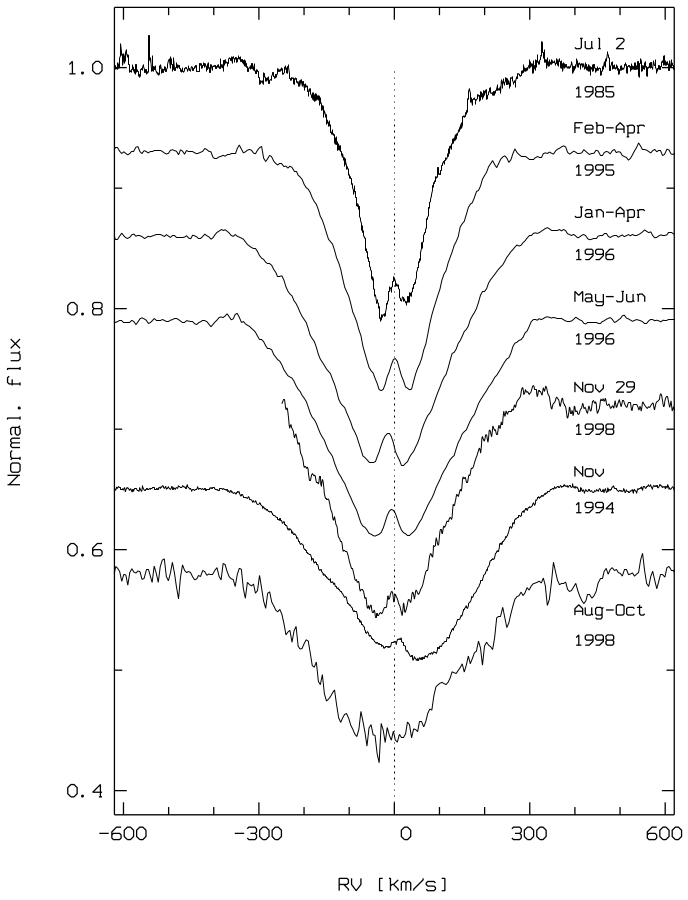}}}
\caption[Temporal variability of the CQE feature in the \ion{He}{I} $\lambda$6678 line of $\epsilon$ Capricorni (B3Ve).]
{Temporal variability of the CQE feature in the \ion{He}{I} $\lambda$6678 line of $\epsilon$ Capricorni (B3Ve). See text for discussion. Reproduced from~\citet{rivinius99}.}
\label{sec_intro_rivinius99_fig3}
\end{figure}

Be stars also often exhibit variation of the Violet-to-Red ($V/R$) peak ratio with time-scales ranging from months to over a decade~\citep[e.g.,][]{rivinius06}. This quantity is given by the ratio between the relative flux (to the continuum) in the blueward and redward emission-peaks maxima (usually in H$\alpha$). The shorter $V/R$ variations (time-scale of months) are understood as arising from binarity effects, while the longer ones (time-scale of years) are well explained by the precession of one-armed density waves in the circumstellar disk~\citep{okazaki91, okazaki97}.\par 

The edge-on Be star $\omicron$ Aquarii, discussed in Fig.~\ref{sec_intro_rivinius_carciofi_fig1}, has never presented longer-term $V/R$ variability~\citep[e.g.,][]{rivinius06}.~\citet{rivinius06} also verified central quasi-emission (CQE) in the \ion{Mg}{II} $\lambda$4481 line of this star. This spectral feature appears in certain line profiles as of \ion{He}{I}, \ion{Fe}{II}, and \ion{Mg}{II}. Despite being shown as a ``bump'' in emission, these features are understood as a pure geometrical effect that can arise (under certain conditions, depending on the disk opacity and size in the continuum) in a (quasi-)Keplerian-rotating disk seen close to the edge-on case~\citep{hanuschik95, rivinius99}. However, as pointed out by~\citet{rivinius06}, it is very unlikely, considering the same period of time, to observe CQEs and variability in the $V/R$ ratio. This happens since the occurrence of this feature requires very low radial velocities in the disk.\par 

Fig.~\ref{sec_intro_rivinius99_fig3} shows the variability of CQE in the \ion{He}{I} $\lambda$ 6678 line of the Be star $\epsilon$ Capricorni (B3Ve). The spectra are ordered to show the temporal variation of strengthen of the CQE feature: strongest CQE in the top (1985) and the absence of CQE in the bottom (1998). As suggested by~\citet{rivinius99}, such a temporal variation of the CQE feature must be related to changes on the disk structure: CQEs in the spectra of Be stars are more likely to happen during the accretion phase of the circumstellar disk.\par

Finally, another quantity of interest to address the temporal changes of Be disks is the equivalent width ($EW$) of H$\alpha$ and also other Balmer lines~\citep[e.g.,][]{dachs86, mennickent91, jones11}.\par 

The line equivalent width, within the interval of wavelength $\lambda_{0}$ and $\lambda_{f}$, is calculated as follows:

\begin{equation}
EW = \int_{\lambda_{0}}^{\lambda_{f}} \left(1 - \frac{F_{\lambda}}{F_{0}}\right) d\lambda,
\end{equation}
where $F_{0}$ is the continuum flux around the spectral line and $F_{\lambda}$ is the line flux.\par 

For emission-lines, $EW$, as defined above, is negative, but in this case sometimes the absolute value (positive) of $EW$ is considered, as in Fig.~\ref{sec_intro_pollmann18} where we show the measured H$\alpha$ equivalent width of the binary Be star 28 Tau (Pleione, B8Vpe) covering about 65 years of observations, showing a high long-term variability in H$\alpha$.\par 

\begin{figure}[t]
\centerfloat
\centerline{\resizebox{1.00\textwidth}{!}{\includegraphics{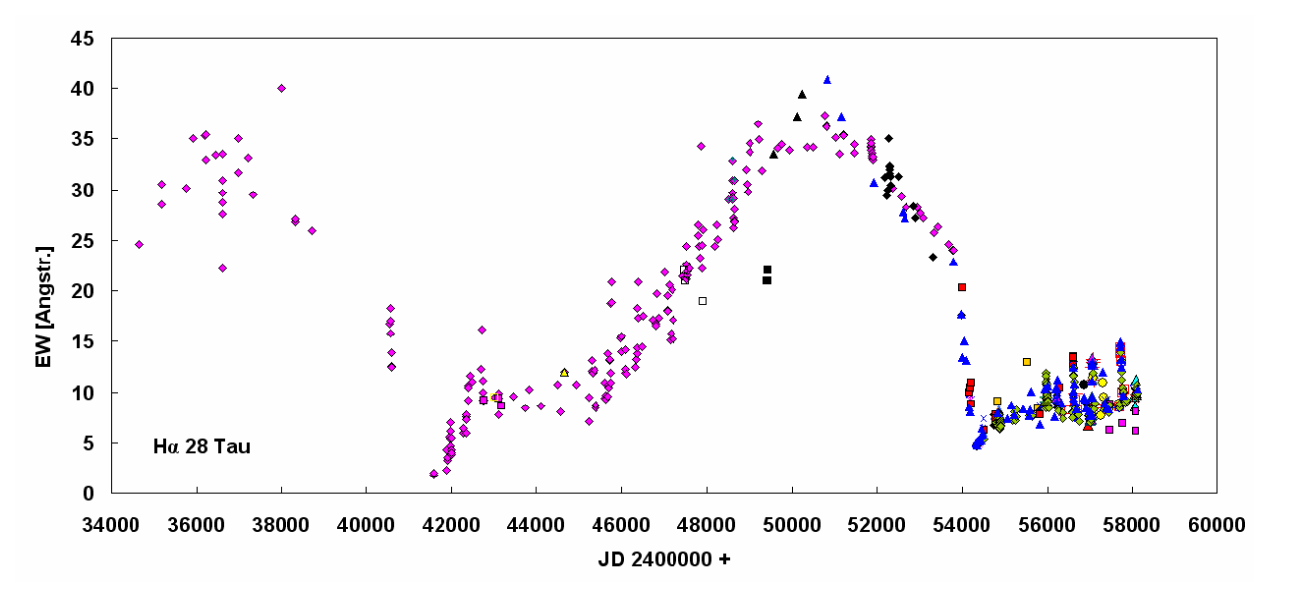}}}
\caption[Long-term variability of the H$\alpha$ equivalent width of 28 Tau (B8Vpe).]
{Long-term variability of the H$\alpha$ equivalent width of 28 Tau (B8Vpe). Observations are taken from about 1953 up to 2018. See text for discussion. Reproduced from~\citet{pollmann18}.}
\label{sec_intro_pollmann18}
\end{figure}

The understanding of long-term variability in H$\alpha$ of 28 Tau is still an open question in literature. As initially proposed by~\citet{hirata07}, the change on the observed H$\alpha$ profile of 28 Tau would arise from a geometric effect due to the disk precession around the central star because of the presence of a secondary component. The observations of~\citet{pollmann18} support this interpretation since these authors found a high correlation between the variation of the central absorption component of H$\alpha$ and the minima of the $V/R$ ratio.\par 

However, based on the analysis of the full width at half maximum (FWHM) in the H$\alpha$ line,~\citet{nemravova10} argued that such a huge variation in H$\alpha$ arises from structural changes on the disk, i.e., passing from a dissipation phase to a growing phase between about 1993 and 2012. Note, from Fig.~\ref{sec_intro_pollmann18}, that the H$\alpha$ equivalent width decreases between 1993 (MJD = 49000) and 2006 (MJD = 54000), and then increases again between 2006 and 2012 (MJD = 56000). Nevertheless, it is conspicuous how the change on the disk density, by a binary companion or by the disk evolution (growth and dissipation phases) affects this observable in the H$\alpha$ line.\par

\subsection{Oe and Ae stars: counterparts to the Be phenomenon?}
\label{sec_intro_Oe_Ae_stars}

As pointed out, in Sect.~\ref{sec_intro_be_phenomenon}, other hot stars with spectral type beyond the B-type can present double-peaked Balmer emission lines. Despite being more rare than in B stars, this is also observed in fast rotators with adjacent type to B stars: late (around O8-9) O-type~\citep[e.g.,][]{negueruela04, goldenmarx16, li18} and early (around A0-1) A-type stars~\citep[e.g.,][]{monin03, bohlender16}. This could suggest that these stars are the high and low-mass analogs of the Be phenomenon, that is, they also have decretion circumstellar disks in rotation due to mass injection from the central star during the main sequence phase.\par

\citet{goldenmarx16} verified a much larger fraction of Oe/O stars in the Small Magellonic Cloud (SMC) of 0.26 $\pm$ 0.04, in comparison with the one found in the Galaxy of 0.03 $\pm$ 0.01. To date, they found the earliest Oe stars (type O6) known in the both galaxies. This is important to better understand the Be phenomenon itself as it concerns the role of stellar winds, which are weaker as the stellar metallicity is reduced, to allow the process of growth and dissipation of circumstellar disks even in such more massive stars (O-type). Based on the results from these authors, this means that in the SMC even O6 stars, which have stronger winds than the later O stars, could form a circumstellar disk due to the less intense winds in comparison with early O stars but in the Galaxy.\par

Furthermore, Ae/An stars are known to present very weak emission in the H$\alpha$ line wings. More recent, the results from~\citet{bohlender16} also support this scenario of counterpart to the Be phenomenon, but as the lower-mass analogs of the Be phenomenon. About 15\% of the stars of their large sample of A-type stars ($\sim$400 stars) show shell or emission features in the H$\alpha$ line. They verified that the incidence of such features decreases toward the earlier A-type stars and also according to lower values of $v \sin i$.\par

Despite these evidences discussed above, this scenario for Oe and Ae/An stars, as being the more and less massive extensions of classical Be stars, is still an open question in literature. For instance, the spectro-polarimetric study of peculiar O stars by~\citet{vink09} detected depolarization ``line effects'' in H$\alpha$ for just one out of five Oe stars of their sample. However, such incidence in classical Be stars is quite larger up to $\sim$55\% as an evidence of their asymmetric circumstellar envelopes~\citep{vink02}. As pointed out by~\citet{rivinius13}, some Ae/An stars should be composed of pre-main sequence objects, as Herbig stars. In this case, their disks are not created by material that is ejected from the stellar surface, but star-forming disks showing infrared excess due to dust~\citep[e.g.,][]{jamialahmadi15}.\par

\section{Circumstellar disks of Be stars}
\label{sec_intro_Be_disks}

\subsection{Formation and dynamics}\label{sec_intro_disk_dynamics}

Different dynamical model tried to explain the disk formation of Be stars, as shown schematically in Fig.~\ref{sec_intro_porter03_fig8}.\par

\begin{figure}[t]
\centerfloat
\centerline{\resizebox{0.35\textwidth}{!}{\includegraphics[angle=270]{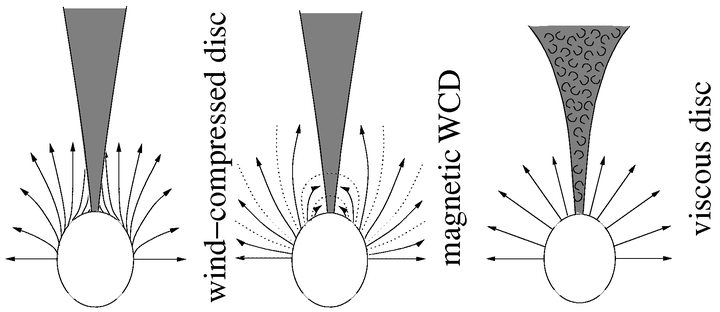}}}
\caption[Schemes of different dynamical models to form disks in Be stars.]
{Schemes of different dynamical models to form disks in Be stars. The stellar wind streamlines are represented by rows. See text for discussion. Reproduced from~\citet{porter03}}
\label{sec_intro_porter03_fig8}
\end{figure}

First, based on the theory of radiative line-driven winds introduced by~\citet{lucy70} and~\citet{castor75} for massive hot stars,~\citet{bjorkman93} presented the first quantitative physical model in order to explain the circumstellar disk formation in Be stars, the so-called wind-compressed disk (WCD) model.\par

To explain the formation of a circumstellar disk, the WCD model just needs a fast rotating star and a radiative line-driven wind. Hence, the wind streamlines are distorted due to fast rotation, forcing the outflow to be confined around the stellar equatorial plane and thus forming a disk. In this case, the mass-loss rate and the terminal velocity are scaled to the stellar rotational velocity $\Omega$ and the co-latitudinal angle $\theta$ as follows:

\begin{equation}
\dot{M}(\Omega, \theta) \propto (1 - \Omega^2 \sin^2 \theta)^{1 - 1/\alpha^{'}},
\end{equation}

\begin{equation}
v_{\infty}(\Omega, \theta) \propto (1 - \Omega^2 \sin^2 \theta)^{1/2},
\end{equation}
where $\alpha'$ is defined from the force multiplier parameter from the CAK-theory as presented in Sect.~\ref{sec_intro_theory_radiative_line_winds}. Thus, this results that $\dot{M}$ is increased, while $v_{\infty}$ is decreased, as $\theta$ tends to $\pi/2$ (stellar equator).\par

Despite being a benchmark work to predict quantitatively the disk dynamics in Be stars, the WCD model is invalid. As discussed in Section \ref{sec_intro_fast_rotation}, the gravity darkening effect highly impacts the stellar properties in the case of fast rotators, as the case of Be stars.~\citet{owocki96} and~\citet{owocki98} showed that the inclusion of non-radial forces (arising from the finite size of star) and gravity darkening lead to a significant reduction of the equatorial mass-loss, and then preventing the formation of a high density disk as predicted by the WCD model.\par

Therefore, just from a theoretical point-of-view, the WCD is ruled out when taking into account more realistic physical assumptions in the hydrodynamical simulations of Be disks. Nevertheless,~\citet{cassinelli02} modified the original WCD model, including dipolar stellar magnetic fields, being then called magnetic WCD model (Fig.~\ref{sec_intro_porter03_fig8}). In this case, in addition to the ram pressure due to the stellar winds, the magnetic field lines also contribute to confine the circumstellar material around the equator. In order to affect the wind structure in Be stars, this model only need a relatively weak magnetic field of the order of $\sim$$10^2$ G.\par 

As discussed in Sect.~\ref{sec_intro_radiative_winds_hr_diagram}, magnetic fields have already been detected in OBA-type stars, but their detection in Be stars proved to be quite elusive. Some studies have reported measurements of weak fields ($\lesssim$ 150 G) in Be stars~\citep[e.g.,][]{yudin11}, while others point out that these measurements are spurious~\citep[][]{bagnulo12}. To date no firm evidence of magnetic fields was found for Be stars. Such an absence of magnetic field in Be stars poses a question if fast rotation would prevent the stability of a dipolar magnetic field in these stars~\citep[e.g.,][]{wade16b, wade18}. In short, this lack of observational evidence of dipolar magnetic fields in Be stars is an important problem to support the magnetic WCD model.\par

Taken at face value, the WCD model implies that the circumstellar disk has an azimuthal velocity field $v_{\phi}(r)$:

\begin{equation}
v_{\phi}(r) = v_{\mathrm{rot}}(r/R_{\star})^{-1},
\end{equation}
where $v_{\mathrm{rot}}$ is the rotational velocity at the base of the disk. This arises from the assumption that the angular momentum is radially conserved through the disk.\par

On the other hand, if we consider that the disk dynamics is driven by gas turbulent viscosity, from the balancing of the gravitational and centrifugal force on an element of the disk, this would imply that the disk rotates in a Keplerian fashion as follows:

\begin{equation}
v_{\phi}(r) = v_{\mathrm{rot}}(r/R_{\star})^{-1/2}.
\label{eq:keplerian}
\end{equation}

To date, all the different types of observables strongly support this last scenario of a Keplerian rotating disk in Be stars.\par 

The first clear direct determination of a Keplerian disk around Be stars was found by~\citet{meilland07a} for $\alpha$ Arae (B3Ve), from the modeling of VLTI/AMBER spectro-interferometric data in the K-band using the radiative transfer code SIMECA~\citep{stee94,stee08}. In addition to several spectroscopic and interferometric works~\citep[see, e.g.,][and references therein]{stee12, rivinius13}, more recent spectro-interferometric analyses, as the VLTI/AMBER surveys of Be stars from~\citet{meilland12} and~\citet{cochetti19}, provided strong evidence for a Keplerian disk rotation in almost all their Be star samples.\par

This fact directly supports the last model presented in Fig.~\ref{sec_intro_porter03_fig8}, the so-called viscous decretion disk (VDD) model~\citep{lee91, okazaki01, bjorkman05}. In the VDD model, given that mass is removed from the stellar surface to the Keplerian orbit at the base of the disk, the material will then be transported outward due to viscous diffusion. This same model is also able to explain the hydrodynamics of the disk in young stellar objects~\citep[e.g.,][]{pringle81, balbus03}, with the only difference that the flow is inward in that case (accretion disk instead of decretion disk in Be stars). Said differently, the disk dynamics of Be stars is viscously-driven and then described by the Navier-Stokes equations~\citep[see, Eq.~3.1 to 3.4 of ][]{carciofi11}.\par 
Clearly, one quantity of interest is the disk volume mass density $\rho(r, z, t)$, which is expressed as a function of position (in cylindrical coordinates) and time. From that, the disk surface mass density $\Sigma(r,t)$ is then defined:

\begin{equation}
\Sigma(r,t) \equiv \int_{-\infty}^{+\infty} \rho(r,z,t) dz.
\label{eq:sigma}
\end{equation}

From the Navier-Stokes equations, we can write a diffusion equation for the disk surface density~\citep[e.g.,][]{pringle81}:

\begin{equation}
\diffp{\Sigma(r,t)}{t} = \frac{1}{r} \frac{\partial}{\partial r} \biggl[\frac{ \frac{\partial}{\partial r} (r^2 \Sigma(r,t) \alpha {c_{s}}^2)  }{ \frac{\partial}{\partial r} (r^2 \Omega)  }\biggr],
\label{eq:diffusion_equation}
\end{equation}
where $c_{s}$ is the disk isothermal sound speed that is given by:

\begin{equation}
c_s = \sqrt{\frac{ k_{B}T  }{\mu m_{H}}},
\label{eq:sound_speek_disk}
\end{equation}
where $k_{B}$ is the Boltzmann constant, $\mu$ is the mean molecular weight of the gas, $m_{H}$ is the hydrogen mass, and $T$ is isothermal electron temperature. $\Omega$ is the angular frequency of disk rotation. Lastly, $\alpha$ is the so-called Shakura-Sunyaev's viscous parameter. The kinematic viscosity $\nu$ of Be disks is typically parameterized in terms of $\alpha$~\citep{shakura73}:

\begin{equation}
\nu = \alpha c_{s} H,
\label{eq:viscosity}
\end{equation}
where H is the isothermal disk scale height:

\begin{equation}
H(r) = H_{0} \left(\frac{r}{R_{\mathrm{eq}}}\right)^{3/2},
\label{eq:disk_height_scale}
\end{equation}
and $H_{0}$ is the scale height at the disk base,

\begin{equation}
H_{0} =  c_{s} R_\mathrm{eq} \, \left( \frac{G M_{\star}}{ R_\mathrm{eq}}      \right)^{-1/2}.
\label{eq:disk_height_scale_base}
\end{equation}

In Eq.~\ref{eq:viscosity}, $\alpha$ is defined between 0 and 1. This parameter can be understood as a scale to the evolution of the disk dynamics: larger values of $\alpha$ mean that the disk growth or dissipate in a shorter time-scale~\citep[see, e.g.,][]{haubois12}, and then the variation of $\alpha$ will affect the observables predicted from the VDD model~\citep[e.g., see Fig.~5 of][]{carciofi11}. Therefore, $\alpha$ is related to the so-called viscous diffusion timescale $t_{\mathrm{diff}}$ given by:

\begin{equation}
t_{\mathrm{diff}} = \frac{r^2}{\nu} = \frac{r^2}{\alpha c_{s} H},
\label{eq:diffusion_timescale}
\end{equation}
expressing the typical time-scale for changes on the density structure due to viscosity. This quantity must be taken with caution as it does not represent the time-scale for the disk growth. As pointed out by~\citet{haubois12}, considering their VDD model simulations with $\alpha$ = 1.0, it is needed about 30 years to reach 95\% of the limit density at 10 $R_{\star}$, while $t_{\mathrm{diff}}$ = 1.7 years at this same distance in the disk. Then the value of $t_{\mathrm{diff}}$ (at a certain distance) is quite smaller than the one to the disk to reach (almost) the steady-state regime.\par

In the particular case of steady-state regime\footnote{Assuming $\diffp{\Sigma(r,t)}{t}$ = 0 and no-null mass injection from the central star to the disk.}, we are able to solve analytically Eq.~\ref{eq:diffusion_equation} for an isothermal disk as follows~\citep{okazaki07}:

\begin{equation}
\Sigma(r) = \frac{\dot{M} v_{\mathrm{orb}} {R_{\star}}^{1/2}}{3 \pi \alpha {c_{s}}^2 r^{3/2} } \left( \sqrt{\frac{R_{0}}{r}} - 1  \right),
\label{eq:sigma_analytical_steady}
\end{equation}
where $\dot{M}$ is the rate of mass injection to the disk and is assumed constant in this derivation for $\Sigma(r)$. $R_{0}$ is an integration constant that is related to the extension of the disk.\par

Due to the complexity of the radial density profile of Be disks, for example, being affected by presence of a binary companion~\citep[][]{panoglou16}, one common approach in the literature is to parameterize $\rho(r,z)$ using a power-law approximation, based on a vertically isothermal disk:

\begin{equation}
\rho(r,z) = \rho_{0}{\left(\frac{R_{\mathrm{eq}}}{r}\right)}^{m} \exp\left( \frac{-z^{2}}{2{H(r)}^{2}}  \right),
\label{eq:rho}
\end{equation}
where $\rho_{0}$ is the volume density at the disk base (at $r = R_{\mathrm{eq}}$ and $z$ = 0). Hence, apart from the stellar parameters\footnote{The temperature $T$, which is used to calculate $H_{0}$ (Eq.~\ref{eq:disk_height_scale_base}) is usually adopted as a fraction of the effective temperature~\citep[e.g.,][]{carciofi06a}}, the density profile is fully described by just two parameters: $\rho_{0}$ and $m$ (the density law exponent).\par

From Eq.~\ref{eq:sigma}, the volume and surface density profiles are related to each other by:

\begin{equation}
\rho(r,z) = \frac{\Sigma(r)}{H(r) \sqrt{2\pi}} \exp\left( \frac{-z^{2}}{2{H(r)}^{2}}  \right).
\label{eq:rho_sigma}
\end{equation}

Conversely, the surface density profile $\Sigma(r)$ can also parameterized by a power-law approximation:

\begin{equation}
\Sigma(r) = \Sigma_{0} \left(\frac{R_{\mathrm{eq}}}{r}\right)^{n},
\label{eq:radial_surface_density}
\end{equation}
where $\Sigma_{0}$ is the surface density at the disk base. Since the scale height $H(r) \propto r^{3/2}$, from Eq.~\ref{eq:rho} the volume ($m$) and surface ($n$) density law exponents are related to each other by\footnote{We point out that some authors define these variables in the opposite way: $m$ as the surface density law exponent and $n$ the volume density law exponent~\citep[e.g.,][]{haubois12}}: 

\begin{equation}
n = m - 1.5.
\label{sec_intro_relation_n_m}
\end{equation}

Finally, the volume and surface densities at the disk base are related as follows:

\begin{equation}
\rho_{0} = \Sigma_{0} \sqrt{\frac{GM_{\star}}{ 2\pi {c_s}^{2} {R_{\mathrm{eq}}}^{3}     }   }.
\label{sec_intro_relation_sigma0_rho0}
\end{equation}

From this simple parameterization for the disk density, models with radial density exponent $m$ = 3.5 mimic the analytical solution for the steady-state regime, when considering a vertical isothermal disk (Eq.~\ref{eq:sigma_analytical_steady}). We emphasize that it is not reliable to characterize the whole disk extension by a single value of radial density law exponent~\citep[e.g.,][]{haubois12}. Nevertheless, circumstellar disks passing by distinguishable phases of growth or dissipation must present quite different values of $m$. A growing disk will have a value of $m$ in the interval of $\sim$3.5-5.0, while $m$ $\sim$1.5-3.5 indicates a dissipation phase~\citep[e.g.,][]{vieira17}.\par

In short, once the material is ejected from the central star to the disk, the VDD model is able to explain the outflow dynamics in the disk in terms of turbulent viscosity. One can note that a clear weakness of this model is the inability to explain the mechanism of mass loss in these stars. As commented above, the VDD formulation requires the assumption of the rate of mass injection from the star to the disk. As already commented in Section \ref{sec_intro_fast_rotation}, the own high rotational rate of Be stars is recognized as a source of force to pull out material from the star, possibly more important for the later Be stars.\par

Lastly, we point out here some possible departures from the VDD model. First, some interferometric studies found deviations from a disk Keplerian rotation (c.f., Eq.~\ref{eq:keplerian}) in Be stars. From modeling spectro-interferometric VLTI/AMBER data (K-band) and the Pa$\beta$ line profile using the code SIMECA,~\citet{meilland07b} determined disk velocity law of $\beta$ = -0.32 $\pm$ 0.1 for $\kappa$ Canis Majoris (B2IVe). On the other hand, analysing AMBER data of better quality for this star, the subsequent analysis from~\citet{meilland12} provided a better agreement with the Keplerian picture from the VDD model, showing $\beta$ = -0.5 $\pm$ 0.2.~\citet{delaa11} also found a value of $\beta$ around -0.3 ($\beta$ = -0.35 $\pm$ 0.05) for the Be star $\psi$ Persei (B3Ve) from the analysis of spectro-interferometric data in the H$\alpha$ line from the CHARA/VEGA instrument. From the analysis of AMBER data,~\citet{cochetti19} found at least one clear departure of $\beta$ = -0.5 for 228 Eri (B2Vne) with $\beta$ = -0.35 $\pm$ 0.05.\par

Furthermore, a series of works about Be stars that present fairly stable disks (in the time-scale of decades) found clear deviations from the canonical value of $m$ = 3.5 for the steady-state regime. For example,~\citet{klement15} found $m$ = 2.9 for the late-type Be star $\beta$ Canis Minoris (B8Ve) and~\citet{mota19} derived $m$ = $2.44^{+0.27}_{-0.16}$ for $\alpha$ Ara (B2Vne). Thus, the radial density exponent is consistently equal or than 3.0 for these Be stars with stable disks. From analysing the temporal variation of the disk density,~\citet{vieira17} identified a slightly extended range of $m$ = 3.0-3.5 for the steady-state regime. This indicates that the standard theory is either incorrect or incomplete, since it predicts that (vertically) isothermal steady-state disks (considering a single star) should have $m$= 3.5. One possibility to explain this fact could rely on both non-isothermal and binarity effects in the disk structure~\citep[see, e.g.,][]{carciofi08, panoglou16, cyr17}.\par


\subsection{Geometry and size}\label{sec_intro_disk_geometry_extension}

As discussed above, the disk density is fairly well described considering a simple structure given by Eq.~\ref{eq:rho_sigma} (power law + vertical Gaussian fall-off). For instance, using radiative transfer models, under the approximation of Eq.~\ref{eq:rho_sigma} in the VDD model,~\citet{touhami11} were able to well explain the spectral energy distribution for a large sample of Be stars (130 objects). As one may expect, the disk density distribution is related to the disk physical extension. As pointed out by~\citet{rivinius13}, this quantity is rather difficult to be derived, unless in the case of a truncated  disk (due to a binary companion), and it has no unambiguous determination for Be stars up to date.\par 

\begin{figure}[t]
\centerfloat
\centerline{\resizebox{0.65\textwidth}{!}{\includegraphics{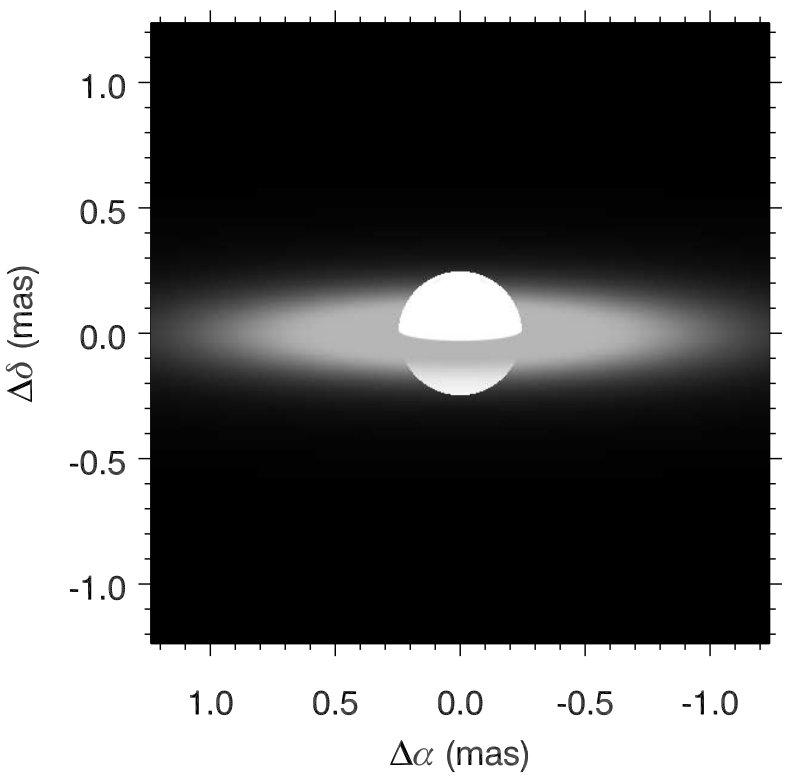}}}
\caption[Model intensity map (projected on the sky) in the K-band from~\citet{gies07} to the analysis of interferometric data of the Be star $\gamma$ Cassiopeiae (B0.5IV).]
{Model intensity map (projected on the sky) in the K-band from~\citet{gies07} to the analysis of interferometric data of the Be star $\gamma$ Cassiopeiae (B0.5IV). In this model, these authors assumed $i$ = \ang{80} for the central star (represented by a uniform disk distribution). The disk density law is assumed by $\rho$ = $7.2\e{-11}$ g cm\textsuperscript{-3} and $n$ = 2.7. See text for discussion. Adapted from~\citet{gies07}.}
\label{sec_intro_gies07_fig2}
\end{figure}

Nevertheless, based on interferometric observations, a series of papers have been presented results for the size of the emitting region of Be disks at certain specific spectral regions~\citep[e.g.,][]{quirrenbach97, gies07, meilland12, touhami13, dalla17, cochetti19}. One challenge imposed by Be stars disks comes from the inability to resolve them using single aperture telescopes. Currently, optical long-baseline interferometry is the only technique able to achieve sufficiently high spatial resolution in order to directly constrain the geometry of Be disks, which are typically extended up to a few milliarcseconds in the visible and infrared regions~\citep[see, e.g.,][]{chesneau12, stee12}. This observational technique will be explained in details in Chapter~\ref{chapter_interferometry}.\par

Since the size of the emitting region of the disk is a proxy to its physical extension, this first quantity will be called the disk size, or disk extension, in the remaining of this thesis. The disk size can be understood in terms of the so-called pseudo-photosphere~\citep[see, e.g.,][]{carciofi06a, carciofi11}. As suggested by the term employed to this region, the circumstellar disk will play like an extension to the stellar atmosphere, as it becomes optically thick, contributing to the line formation region (and to the continuum). The pseudo-photosphere is then defined as the circumstellar disk region that is optically opaque to the radiation in the continuum or in line, being wavelength dependent.\par

Fig.~\ref{sec_intro_gies07_fig2} shows the intensity image in the K-band, projected on the sky, from simple radiative transfer models~\citep[based on the approach of][]{hummel00} of~\citet{gies07} to interpret interferometric data. Here, these authors also employed the density law in the disk given by Eq.~\ref{eq:rho_sigma} and they assume a simple uniform disk intensity distribution for the central star. For this model, close to the edge-on case ($i$ = \ang{80}), one can clearly see how the disk intensity falls toward larger distances, following approximately a Gaussian distribution, as expected since the mass density law in the disk is assumed according to Eq.~\ref{eq:rho_sigma} in the calculations by these authors.\par 

\begin{figure}[t]
\centerfloat
\centerline{\resizebox{0.65\textwidth}{!}{\includegraphics{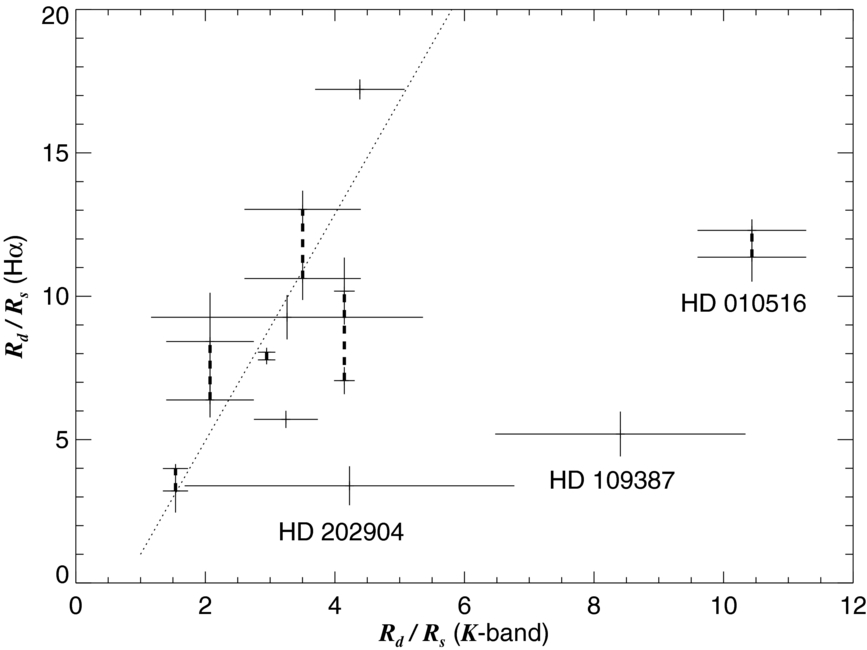}}}
\caption[Comparison of Be disks sizes derived in the H$\alpha$ line and in K-band from the CHARA Array interferometric survey of~\citet{touhami13}.]
{Comparison of Be disks sizes derived in the H$\alpha$ line and in K-band from the CHARA Array interferometric survey of~\citet{touhami13}. The linear fit to the data is shown in dotted line. Note that the disk extension is usually larger in the H$\alpha$ line than in the K-band region. Reproduced from~\citet{touhami13}.}
\label{sec_intro_touhami13_fig12}
\end{figure}

In comparison with radiative transfer calculations, geometric models are a much simpler approach to estimate the disk extension based on fitting interferometric observables. In this case, analytical functions are used to represent the intensity contribution from the central star and from the circumstellar disk. Typically, the disk intensity distribution is modelled as following a Gaussian distribution~\citep[see, e.g,][]{touhami13}. Thus, one useful quantity to estimate the disk extension is the FWHM of a Gaussian distribution, expressing the region of the disk where most part ($\sim$80\%) of emission is formed in a certain wavelength channel~\citep[see, e.g.,][]{delaa11, meilland12, cochetti19}. A comprehensive compilation of literature results for Be disks sizes, from the H$\alpha$ line up to the radio region at 2 cm, is found in Tables 1 and 2 of~\citet{rivinius13}.\par

\begin{figure}[t]
\centerfloat
\centerline{\resizebox{1.00\textwidth}{!}{\includegraphics{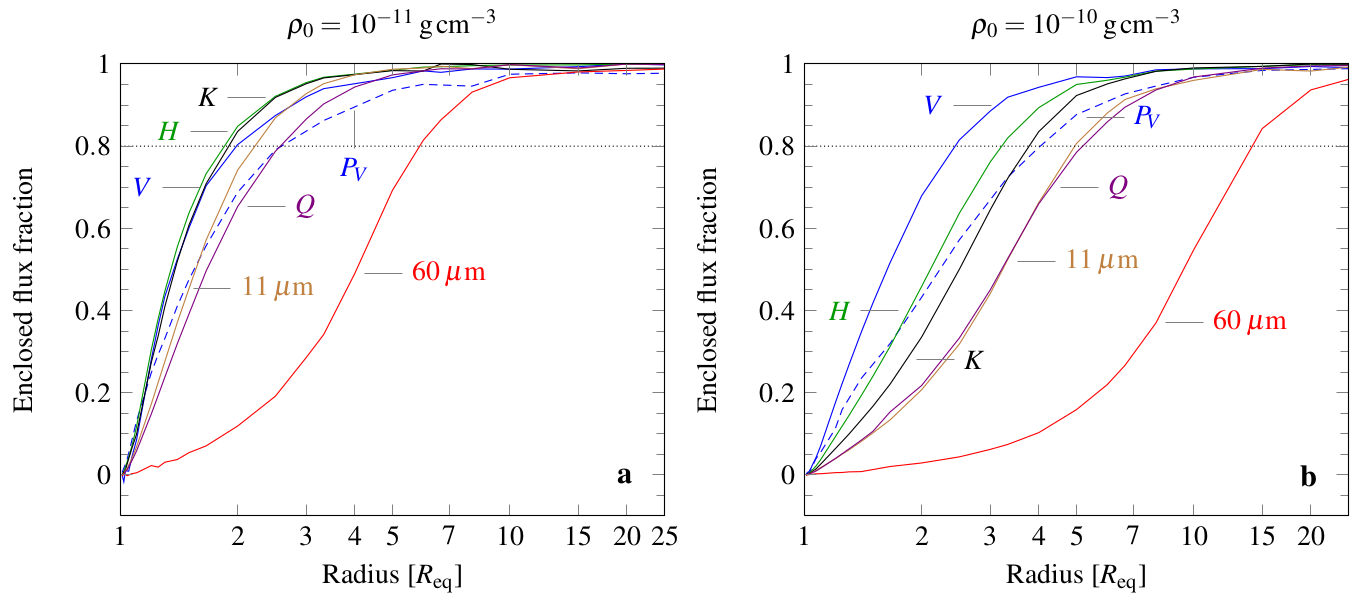}}}
\caption[Theoretical Be disk formation at different wavelength regions from calculations with code HDUST.]
{Theoretical Be disk formation at different wavelength regions from calculations with code HDUST. Results from model with have different base disk densities of $\rho_{0}$ = $10^{-11}$ g cm\textsuperscript{-3} (left panel) and $\rho_{0}$ = $10^{-11}$ g cm\textsuperscript{-3} (right panel). For the HDUST model with higher $\rho_{0}$, $\sim$ 50\% of the disk emitting region is formed within $\sim$ 10 $R_{\mathrm{eq}}$ in 60 $\mu$m. See text for discussion. Reproduced from~\citet{rivinius13}.}
\label{sec_intro_rivinius13_fig2}
\end{figure}

The understanding of Be disks extension at different wavelength regions is still an open issue in the literature. For example,~\citet{gies07} derived the angular sizes of four Be stars ($\gamma$ Cas, $\phi$ Per, $\zeta$ Tau, and $\kappa$ Dra) in the K-band using simple geometric models to fit interferometric data measured with the CHARA/CLASSIC instrument. They showed that the disks of these stars were significantly larger (by up to $\sim$1.5-2.0 times) in the H$\alpha$ line than in near-infrared region (K-band, enclosing the Br$\gamma$ line). These authors explained that in terms of a larger opacity in H$\alpha$ than in the K-band region, resulting in a more extended disk at lower densities toward larger distances. Such a trend of larger disk sizes in H$\alpha$ than in the K-band is also verified by the interferometric Be survey of of~\citet{touhami13} as shown in Fig.~\ref{sec_intro_touhami13_fig12}. Here,~\citet{touhami13} compare the disk size ($R_{d}/R_{s}$) in the H$\alpha$ line to their results found in the K-band. This quantity for the disk size is defined as the ratio between the angular size of the Gaussian ellipse (representing the disk) and the uniform disk (representing the star) in their best-fit geometric models. Apart from a few cases, one clearly sees that the most part of their Be star sample show a quite larger disk extension in H$\alpha$ than in K-band with the highest discrepancy up to $\sim$4 times.\par

On the other hand, some studies have found larger or similar Be disk extension in Br$\gamma$, when compared with H$\alpha$. Using the code SIMECA,~\citet{stee01} theoretically investigated the formation region of different lines in Be disks based on typical stellar and disk parameters for both early and late-type Be stars. Interestingly, their simulations support that the disk size can be up 2 times larger in the Br$\gamma$ line than in the H$\alpha$ line. Moreover,~\citet{carciofi11} using the radiative transfer code HDUST found that Be disks can have a quite similar V- and K-bands formation regions.\par

Fig.~\ref{sec_intro_rivinius13_fig2} shows the formation region of the disk emission at different spectral bands using HDUST. Here,~\citet{rivinius13} presented their results from HDUST models with different base disk densities of $\rho_{0}$ = $10^{-11}$ g cm\textsuperscript{-3} and $\rho_{0}$ = $10^{-10}$ g cm\textsuperscript{-3}. One sees that the formation regions of the V and K-bands are quite similar in the lower density model, while the denser model shows a larger emitting region in the K-band. Moreover, both cases show a larger disk emitting region toward the far-infrared region (60 $\mu$m) in comparison with the visible and near-infrared regions. Considering the higher density model, about the half of the disk emission is formed at 10 $R_{\mathrm{eq}}$ in 60 $\mu$m, while the same fraction of flux is reached at less than 2 $R_{\mathrm{eq}}$ in the V-band. Overall, these results can be interpreted in terms of the definition of the disk pseudo-photosphere. One may expect to have a larger pseudo-photosphere, at a certain wavelength, for the model with higher disk density. Indeed, we see that at all the considered wavelength regions, the disk formation is shifted toward larger distances due to a larger pseudo-photosphere of the disk in the model with higher density.\par

Lastly, more recently,~\citet{cochetti19} determined the disk size for a large sample of Be stars (26 objects) using a kinematic model to interpret their VLTI/AMBER data. This model is described in details in Sect.~\ref{sec_analytical_modeling_kinematic_models}. However, when comparing their results to the ones found by previous studies but in H$\alpha$, these authors could not verify a clear correlation between the disk size in the H$\alpha$ and Br$\gamma$ lines. As pointed out by these authors, it is important to constrain the disk extension from the analysis of quasi-contemporaneous observations in both spectral regions since the circumstellar environments of Be stars can be highly variable in a short period of time.\par


\section{Outline of this thesis}\label{intro_objetives}

As discussed above, despite the success of the CAK-theory to explain different observational features of winds of massive hot stars, further efforts are still needed to better understand their fundamental parameters. As pointed out in Sect.~\ref{sec_intro_radiative_winds_hr_diagram}, certain classes of O-type stars show large discrepancies between the predicted and measured (from spectroscopy) mass-loss rates (weak winds). In addition, it is still hard to conciliate the results for the mass-loss rates of massive hot stars obtained at different spectral regions (as UV, visible, and infrared).\par

Regarding Be stars, the VDD theory is also able to reproduce several types of observables for these stars, but the mechanisms underlying the mass loss process on Be stars are still not very well understood. Our discussion in Sect.~\ref{sec_intro_disk_dynamics} evidences the importance of a more detailed characterization on the properties of Be stars, as their rotational rate and disk physical parameters, to draw a better picture of the Be phenomenon.\par

In this thesis, I address these questions above, investigating both the physical properties of massive hot stars with radiative line-driven winds and rotating equatorial disks.\par 

In Chaps.~\ref{chapter_spectroscopy} and \ref{chapter_interferometry}, I discuss the two main observational techniques used in this thesis: spectroscopy and spectro-interferometry. Chap.~\ref{chapter_radiative_transfer_modeling} discusses some basic concepts of radiative transfer modeling, focusing on the two radiative transfer codes used in this thesis to interpret the spectroscopic and interferometric datasets: CMFGEN and HDUST. Lastly, less complex modeling tools to interpret interferometric data are also discussed in Chap.~\ref{chapter_analytical_models}.\par

The results found in this thesis are discussed in Chap.~\ref{chapter_results}. In Sect.~\ref{sec_results_ostars}, I discuss my UV and visible spectroscopic study of giants O-type stars, in the context of the so-called weak wind phenomenon, using CMFGEN to determine their stellar and wind parameters. In Sect.~\ref{sec_results_pcygni}, I discuss my collaboration using CMFGEN to interpret first H$\alpha$-band intensity interferometric observations of the LBV star P Cygni, and then to estimate its distance. Lastly, in Sect.~\ref{sec_results_omicron_aquarii}, I discuss my CHARA/VEGA and VLTI/VLTI spectro-interferometric study on the Be star $\omicron$ Aquarii. For this purpose, I used modeling tools of increasing complexity to constrain both the morphology, kinematics and physics of its disk and central star. My ongoing and near-future studies about winds and disks of massive hot stars are discussed in Chap.~\ref{chapter_perspectives}.\par 

In the end, Chap.~\ref{chapter_conclusions} gives an overview on the main findings of this thesis and some broader perspectives to study the environments of massive hot stars.\par

\cleardoublepage
\pagestyle{fancy}

\chapter{Stellar spectroscopy}
\label{chapter_spectroscopy}
\minitoc


\section{Stellar classification}
\label{sec_spectro_stellar_classification}

Since the discovery of spectral lines in the solar spectrum by Joseph von Fraunhofer in 1814\footnote{In fact, William Wollaston was the first, in 1802, to report absorption lines in the solar spectrum, but Fraunhofer performed a deeper analysis of the solar spectrum, identifying more than 500 different spectral absorption lines: afterward named as the Fraunhofer lines~\citep[e.g., see page 330 of][]{pannekoek61}.}, a series of schemes have been developed to classify the stellar spectra, mainly based on the observation of spectral lines in the visible region, to draw a better picture on the physics underlying the stellar atmospheres.\par 

One of the first systematic studies on stellar classification was performed by Angelo Secchi in the middle of the 19th century. He classified the spectra of about 4000 stars, and then introducing the so-called Secchi classes that grouped the stars as early-type (class I, showing strong hydrogen lines) and late-type (class II, showing lines of metals and weak hydrogen lines). It is interesting to note that the terms early-type (for hotter stars) and late-type (for cooler stars) were originally coined due to a wrong understanding about stellar evolution (based on the Kelvin-Helmholtz mechanism\footnote{At the beginning of the 20th century, it was thought that the Kelvin-Helmholtz mechanism explained the stellar radiation.~\citet{eddington26} was the first to correctly propose that nuclear fusion is the source of the stellar radiation.}), meaning that the cooler stars would be necessarily more evolved than the hotter ones. Secchi included other classes, as III (modernly, M-type stars) and IV (carbon stars). Also interestingly, Secchi was the first to observe the spectrum (in the H$\beta$ line) of a Be star, $\gamma$ Cassiopeiae, in 1866~\citep{secchi66}. More generally, Secchi grouped these stars, which display Balmer emission lines, as being of class V~\citep[e.g., see pages 450-452 of][]{pannekoek61}.\par

More fundamental systems were developed at Harvard College Observatory from 1885 until the beginning of the 20th century, namely, the Harvard classification system, that was led by the astronomers Williamina Fleming and Annie Jump Cannon. The letters employed in the Harvard system resulted in the current classification of O-B-A-F-G-K-M-types. Until her death in 1941, Cannon had classified up to about 395000 stellar spectra. For a more comprehensive overview on the history of stellar classification, we refer the reader to 
Chap.~40 of~\citet{pannekoek61} and Chap.~1 of~\citet{gray09}.\par

Today, the most used system is the so-called Yerkes or Morgan-Keenan-Kellman (MKK) system~\citep{morgan43}. It was updated with inclusion of standard stars, in order to define prototypes of the spectral classes, by the Morgan Keenan (MK) system~\citep{johnson53}. Basically, this is a two-dimensional scheme with the stars grouped by the canonical spectral types, O, B, A, F, G, K, and M, and numerical suffix from 0 to 9. The second dimension relies on luminosity class of the star: V (dwarfs), IV (sub-giants), III (giants), II (bright giants), I (supergiants\footnote{As for the giant classes, the luminosity class I is sub-divided in Ia (more luminous) and Ib (less luminous) supergiants. It is also extended to class 0 for the hypergiant stars~\citep{keenan71, keenan73}.}). The luminosity class was introduced in order to take into account stars with approximately the same effective temperature in the HR diagram, but with different values of luminosity. In addition, several spectral qualifiers have been used, such as ``e'' for emission line stars (including classical Be stars). We refer the reader to Table 3 of~\citet{sota11} for an extensive list of the stellar qualifiers.\par

Despite having other meanings, one can see that the nomenclature of letters and roman numerals used in the Morgan-Keenan system is inherited from previous works on stellar classification. The MK system expresses different physical properties of the stars. The spectral types from O to M correspond to a decreasing sequence of effective temperature, as firstly understood by the benchmark work of~\citet{payne25}. The Roman numerals correspond to a luminosity sequence, more luminous (and presumably more evolved) stars toward the class I. Fixing the luminosity class V (beginning of the H-burning phase), the dwarf stars also correspond to a sequence of higher stellar radius and mass toward the hotter objects (see Table \ref{table_parameters_OB_dwarfs} for the OB-type dwarfs).\par

The original MK system have been revised and extended by a series of works~\citep[e.g.,][]{morgan73, morgan78, kirkpatrick05}. With respect to O-type stars,~\citet{walborn71} introduced the O3 class to the MK system. Originally, the MK system had the types O5 and O9, as the earliest and the latest ones for O stars, respectively. Later, the spectral types O3.5 and O2 were defined by~\citet{walborn02}, mainly based on the relative ratio strength between \ion{N}{III} and \ion{N}{IV} emission-lines. To date, the O2-type class is the earliest one in the MK system, corresponding to main sequence stars with the highest values of effective temperature, $T_{\mathrm{eff}}$ $\sim$ 55000 K~\citep{walborn04, mokiem07a}.\par 

In addition to the canonical MK system, other spectral types have been taken into account, as the type W for classical Wolf-Rayet stars~\citep[e.g.,][]{smith68}, presented in Figs.~\ref{sec_intro_ekstrom13_table1_conti_scenario} and \ref{sec_intro_groh14_fig3_e_evol_massloss_mass}. Due to their strong winds, Wolf-Rayet stars have hydrogen depleted atmospheres, and thus very weak, or absent, hydrogen lines in their visible spectrum, which is mainly composed of broad and strong emission-lines due to helium and metals as CNO ~\citep[e.g., see Fig.~1 of][]{crowther07}. WR stars are classified as: WN (strong N lines), WC (strong C lines and absence of N lines), and more rarely, WO (both strong C and O lines)~\citep[e.g.,][]{vanderhucht81, figer97}.\par 

From our discussion in Sect.~\ref{sec_intro_evol_stages}, these subtypes of WR stars correspond to distinguish evolutionary stages of massive stars. From Fig.~\ref{sec_intro_groh14_fig3_e_evol_massloss_mass}, notice the predicted sequence WN $\longrightarrow$ WC $\longrightarrow$ WO by~\citet{groh14} for a non-rotating star model with initial mass of 60 $\mathrm{M_{\odot}}$. This comes from the on-going process of mass loss by winds that strip the material of the stellar atmosphere, and then revealing the chemical products created by different core-burning phases. The subtype WN/C is understood as an intermediate evolutionary phase between WN and WC stars~\citep{conti89}. Interestingly, due to their high effective temperature, low-mass stars at very late evolutionary stages, as CSPNs (see Sect.~\ref{sec_intro_theory_radiative_line_winds}), present visible spectra similar to the one of massive stars. In particular, hydrogen-deficient CSPNs essentially show WR-type spectra and then are classified as [WR] stars\footnote{In this case, brackets are used to indicate that they are low-mass stars, instead of classical WR stars~\citep{vanderhucht81}.}.\par

\begin{figure}[t]
\centerfloat
\centerline{\resizebox{0.60\textwidth}{!}{\includegraphics{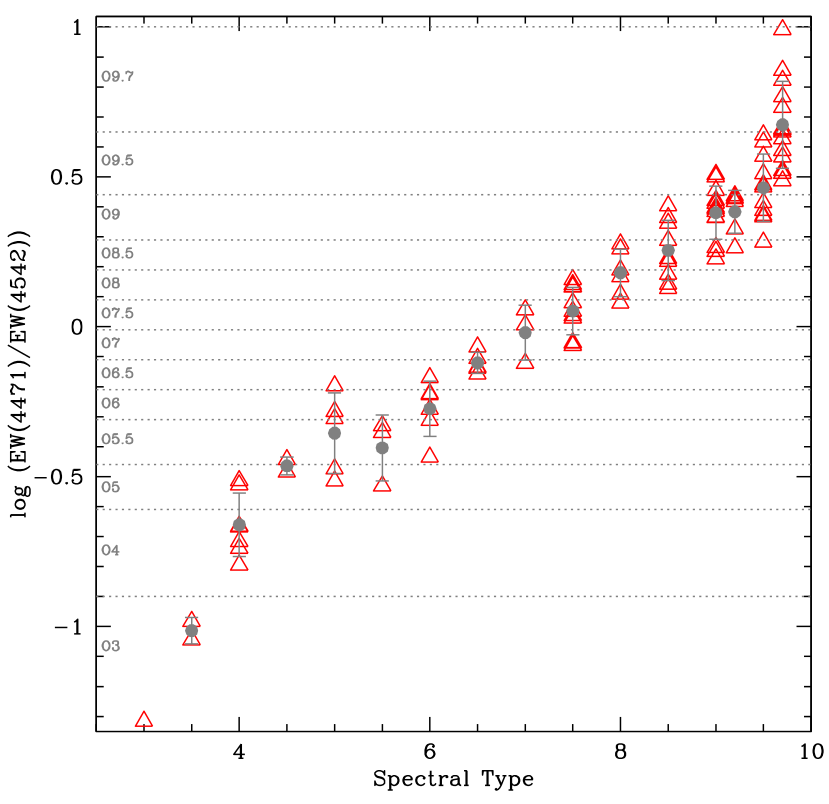}}}
\caption[Ratio between the equivalent width measured in \ion{He}{I} $\lambda$4471 and \ion{He}{II} $\lambda$4542 as a function of spectral type of O stars (105 objects).]
{Ratio between the equivalent width measured in \ion{He}{I} $\lambda$4471 and \ion{He}{II} $\lambda$4542 as a function of spectral type of O stars (105 objects). The ratio for each star is shown in red triangle and the mean values are shown in gray point within the error bars. The dashed gray lines indicate the limits of the ratio according to the spectral type. Overall, the ratio $EW$(\ion{He}{I} $\lambda$4471)/$EW$(\ion{He}{II} $\lambda$4542) increases toward O-type stars with lower effective temperature, toward the type O9.7. Reproduced from~\citet{martins18}.}
\label{sec_spectro_martins18_fig1}
\end{figure}

With respect to the other evolved stages of massive stars, RSGs show spectral type M or K (around F, G, A-types for yellow or warm supergiants) with luminosity class I or 0~\citep[e.g.,][]{neugent10, messineo18}. On the other hand, LBV stars cover a larger range of spectral types, from types O and B ($T_{\mathrm{eff}}$ $\sim$ 12000-30000 K), during the quiescent phase, to the types G and F ($T_{\mathrm{eff}}$ $\sim$ 7000K) during the eruptive phase~\citep[e.g.,][and reference therein]{humphreys94}. In short, it can be a hard task to distinguish them from other evolved stars that also occupy similar regions in the upper HR diagram, as yellow supergiants and also B[e] supergiants~\citep[e.g.,][]{humphreys03, humphreys17}.\par

The spectral type classification of O stars is mainly based on the relative strength line of both neutral and ionized helium. In particular, \ion{He}{I} $\lambda$4471 and \ion{He}{II} $\lambda$4542 are the most important ones as spectral type criteria, which is supplemented by other helium lines such as \ion{He}{I} $\lambda$4144, \ion{He}{II} $\lambda$4200, and \ion{He}{I} $\lambda$4388~\citep[e.g.,][]{plaskett31, abt68, conti71, mathys89, walborn90, sota11}.\par

Fig.~\ref{sec_spectro_martins18_fig1} presents the ratio between the equivalent width ($EW$) calculated in \ion{He}{I} $\lambda$4471 and \ion{He}{II} $\lambda$4542, as a function of spectral type, from O9.7 to O3. Here, there is no distinction among the luminosity classes, covering from the class V to Ia. The ratio $EW$(\ion{He}{I} $\lambda$4471)/$EW$(\ion{He}{II} $\lambda$4542) increases for O-type stars with lower effective temperatures. This can be understood since the \ion{He}{I} $\lambda$4471 intensity decreases, while the \ion{He}{II} $\lambda$4542 intensity increases, toward the hotter O stars, resulting from the ionization balance of \ion{He}{I} and \ion{He}{II}, as shown in Fig.~\ref{sec_spectro_gray09_fig3_1} (dwarf O stars). From Fig.~\ref{sec_spectro_gray09_fig3_1}, one sees that the \ion{He}{I} $\lambda$4144, \ion{He}{I} $\lambda$4388, and \ion{He}{I} $\lambda$4471 lines vanish for the objects earlier than O4-3.5 due to the high degree of ionization in the atmosphere ($T_{\mathrm{eff}}$ $\gtrsim$ 43500 K).\par

\begin{figure}[t]
\centerfloat
\centerline{\resizebox{0.65\textwidth}{!}{\includegraphics{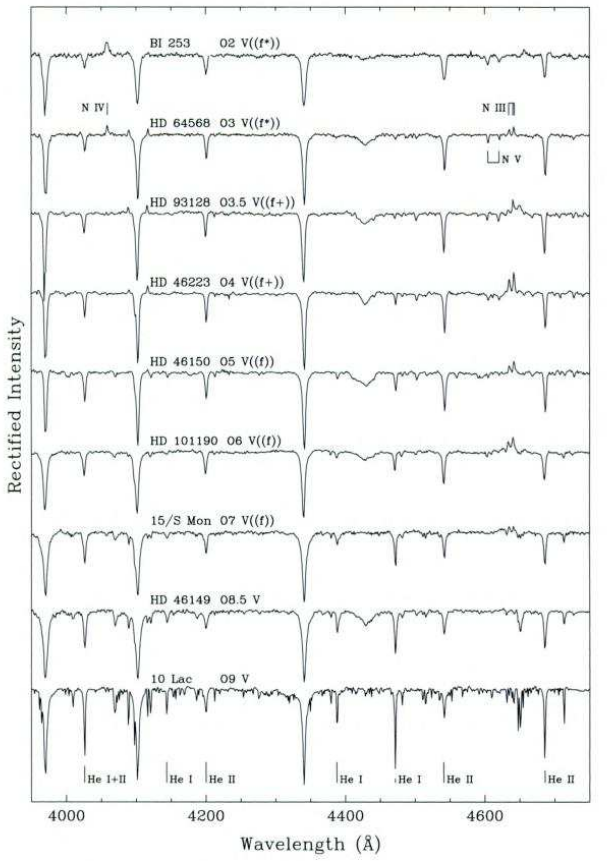}}}
\caption[Visible spectra ($\sim$4000-4700 {\AA}) of O dwarfs (class V) from O9V to O2V.]
{Visible spectra ($\sim$4000-4700 {\AA}) of O dwarfs (class V) from O9V to O2V. Note how the \ion{He}{II} $\lambda$4542 line strengthen in absorption toward the earlier objects, while \ion{He}{I} $\lambda$4471 line strengthen in absorption toward the later objects. See text for discussion. Reproduced from~\citet{gray09}.}
\label{sec_spectro_gray09_fig3_1}
\end{figure}

On the other hand, Fig.~\ref{sec_spectro_gray09_fig3_1} also shows that the \ion{He}{II} $\lambda$4686 intensity tends to decrease as the effective temperature increases. Indeed, \ion{He}{II} $\lambda$4686 is not very useful as a spectral type criteria for O stars but can be used as a luminosity class criterion~\citep[e.g,][]{martins18}. \ion{He}{II} $\lambda$4686 is quite sensitive to the wind parameters (as the mass-loss rate and clumping factor), and thus used as a diagnostic to determine, or investigate theoretically, the wind parameters of O stars~\citep[e.g.,][]{martins11, sundqvist11, tramper14}. This line goes from absorption, as shown in Fig.~\ref{sec_spectro_gray09_fig3_1}, to emission in the more luminous O stars as the early-type supergiants, which have stronger winds. For example, see Fig.~A.6 of~\citet{bouret15}, showing the visible spectrum of HD 16691 (O4If\footnote{The spectral qualifier ``f'' means that \ion{N}{III} $\lambda$4634-4640-4642 and \ion{He}{II} $\lambda$4686 are in emission.}).\par

\begin{figure}[t]
  \centering
  \begin{adjustbox}{minipage=\textwidth,scale=0.65}
  \includegraphics[width=1.00\textwidth]{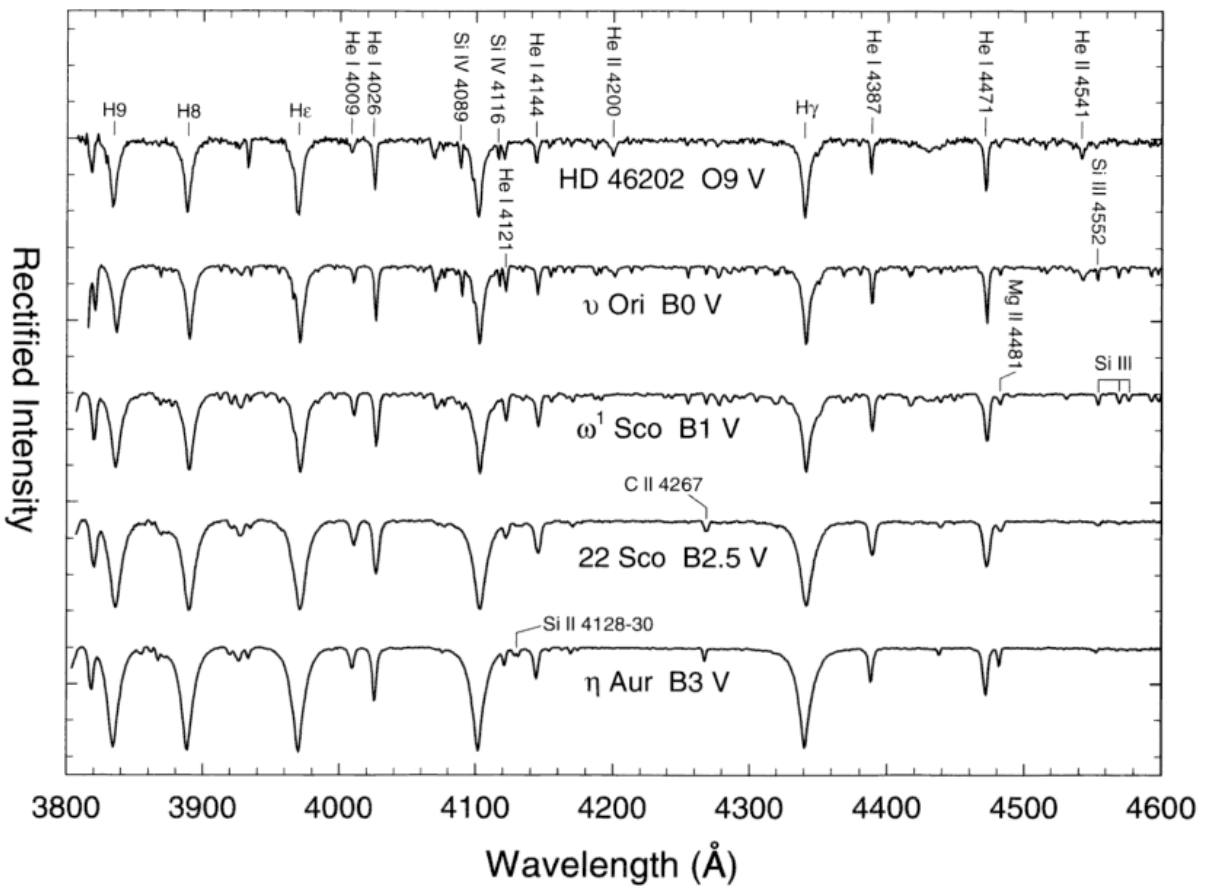}  
  \includegraphics[width=1.00\textwidth]{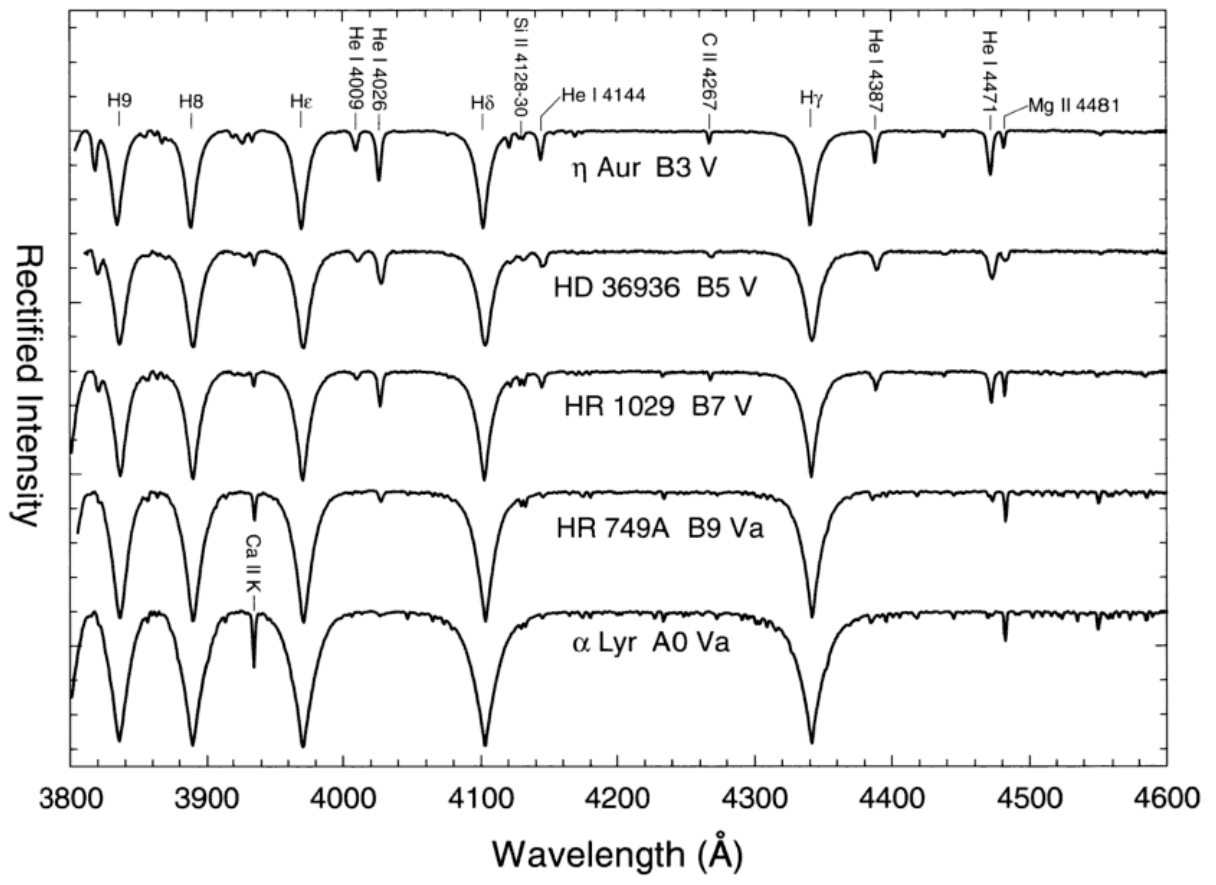}
  \end{adjustbox}
  \caption[Visible spectra ($\sim$ 3800-4600 {\AA}) from O9V to A0V stars, covering the entire range of B dwarf stars (B0V-B9V).]
  {Visible spectra ($\sim$3800-4600 {\AA}) from O9V to A0V stars, covering the entire range of B dwarf stars (B0V-B9V). Note that the \ion{He}{II} $\lambda$4542 line becomes weaker toward the later spectral type, vanishing in the spectrum of B1 stars. Also see how the Balmer lines become stronger from O stars toward the A0 stars. See text for discussion. Reproduced from~\citet{gray09}.}
\label{sec_spectro_gray09_fig4_1_2}
\end{figure}

Originally, in the MKK and MK systems, the distinction between O- and B-type stars was set by the absence of lines due to ionized helium in the spectrum of B0 stars. In this case, the B spectral class was defined to encompass stars showing \ion{He}{I} lines, but not \ion{He}{II} lines, in the visible region~\citep[e.g.,][]{morgan43}. Nevertheless, modern spectroscopic data revealed (weak) \ion{He}{II} lines in the B-class around the B0-type~\citep[e.g.,][]{walborn08}.\par 

The visible spectra of stars with spectral types from O9V to A0V are shown in Fig.~\ref{sec_spectro_gray09_fig4_1_2}. This covers practically the same wavelength region as in Fig.~\ref{sec_spectro_gray09_fig3_1}, but extending the spectral type toward objects later than O9V. One can clearly see \ion{He}{II} $\lambda$4542 in the spectrum of the shown B0V star ($\upsilon$ Orionis), but not for later-types. The intensity of neutral helium lines decreases from the type O to B, and the \ion{He}{I} lines vanish around the type A0. Due to lower values of effective temperature, spectral features in the visible spectrum of metallic ions, mainly, Si, C, N, and O, are used as spectral type criteria for B stars. In particular, the relative ratio strengths between silicon lines at different ionization stages is important for that since they are quite sensitive to the variation of effective temperature, such as \ion{Si}{IV} $\lambda$4089, \ion{Si}{II} $\lambda$4124-4131, and \ion{Si}{III} $\lambda$4552-4567-4574~\citep[e.g.,][]{kilian91a, kilian91b}.\par 

Besides the lines due to helium and metallic ions, the hydrogen Balmer lines also are rather important as a spectral type criteria. Considering the dwarf class, and a pure photospheric contribution, the Balmer lines strengthen from the O-type through the B-type, reaching a maximum around the type A0-A2, and then becoming weaker for later-types. In a qualitative way, the atmosphere of O stars are so hot that most part of hydrogen is either ionized or at high energy states. On the other hand, most part of hydrogen is in the ground energy state in M-type stars~\citep[$T_{\mathrm{eff}}$ $\sim$ 2000-4000 K,][]{allard95}. The maximum fraction of hydrogen atoms populating the second energy level is reached for $T_{\mathrm{eff}}$ $\sim$ 10000-9000 K, corresponding to A0-A2-type stars. For a quantitative discussion, in the framework of the LTE approximation, see, for example, Sect.~7.2 of~\citet{vitense89}.\par

For more details concerning the spectral classification of O stars in the visible, we refer the reader to a series of papers from the Galactic O-Star Catalog\footnote{Available at \url{https://gosc.cab.inta-csic.es/}}~\citep[GOSC][]{maiz13} and the Galactic O-star Spectroscopic Survey~\citep[GOSSS][]{maiz11}. To date, these projects provide detailed classification for about 590 O stars~\citep[see,][]{sota11, sota14, maiz16, maiz18b}. In addition to the visible region, the morphology of ultraviolet lines are also used as spectral classification of OB-type stars~\citep[see, e.g.,][]{walborn85, rountree91, walborn95}. We refer the interested reader to~\citet{heck87} for a detailed discussion on ultraviolet spectral classification both for hot and cool stars.\par

\section{Line formation in the wind: P-Cygni profiles}
\label{sec_pcygni_profiles}

The spectral lines formed in the stellar wind can be found in pure-absorption, pure-emission, or with components in blueward absorption and redward emission. As discussed in Sect.~\ref{sec_intro_theory_radiative_line_winds}, the latter is the so-called P Cygni line profile. These profiles are unequivocal spectral signatures of strong winds around hot massive stars. We point out that the same cannot be stated for pure-emission lines. They can be mainly formed in the stellar atmosphere, for example, as the \ion{N}{III} $\lambda \lambda$ 4634, 4640, 4641 lines that are found in emission in Of stars~\citep{mihalas73}.\par 

Similarly, inverse P Cygni profiles, that is, redward absorption and blueward emission components, are unequivocal spectral signatures of inward flows toward the central star. Both P Cygni and inverse P Cygni line profiles are found in the spectrum of pre-main sequence objects, T-Tauri and Herbig Ae/Be stars, due to a complex interplay between outward (winds) and inward (accretion disk) flows found in these objects~\citep[e.g.,][]{alencar00, deleuil04, cauley14}.\par

P Cygni profiles were first observed in the spectrum of the LBV star P Cygni, showing a large number of them (up to about 140) in the visible and near-infrared regions~\citep[e.g.,][]{struve35, struve39, beals53, burbidge55}. The understanding of the nature of these profiles comes from the 1930s, with~\citet{mccrea29} proposing that P Cygni profiles observed in novae arise from a expanding shell. This suggestion was followed by a series of studies by Carlyle Smith Beals~\citep{beals29, beals30, beals32, beals35}, explaining that such spectral feature, also observed in the spectrum of WR stars and $\eta$ Carinae (another LBV star), originates from ``continuous ejection of gaseous material from the star''~\citep{beals29}. Later,~\citet{beals53} published an extensive catalogue of early-type stars, covering spectral types from O5 to A4, which showed P Cygni-like spectra in the visible region. 

Followed by other observational studies~\citep[e.g.,][]{wilson58, underhill58}, these results supported that strong winds are indeed a common phenomenon among early-type stars, instead of an anomalous or very particular phenomenon of some stars as P Cygni. Soon after, this picture was confirmed by the very first UV spectroscopic observations from~\citet{morton67a, morton67b}, which revealed strong and broad P Cygni profile of \ion{Si}{IV}, \ion{C}{V}, and \ion{N}{V}, in the UV spectrum ($\sim$1200-1600 {\AA}) of OB supergiants, arising from high values of effective temperature (highly ionized wind) and also of mass-loss rate and terminal velocity. Subsequently, UV spectroscopic observations were technically improved by means of the Copernicus and IUE satellites that allowed more systematic analysis of early-type stars~\citep[e.g.,][]{underhill75, morton76, snow76, henrichs83, prinja86, howarth89}. Regarding the IUE studies about early-type stars, the interested reader can see the reviews of~\citet{cassinelli87},~\citet{willis87}, and~\citet{snow1987} about OB, WR, and Be stars, respectively. In short, all these spectroscopic studies discussed above set the conditions to develop and verify the radiative line-driven wind theory from~\citet{lucy70} and~\citet{castor75}.\par

In a qualitative way, the formation of P Cygni line profiles is understood in terms of the flux contribution from different regions in the stellar wind, apart from the photospheric contribution to the line formation in a certain rest-frame frequency $\nu_{0}$ (likely forming a pure-absorption line). For that, we show a geometric scheme (star plus wind) in Fig.~\ref{sec_spectro_bohm_vitense_vol2_fig16_6}. Here, we consider the simplest case of a spherically symmetric expanding shell around the star, following the wind velocity law as described in Eq.~\ref{eq:beta_law}.\par 

\begin{figure}[t]
\centerfloat
\centerline{\resizebox{0.40\textwidth}{!}{\includegraphics{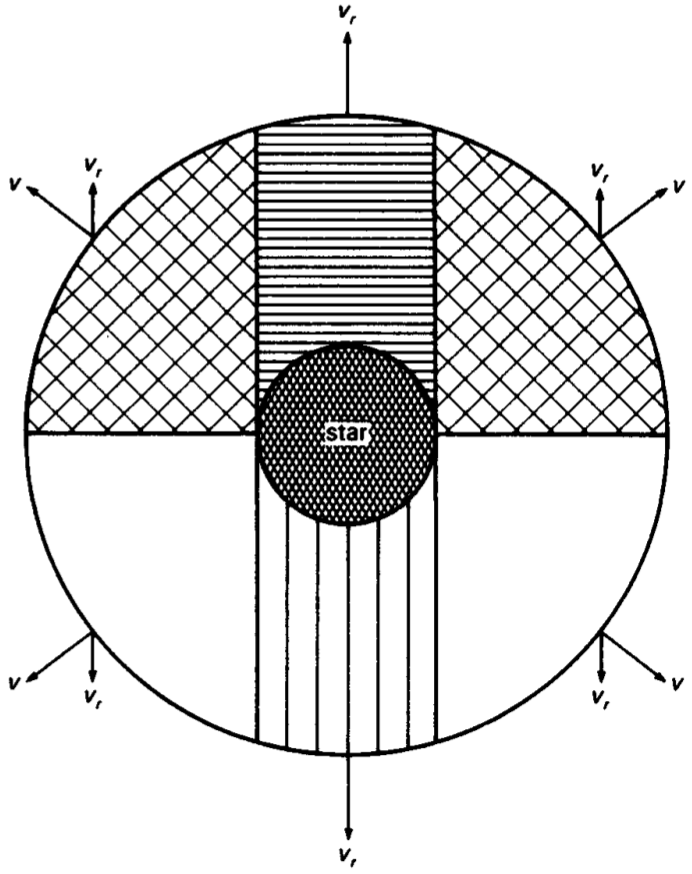}}}
\caption[Geometric scheme for the formation of P Cygni profiles.]
{Geometric scheme for the formation of P Cygni profiles. The star is represented by the central dark region that is surrounded by a expanding shell (wind). $V$ denotes the wind velocity and $V_{r}$ is the radial component, which is parallel to the line of sight of the observer. An asymmetric P Cygni profile is then formed by the superposition of the radiation arising from all these regions. Examples of observed P Cygni profile in the spectrum of O-type stars was previously shown in Fig.~\ref{sec_intro_uv_spectrum}, and Fig.~\ref{sec_spectro_pcygni_profiles_ip} shows examples of theoretical P Cygni profiles in  \ion{C}{IV} $\lambda \lambda$ 1548,1550. Reproduced from~\citet{vitense89}.}
\label{sec_spectro_bohm_vitense_vol2_fig16_6}
\end{figure}

From the Doppler effect, both the absorption and emission components of a certain P Cygni line profile (transition with rest-frame frequency $\nu_{0}$), arise from the interaction between the wind and photons emitted from the star with frequencies higher or equal to $\nu_{0}$. At the distance $r$ in the wind, a photon has frequency $\nu_{\mathrm{obs}}$:

\begin{equation}
\nu_{\mathrm{obs}} = \nu_{0} -  \nu_{0}\frac{v(r)}{c},
\end{equation}
as observed from the stellar surface (rest-frame). Hence, photons emitted from the star with $\nu'_{0}$:
\begin{equation}
\nu'_{0} = \nu_{0} + \nu_{0}\frac{v(r)}{c},
\end{equation}
will contribute at $r$ to the transition with frequency $\nu_{0}$, and thus $\nu'_{0}$ ranges from $\nu_{0}$ ($v(r = \mathrm{R_{\star}})$ $\sim$ 0) up to $\nu_{0}$ + $v_{\infty}/c$.\par

First, the absorption component of the P Cygni profile originates from the scattering of photons in the wind region in front of the star, toward the line of the sight of the observer (vertical dashed area in Fig.~\ref{sec_spectro_bohm_vitense_vol2_fig16_6}). The redward part of the absorption component is formed close to the stellar surface ($v(r)$ close to zero so $\nu'_{0}$ close to $\nu_{0}$), while the blueward one is formed in the outermost wind region (up to $v_{\infty}$). If the wind column density is high enough so the absorption component can be found saturated, that is, having null flux.\par 

The wind region represented by the white and cross-hatched areas in Fig.~\ref{sec_spectro_bohm_vitense_vol2_fig16_6} will thus contribute to the emission component due to the scattering of photons toward the observer. In this case, note that the emission arises from distances in the wind with velocities, with respect to the observer, ranging from -$v_{\infty}$ to $v_{\infty}$, due to the contribution from both wind hemispheres. This same effect does not happen regarding the formation of the absorption component because a portion of the wind region, in the line of sight of the observer, is occulted by the star (horizontal dashed line), and then this region of the wind does not contribute to the observed P Cygni profile.\par 

\begin{figure}[t]
\centerfloat
\centerline{\resizebox{0.90\textwidth}{!}{\includegraphics{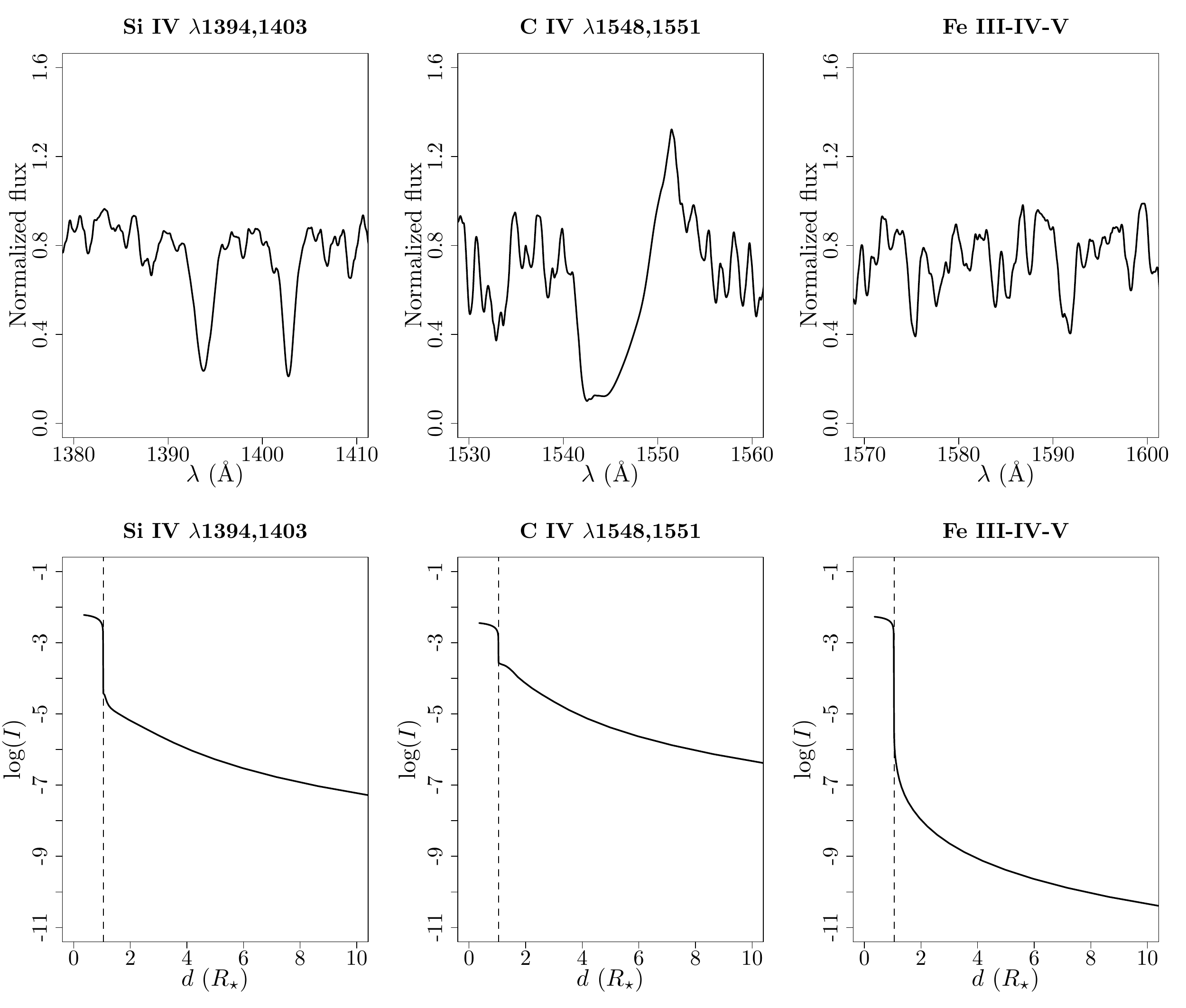}}}
\caption[Comparison of UV line formation through the wind extension.]
{Comparison of UV line formation through the wind extension. The top panels show synthetic line profiles in the UV, calculated by myself, using CMFGEN with parameters as follows: $\log (L_{\star}/L_{\odot})$ = 5.12, $T_{\mathrm{eff}}$ = 30600 K, $\log g$ = 3.50, $R_{\star}$ = 13 $\mathrm{R_{\odot}}$, $M_{\star}$ = 20 $\mathrm{M_{\odot}}$, $\dot{M}$ = $1.5\e{-9}$ $\mathrm{M_\odot}$ yr\textsuperscript{-1}, and $v_{\infty}$ = 1300 km s\textsuperscript{-1}. The bottom panels show the intensity, within the same spectral regions indicated in the top panels, as a function of radial distance from the center of the star. The dashed lines indicate the start of the wind region in the model. Intensity unit is in erg/($\mathrm{cm}^2$ s Hz sr). Note the large drop of intensity at the base of the wind for the spectral region encompassing mainly just Fe lines. See text for discussion.}
\label{sec_spectro_pcygni_profiles_ip}
\end{figure}

From adding the contributions of all these different parts of the wind to the emerging spectrum, an asymmetric P Cygni profile is formed, showing a blueshifted absorption component and a emission component with a peak close to the rest-frame frequency (slightly shifted to the red part of the spectrum), as shown for \ion{C}{IV} $\lambda \lambda$ 1548,1551 in Fig.~\ref{sec_spectro_pcygni_profiles_ip}.\par 

Since P Cygni profiles are formed by scattering, when radiative or collisional processes are dominant in the wind, its absorption component will be thus filled-in due to the photons created in the wind (``true emission''). Thus, instead of forming a P Cygni profile, these processes could result in the formation of a pure-emission line. The resonance transitions, such as, \ion{P}{V} $\lambda \lambda$ 1118,1128, \ion{N}{V} $\lambda$ 1240, \ion{Si}{IV} $\lambda \lambda$ 1394,1403, and \ion{C}{IV} $\lambda \lambda$ 1548,1550, for O and early B stars, are expected to show the strongest P Cygni profiles in the ultraviolet since these transitions have short lifetimes and then occurring more frequently. On the other hand, due to lower values of effective temperature, resonance lines produced by singly ionized metals, such C, Mg, Al, Si, and Fe are relevant in the formation of P Cygni in late B and early A supergiants.\par

The discussion above is only a qualitative picture of P Cygni profiles. For instance, since the ionization fractions are expected to change through the wind, depending on the stellar and wind parameters, a certain metal ion, which produces a certain line, may not occur up to the largest distance in the wind (where $v = v_{\infty}$). This would result in a less broad P Cygni profile. A deeper discussion about the formation of P Cygni profiles can be found in Sect.~2.2 of~\citet{lamers99}.\par

To illustrate, in a quantitative way, the formation of P Cygni profiles in the winds of hot stars, Fig.~\ref{sec_spectro_pcygni_profiles_ip} shows synthetic lines from a radiative transfer model calculated using code CMFGEN (Sect.~\ref{sec_radiative_transfer_modeling_cmfgen}): around \ion{Si}{IV} $\lambda \lambda$ 1394,1403, \ion{C}{IV} $\lambda \lambda$ 1548,1551, and also a narrow spectral region that is mainly composed of lines of iron group elements, specially \ion{Fe}{IV} (considering the parameters of this model).\par

For each one of the presented wavelength regions, we also show the intensity as a function of distance from the center of star. One clearly sees that \ion{C}{IV} $\lambda \lambda$ 1548,1551 forms a fully developed P Cygni profile, while \ion{Si}{IV} $\lambda \lambda$ 1394,1403 is in absorption due to the low value of mass-loss rate adopted in this model ($\dot{M}$ = $1.5\e{-9}$ $\mathrm{M_\odot}$ yr\textsuperscript{-1}), considering the set of stellar parameters here (with $T_{\mathrm{eff}}$ $\sim$ 30000 K). Nevertheless, even in this case, \ion{Si}{IV} $\lambda \lambda$ 1394,1403 shows a significant line intensity through the wind extension, unlike the intensity of the iron lines that suddenly drops at the base of the wind since these lines are mainly formed in the stellar atmosphere. On the other hand, in advance of discussion, one sees, from Fig.~\ref{sec_spectro_efeito_mdot_uv_halpha_infrared}, that fully developed P Cygni profiles are also formed in \ion{Si}{IV} $\lambda \lambda$ 1394,1403, but when considering a much larger value of mass-loss rate in the model ($\dot{M}$ = $2.5\e{-7}$ $\mathrm{M_\odot}$ yr\textsuperscript{-1}).\par 

\section{Multi-wavelength line diagnostics}
\label{sec_spectro_line_diagnostics}

Due to their high effective temperature ($T_{\mathrm{eff}} \gtrsim 10000$ K), hot stars emit most part of their radiative energy in the UV region, resulting in highly ionized environments around these objects. Their UV spectra are composed of a large number of narrow photospheric lines (often called the UV iron forest) and a few strong resonance lines, which are mainly formed in the wind. As discussed in Sect.~\ref{sec_intro_theory_radiative_line_winds}, in general, the iron lines are rather relevant to the radiative acceleration in the wind sub-sonic region (that is, velocity lower than the sonic point of the wind\footnote{See footnote 4 in Sect.~\ref{sec_intro_theory_radiative_line_winds}}), while the contribution from C, N, O, and Si to the radiative acceleration increases toward larger distances from the stellar surface. Moreover, hydrogen recombination lines, such as, H$\alpha$ (visible), Br$\gamma$ (K-band), and Br$\alpha$ (L-band), are also naturally important diagnostics for the determination of the stellar and wind parameters for a large fraction of hot stars (which are not H-depleted).\par

Quantitative spectroscopy relies on the determination of the stellar and wind parameters by modeling the observed spectrum by means of radiative transfer models, as calculated using the code CMFGEN, previously discussed in Figs.~\ref{sec_intro_uv_spectrum} and \ref{sec_spectro_pcygni_profiles_ip}. In Table \ref{table_line_diagnostics}, we summarize the principal line diagnostics in the ultraviolet, visible, and infrared regions, used to determine the effective temperature ($T_{\mathrm{eff}}$), surface gravity ($\log g$), mass-loss rate ($\dot{M}$), terminal velocity ($v_{\infty}$), and the wind clumping ($f_{\infty}$) of O-type, WR, and also B supergiant stars.\par

\begin{table}[t]
\caption{\label{table_line_diagnostics} Summary of the main ultraviolet, visible, and infrared line diagnostic used to determine the photospheric and wind parameters of massive stars, in particular, O-type, WR, and B supergiant stars. Adapted from~\citet{martins11}.}
\centering
\begin{adjustbox}{width=1.00\textwidth}
\begin{tabular}{lccc}
\toprule
\toprule
\multicolumn{1}{l}{Parameter} & \multicolumn{1}{c}{Ultraviolet} & \multicolumn{1}{c}{Visible} & \multicolumn{1}{c}{Infrared} \\
\midrule 

$T_{\mathrm{eff}}$ & \makecell{Iron forest \\ (\ion{Fe}{III-IV-V-VI})} & \makecell{\ion{He}{I} $\lambda$4471 and \ion{He}{II} $\lambda$4542 \\ \ion{Si}{II} $\lambda$4124, \ion{Si}{III} $\lambda$4552, and \ion{Si}{IV} $\lambda$4116} &   \ion{He}{I} $\lambda$2.112 and \ion{He}{II} $\lambda$2.189\\

\midrule

$\log g$ & --------- &  H$\alpha$, H$\beta$, H$\gamma$, and H$\gamma$ &  Br$\gamma$\\

\midrule 

$\dot{M}$ & \makecell{\ion{P}{V} $\lambda \lambda$ 1118,1128 \\ \ion{N}{V} $\lambda$1240 \\ \ion{Si}{IV} $\lambda \lambda$1394,1403 \\ \ion{C}{IV} $\lambda \lambda$ 1548,1550 \\ \ion{N}{IV} $\lambda$1718 } & H$\alpha$ and \ion{He}{II} $\lambda$4686  & Pf$\gamma$, Br$\alpha$, and Br$\gamma$ \\

\midrule

$v_{\infty}$ & \makecell{\ion{P}{V} $\lambda \lambda$ 1118,1128 \\ \ion{N}{V} $\lambda$1240 \\ \ion{Si}{IV} $\lambda \lambda$1394,1403 \\ \ion{C}{IV} $\lambda \lambda$ 1548,1550 \\ \ion{N}{IV} $\lambda$1718 } & \makecell{H$\alpha$, H$\beta$, H$\gamma$, and \ion{He}{I} $\lambda$4471 \\ (\textit{if strong winds})}  & \makecell{ \ion{He}{I} $\lambda$2.058, \ion{He}{I} $\lambda$2.112, and Br$\gamma$ \\ (\textit{if strong winds})} \\

\midrule

$f_{\infty}$ & \makecell{\ion{P}{V} $\lambda \lambda$ 1118,1128 \\ \ion{O}{V} $\lambda$1371 \\ \ion{N}{IV} $\lambda$1718} & H$\alpha$ and \ion{He}{II} $\lambda$4686 & Br10 and Br11\\

\bottomrule
\end{tabular}
\end{adjustbox}
\end{table}

In the following, we focus our discussion on the stellar effective temperature and the wind mass-loss rate. For that, we use state-of-the-art radiative transfer models, calculated with the code CMFGEN. Further details on multi-wavelength quantitative spectroscopy of hot stars can be found in~\citet{martins11}, extending the discussion of this section to the analysis of surface magnetic field, surface chemical abundance, projected rotational velocity, and macroturbulence.\par

\begin{figure}[t]
\centerfloat
\centerline{\resizebox{0.50\textwidth}{!}{\includegraphics{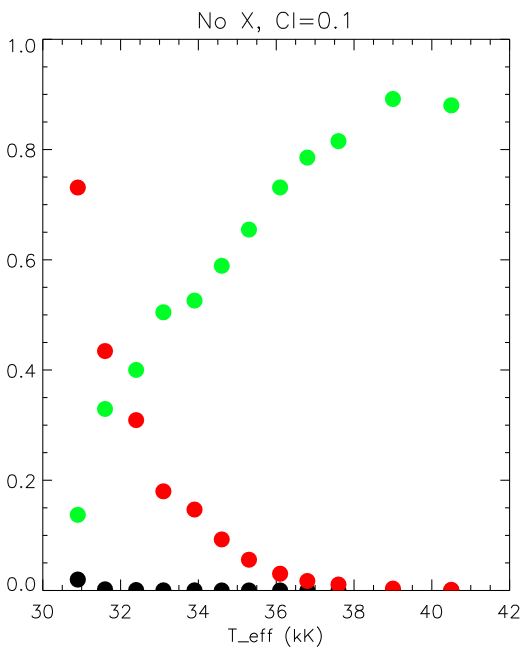}}}
\caption[CMFGEN ion fraction of silicon as a function of effective temperature for the parameter space of O supergiant stars.]
{CMFGEN ion fraction of silicon as a function of effective temperature for the parameter space of O supergiant stars. Black: \ion{Si}{III}. Red: \ion{Si}{IV}. Green: \ion{Si}{V}. In the top, it is indicated the level of wind clumping ($f_{\infty}$ = 0.1) and the absence of X-Rays in the models. Adapted from~\citet{austin11_thesis}.}
\label{sec_spectro_ion_fractions_silicon_supergiants}
\end{figure}

The effective temperature is one of the main stellar parameters affecting the wind properties. This happens since this parameter impacts the wind ionization structure. In Fig.~\ref{sec_spectro_ion_fractions_silicon_supergiants}, we show the ionization fraction of silicon, calculated with CMFGEN, according to the variation of the effective temperature. Here, the model parameter space is set typically for O supergiants, as investigated by~\citet{austin11_thesis}. In this case, the fraction of \ion{Si}{IV} increases toward lower values of $T_{\mathrm{eff}}$. Note how practically all silicon is found in the higher ionization stage considered here (\ion{Si}{V}) for higher values of effective temperature (up to $T_{\mathrm{eff}}$ $\sim$ 40000 K).\par

Looking the synthetic spectra, in advance, as shown in Fig.~\ref{sec_spectro_efeito_teff_uv_visible}, one sees how the \ion{Si}{IV} $\lambda \lambda$1394,1403 is significantly affected by varying the effective temperature. In this case, when considering a lower value of $T_{\mathrm{eff}}$, closer to 30000 K, \ion{Si}{IV} $\lambda \lambda$1394,1403 shows a larger line width, resulting from a larger wind contribution. In short, this means that reliable estimations for the wind mass-loss rate depend on accurate constrains on $T_{\mathrm{eff}}$.\par

In a more quantitative way, this can be understood since the intensity of the line profile due to a certain atomic specie (ion), that is related to the optical depth for such wavelength region, depends both on the mass-loss rate and the ion fraction. For instance, considering \ion{Si}{IV}, under the Sobolev approximation\footnote{See footnote 6 in Sect.~\ref{sec_intro_theory_radiative_line_winds}.} the optical depth (in this case, denoted as $\tau_{\mathrm{\ion{Si}{IV}}}$) is then directly proportional to the mass-loss rate, the chemical abundance of abundance of silicon ($\epsilon_{\mathrm{Si}}$), and to the \ion{Si}{IV} fraction ($q_{\mathrm{\ion{Si}{IV}}}$): $\tau_{\mathrm{\ion{Si}{IV}}}$ $\propto$ $\dot{M} q_{\mathrm{\ion{Si}{IV}}} \epsilon_{\mathrm{Si}}$.\par

From Section \ref{sec_spectro_stellar_classification}, it is conspicuous that the main diagnostics of the effective temperature are based on the ionization balances of helium and silicon lines in the visible spectrum. The lines of \ion{He}{I} $\lambda$4471 and \ion{He}{II} $\lambda$4542, and  \ion{Si}{II} $\lambda$4124, \ion{Si}{III} $\lambda$4552, and \ion{Si}{IV} $\lambda$4116 are particularly important to the determine the $T_{\mathrm{eff}}$ of O- and B-type stars, respectively.\par

In Fig \ref{sec_spectro_efeito_teff_uv_visible}, we compare the synthetic spectra from two different CMFGEN models, covering the UV ($\sim$1200-2000 {\AA}) and visible regions ($\sim$4050-4750 {\AA}). The parameter values of these models are the same than the ones shown in Fig.~\ref{sec_spectro_pcygni_profiles_ip}, but with $T_{\mathrm{eff}}$ = 27500 K and $T_{\mathrm{eff}}$ = 37500 K.\par

\begin{figure}[t]
\centerfloat
\centerline{\resizebox{1.00\textwidth}{!}{\includegraphics{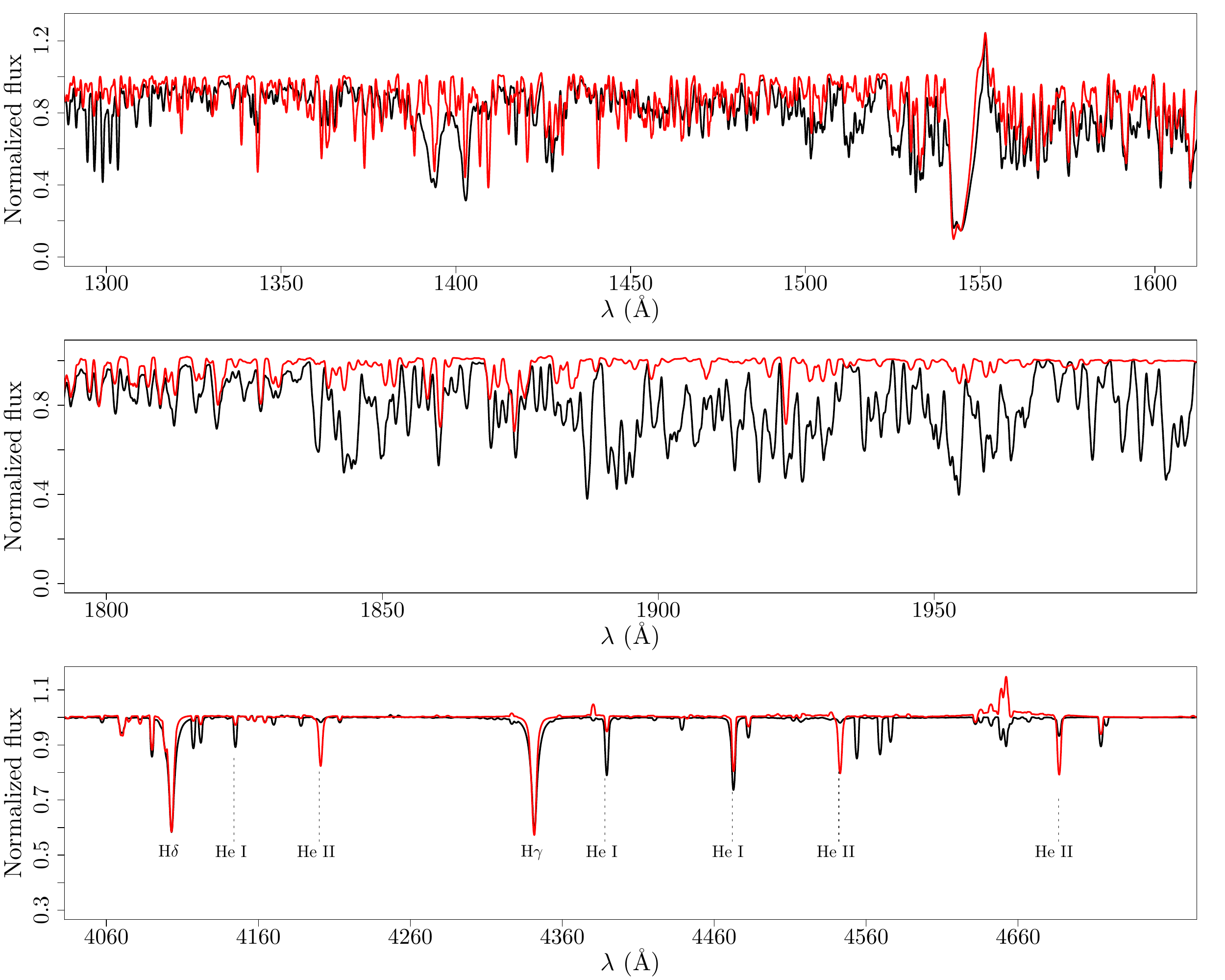}}}
\caption[Effect of varying the effective temperature on the UV and visible regions.]
{Effect of varying the effective temperature on the UV and visible regions. Synthetic spectra are shown for two CMFGEN models, calculated by myself, with different values of effective temperature: $T_{\mathrm{eff}}$ = 27500 K (black) and $T_{\mathrm{eff}}$ = 37500 K (red). All the other parameters are fixed according to the model shown in Fig.~\ref{sec_spectro_pcygni_profiles_ip}. See text for discussion.}
\label{sec_spectro_efeito_teff_uv_visible}
\end{figure}

As one can see from Fig.~\ref{sec_spectro_efeito_teff_uv_visible}, such a high contrast between the synthetic spectra, both in the UV and visible, is due to the high difference in $T_{\mathrm{eff}}$, namely, 10000 K. In particular, in the UV, the highest effects are seen in the iron forest around 1850-2000 {\AA} (in this case, mainly composed of \ion{Fe}{III} lines). In the visible, the strongest effects are seen in the \ion{He}{I} $\lambda$4471 and \ion{He}{II} $\lambda$4542 lines, as discussed above, as the parameters of these models are tuned for the analysis of O-type stars. The $T_{\mathrm{eff}}$ of O stars is generally determined from such quantitative spectroscopic analysis, with uncertainties of $\sim$1000-2000 K, from the fit to the visible \ion{He}{I-II} lines, and somewhat larger values of $\sim$2000-4000 K from the fit purely based on the iron forest in the ultraviolet~\citep[e.g.,][]{repolust04, martins04, holgado18}.\par

Overall the $T_{\mathrm{eff}}$ derived, in separate way, from the UV iron forest lines and the visible helium lines agrees fairly well~\citep[e.g., using the code CMFGEN as in][]{hillier03, martins05_weakwinds}. However, significant differences are still reported in the literature. For example,~\citet{bianchi02} derived quite lower values of $T_{\mathrm{eff}}$ of O stars from the ultraviolet, of about $\sim$6000 K, when compared with the values derived from the visible spectrum. This points out the current need of a multi-wavelength approach in order to better constrain such a fundamental stellar parameter, allowing us to achieve lower uncertainties on the determination of the wind properties, as discussed above.\par

\begin{figure}[t]
\centerfloat
\centerline{\resizebox{0.70\textwidth}{!}{\includegraphics{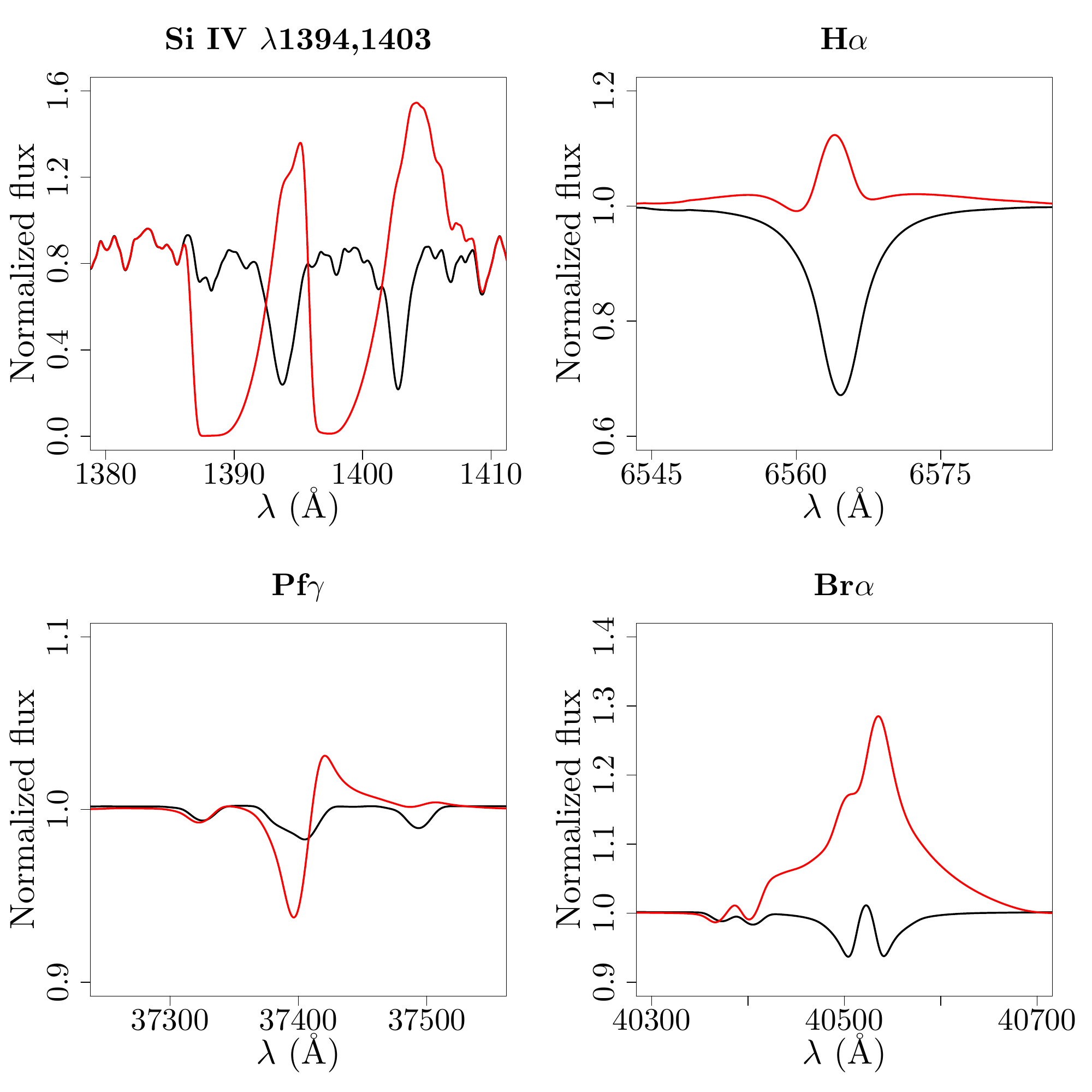}}}
\caption[Effect of varying the mass-loss rate on the UV (\ion{Si}{IV} $\lambda \lambda$ 1394,1403), visible (H$\alpha$), and infrared (Pf$\gamma$ and Br$\alpha$) regions.]
{Effect of varying the mass-loss rate on the UV (\ion{Si}{IV} $\lambda \lambda$ 1394,1403), visible (H$\alpha$), and infrared (Pf$\gamma$ and Br$\alpha$) regions. Synthetic spectra are shown for two CMFGEN models, calculated by myself, with different values of $\dot{M}$: $\dot{M}$ = $1.5\e{-9}$ $\mathrm{M_\odot}$ yr\textsuperscript{-1} (black) and $\dot{M}$ = $2.5\e{-7}$ $\mathrm{M_\odot}$ yr\textsuperscript{-1} (red). All the other parameters are fixed according to the model shown in Fig.~\ref{sec_spectro_pcygni_profiles_ip}. Note how all these diagnostic lines for the mass-loss rate are highly affected by just varying this parameter in the models.}
\label{sec_spectro_efeito_mdot_uv_halpha_infrared}
\end{figure}

Regarding the wind parameters, the line diagnostics for the mass-loss rate and the terminal velocity are essentially the same in the UV region. However, as discussed in Sect.~\ref{sec_pcygni_profiles}, depending on the stellar and wind parameters, it is possible to observe saturated P Cygni profiles (mainly in the objects of stronger winds), and in this case, the terminal velocity can be unambiguously derived by adjusting the blueward extension of their absorption component, while the wind mass-loss rate will affect highly the P Cygni intensity overall, and thus its morphology. As shown in Table \ref{table_line_diagnostics}, the hydrogen and helium lines, found in the visible and infrared, are useful to estimate the terminal velocity of objects with strong stellar wind, as WR and early B-supergiant stars, showing $\dot{M}$ as higher as $10^{-5}$ $\mathrm{M_\odot}$ yr\textsuperscript{-1}. Furthermore, these lines are also used as diagnostics for the determination of the wind mass-loss rate.\par 

In Fig.~\ref{sec_spectro_efeito_mdot_uv_halpha_infrared}, we compare synthetic profiles in the UV, visible, and infrared, computed from two CMFGEN models, in this case, only varying the mass-loss rate: $\dot{M}$ = $1.5\e{-9}$ $\mathrm{M_\odot}$ yr\textsuperscript{-1} (black) and $\dot{M}$ = $2.5\e{-7}$ $\mathrm{M_\odot}$ yr\textsuperscript{-1} (red). As in the previous example, for the effective temperature, all the other parameters are fixed with the same values as those of the model presented in Fig.~\ref{sec_spectro_pcygni_profiles_ip}. These lines shown here are one of the main mass-loss diagnostics presented in Table \ref{table_line_diagnostics}, and it is conspicuous their high sensitive to the variation of the mass-loss rate (of about 2 orders of magnitude).\par

We point out that UV resonance lines, such as \ion{Si}{IV} $\lambda \lambda$ 1394,1403, forming P-Cygni profiles in the model with the highest $\dot{M}$ in Fig.~\ref{sec_spectro_efeito_mdot_uv_halpha_infrared}, respond differently to the change in $\dot{M}$ than the hydrogen recombination lines. As resonance lines, their intensity ($I$) is directly proportional to the local density in the wind, i.e., $I$ $\propto$ $\rho$. On the other hand, the intensity of recombination lines, as H$\alpha$, Pf$\gamma$, and Br$\gamma$, shown in Fig.~\ref{sec_spectro_efeito_mdot_uv_halpha_infrared}, depends on the square of $\rho$, $I$ $\propto$ $\rho^{2}$, since the formation of these lines requires the interaction of two particles (an electron and an ion).\par 

This fact above is relevant with respect to the estimation of the $\dot{M}$ of massive stars with weaker winds, having $\dot{M}$ $\sim$ $10^{-10}$-$10^{-9}$ $\mathrm{M_\odot}$ yr\textsuperscript{-1}. Despite the less accurate modeling of their ionization structure, in comparison to the H lines, the UV resonance lines better trace $\dot{M}$ for low-density wind regimes, as the H and He recombination lines are typically formed in the wind region closer to the stellar surface (larger wind density) than the UV resonance lines. However, recombination lines will be more useful as mass-loss diagnostics of stronger (denser) winds, showing $\dot{M}$ $\sim$ $10^{-8}$--$10^{-7}$ $\mathrm{M_\odot}$ yr\textsuperscript{-1}~\citep[see, e.g.,][]{puls08, marcolino09}.\par

A multi-wavelength quantitative spectroscopic approach, that is, using spectral lines formed in different wavelength regions, to better constrain the real rates of mass-loss in massive stars, is one of the main objective of this thesis. This issue will be discussed in details, in Sect.~\ref{sec_results_ostars}, when presenting our results for the wind parameters of O-type stars, concerning the so-called weak wind problem, firstly discussed in Sect.~\ref{sec_intro_radiative_winds_hr_diagram}.\par
\pagestyle{empty}
\cleardoublepage
\pagestyle{fancy}

\chapter{Optical long-baseline stellar interferometry}
\label{chapter_interferometry}
\minitoc

\section{Why we need high angular resolution observations?}
\label{sec_interf_hra}

In the previous chapter, we highlighted the power of spectroscopy to characterize the stellar and circumstellar properties of massive hot stars.\par 

However, due to the very low values of stellar angular size, we usually are not able to retrieve information about the spatial intensity distribution of the star and of its environments projected across the sky, purely based on the analysis of line profiles. Nevertheless, one possibility, using spectroscopic data, in order to indirectly resolve the structure of stellar surfaces relies on the so-called Doppler imaging technique~\citep[e.g.,][]{vogt87}. On the other hand, this technique works optimally for fast rotators, with intermediate values of inclination angle, preferentially between $\sim$\ang{20}-\ang{70}, arising ambiguities on the polar locations for stars seen close to edge-on~\citep[e.g.,][]{ strassmeier02}.\par

Spatial resolution means the ability to resolve, that is, distinguish, different points of the observed object's image. The angular resolution provided by a single-mirror telescope depends on its mirror's diameter ($D$) and the observing wavelength ($\lambda$), being usually approximated by the so-called Rayleigh criterion:

\begin{equation}
\theta^{\mathrm{diffraction}}_{\mathrm{res}} = 1.22\frac{\lambda}{D} \sim \frac{\lambda}{D}.
\label{eq:power_spatial_resolution}
\end{equation}

This simple estimation for the angular resolution of a telescope comes from the diffraction limit imposed by the interference of the incident light wavefronts of an unresolved object across the telescope aperture. Considering a uniformly-illuminated circular aperture, with diameter $D$, the diffraction pattern is described by an Airy disk, with brightness, $I(\rho)$, given as follows:

\begin{equation}
I(\rho) = \left(\frac{\pi D^{2}}{4}\right)^{2} \left[\frac{2 J_{1}(\pi \rho D)}{\pi \rho D}\right]^{2},
\label{eq:airy_disk}
\end{equation}
where $\rho$ is the radial distance from the center of the Airy disk and $J_{1}$ is the first order Bessel function of the first kind~\citep[e.g., see Chap.~9 of][]{abramowitz72}.\par 

As shown in Fig.~\ref{sec_spectro_airy_disk}, the Airy disk pattern has a central bright region with a diameter of $\sim$2.44$\frac{\lambda}{D}$. Thus, the factor 1.22, in Eq.~\ref{eq:power_spatial_resolution}, which is the radius of the Airy disk's first zero (first dark ring), expresses the ability to separate two point source images at the focal plane of the telescope.\par

\begin{figure}
\centerfloat
\centerline{\resizebox{0.85\textwidth}{!}{\includegraphics{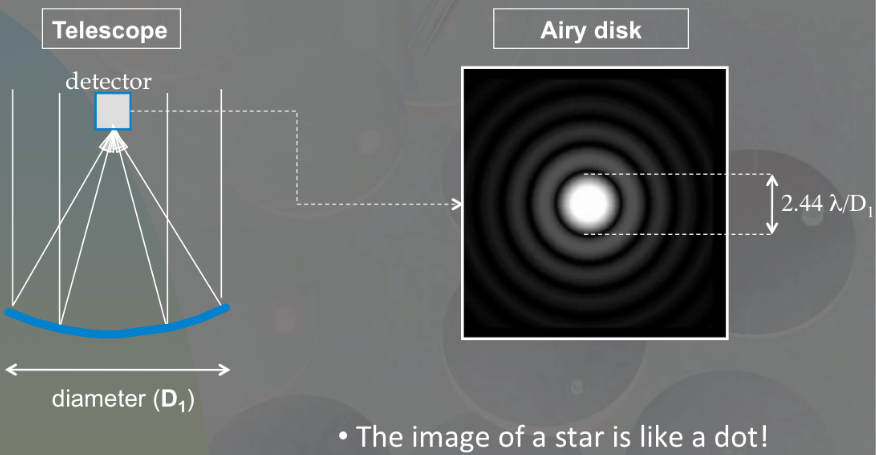}}}
\caption[Schematic of an Airy disk, formed by the phenomenon of diffraction, arising from the incidence of the light wavefronts into a telescope (round pupil).]
{Schematic of an Airy disk, formed by the phenomenon of diffraction, arising from the incidence of the light wavefronts into a telescope (round pupil). The light wavelength is given by $\lambda$ and the telescope principal mirror by $D_1$. The Airy disk's central bright region has a diameter of $\sim$2.44$\lambda / D$. See text for discussion. Reproduced from~\citet{surdej19}.}
\label{sec_spectro_airy_disk}
\end{figure}

As pointed out, in Fig.~\ref{sec_spectro_airy_disk}, the Airy disk pattern results from the observation of an unresolved object, a point source, describing then the ideal (i.e., without instrumental and atmospheric effects) point spread function (PSF) of a circular single-mirror telescope. Even considering a single-mirror telescope, we stress that the PSF changes according to the employed aperture geometry for the telescope pupil~\citep[e.g., as shown in Fig.~2 of][]{millour14}.\par

We also stress that the angular resolution of a telescope, as approximated by Eq.~\ref{eq:power_spatial_resolution}, does not take into account the optics wavefront aberrations nor atmospheric turbulence effects on the focal plane image, which can drastically reduce the telescope's capacity to spatially resolve an object. To overcome this problem, adaptive optics systems have been extensively developed since the end of the 1990's in order to reduce the image blur due to the atmospheric turbulence. For example, in the visible spectral region, without the use of adaptive optics, the spatial resolution of a 10-m telescope would be equivalent of a 10-cm mirror telescope. Details about this technique can be found in the reviews of~\citet{beckers93} and~\citet{davies12}.\par

Nevertheless, the current generation of 10-m class telescopes are only able to resolve the nearest supergiant stars. For instance, we consider Achernar that is the brightest (V = 0.46) and nearest Be star with a distance of 42.8 pc~\citep{vanleeuwen07}. Achernar has an equatorial angular diameter of $\sim$1.99 mas, considering the linear equatorial radius of $\sim$9.16 $R_{\odot}$, and a mean angular diameter of 1.77 mas when taking into account its geometrical oblateness~\citep{domiciano14}. From Eq.~\ref{eq:power_spatial_resolution}, considering $\lambda$ = 6563 {\AA} (centered at the H$\alpha$ line) and a 8-m telescope, as the Unit Telescopes at the ESO/VLT facility, we find that $\theta_{\mathrm{res}}$ $\sim$ 0.02 arcsecond, that is, 20 milliarcseconds. This is one order of magnitude larger than the angular diameter of Achernar projected onto the sky.\par

Thus, even considering the nearest Be star, this shows the impossibility to obtain finer spatial information of its surface, using such a large single-mirror telescope as provided by the ESO Unit Telescopes. In addition, as discussed in Sects.~\ref{sec_intro_Be_stars} and \ref{sec_intro_Be_disks}, Be stars possess transient or stable (within a certain time-scale) circumstellar disks with emitting envelope extension reaching up to a few milliarcseconds in angular size. Therefore, in addition to spectral resolution, this evidences the necessity of spatial resolution in order to constrain the physical properties of Be stars, and more generally of massive stars, and their environments (composed of winds, disks, or both of them).\par

\begin{figure}[t]
\centerfloat
\centerline{\resizebox{1.00\textwidth}{!}{\includegraphics{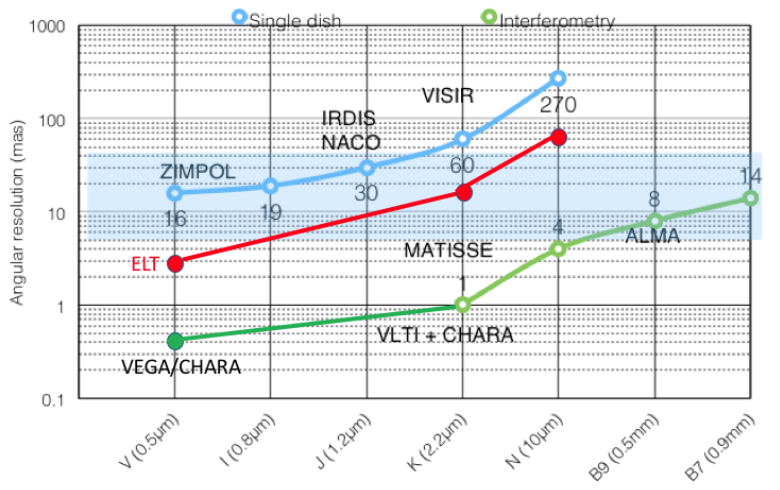}}}
\caption[Comparison between the angular resolution (in milliarcsecond) provided by current single-mirror telescope (blue line; ZIMPOL, IRDIS, VISIR instruments) and interferometric facilities (green line; VEGA/CHARA, MATISSE, VLTI + CHARA instruments, and ALMA), as a function of wavelength.]
{Comparison between the angular resolution (in milliarcsecond) provided by current single-mirror telescope (blue line; ZIMPOL, IRDIS, VISIR instruments) and interferometric facilities (green line; VEGA/CHARA, MATISSE, VLTI + CHARA instruments, and ALMA), as a function of wavelength. The blue region indicates a similar range of angular resolution that is achieved by both the current facilities and the future ones, as the Extremely Large Telescope (ELT, 39-m telescope) in red line. See text for discussion. Reproduced from~\citet{mourard19}.}
\label{sec_spectro_single_interf_res}
\end{figure}

To date, interferometry is the only technique that enables to overcome the limitations of angular resolution imposed by the aperture size of current single-mirror telescopes, and allowing us to resolve the surface and the environment of massive stars, which typically have angular sizes as low as $\sim$1 and 10 mas, respectively. In this case, by combing the light-beams from different single-mirror telescopes in order to obtain interference fringes, the angular resolution can now be approximated by the distance $B_{\mathrm{proj}}$ of a pair of telescopes, projected onto the sky in the direction of the observed object, that is, the projected baseline length, replacing $D$ in Eq.~\ref{eq:power_spatial_resolution} by:

\begin{equation}
\theta^{\mathrm{interf}}_{\mathrm{res}} \sim \frac{\lambda}{B_{\mathrm{proj}}}.
\label{eq:power_spatial_resolution_interferometer}
\end{equation}

We thus achieve a spatial resolution as provided by the diffraction-limit of a ``synthesized telescope'' with its mirror diameter equal to the projected baseline length of the telescope array.\par 

Fig.~\ref{sec_spectro_single_interf_res} compares the spatial resolution provided by different single-mirror telescopes and by interferometric facilities, as a function of wavelength, from the visible to the millimetric region. First, we note how the ability to spatially resolve is typically reduced toward longer wavelength (in agreement with Eqs.~\ref{eq:power_spatial_resolution} and \ref{eq:power_spatial_resolution_interferometer}). Moreover, to date, one sees that the highest angular resolution, in the visible band (including the H$\alpha$ line) are obtained using the CHARA/VEGA instrument, reaching a maximum resolution of $\sim$0.3-0.4 mas. Meanwhile, the CHARA (such as CLASSIC) and VLTI instruments (such as MATISSE) provide the highest ones in the near-infrared (including the Br$\gamma$ line) and in the mid-infrared (including the Br$\alpha$ line).\par

Fig.~\ref{sec_spectro_single_interf_res} also shows the possible synergies between the single-mirror telescopes and interferometric facilities. For example, the ELT array will be able to obtain similar value of angular resolution ($\sim$3 mas) in the visible than the current VLTI/MATISSE instrument in the N-band, as well as the Atacama Large Millimeter Array (ALMA), allowing us to compare the  properties of the objects studied then under the same level of spatial resolution in different spectral channels (visible, mid-infrared, and sub-millimeter regions).\par


\section{Historical overview}
\label{sec_interf_historical_overview}

Optical long-baseline interferometry (OLBI) means the use of optical systems to combine the light received from separate pairs of telescopes in order to obtain fringe patterns. As pointed out by~\citet{monnier03}, the term ``optical'' used here indicates the implementation of such optical system to manipulate the collected light, instead of indicating the wavelength domain of the observations. Currently, OLBI is limited in the spectral channels from the optical/visible region ($\sim$400 nm) to the mi-infrared ($\sim$10 $\mu$m). Further details on the history of stellar interferometry can be found in~\citet{lawson00} and~\citet{lena14}.\par

The basis for stellar interferometry comes from the demonstration of the wave properties of the light by Thomas Young in 1801. Fig.~\ref{sec_interf_young_double_slit_experiment} shows the basic scheme of the so-called Young's double slit experiment. Under certain condition, as it will be discussed in more details in Sect.~\ref{sec_interf_elementary}, an interference pattern composed of bright and dark fringes arises from the interaction of the light waves originating from the two slits.\par 

As indicated in Fig.~\ref{sec_interf_young_double_slit_experiment}, each point of a light wavefront works like his own source of spherical waves, following the Huygens-Fresnel principle, originally proposed by Christiaan Huygens in 1678. This allows to explain an interference pattern resulting from the combination of overlapping light waves, which are originated from the two slits. Considering the simplest case of a monochromatic light source, that is, fixing the light wavelength, the angular separation between the fringes can then be expressed as a function of distance between the two slits. For a certain constant distance from the slits to the pattern screen, a larger distance between the slits implies larger separation between the interference fringes, a lower contrast of the fringes.\par 

\begin{figure}
\centerfloat
\centerline{\resizebox{1.00\textwidth}{!}{\includegraphics{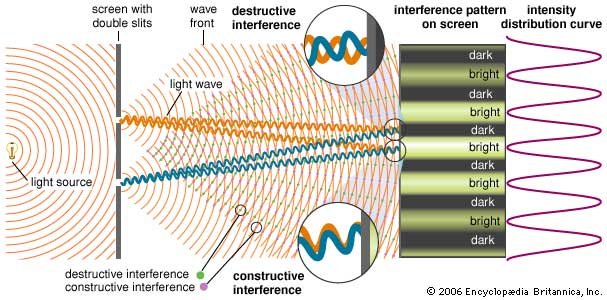}}}
\caption[Scheme for the propagation of light wavefronts through a double-slit screen, the Young's double-slit experiment.]
{Scheme for the propagation of light wavefronts through a double-slit screen, the Young's double-slit experiment. For simplification, the light source is considered monochromatic with wavelength $\lambda$. By the Huygens–Fresnel principle, each point of a light wavefront behaves itself as a own source of spherical waves. The light interference happens according to the difference in phase between different wavefronts. Complete constructive interference happens when the path difference is equal to $\lambda$, or a integer multiple of that, while complete destructive interference happens when the path difference is equal to $\lambda$/2, or a integer multiple of that. Interference fringes (bright and dark bands) are then formed by the superposition of the overlapping light waves propagating toward the screen. Source: \url{https://www.britannica.com/}.}
\label{sec_interf_young_double_slit_experiment}
\end{figure}

Based on the Young's double-slit experiment, Hippolyte Fizeau was the first to propose, in 1868, the measurement of stellar diameters by the observation of interference fringes, using slits in front of the telescope principal mirror. Until then, it was considered an impossible task to measure stellar diameters for stars other than the Sun. As figured out by Fizeau, the contrast of the fringes will be a function of both the angular diameter of the star and the slits separation.\par

\begin{table}[t]
\caption{\label{table_results_interf} Summary of some very first visible and infrared interferometric studies using two-telescope configuration. Reproduced from~\citet{lena14}.}
\centering
\renewcommand{\arraystretch}{1.25}
\begin{adjustbox}{width=1.00\textwidth}
\begin{tabular}{lccccc}
\toprule
\toprule
\multicolumn{1}{l}{Year} & \multicolumn{1}{c}{Instrument/baseline} & \multicolumn{1}{c}{Location} & \multicolumn{1}{c}{Observation} & \multicolumn{1}{c}{Wavelength} & \multicolumn{1}{c}{Resolution}  \\
\midrule 

1970 & 6 m & \makecell{Pulkovo \\ (Crimea)} & Orbit of Capella & Visible & $\sim$1 mas \\

1974 & I2T/12 m & Calern (France) & Vega & Visible & 1 mas \\

1974 & ISI/5.5 m & \makecell{Kitt Peak \\ (Arizona)} & Mercury & 10.6 $\mu$m & 200 mas \\

1986 & I2T/8 to 25 m & Calern (France) & $\gamma$ Cas envelope & H$\alpha$ & 1.4 mas \\

1986 & SUSI/11 m & \makecell{Narrabri \\ (Australia)} & Sirius diameter & Visible & 1 mas \\

1986 & COAST/6 m & Cambridge (UK) & Orbit of Capella & Visible & 10 mas \\

1988 & MarkIII/12 m & \makecell{Mt. Wilson \\ (California)} & Diameter of 4 stars & Visible & 1 mas \\

\bottomrule
\end{tabular}
\end{adjustbox}
\end{table}

The first successful implementation of this method was the measurement of the diameters of the Jupiter's moons (angular size of $\sim$1 arcsecond) by~\citet{michelson91}, using the 12-inch refractor at Lick Observatory. Three decades later,~\citet{michelson21} finally managed to measure the first angular diameter of a star (other than the Sun) using the 2.5-m refractor at Mount Wilson Observatory in California. To increase the angular resolution, they mounted a 20-feet beam (6 m) equipped with mirrors over the telescope structure. Thanks to this modification, they measured the angular diameter of the red supergiant star Betelgeuse, finding an angular size of $\sim$47 mas with an uncertainty of about 10\%.\par

Due to both instrumental and atmospheric effects, later interferometric works had a quite limited quality in measuring fringes for a larger sample of stars. This drastically limited the pioneer studies of stellar interferometry to longer wavelengths, for which the manipulation of the light to obtain interference is easier than in the visible/infrared. From 1940's to the beginning of the 1970's, most part of the interferometric works were performed using radio interferometers, for example, using the Five Kilometer Array~\citep{ryle50}. This happens since the coherence time-scale, allowing interference, is dependent on the light wavelength. In the practice, for grounded-based observations, resulting from the atmospheric turbulence, the coherence time-scale is limited to very short value of $\sim$$10^{0}$-$10^{1}$ ms in the visible, when compared with the radio domain, reaching up to several minutes~\citep[e.g., see Table 2 of][]{monnier13}.\par

In 1974, Antoine Labeyrie was the first to directly combine coherently the visible light received by separate telescopes in order to measure interference fringes, using the \textit{Interféromètre à 2 Télescopes} (I2T) interferometer (Plateau de Calern, France), providing an upper limit of 5 mas for the stellar diameter of Vega~\citep{labeyrie75}. This ground-breaking work established the modern technique of OLBI, and then reviving the stellar interferometry at the visible and infrared wavelengths.\par 

Still using the I2T interferometer, \cite{thom86} was the first to spatially resolve the circumstellar environment of a classical Be star, $\gamma$ Cassiopeiae, measuring an extension of 3.6 mas in the H$\alpha$ line. Still observing in H$\alpha$,~\citet{mourard89} was one of the first spectro-interferometric studies to find evidences of a rotating disk in $\gamma$ Cassiopeiae, with the \textit{Grand Interféromètre à 2 Télescopes} (GI2T) interferometer, also installed on Plateau de Calern.~\citet{stee95} were the first to interpret GI2T spectro-interferometric observations of $\gamma$ Cassiopeiae using a radiative transfer model (code SIMECA). Later,~\citet{vakili97} provided the first spectro-interferometric observations of the LBV star P Cygni, using the GI2T interferometer, in the H$\alpha$ and \ion{He}{I} $\lambda$6678 emission lines. In addition to constraining the extension of P Cygni' wind in these lines, these authors found evidences of inhomogeneities in its winds based on differential phase measurements in H$\alpha$. Some of these early results, together with other remarkable ones found using two-telescope interferometry, are listed in Table \ref{table_results_interf}.\par


\section{Elementary concepts of OLBI}
\label{sec_interf_elementary}

Fig.~\ref{sec_interf_monnier03_fig10} shows the principal components of an optical interferometer. From that, we point out that the capability to spatially resolve the object is dependent on the projected baseline length, $B_{\mathrm{proj}}$, defined as follows:

\begin{equation}
B_{\mathrm{proj}} = B \sin{\theta},
\end{equation}
where $B$ is the linear baseline length between the pair of telescopes, and $\theta$ is the angle formed between the baseline vector, $\hat{B}$, and the line of sight toward the object. If the baseline is perpendicular to the line of sight, we have $B_{\mathrm{proj}}$ = $B$.\par

Moreover, one sees that a geometrical delay, $\tau_{g}$ (time unit), naturally arises from interferometry since the light wavefronts are collected at different times by each telescope forming the array, being given as follows:

\begin{equation}
\tau_{g} = \frac{\vec{B} \,.\, \hat{s}}{c}, 
\label{eq:geometrical_delay}
\end{equation}
where $\vec{B}$ is the baseline vector and $\vec{s}$ is a unit vector in the direction of the source. This quantity, $\tau_{g}$, is related to the optical path difference (OPD) in the interferometer, OPD = $\tau_{g} c$.\par

The optical system used to send the light to the beam combination facility, where the interference is indeed performed, is represented in the bottom of the schematic in Fig.~\ref{sec_interf_monnier03_fig10}. At first glance, it may sound an easy task, but several technical issues are related to this process. For instance, it is necessary to transport the collected light through nearly vacuum pipes, or to employ a dispersion compensator, in order to overcome the effect of spectral dispersion due to the long distances that the light must travel between the telescope and the beam combiner~\citep[e.g.,][]{tenbrummelaar95}.\par

\begin{figure}[t]
\centerfloat
\centerline{\resizebox{0.75\textwidth}{!}{\includegraphics{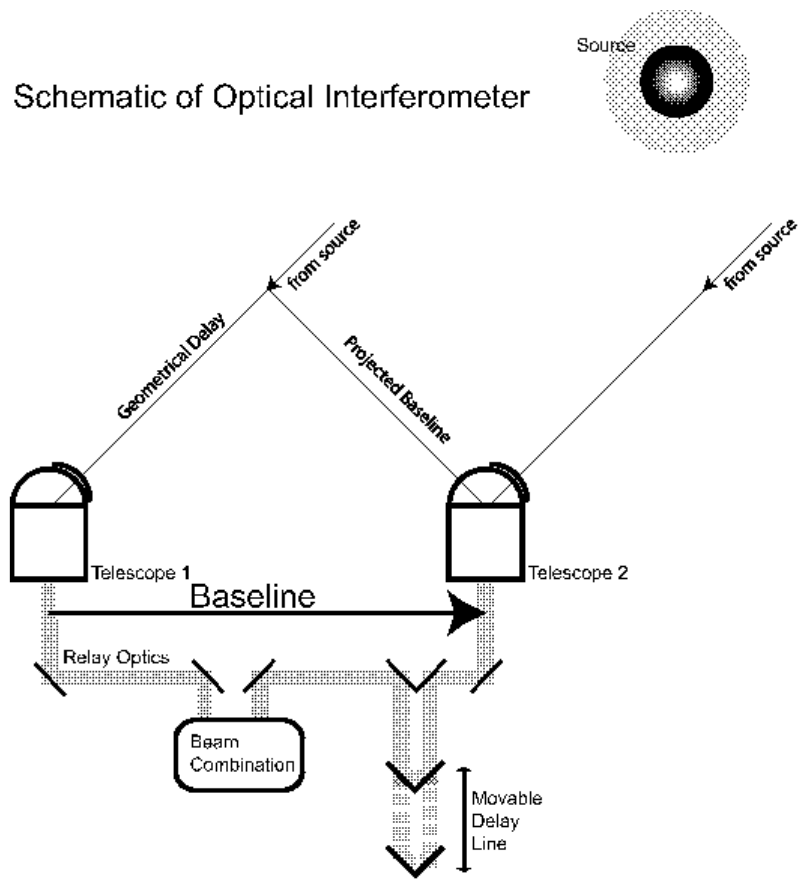}}}
\caption[Schematic representing the basic elements of an optical long-baseline interferometer: the telescopes and optical subsystems (relay optics, delay line, and beam combination).]
{Schematic representing the basic elements of an optical long-baseline interferometer: the telescopes and optical subsystems (relay optics, delay line, and beam combination). See text for discussion. Reproduced from~\citet{monnier03}.}
\label{sec_interf_monnier03_fig10}
\end{figure}

In addition, the optical system also contains movable delay lines, which are needed to compensate the geometrical delay, as discussed above, originating in the sidereal motion of the source, as shown in Eq.~\ref{eq:geometrical_delay}. This change in the optical path difference is rather important concerning the own phenomenon of light interference, in order to allow the formation of fringes, as it is discussed below.\par

As stated in the previous section, an interferometer measures the interference fringe pattern produced by the combination of the light-beams received by a pair of telescopes. Said differently, an interferometer is then sensitive to the degree of coherence of the light wavefronts received by a pair of telescopes.\par

The concept of coherence is very useful in order to understand the phenomenon of interferometry. If two light wavefronts show the same phase difference, the same wavelength, and the same waveform, they are said to be coherent. Ideally, light interference pattern happens under such condition. For this reason, the Young's double-slit experiment works optimally (to produce interference patterns) using a laser source, a highly coherent light source, since its radiation beam is highly directional, that is, spatially narrow, and also nearly monochromatic. On the other hand, a conventional light bulb works like a highly incoherent source of light since it is composed of a wide range of wavelengths (in the visible region, $\sim$3000-7000 {\AA}) and also emitted in many directions away from the source.\par 

For instance, as a thought experiment, image raindrops falling on a lake. Here, each raindrop works like a source of (mechanical) waves in the water, and thus, at each point of the lake, the resulting wavefronts are formed by the superposition of all these wave sources. Since the raindrops reach the lake in a random way, both in space and time, one might expect that the wavefronts, at different points on the lake, are highly uncorrelated, not showing a constant phase difference. In this case, we say that the wave sources are incoherent.\par 

\begin{figure}[t]
\centerfloat
\centerline{\resizebox{1.00\textwidth}{!}{\includegraphics[angle=0]{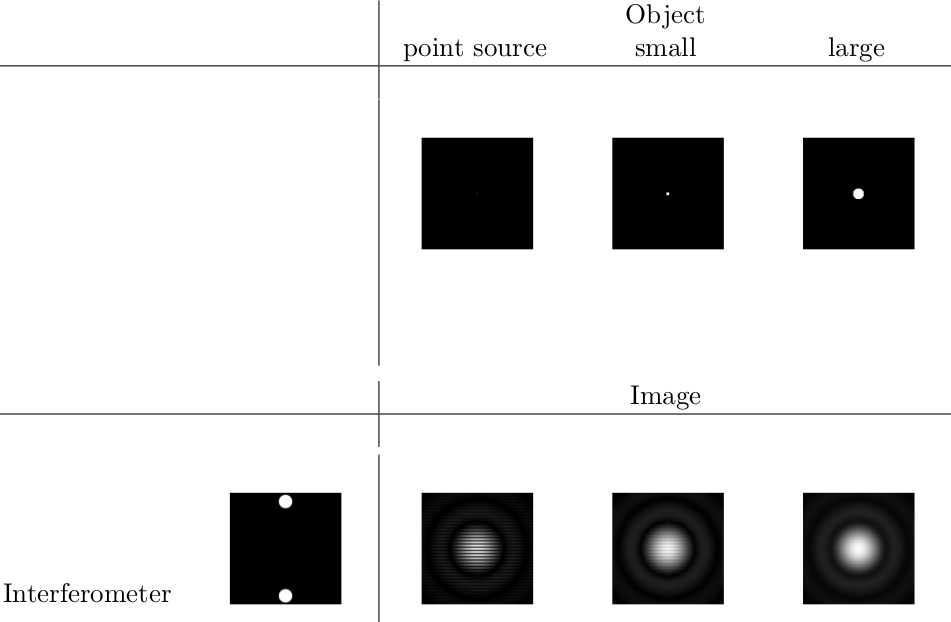}}}
\caption[Simulated fringe patterns for objects of different sizes, observed using a two-telescope interferometer.]
{Simulated fringe patterns for objects of different sizes, observed using a two-telescope interferometer. In the first column, we show the ideal case of a completely spatially unresolved source (point source). Note how the contrast of the interferometric fringes changes due to the change on the degree of coherence of the light, between the two telescopes, as the object's size increases and begins to be resolved by the interferometer. See text for discussion. Adapted from~\citet{millour14}}
\label{sec_interf_millour14_fig3}
\end{figure}

From that, we point out that the concept of coherence works in the domain of space and time. Spatial coherence measures the correlation between the wave phase at different points along the transverse direction of the light propagation, and then stating how uniform is the phase wave. Ideally, a point source produces waves that are perfectly coherent in the spatial domain. Moreover, this means that some level of incoherence is always introduced when we are able to spatially resolve the object (as a star), as the different parts of the (resolved) object begins to work independently, thus incoherent sources of light, lacking a constant phase relationship between the waves emitted by them.\par

Returning to the Young's double-slit experiment, interference fringes are lost when the distance between the pinholes are larger than a certain value $r_{c}$:

\begin{equation}
r_{c} = \frac{\lambda}{\theta},
\end{equation}
where $\theta$ is the angular diameter of the source. This critical value, $r_{c}$, is called the coherence distance of the source~\citep[e.g., see page 95 of][]{labeyrie06}.\par 

As discussed in Sect.~\ref{sec_interf_historical_overview}, the implementation of this method, in order to measure stellar diameters, founded the field of stellar interferometry at the end of the 19th century.\par 

In Fig.~\ref{sec_interf_millour14_fig3}, we show simulations of fringe patterns measured by a two-telescope interferometer, for different cases of spatial resolution, a completely unresolved source (i.e., point source), and also other cases with larger objects up to a fully resolved one. One sees here how the fringes are blurry, showing a lower contrast, as the object is resolved, associated to the decrease in the level of coherence of light collected by the two telescopes (due to the increase of the object's size projected onto the sky). Conversely, considering an object with fixed angular size, we are able to measure fringes produced under different degrees of light coherence by varying the baseline length and orientation.\par

\begin{figure}[t]
\centerfloat
\centerline{\resizebox{0.60\textwidth}{!}{\includegraphics{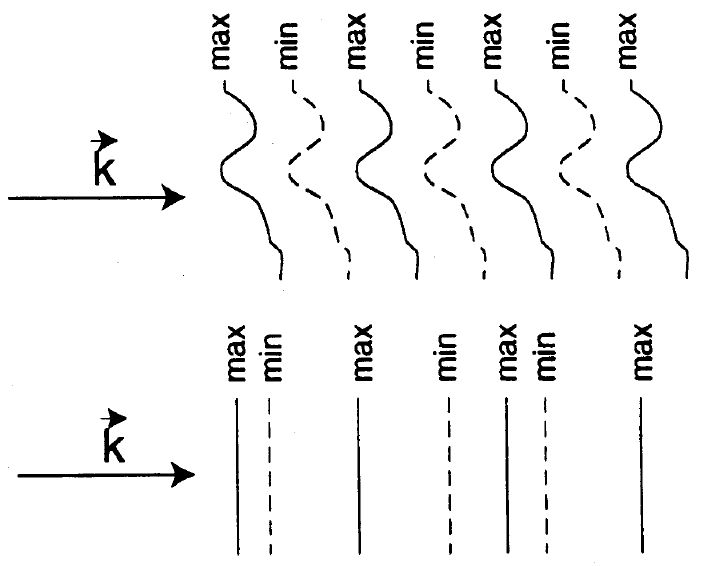}}}
\caption[Schematic of two different cases of spatial and temporal coherence.]
{Schematic of two different cases of spatial and temporal coherence. The vector $\hat{k}$ sets the direction of light propagation. In the top, there are shown highly temporally coherent waves, but also highly incoherent in the space domain. In the bottom, the opposite case is illustrated: the waves are highly coherent in the space domain, but highly incoherent in the time domain. Source: \url{https://www.brown.edu}.}
\label{sec_interf_spatial_temporal_coherence}
\end{figure}

In Sect.~\ref{sec_interf_young_double_slit_experiment}, the Young's double-slit experiment was discussed assuming a monochromatic light source, that is, emitting light described, ideally, by just one given value of wavelength. For sure, this description is quite unrealistic for natural light sources, since most astrophysical objects are indeed polychromatic sources of light, emitting a non-monochromatic wave packet (or also called as wave train).\par

This fact above is important concerning the concept of coherence since we need to deal now with wavefronts that are composed by multiple frequencies. Temporal coherence measures the correlation between the wave phase at different points in the own direction of the light propagation. In Fig.~\ref{sec_interf_spatial_temporal_coherence}, we illustrate somewhat extreme cases of coherence: spatially coherent and temporally incoherent waves, and the opposite case, temporally coherent and spatially incoherent waves. Thus, some level of temporal incoherence is always introduced by the polychromatic nature of the light sources because the phase waves are displaced, in the direction of the light propagation, due to the different values of wavelengths that result in the wave packet. In short, the degree of temporal coherence of the source is related to its spectral purity, and, for sure, to the instrumental spectral bandwidth of the observations.\par

In the practice, polychromatic light can be well described by a spectral band of wavelength width $\Delta \lambda$, centered at a certain value of wavelength $\lambda_{0}$ (the maximum peak of the light intensity). One characteristic time-scale, $\tau_{c}$, can then be defined in order to allow temporal coherence:

\begin{equation}
\tau_{c} = \frac{1}{\Delta \nu} =  \frac{\lambda_{0}^{2}}{c \Delta \lambda},
\end{equation}
where $c$ is the light speed in vacuum and $\Delta \nu$ is the frequency width corresponding to $\Delta \lambda$.\par

Physically, this quantity describes the effective oscillation time of a wave packet composed of non-monochromatic light. From that, the length-scale for temporal coherence, $l_{c}$, associated to $\tau_{c}$, is then defined as follows:

\begin{equation}
l_{c} = c \tau_{c} = \frac{\lambda_{0}^2}{\Delta \lambda},
\end{equation}
being called as the coherence length.\par

These characteristic scales above are rather important concerning the operation of a interferometer. Looking the parameter $l_{c}$, the geometrical delays between the beams, that is, the OPD, must be set to values lower than $l_{c}$ to allow the occurrence of fringe patterns. This shows the importance of the movable delays lines, discussed above (Fig.~~\ref{sec_interf_monnier03_fig10}), in order to maintain the optical path difference, that changes with the time, lower than such critical length-scale, during the observations.\par


\section{Light coherence}

In the following section, we deal with the fundamental interferometric quantities, which are extracted from the measurement of fringes. For that, firstly, a more quantitative description of light coherence is needed. We consider the electric field $\vec{E}$ component of an electromagnetic wave, in 3 dimensions, as follows:

\begin{equation}
\vec{E}(\vec{r},t) = \vec{E}_{0} \exp[i(\vec{k}.\vec{r} - wt)].
\end{equation}

Here, the position $\vec{r} = x\hat{x} + y\hat{y} + z\hat{z}$, in Cartesian coordinates, $|\vec{E}_{0}|$ is the amplitude of the electric field, $|\vec{k}|$ = $k$ = $2\pi / \lambda$ is the wavenumber, $w$ = $2\pi c / \lambda$ is the angular frequency, that is, $w$ = $c k$, and $\lambda$, as usual, is the light wavelength.\par

The light intensity $I$, as measured by a detector, is then given by the superposition of the wavefronts coming from different positions $\vec{r}_{i}$:

\begin{equation}
I = \Bigg \langle  \norm{\sum_{i} \vec{E}(\vec{r}_{i}, \, t - \tau_{i})}^{2} \Bigg \rangle,  
\label{eq:light_intensity}
\end{equation}
where $\tau_{i}$ is the delay time of the wavefronts coming from $\vec{r}_{i}$, and the use of brackets denotes the temporal average of the electric fields $\vec{E}$ over $t$. As defined above, for sure, the light intensity is a real number.\par

The coherence function, $\Gamma_{i,j}$, quantifies the correlation between the light wavefronts $\vec{E}$ received from each pair of positions, denoted by $\vec{r}_{i}$ and $\vec{r}_{j}$, with a delay time $\tau$:

\begin{equation}
\Gamma_{i,j}(\tau) = \Big \langle \vec{E}(\vec{r}_{i}, \, t - \tau) \vec{E}^{*}(\vec{r}_{j}, t) \Big \rangle,
\label{eq:coherence_function1}
\end{equation}
where the use of asterisk denotes the complex conjugate.\par

Therefore, using Eq.~\ref{eq:light_intensity}, the light intensity is expressed as a function of degree of coherence $\Gamma_{i,j}$, as follows:

\begin{equation}
I = \sum_{i} \Gamma_{i,i}(0) + \sum_{i,j} \big[\Gamma_{i,j}(\tau_{i} - \tau_{j}) + \Gamma_{i,j}^{*}(\tau_{i} - \tau_{j})\big].
\label{eq:light_intensity_coherence}
\end{equation}

Considering the simplest case of two wavefronts, received by a single pair of telescopes of an interferometer, denoted here by the indices ``1'' and ``2', Eq.~\ref{eq:light_intensity_coherence} takes the form of:

\begin{equation}
I = I_{1} + I_{2} + \Gamma_{1,2}(\tau) +  \Gamma_{1,2}^{*}(\tau),
\label{eq:light_intensity_coherence2}
\end{equation}
being the delay $\tau = \tau_{i} - \tau_{j}$ from Eq.~\ref{eq:light_intensity_coherence}. In addition, in this case, the correlation function, Eq.~\ref{eq:coherence_function1}, can be simplified as:

\begin{equation}
\gamma_{1,2}(\tau) = \frac{\Gamma_{1,2}(\tau)}{\Gamma_{1,1}(0) + \Gamma_{2,2}(0)}.    
\label{eq:coherence_function2}
\end{equation}

Here, the correlation function, $\gamma_{1,2}$, is normalized by the total intensity that is collected by the array of two telescopes: $I_{1} + I_{2} = \Gamma_{1,1}(0) + \Gamma_{2,2}(0)$.\par

Lastly, from Eqs.~\ref{eq:light_intensity_coherence2} and \ref{eq:coherence_function2}, the intensity $I$ can be written as follows (considering a 1D interferograph):

\begin{equation}
I(x) = [I_{1}(x) + I_{2}(x)][1 + \Re{(\gamma_{1,2}(\tau))}].
\label{eq:interferometric_equation}
\end{equation}


\section{Interferometric quantities}
\label{sec_interf_quantities}

\subsection{The Zernike-van Cittert theorem}
\label{sec_interf_quantities_zernike}

Eq.~\ref{eq:interferometric_equation} is often called the interferometric equation, that is, describing the interferometric fringe patterns formed by each array of two telescopes.\par 

Taking the real part of the correlation function $\gamma_{1,2}(\tau)$, in the interferometric equation, we have:

\begin{equation}
\Re{(\gamma_{1,2}(\tau))} = V \cos{(\frac{2\pi x}{\lambda} + \phi)},
\label{eq:real_part_correlation_function}
\end{equation}
where $V$ denotes the modulus of $\gamma_{1,2}$, $V = |\gamma_{1,2}|$, and $\phi$ is the phase of $\gamma_{1,2}$. Note that the fringe pattern itself is related to the cosine term in Eq.~\ref{eq:real_part_correlation_function}.\par

Typically, the quantity $\gamma_{1,2}$, a complex number, is called as the (normalized) complex visibility, often denoted by $\widetilde{V}$. In a more general way, the complex visibility can be written as follows:

\begin{equation}
\widetilde{V}(u, v, \lambda) = |\widetilde{V}(u, v, \lambda)| \exp{[i\phi(u, v, \lambda)]},
\label{eq:complex_visibility}
\end{equation}
where $u$ and $v$ are the spatial frequencies of the projected baseline ($B_{\mathrm{proj}})$:

\begin{equation}
\frac{B_{\mathrm{proj}}}{\lambda} = \sqrt{u^2 + v^2}.
\label{eq:spatial_frequencies}
\end{equation}

First, as written in Eq.~\ref{eq:complex_visibility}, the complex visibility is explicitly dependent on the observing wavelength. Moreover, note that the spatial frequencies, defining the called $uv$ plan of the observations, are themselves dependent on $\lambda$ (Eq.~\ref{eq:spatial_frequencies})·\par

The real quantities $V$ and $\phi$ are the visibility amplitude and visibility phase of the fringe pattern. However, they are often simply called in literature, respectively, visibility and phase, and these terms are also used in the remaining of this thesis.\par 

The discussion above clarifies why we stated, in the previous section, that an interferometer is sensitive to the degree of coherence of the light, which is received by the telescope array. Thus, these are the most fundamental quantities -- the visibility and phase -- that can be retrieved from the measurement of a fringe pattern using an OLBI.\par

Concerning the fringe pattern, while the phase is related of the position of the fringes, the visibility is a measurement of the fringe contrast. It was firstly defined by~\citet{michelson90} as follows:

\begin{equation}
V = |\gamma_{1,2}| = |\widetilde{V}| = \frac{ I_{\mathrm{max}} - I_{\mathrm{min}}}{I_{\mathrm{max}} + I_{\mathrm{min}}},
\label{eq:def_visibility_norm}
\end{equation}
where $I_{\mathrm{max}}$ is the maximum fringe intensity (related to the bright bands) and $I_{\mathrm{min}}$ is the minimal one (related to the dark bands). Note that, as defined in Eq.~\ref{eq:def_visibility_norm}, the visibility is normalized as a function of intensity, ranging between 0 and 1.\par

Here, the ideal case of $V$ = 1 means an object that is fully unresolved by the interferometer, that is, ideally, a monochromatic point source. The other ideal case, $V$ = 0, means that the object is fully resolved by the interferometer, and then the interferometric fringes completely disappear. For example, see, again, Fig.~\ref{sec_interf_millour14_fig3}. From that, the visibility of the star is dropping, in the different cases, shown from the left to right panels, as the degree of spatial resolution is increased (in this case, due to the change of the object's angular size on the sky).\par

Up to this point, it must be clear that the fringe pattern encodes information on the spatial intensity distribution of the object (projected onto the sky). We can say that the essential goal of OLBI is to measure fringes under different degrees of coherence, showing visibilities between the two ideal cases, as described above, in order to obtain finer spatial information on the intensity distribution of the object. However, how could we retrieve such information from the measurement of the complex visibility?\par

The ``ace in the hole'' of interferometry is the so-called Zernike-van Cittert theorem~\citep{cittert34, zernike38}:

\begin{equation}
\widetilde{V}(u,v)_{\lambda} = \frac{\mathlarger{  \int \int S(\alpha,\delta)_{\lambda} \exp{[-2i\pi(u\alpha + v\delta)]} \, d \alpha \, d \delta }}{S^{\mathrm{total}}_{\lambda}},
\label{eq:van_cittert_theorem}
\end{equation}
where $\alpha$ and $\delta$ are the usual angular coordinates in the sky, and $S(\alpha,\delta)_{\lambda}$ is the brightness distribution of the source (projected onto the sky).\par

Thus, the Zernike-van Cittert theorem relates the complex visibility to the Fourier transform of the source brightness distribution. Also note that, in principle, given $\widetilde{V}(u,v)_{\lambda}$, we are thus able to retrieve $S(\alpha,\delta)_{\lambda}$ from the inverse Fourier transform of $\widetilde{V}(u,v)_{\lambda}$\footnote{Despite the apparent simplicity of this description, in the practice, it is indeed a hard task to perform an image reconstruction of the observed source, both from an observational and modeling point-of-views. For instance, see~\citet{millour12},~\citet{thiebaut17}, and references therein.}.\par

Note that, as defined above, the complex visibility is normalized by the total source brightness:

\begin{equation}
S^{\mathrm{total}}_{\lambda} = \int \int S(\alpha,\delta)_{\lambda} d \alpha \, d \delta.
\end{equation}

As explicitly set in Eq.~\ref{eq:van_cittert_theorem}, the Zernike-van Cittert theorem works for each value of wavelength. Apart from the dependence on $\lambda$ from the spatial coordinates $u$ and $v$ in the Fourier plan (Eq.~\ref{eq:spatial_frequencies}), $S(\alpha,\delta)$ can be highly dependent on $\lambda$, so $\widetilde{V}$ must also change with respect to the wavelength. For simplicity, this dependence of the Zernike-van Cittert theorem on the wavelength will be omitted in the remaining of this section.\par

Besides the own technical issues related to the quality of the fringe measurements, due to instrumental and atmospheric effects, one of the biggest challenges of interferometric studies relies on filling the $uv$ plan in order to retrieve finer spatial information about the object, covering different values of spatial frequencies (in the Fourier domain). This means recording fringes under different degrees of coherence. In practice, this task requires multiple observations performed using different arrays configurations (different baseline lengths and orientations on the sky).\par

\begin{figure}[t]
\centerfloat
\centerline{\resizebox{0.35\textwidth}{!}{\includegraphics{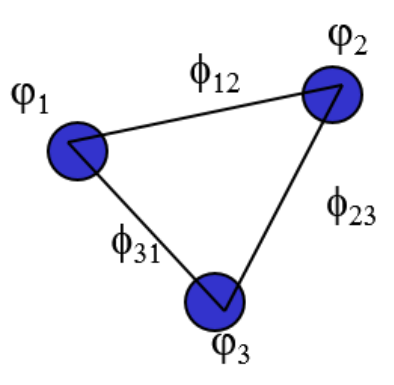}}}
\caption[Schematic of a triplet of telescopes used for measuring closure phase.]{Schematic of a triplet of telescopes used for measuring closure phase. The atmospheric phase, introduced in the fringe measurement, on each telescope, is indicated by $\phi_{1}$, $\phi_{2}$, and $\phi_{3}$. The true phase of the object is also indicated by $\Phi_{12}$, $\Phi_{23}$, and $\Phi_{31}$. See text for discussion. Reproduced from~\citet{mourard19}.}
\label{sec_interf_mourard19_fig97}
\end{figure}


\subsection{Closure phase}

We point out that the true information about the object's phase is indeed lost, when measuring fringes with ground-based telescopes.\par 

This happens due to the alteration or disturbance of the wavefronts by the atmosphere, as the atmospheric turbulence displaces the received wavefronts, destroying the object phase signal. This is known as the piston effect. Fortunately, it is possible to partially overcome this issue by the measurement of another quantity, the closure phase, which is related to the phase, but measured using different triplets of telescopes (Fig.~\ref{sec_interf_mourard19_fig97}).\par 

For this purpose, it is necessary to employ an array composed of at least three telescopes, as indicated in Fig.~\ref{sec_interf_mourard19_fig97}. From that, $\Phi_{12}$, $\Phi_{23}$, and $\Phi_{31}$ would correspond to the object fringe phases that would be measured by each pair of telescopes without atmospheric effects. The phases that are, in fact, measured by each baseline, denoted here by $\psi$, are then:

\begin{equation}
\begin{aligned}
\psi_{12} = \Phi_{12} + \phi_{1} - \phi_{2},
\\
\psi_{23} = \Phi_{23} + \phi_{2} - \phi_{3},
\\
\psi_{31} = \Phi_{31} + \phi_{3} - \phi_{1},
\end{aligned}
\label{eq:closure_phase1}
\end{equation}
where $\phi$ is the piston effect introduced by the atmospheric turbulence in wavefronts collected by each telescope.\par

From Eq.~\ref{eq:closure_phase1}, one sees that the piston effect can be cancelled by the sum of $\psi_{12}$, $\psi_{23}$, and $\psi_{31}$, and thus, defining the closure phase $CP$:

\begin{equation}
CP = \psi_{12} + \psi_{23} + \psi_{31},
\label{eq:closure_phase2}
\end{equation}
that is, a certain triplet configuration is associated to one measurement of closure phase.\par 

This interferometric quantity is important, for example, to the study of circumstellar environments, as it provides information on the amount of asymmetry of the object. Considering that the object is at least partially resolved, the measurement of $CP$ traces the photocenter of the object projected onto the sky~\citep[e.g., see][]{monnier03_closure_phases}. For instance, considering a center-symmetric intensity map, as arising from a spherically symmetric outflow, the measured $CP$ would be null (equivalently, equal to -\ang{180} or +\ang{180}). On the other hand, in this case, any departure from a center-symmetric intensity map would result in non-zero closure phases (ranging between -\ang{180} and +\ang{180}). Further details on the mathematical description of closure phases can be found in~\citet{monnier00} and~\citet{monnier03_closure_phases}.\par 


\subsection{Visibility calibration}

Concerning the visibility of a certain target, we are able to estimate its true, or absolute, visibility by also observing close stars (in angular size) to the target, in order to overcome atmospheric and instrumental effects, using the called transfer function $TF$:

\begin{equation}
TF = \frac{V^{\mathrm{cal}}_{\mathrm{obs}}}{V^{\mathrm{cal}}_{\mathrm{true}}}, \, 0 \leq TF \leq 1, 
\label{eq:transfer_function}
\end{equation}
where $V^{\mathrm{cal}}_{\mathrm{obs}}$ is the measured visibility of the calibrator star and $V^{\mathrm{cal}}_{\mathrm{true}}$ is its (true) absolute visibility.

The estimation of the true visibility of the target, $V^{\mathrm{target}}_{\mathrm{true}}$, is then given by:

\begin{equation}
V^{\mathrm{target}}_{\mathrm{true}} = \frac{V^{\mathrm{target}}_{\mathrm{obs}}}{TF},
\label{eq:true_visibility}
\end{equation}
where $V^{\mathrm{target}}_{\mathrm{obs}}$ is the measured visibility of the target object.\par 

Note that $V^{\mathrm{cal}}_{\mathrm{true}}$ is a prior information to be used in Eq.~\ref{eq:transfer_function}. Thus, it is a good practice to select (nearly) unresolved sources, that is, $V^{\mathrm{target}}_{\mathrm{true}}$ $\sim$ 1, for a certain interferometric configuration, or with a very accurate determination of angular diameter. To further details on interferometric calibration, we refer the interested reader to \cite{percheron08}.\par

\section{Spectro-interferometry}
\label{sec_interf_spectro_interferometry}

Spectro-interferometry, also called differential interferometry~\citep{petrov89}, or self-phase-referencing interferometry~\citep{woilez12}, means that the measured fringes patterns are dispersed along the spectral dimension using a spectrograph. This technique was first used with the I2T interferometer by~\citet{koechlin79} to detect differential phase effects on Capella..\par

In Fig.~\ref{sec_interf_weigelt07_fig1}, we show the interference fringes of the LBV star $\eta$ Carinae observed in high spectral resolution mode ($R$ = 12000) with the VLTI/AMBER instrument in the K-band ($\sim$2.156-2.175 $\mu$m, including the Br$\gamma$ line). One clearly sees how the contrast of the fringes is lower in Br$\gamma$ (brighter region), that is, showing a lower value of visibility, when compared with the close-by continuum, due to the larger flux contribution from its strong wind in Br$\gamma$ than in the continuum. Indeed, looking the near-infrared spectrum of $\eta$ Carinae, one can find a strong and broad Br$\gamma$ emission-line, indicating then a significant line formation through the wind region~\citep[e.g., see Fig.~2 of][]{weigelt07}.\par

It is important to stress that spectro-interferometry enables to access much more information than the one provided by single-band interferometry, but performed sequentially at different spectral bands. Spectro-interferometry introduces new quantities: the differential visibility and differential phase (including here the closure phase).\par

\begin{figure}[t]
\centerfloat
\centerline{\resizebox{0.50\textwidth}{!}{\includegraphics{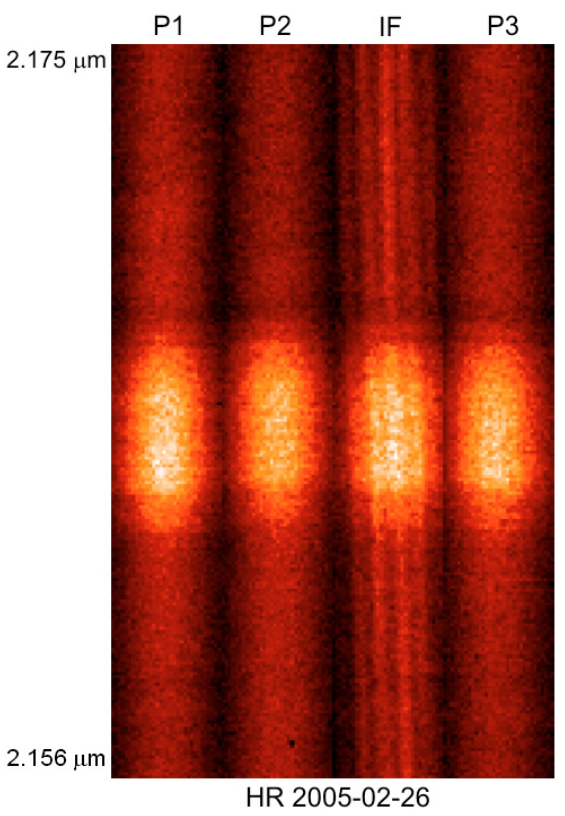}}}
\caption[Spectrally dispersed ($\sim$2.156-2.175 $\mu$m) fringes of $\eta$ Carinae measured with VLTI/AMBER in high spectral resolution mode (HR, $R$ = 12000) at 26 February 2005.]
{Spectrally dispersed ($\sim$2.156-2.175 $\mu$m) fringes of $\eta$ Carinae measured with VLTI/AMBER in high spectral resolution mode (HR, $R$ = 12000) at 26 February 2005. The spectrally dispersed fringe signal is shown in IF, while the photometric calibration signals from the three telescopes are shown from P1 to P3. Note that the bright regions are associated with the Doppler-broadened Br$\gamma$ emission (2.166 $\mu$m). Reproduced from~\citet{weigelt07}.}
\label{sec_interf_weigelt07_fig1}
\end{figure}

Considering a narrow region in wavelength, one might expect that the atmospheric effects on the wavefronts are roughly the same. Thus, we are able to retrieve information of the object, in a certain spectral line, relatively to other reference measurement simultaneously observed, at a wavelength close to the line. Usually, the continuum region, close to the line, is chosen as the wavelength of reference in order to calculate the differential quantities (see Fig.~\ref{sec_inter_meilland14_fig8_9}). Nevertheless, this choice can have some caveats, for example, a Be disk can contribute up to 20\% of the total continuum emission, being thus relativity significant when compared with the flux contribution from the central star~\citep[see, e.g.,][]{stee94}. In advance of discussion, we stress, however, that this problem does not stand in our study of the Be star $\omicron$ Aquarii, showing a flux contribution from the disk in the H$\alpha$ close-by continuum of 0 up to 6\%~\citep[see Table 2 of][]{paperIII}.\par 

Furthermore, having measurements of differential visibilities, and also of absolute visibilities (as given by Eq.~\ref{eq:true_visibility}), for example, observed at a close-by continuum region to the spectral line, we are thus able to retrieve information on the absolute visibility of the object with spectral resolution.\par

As pointed out in Sect.~\ref{sec_interf_hra}, interferometric observations are needed in order to measure the geometry and extension of the stellar surfaces and their environments. Attempting this task is much harder, or simply impossible, using single-mirror telescopes, being marginally feasible just for the nearer and brighter objects (such as red giants and AGB stars), and using others high angular resolution techniques, as with the instruments of VLT/SPHERE~\citep[see, e.g.,][]{kervella15, ohnaka16, khouri16}. A detailed overview on the SPHERE instruments can be found in~\citet{beuzit19}.\par

Combining simultaneously the power of spectroscopy and interferometry, spectro-interferometry is one of the most suitable observational methods for the analysis of stellar environments. In addition to information about the spatial distribution of their environments, as provided by non-spectrally-dispersed interferometry, this technique also allows us to probe its kinematics by observing certain spectral lines, for example, as H$\alpha$, Br$\gamma$, and Br$\gamma$. Furthermore, as discussed in Sect.~\ref{sec_spectro_line_diagnostics} (see, again, Table \ref{table_line_diagnostics}), these lines are the most sensitive to probe the physical conditions (as the local density that is related to the mass-loss rate) of the atmospheres and environments of massive hot stars.\par

For example, we show, in Fig.~\ref{sec_inter_meilland14_fig8_9}, three different scenarios for a stellar environment, that is, having different kinematics: a purely-rotation Keplerian disk, a purely-expanding disk (i.e., an equatorial outflow), and a mixed case between these extreme cases. These models are calculated using a kinematic code presented in details in~\citet{meilland12} and~\citet{cochetti19}. In short, in this case, as shown in Fig.~\ref{sec_inter_meilland14_fig8_9}, the intensity map, representing the flux contribution due to the circumstellar environments is assumed as a simple uniform disk following these kinematics mentioned above. Details on this kinematic code will be properly addressed in Sect.~\ref{sec_analytical_modeling_kinematic_models}.\par

\begin{figure}
  \begin{adjustbox}{minipage=\textwidth,scale=1.00}
  \centering
  \includegraphics[width=1.00\columnwidth]{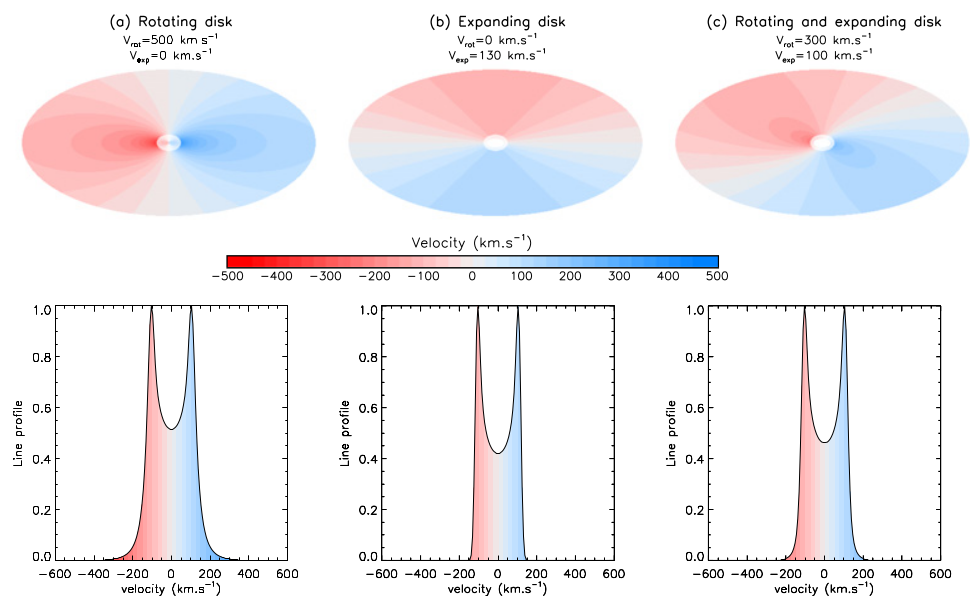}
  \medskip
  \includegraphics[width=1.00\columnwidth]{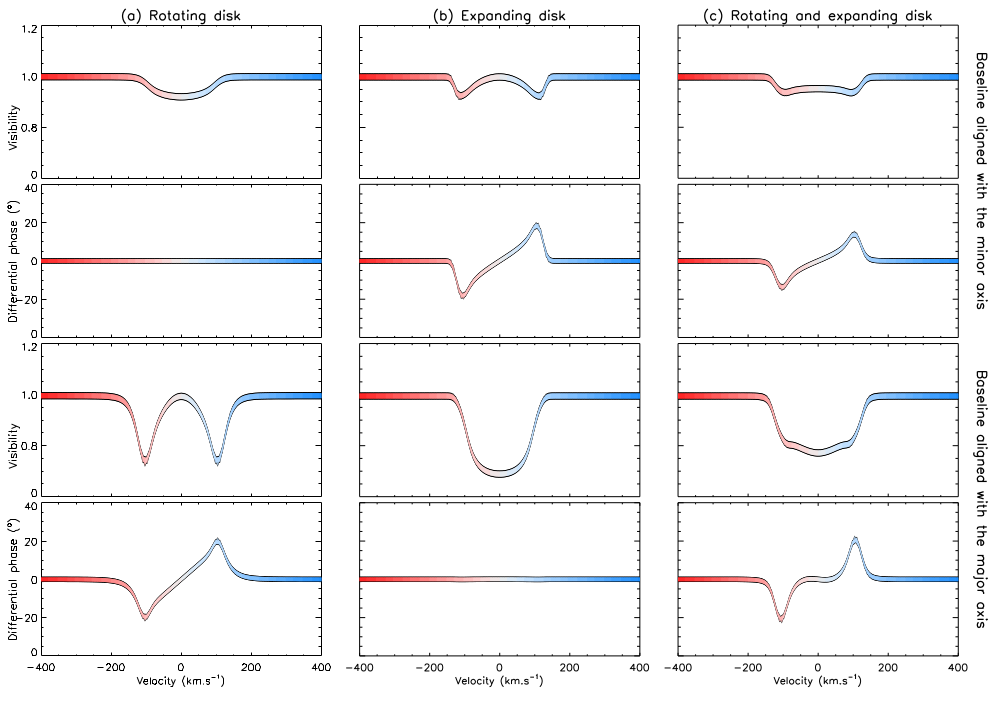}
  \end{adjustbox}
  \caption[Different scenarios for the kinematics of a circumstellar disk, calculated using a kinematic model: purely-rotating disk, purely-expanding disk, and a hybrid case, respectively, from the left to the right panels.]{Different scenarios for the kinematics of a circumstellar disk, calculated using a kinematic model: purely-rotating disk, purely-expanding disk, and a hybrid case, respectively, from the left to the right panels. The top panels show the iso-velocity maps for these different models. For each one of these models, the synthetic H$\alpha$ line profile is shown in the second row. The synthetic differential interferometric quantities, the visibility and phase, are also shown, but for two extreme cases of interferometer baseline alignment, with respect to the circumstellar disk major-axis. See text for discussion. Reproduced from~\citet{meilland14}.}
\label{sec_inter_meilland14_fig8_9}
\end{figure}

From Fig.~\ref{sec_inter_meilland14_fig8_9}, one sees that the morphology of the observed synthetic line profiles, calculated in H$\alpha$, are very similar among all the different kinematical scenarios, apart from the line wings. Nevertheless, it is important to stress that the H$\alpha$ wings of hot stars, as the Be stars, can be highly affected by non-kinematic line-broadening, as due to non-coherent scattering~\citep[e.g.,][]{hummel92, delaa11}. Moreover, the emission line profile will be affected by changing the geometry representing the circumstellar disk in the model, for example, a Gaussian distribution, instead of the simplest case of a uniform disk.\par 

In short, disentangling the kinematics of these environments, only based on spectroscopy, is clearly a difficult task.\par 

In addition to the spectral line profiles, Fig.~\ref{sec_inter_meilland14_fig8_9} also shows the predicted differential interferometric observables, the differential visibility and phase, in the H$\alpha$ line. Here, two different cases are shown, one for an interferometry baseline aligned along the disk major-axis, and other for baseline perpendicular to the disk major-axis. From that, one clearly sees, due to the ability of spatial resolution, how both the differential visibility and phase are sensitive to different considered kinematics, and to the different baseline configuration with respect to the object. For instance, considering the interferometer baseline perfectly aligned along the disk major-axis (purely-expanding disk case, middle panel of Fig.~\ref{sec_inter_meilland14_fig8_9}) results in null value of phase since the projected velocity of the envelope is null through all the disk extension, being then the photocenter of the object unshifted in the sky regardless the value of wavelength. \par

\begin{table}[t]
\begin{adjustbox}{width=1.00\textwidth}
\begin{threeparttable}
\caption{\label{table_spectro_interf_instruments} Summary of telescope arrays with spectro-interferometric instruments. Adapted from~\citet{hadjara17}.}\
\centering
\renewcommand{\arraystretch}{1.10}
\begin{tabular}{lccccc}
\toprule
\toprule
\multicolumn{1}{l}{Array} & \multicolumn{1}{c}{Location} & \multicolumn{1}{l}{Year} & \multicolumn{1}{c}{\textnumero \,of telescopes} & \multicolumn{1}{c}{$B_{\mathrm{max}}$ (m)} & \multicolumn{1}{c}{$\lambda$ ($\mu$m)}  \\
\midrule 

NPOI\tnote{a} & Anderson Mesa (USA) & 1994-- & 6 & 435 & 0.45-0.85 (B-I) \\

CHARA\tnote{b} & Mount Wilson (USA) & 2001-- & 6 & 331 & 0.45-2.50 (B-K) \\

VLTI\tnote{c} & Cerro Paranal (Chile) & 2001-- & 4 & 202 & 1.2-13.0 (J-N) \\

ISI\tnote{d} & Mount Wilson (USA) & 1988-- & 3 & 70 & 8-13 (N) \\

SUSI\tnote{e} & Narrabri (Australia) &1991-- & 2 & 640 & 0.4-0.9 (B-I) \\

\textsuperscript{\textdagger}PTI\tnote{f} & Mount Palomar (USA) & 1995--2008  &2 & 110 & 1.5-2.4 (H-K) \\

\textsuperscript{\textdagger}KECK-I\tnote{g} & Mauna Kea (USA) & 2003-2012 & 2 & 85 & 2.2-10.0 (K-N) \\

\bottomrule
\end{tabular}

\begin{tablenotes}
\item[a] Navy Precision Optical Interferometer.
\item[b] Center for High Angular Resolution Astronomy.
\item[c] Very Large Telescope Interferometer.
\item[d] Berkeley Infrared Spatial Interferometer.
\item[e] Sydney University Stellar Interferometer.
\item[f] Palomar Testbed Interferometer.
\item[g] Keck Interferometer. 
\item[\textdagger] Discontinued.  

\end{tablenotes}

\end{threeparttable}
\end{adjustbox}

\end{table}


\section{Spectro-interferometric instruments}

Table \ref{table_spectro_interf_instruments} summarizes the current (and recently discontinued) observational facilities with high spectral resolution spectro-interferometric instruments. In addition to their location and operating date, the number of telescope configuration, maximum baseline length, and the wavelength range (and the corresponding photometric bands) are also shown here. In particular, we point out that the CHARA and VLTI Arrays host the current state-of-the-art spectro-interferometric instruments.\par 

In the following sections, we briefly describe some of the instruments hosted on these arrays, in particular, the CHARA/VEGA and VLTI/AMBER spectro-interferometric instruments, since large sets of data observed with these instruments are analysed in this thesis (Sect.~\ref{sec_results_omicron_aquarii}).\par

\subsection{The CHARA array}
\label{sec_interf_chara}

The Center for High Angular Resolution Astronomy (CHARA\footnote{\url{http://www.chara.gsu.edu/}}) is an optical-interferometric array, owned by Georgia State University, located at Mount Wilson Observatory, California (USA).\par 

CHARA is composed of six 1-m class telescopes, forming a Y-shape array, as shown in Fig.~\ref{sec_interf_chara_array}, in order to optimize the $uv$ plan coverage. The light from the telescopes are sent through vacuum pipes to a recombining laboratory equipped with fixed and movable delay lines. In total, there are 15 different baseline configurations, formed by each pair of two telescopes, being the S2-S1 configuration the shortest with $\sim$34 m, while E2-S1 is the largest one with $\sim$331 m (see the maximum baseline length in Table \ref{table_spectro_interf_instruments}). A full list describing the CHARA baseline configuration can be found in Table 1 of~\citet{tenbrummelaar05}. Further technical details on the elements composing the CHARA Array also are described in~\citet{tenbrummelaar05}.\par

\begin{figure}[t]
\centerfloat
\centerline{\resizebox{0.85\textwidth}{!}{\includegraphics{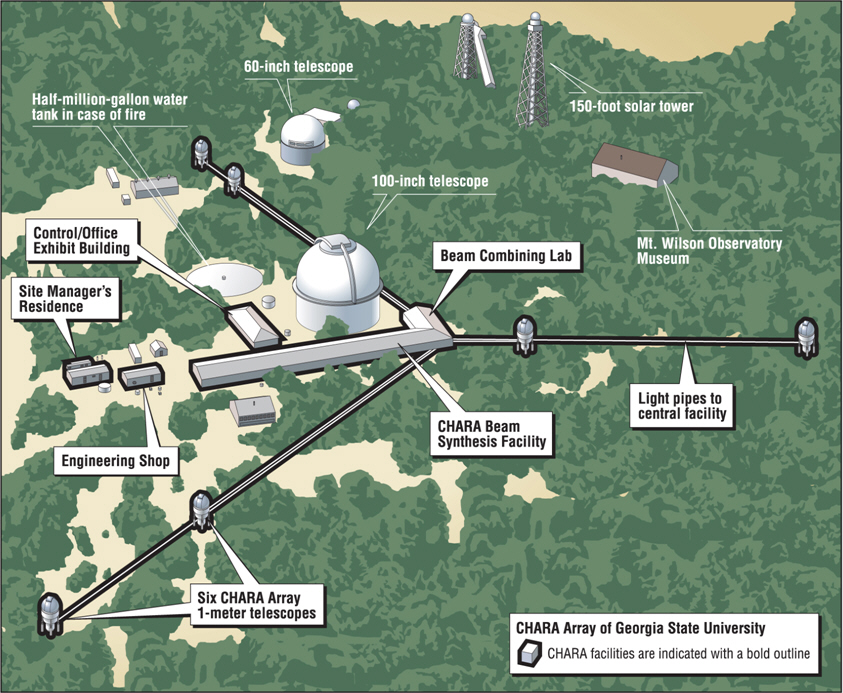}}}
\caption[Schematic layout of the CHARA Array, installed on the Mount Wilson Observatory (California, USA).]{Schematic layout of the CHARA Array, installed on the Mount Wilson Observatory (California, USA). Source: \url{http://www.chara.gsu.edu/}.}
\label{sec_interf_chara_array}
\end{figure}

Currently, six spectro-interferometric are working at CHARA, operating in the near-infrared or in visible region, namely: 

\begin{enumerate}[label=(\roman*)]
\setlength\itemsep{1em}

\item CLASSIC and CLIMB: JHK-bands~\citep[][]{tenbrummelaar13}

\item JouFLU\footnote{Acronym for ``Jouvence of FLUOR'', being the Fiber Linked Unit for Optical Recombination (FLUOR) an older near-infrared instrument (see references above).}: K-band~\citep[][]{foresto03, scott13}

\item MIRC-X\footnote{An upgrade of the Michigan Infrared Combiner (MIRC) instrument (see references above).}: JH-bands~\citep[][]{monnier04, anugu18}

\item PAVO\footnote{Acronym for ``Precision Astronomical Visible Observations''.}: 6300-9500 {\AA}~\citep{ireland08}

\item VEGA\footnote{Acronym for ``Visible spEctroGraph and polArimeter''.}: 4800-8500 {\AA}~\citep{mourard09}

\end{enumerate}

Jointly, these instruments were used, from 2005 to date, in about 90 refereed articles, ranging from the study of circumstellar disks in low and massive stars up to the characterization of exoplanets.\par

Among other on-going instrumental projects, we point out the Fibered spectrally Resolved Interferometer -- New Design (FRIEND) instrument, a new visible beam-combiner, covering the spectral interval of $\sim$6200-8500 {\AA}~\citep{martinod18}. This spectro-interferometric instrument was designed to overcome certain limitations of the VEGA instrument, as  unreliable estimations of closure phase~\citep[see, e.g.,][]{mourard12}. It was mainly a test instrument for the future six-telescope visible combiner SPICA at CHARA~\citep{mourard17}.\par


\subsection{CHARA/VEGA}
\label{sec_interf_vega}

Despite theses limitations, VEGA~\citep[Visible spEctroGraph and polArimeter,][]{mourard09} is a unique instrument with respect to the capability of combining both high spatial resolution (up to $\sim$0.3mas) and high spectral resolution (up to $R$ = 30000 in the high resolution mode) in the visible domain. Currently, VEGA can combine simultaneously up to four beams operating at different wavelengths from 4500 {\AA} to 8500 {\AA}, that is, including the H$\alpha$ line, being particularly important to perform spectro-interferometric studies of massive hot stars and their environments. As a non-exhaustive list of works on these objects, based on VEGA observations, see~\citet{chesneau10},~\citet{meilland11},~\citet{delaa11},~\citet{stee12a}, and \cite{mourard15}, for example\footnote{A full list of refereed articles, using the VEGA instrument, can be found in the web-site: \url{https://www.oca.eu/fr/publications-vega}.}.\par

Together with PAVO, mentioned above, VEGA is one of the two visible spectro-interferometric instruments on CHARA. One clear advantage of VEGA, with respect to PAVO ($R$ = 30), relies on its high instrumental spectral resolution ($R$). To date, it is the only instrument operating at CHARA with a spectral resolution high enough to resolve narrow spectral features such as atomic and molecular lines, in short, offering three spectral modes: $R$ = 1000 (LR, low resolution mode), $R$ = 6000 (MR, medium resolution mode), and $R$ = 30000 (HR, high resolution mode).\par


\subsection{The VLTI array}
\label{sec_interf_vlti}

The Very Large Telescope Interferometer (VLTI) is the interferometric mode of the Very Large Telescope (VLT), located at Cerro Paranal Observatory in the Atacama desert. It is owned by the European Southern Observatory (ESO), an intergovernmental consortium, founded on 1962, actually composed of 16 state-members\footnote{Further details can be found in \url{https://www.eso.org/public/about-eso/memberstates/}.}. Fig.~\ref{sec_interf_vlt_view} shows an aerial view of the ESO/VLT facility. The interested reader can see~\citet{blaauw91} for a historical overview of ESO.

Despite the initial objection, during the 1970's, of the astronomical community to the implementation of a large interferometric facility, the same was rapidly encouraged by successful works with optical long-baseline interferometers in the 1980's, for example, as with the GI2T (Plateau de Calern, France), a direct upgrade from the pioneer I2T interferometer. As pointed in Sect.~\ref{sec_interf_historical_overview}, the GI2T was the first interferometric facility used to successfully probe the circumstellar environments of hot stars. In short, ESO approved the construction of the VLT project, also including a interferometric model (VLTI), in 1987~\citep[see Sect.~5 of][]{lena14}.\par

\begin{figure}[t]
\centerfloat
\centerline{\resizebox{1.00\textwidth}{!}{\includegraphics{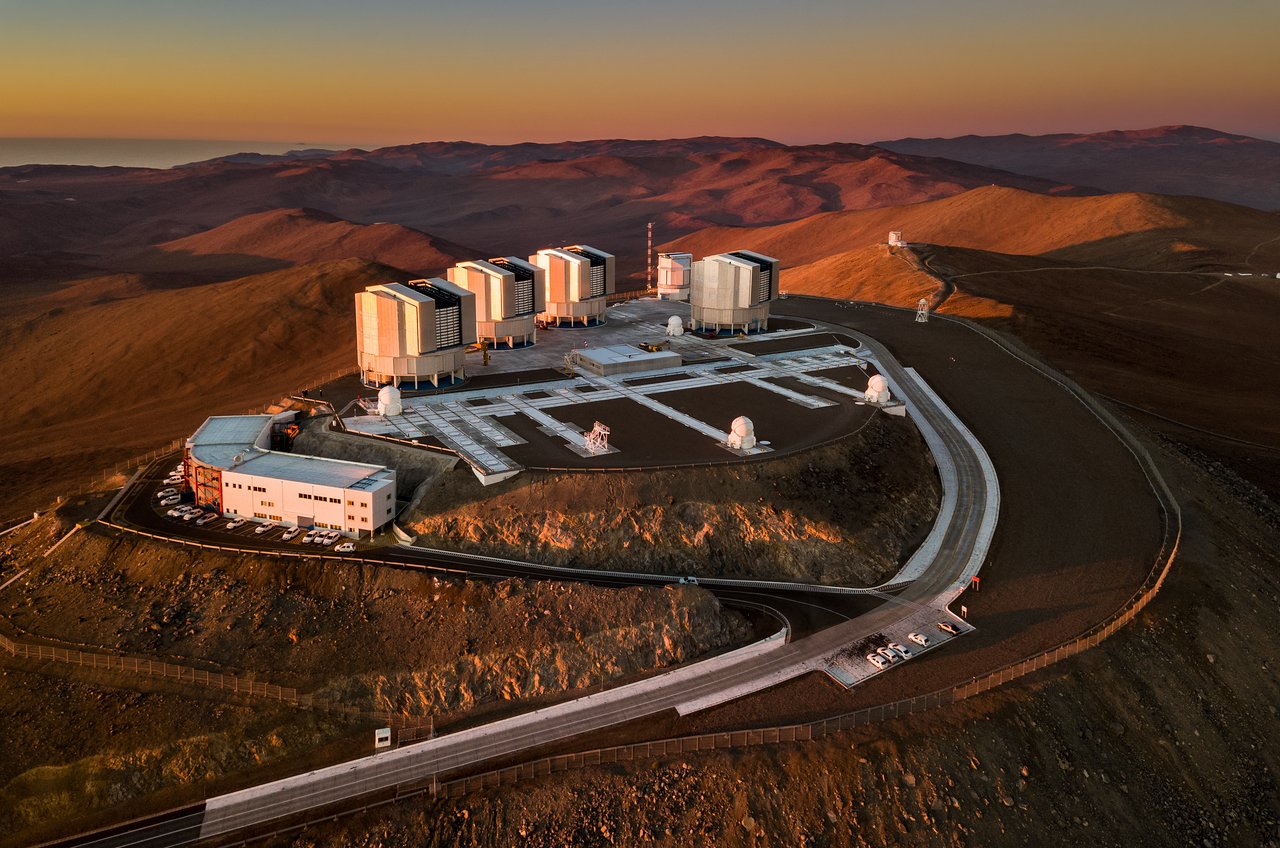}}}
\caption[Aerial view of the Very Large Telescope's observing platform, installed at the Cerro Paranal Observatory (Atacama desert, Chile).]{Aerial view of the Very Large Telescope's observing platform, installed at the Cerro Paranal Observatory (Atacama desert, Chile). The 4 fixed 8-m telescopes and the 4 movable 1.8-m telescopes are shown here. Source: \url{https://www.eso.org/}.}
\label{sec_interf_vlt_view}
\end{figure}

The VLTI Array is composed of four fixed 8.2-m telescopes, called as Unit Telescopes (UTs), in addition to four movable 1.8-m telescopes, called as Auxiliary Telescopes (ATs). A layout of the VLTI baselines is shown in Fig \ref{sec_interf_vlti_array}. The shortest baseline configuration (B0-C0) has about 8 m, while the largest one has about 202 m (B5-J6). One clear advantage of employing movable telescopes for interferometry is to increase the number of possible baseline configurations, reaching up to 518 different configurations with VLTI in total. Then the  ability of covering the $uv$ plan is drastically improved in this case, when compared with facilities using fixed telescopes as CHARA. A complete list of the baselines on VLTI can be found in the web-site for the ``VLTI Station Position Technical Data'', hosted by ESO: \url{https://www.eso.org/observing/etc/doc/viscalc/vltistations.html}.

The VINCI beam-combiner started to be built in 1998 and it was the first VLTI instrument measuring interference fringes in 2001, in the near-infrared (K-band). This was an instrument designed to test and tune the performance of the VLTI Array~\citep[see, e.g.,][]{kervella00, glindemann02}. Despite being essentially a test instrument for VLTI, observations with VINCI resulted in a significant number of studies. Regarding the photosphere and environments of hot stars, for example, see~\citet{domiciano03},~\citet{domiciano05},~\citet{kervella07}.\par

\begin{figure}[t]
\centerfloat
\centerline{\resizebox{0.90\textwidth}{!}{\includegraphics{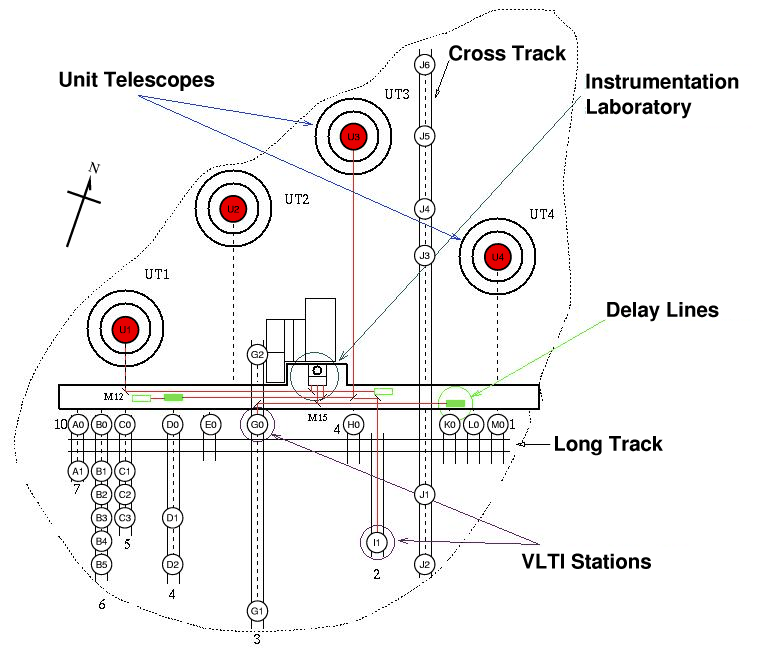}}}
\caption[Schematic layout of the VLTI Array. The Unit Telescopes (UT) are indicated in red, fixed 8-m telescopes, with stations named from U1 to U4.]{Schematic layout of the VLTI Array. The Unit Telescopes (UT) are indicated in red, fixed 8-m telescopes, with stations named from U1 to U4. The possible positions of the Auxiliary Telescopes (AT), 1.8-m telescopes, have stations named from A0 to M0. Source: \url{https://www.eso.org/}.}
\label{sec_interf_vlti_array}
\end{figure}

From that, followed the instruments MIDI (mid-infrared, N-band) in 2003 and AMBER~\citep[JHK-bands,][]{petrov07} in 2004. They are called as the VLTI first generation of interferometric instruments. For a summary of their characteristics, for example, see Table 2 of~\citet{richichi05}. The VLTI second generation is composed by GRAVITY~\citep[K-band,][]{gravity_collaboration17} and MATISSE~\citep[LMN-bands,][]{lopez14} instruments, with first light very recently in 2016 and 2018, respectively.\par

In summary, these instruments were employed in a large number works, to date, up to about 460 referee articles in total. From them, just PIONIER\footnote{It is a 4-telescope visitor spectro-interferometric instrument at VLTI~\citep[H-band,][]{lebouquin11}, constituting a gap between the first and second VLTI generations.}, GRAVITY, and MATISSE are still working, being the others decommissioned. Also, note that all the instruments installed on VLTI operate, or operated, in the near-infrared (JHK-bands), such as AMBER and GRAVITY, or longer wavelengths up to the mid-infrared region, such as MATISSE (LMN-bands) and MIDI (N-band).\par

In addition to these references above, as a non-exhaustive list of work on hot stars with VLTI instruments, we also refer the reader to~\citet{meilland07a},~\citet{meilland07b},~\citet{domiciano07},~\citet{meilland12},~\citet{weigelt16}, and~\citet{cochetti19}, being all these latter ones based on VLTI/AMBER observations.\par


\subsection{VLTI/AMBER}
\label{sec_interf_amber}

The Astronomical Multi-BEam combineR~\citep[AMBER,][]{petrov07} was a three beam combiner operating in the H- and K-bands, with three spectral resolution models: $R$ = 35 (LR), $R$ = 1500 (MR), and $R$ = 12000 (HR). It was the instrument installed at VLTI offering the highest spectral resolution thanks to its HR mode, and thus the most adapted to study gaseous circumstellar environment in emission lines.\par 

Despite being a decommissioned instrument, by the end of 2017, a large number of observations performed with AMBER are still used in very recent works about circumstellar environments~\citep[e.g.,][]{koutoulaki18, cochetti19, adam19, ohnaka19, hadjara19}, and it yielded the largest number of articles among the VLTI instruments to date.\par


\subsection{Comparison between CHARA and VLTI}

With respect to CHARA (Sect.~\ref{sec_interf_chara}), we point that one caveat of VLTI (Sect.~\ref{sec_interf_vlti}) relies on its shorter maximum baseline length, of about 202 m, allowing us to reach lower angular resolutions than the ones with the CHARA Array (baselines up to $\sim$331 m). Furthermore, currently, there are no spectro-interferometric instrument, operating at the visible region, installed on the VLTI Array. As discussed above, to date, all the VLTI instruments work from the near- to the mid-infrared region. For instance, see the interesting science cases pointed out by~\citet{millour18} for a future visible spectro-interferometric instrument on VLTI.\par 

On the other hand, VLTI offers larger (1.8-m and 8-m mirror), and also movable telescopes, instead of the 1-m fixed telescopes on CHARA. As pointed out above, the usage of the movable auxiliary telescopes of VLTI allows us to reach up to 518 different baseline configurations, in instead of the 15 baselines with CHARA. For sure, this results in a better coverage of the $uv$ plan. In addition, VLTI has shorter baselines, of about 8m, the ones on CHARA, of about 34 m, and then mapping smaller spatial frequencies in the Fourier space.\par 

In conclusion, the spectro-interferometric instruments currently mounted on CHARA and VLTI are thus complementary in order to perform more detailed interferometric studies on hot stars, in the framework of a multi-wavelength approach, covering both the visible and the near-infrared regions, at different levels of angular resolution. This complementary task, relying both on CHARA and VLTI observations, is part of the objective of this thesis, when studying the surface and environment of the Be star $\omicron$ Aquarii (Sect.~\ref{sec_results_omicron_aquarii}).\par

\pagestyle{empty}
\cleardoublepage
\pagestyle{fancy}

\chapter{Radiative transfer modeling}
\label{chapter_radiative_transfer_modeling}
\minitoc

\section{Elementary concepts of radiative transfer}
\label{sec_radiative_transfer_modeling_elementary_concepts}

Stellar radiation flux contains information on the physics and chemistry of the matter composing the photosphere and the circumstellar environment, as the emergent radiation field results from the interaction of the photons with both of them. The goal of radiative transfer modeling is to explain such radiation field in terms of (almost) fundamental parameters, for the photosphere, such as the stellar luminosity, effective temperature, surface gravity, and for the circumstellar environment, such as the mass-loss rate, the density, and the velocity field.\par 

This is usually a hard task to be performed. As result of the interaction photon-particle, both the radiation and gas fields are modified in the end, being thus a coupled problem to be solved. This means that, in order to predict, as precisely as possible, the stellar radiation, a stellar atmosphere model must also describe the state of the gas, that is, to know how are distributed the atoms over bound and free states, and thus requiring knowledge on the thermodynamics and hydrodynamics of the gas.\par 

Before setting the radiative transfer problem itself, some quantities of interest are discussed in the following. The specific (or also called monochromatic) intensity is defined as follows:

\begin{equation}
I_{\nu} = \frac{dE_{\nu}}{dA \, dt \, d\Omega \, d\nu}, 
\label{eq:specific_intensity}
\end{equation}
and thus describing the amount of energy $dE$ going through an surface $dA$ from light beams with frequency ranging between $\nu$ and $\nu$ + $d\nu$, forming an solid angle $d\Omega$, per unit time $dt$. Besides the own dependence on the light's frequency, also note that the intensity depends on the spatial direction due to the dependence on the solid angle.\par

Another very useful related quantity is the specific flux $F_{\nu}$:

\begin{equation}
F_{\nu} =  \int_{0}^{2\pi} \int_{0}^{\pi} I_{\nu}(\theta,\phi) \cos{{\theta}} \, \sin{{\theta}} d\theta d\phi,
\label{eq:specific_flux}
\end{equation}
where $\theta$ and $\phi$ are the usual spherical coordinates. This is the net amount of energy that is emitted by the star at the frequency $\nu$.\par

Formally, the radiative flux comes from the moments of the specific intensity:

\begin{equation}
M_{\nu}^{n} = \frac{1}{2} \int_{-1}^{+1} I_{\nu}  \, \mu^{n}  \, d\mu,    
\label{eq:moments_intensity}
\end{equation}
where $M_{\nu}^{n}$ is the moment of n-th order and $\mu$ = $\cos{\theta}$ (Fig.~\ref{sec_modeling_radiative_transfer_plane_parallel}). The so-called Eddington Flux, $H_{\nu}$, is then derived from first moment of the intensity  ($M_{\nu}^{1}$):

\begin{equation}
H_{\nu} = \frac{1}{2} \int_{-1}^{+1} I_{\nu} \, \mu \, d\mu,
\end{equation}
and is related to the specific flux (Eq.~\ref{eq:specific_flux}) by:

\begin{equation}
F_{\nu} = 4\pi H_{\nu}.
\label{eq:relation_mono_eddington_flux}
\end{equation}

Another very useful quantity, which is used for solving the radiative transfer problem~\citep[e.g., see Eq.~2 of][]{hillier98}, is derived from the zero-order intensity moment ($M_{\nu}^{0}$):

\begin{equation}
J_{\nu} = \frac{1}{2} \int_{-1}^{+1} I_{\nu} \, d\mu,
\end{equation}
being $J_{\nu}$ the mean intensity, at a certain position and wavelength, evaluated over all the directions.\par

Note that the spatial information -- how the energy flow is angularly distributed ($\theta$,$\phi$) -- is lost in the definition of flux, in Eq.~\ref{eq:specific_flux}, when compared with Eq.~\ref{eq:specific_intensity}, due to the integration on solid angle. This means that the flux is the most natural quantity to be addressed by (unresolved) spectroscopy, being this the case for stars, while interferometry is sensible to the intensity distribution of the star (across the sky).\par

Furthermore, unlike the intensity, the flux is dependent on the distance to the source. Thus, by the energy conservation and the Stefan-Boltzmann theorem (Eq.~\ref{eq:stefan_boltzmann}):

\begin{align}
\label{eq:energy_conservation_obs_flux}
\begin{split}
L_{\star} = 4\pi R_{\star}^{2} F = 4\pi d^{2} F_{\mathrm{obs}},
\\
F_{\mathrm{obs}} =  \left(\frac{R_{\star}}{d}\right)^2 F,
\end{split}
\end{align}
where, as usual, $L_{\star}$ and $R_{\star}$ are the stellar bolometric luminosity and radius, $F$ is the emergent flux at the stellar surface, and $F_{\mathrm{obs}}$ the observed flux at the distance $d$ to the star.\par 

In Eq.~\ref{eq:energy_conservation_obs_flux}, since we are dealing with the bolometric luminosity $L_{\star}$, the flux quantities are integrated in frequency:

\begin{equation}
F = \int_{0}^{\infty} F_{\nu}\,d\nu.
\end{equation}

In the most general form, the equation of radiative transfer can be expressed as follows~\citep[see Eq.~26 in Chap.~2 of][]{mihalas78}:

\begin{equation}
 \frac{1}{c} \frac{\partial I_{\nu}}{\partial t} + \vec{\nabla} . (I_{\nu} \vec{n}) = \eta_{\nu} - \chi_{\nu}I_{\nu},
 \label{eq:equation_radiative|_transfer_general_form}
\end{equation}
where $\vec{n}$ is the vector setting the position where the radiative field is evaluated.

\begin{figure}[t]
\centerfloat
\centerline{\resizebox{0.70\textwidth}{!}{\includegraphics{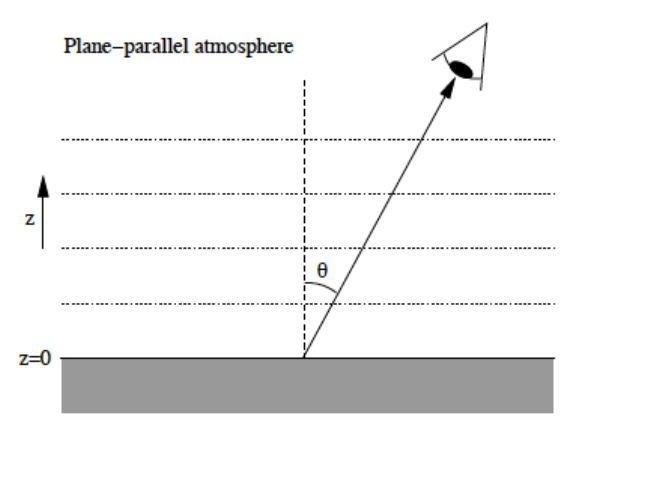}}}
\caption[Schematic of a plane-parallel geometry.]{Schematic of a plane-parallel geometry. The position is set by the distance $z$ and the angle $\theta$ that is formed between the normal vector (vertical dashed line) to the surface of the stellar atmosphere and the light array propagation (toward an observer).}
\label{sec_modeling_radiative_transfer_plane_parallel}
\end{figure}

The quantities $\eta_{\nu}$ and $\chi_{\nu}$ are the emission and the absorption coefficients, respectively. These coefficients share a common origin, in the way that they are related to Eq.~\ref{eq:specific_intensity}, but stating, respectively, the quantity of energy that is added or removed to a light beam through the medium. Note that they are themselves dependent on the frequency.\par

The emission coefficient $\eta_{\nu}$ is defined as follows:

\begin{equation}
dE_{\nu} = \eta_{\nu} \, dV \, d\Omega \, d\nu \, dt, 
\end{equation}
having dimensions of erg cm\textsuperscript{-3} sr\textsuperscript{-1} Hz\textsuperscript{-1} s\textsuperscript{-1} (CGS system).\par 

The absorption coefficient $\chi_{\nu}$ is given by:

\begin{equation}
dI_{\nu} = -\chi_{\nu}I_{\nu}ds,
\end{equation}
where $ds$ is the distance travelled by the light beam, and $\chi_{\nu}$  has dimension of cm\textsuperscript{-1} in the CGS system. Usually the absorption coefficient is also called as the opacity of the medium. One useful quantity, related to the absorption coefficient, is the mean free path:

\begin{equation}
l_{\nu} = \frac{1}{\chi_{\nu}}.    
\label{eq:mean_free_path}
\end{equation}

Eq.~\ref{eq:mean_free_path} sets the typically distance that a photon travels (freely) within consecutive processes of interaction with the gas particles of the atmosphere.\par

In addition, another important quantity, but related to the opacity $\chi_{\nu}$ of the medium, is the optical depth $\tau_{\nu}$:

\begin{equation}
\tau_{\nu} = \int_{0}^{s} \chi_{\nu} ds^{\prime}, 
\label{eq:optical_depth}
\end{equation}
being evaluated through a distance $s$. This quantity is useful as it depends both of the distance length travelled by the light and of the medium opacity. When $\tau_{\nu}$ $>>$ 1, the medium is said to be optically thick (at a certain value of frequency), while a optically thin medium has $\tau_{\nu}$ $<<$ 1.\par

We are able to simplify the radiative transfer equation (Eq.~\ref{eq:equation_radiative|_transfer_general_form}) considering the case of a stationary atmosphere, and thus assuming that:

\begin{equation}
\frac{\partial I_{\nu}}{\partial t} = 0.
\end{equation}

Furthermore, other important simplification is provided when assuming a plane-parallel geometry for the atmosphere. Fig.~\ref{sec_modeling_radiative_transfer_plane_parallel} shows a schematic of a plane-parallel geometry. The atmosphere is composed of parallel planes and a constant value $\theta$, which is formed between the light ray (seen toward the observed) and the normal to the surface of the atmosphere. Considering the usual Cartesian coordinates, and following the schematic of Fig.~\ref{sec_modeling_radiative_transfer_plane_parallel}, we thus have:\par 

\begin{equation}
\frac{\partial I_{\nu}}{\partial x} = \frac{\partial I_{\nu}}{\partial y} = 0,
\end{equation}
and
\begin{equation}
\vec{n_{z}} . \vec{n} = \cos{\theta} = \mu.
\end{equation}

This means that, for a certain direction (given by $\cos{\theta}$), the solution of the radiative transfer equation is dependent only of the variable $z$. Then the left hand of Eq.~\ref{eq:equation_radiative|_transfer_general_form} is simplified to:

\begin{equation}
\mu \diff{I_{\nu}}{z} = \eta_{\nu} - \chi_{\nu}I_{\nu}.
\label{eq:radiative_transfer_plane_parallel1}
\end{equation}

Usually the radiative transfer equation is expressed as a function of the optical depth. For that, it is introduced the called source function, $S_{\nu}$, defined in terms of the emission and absorption coefficients, as follows:

\begin{equation}
S_{\nu} =  \frac{\eta_{\nu}}{\chi_{\nu}}.
\label{eq:source_function}
\end{equation}

Physically, the source function can be interpreted as the amount of energy emitted along a photon mean free path (Eq.~\ref{eq:mean_free_path}).\par

From Eq.~\ref{eq:optical_depth}, we have that:

\begin{equation}
d\tau_{\nu} = \chi_{\nu} ds,
\label{eq:element_optical_depth}
\end{equation}
being $s = z/\mu$

Thus, from Eqs. \ref{eq:source_function} and \ref{eq:element_optical_depth}, the equation of radiative transfer for the stationary and plane-parallel case is then reformulated as follows~\citep[Eq.~1.27 of ][]{rybicki86}:

\begin{equation}
\diff{I_{\nu}}{\tau_{\nu}} = S_{\nu} - I_{\nu},    
\label{eq:radiative_transfer_plane_parallel2}
\end{equation}
being $I_{\nu} = I_{\nu}(\tau_{\nu}, \theta$).\par

Consider that the light array travels through the atmosphere from a optical depth $\tau_{2}$ to $\tau_{1}$, being $\tau_{2} > \tau_{1}$\footnote{For simplicity, when specifying a certain value of optical depth, we remove the $\nu$ index, but we keep using this index when expressing the variable ($\tau_{\nu}$).}. Thus, the formal solution of Eq.~\ref{eq:radiative_transfer_plane_parallel2} is given as follows:

\begin{equation}
I_{\nu}(\tau_{1},\mu) = I_{\nu}(\tau_{2}, \mu)\exp{[-(\tau_{2} - \tau_{1})/\mu]} + \frac{1}{\mu} \int_{\tau_{1}}^{\tau_{2}} S_{\nu}(\tau_{\nu}) \exp{[-(\tau_{\nu}^{\prime} - \tau_{1})/\mu]} d\tau_{\nu}^{\prime}.
\label{eq:solution_radiative_transfer_plane_parallel}
\end{equation}

Note that the solution provided by Eq.~\ref{eq:solution_radiative_transfer_plane_parallel} is not valid for an extended atmosphere, as formed by winds and disks on massive hot stars, having an extension much larger than $R_{\star}$ (up to $10^{1}$-$10^{2}$ $R_{\star}$). However, it is a good approximation to model the phostophere (hydrostatic region) since its extension is very small in comparison with $R_{\star}$. As will be latter discussed (Sect.~\ref{sec_radiative_atmosphere_codes_hot_star}), atmosphere models for hot stars must solve the equation of radiative transfer using a spherical geometry, instead of plane-parallel geometry, in order to to able to explain the line-formation in the circumstellar envelope.\par 


Also note that, as formulated in Eq.~\ref{eq:radiative_transfer_plane_parallel2}, the solution for the radiative transfer problem is specified by knowing the source function $S_{\nu}(\tau_{\nu})$. In this case, one important simplification is provided in terms of the so-called hypothesis of local thermodynamic equilibrium (LTE).\par 

In LTE, we assume that the source function, at any point of the atmosphere, is given by the Planck distribution $B_{\nu}(T(\tau_{\nu}))$:

\begin{equation}
S_{\nu}(\tau_{\nu}) = B_{\nu}(T(\tau_{\nu})) = \frac{2h\nu^3}{c^2} \, \frac{1}{\exp{[h\nu/k_{\mathrm{B}}T]} - 1},   
\label{eq:local_thermodynamic_equilibrium}
\end{equation}
being $k_{\mathrm{B}}$ the constant Boltzmann and $T$ the (local) temperature in the atmosphere.\par

As stated above, $T = T(\tau_{\nu})$, that is, this assumption works locally in the atmosphere, being valid at each point in the atmosphere. It does not states that the entire extension of atmosphere is modeled as having a unique value of temperature. Quite the contrary, in general, the temperature decreases outwards the photosphere. Indeed, as shown in Eq.~\ref{eq:flux_radiative_equilibrium} (LTE case), the radiative flux is dependent on the radial gradient of the temperature.\par

In the end, one sees how this latter hypothesis simplifies the solution of the radiative transfer problem. From Eq.~\ref{eq:local_thermodynamic_equilibrium}, the source function (at a certain depth point) only depends on temperature. Thus, by knowing the temperature structure of the atmosphere, Eq.~\ref{eq:solution_radiative_transfer_plane_parallel} can be solved. This means that the state of the gas is set by the Maxwell-Boltzmann and Saha distributions, described, locally, by a single value of temperature. The LTE approximation works well when the gas density of the atmosphere is high enough, meaning a larger opacity, and then a shorter mean free path of the gas particles and photons. In good approximation, this physical condition ensures that both the radiation and gas fields share, locally, the same value of temperature (holding for the Planck,  Maxwell-Boltzmann, and Saha distributions).\par

For instance, LTE is a reasonable approach to model the solar photosphere: it is well-known that a LTE radiative transfer modeling is able to reproduce well the solar spectrum~\citep[e.g.,][]{holweger67, kurucz91, kurucz05}. Nevertheless, even regarding the solar spectrum, an non-LTE approach is mandatory for the analysis of certain spectral lines, being particularly important for accurate determinations of chemical abundances~\citep[e.g., see][and references therein]{grevesse10, grevesse12, amarsi20}.\par

However, it is a complete different scenario when dealing with the atmospheres of massive hot stars. Even considering the photospheric region, these stars have much lower values of mass density, when compared with main sequence low-mass stars, such as the Sun.\par 

For instance, massive O-type stars have $\mathrm{M_{ZAMS}}$ $\gtrsim$ 8 $\mathrm{M_\odot}$, but, at the same, radius much larger than the Sun, up to 14 $\mathrm{R_{\odot}}$ (for a O3 dwarf, Table \ref{table_parameters_OB_dwarfs}). As a very crude estimation, the mass density is linearly proportional to the mass, but inversely proportional to the cube of the radius. Thus, following this approximation, the density in the photosphere of a O3 dwarf is lower by a factor of $10^{2}$ than the solar case. In this case, LTE is a very bad hypothesis to model the phostophere, as the radiative processes begins to be more important than the collisional processes in the transport of energy. Furthermore, the winds and disks around these stars also can not be modelled under LTE, as the density quickly drops in the extended atmosphere in comparison with the hydrostatic region (photosphere).\par 

In conclusion, the radiative transfer problem for massive hot stars must be treated, both for the phostophere and extended atmosphere, without the assumption of LTE, that is, requering a non-LTE treatment. This means that the equation describing the radiation field and the state of the gas must be treated simultaneously. Despite being valid for the particular case of plane-parallel geometry, one sees, from Eq.~\ref{eq:radiative_transfer_plane_parallel2}, that the radiative transfer problem itself is dependent on the state of the gas (that affects the source function). On the other hand, the LTE approximation makes the source function equal to a very simple expression (Eq.~\ref{eq:local_thermodynamic_equilibrium}).\par

In the following section, we discuss the principal atmosphere model codes for massive hot stars.\par


\section{Overview on radiative transfer codes for hot stars}
\label{sec_radiative_atmosphere_codes_hot_star}

Non-LTE radiative transfer modeling is a central part of this thesis, allowing us to perform a quantitative analysis of both spectroscopic and (spectro-)interferometric data, in terms of fundamental parameters of the photosphere and the circumstellar environment. Despite being widely used for spectroscopic analysis, the use of radiative transfer codes, such as CMFGEN and HDUST, to interpret multi-wavelength spectro-interferometric observations is however still somewhat scarce and quite recent in the literature~\citep[e.g.,][]{meilland07a, meilland07b, domiciano04, domiciano05, chesneau10, chesneau12}.\par 

This modern and physically consistent methodology was also employed in this thesis, combining both spectroscopic and interferometric observations to extract the physical properties of the photospheres of hot stars and their circumstellar environments. I worked using radiative transfer models on all the topics of this thesis. I used a pre-calculated grid of HDUST models to interpret multi-wavelength spectro-interferometric data (CHARA/VEGA and VLTI/AMBER data) of the Be star $\omicron$ Aquarii (Sect.~\ref{sec_results_omicron_aquarii}). In addition, I worked using the code CMFGEN to perform a detailed multi-wavelength (UV and visible) spectroscopic modeling of O-type stars (Sect.~\ref{sec_results_ostars}), and also, when collaborating to model recent H$\alpha$ intensity interferometry measurements of P Cygni (Sect.~\ref{sec_results_pcygni}). A detailed discussion on CMFGEN and HDUST (and the BeAtlas grid) are provided in Sect \ref{sec_radiative_transfer_modeling_cmfgen} and \ref{sec_radiative_transfer_modeling_hdust}.\par 

Nevertheless, we stress that other radiative transfer codes are used in the literature for the analysis of hot stars. For instance, the code Postdam Wolf-Rayet (PoWR, Sect.~\ref{sec_intro_theory_radiative_line_winds}) provides a similar treatment to the one by CMFGEN (Table \ref{table_radiative_transfer_codes}), being also well suited to model the extended atmosphere of hot and massive stars, such as O-type and Wolf-Rayet stars~\citep[see, e.g.,][and references therein]{hamann08, maryeva12}. In comparison with HDUST, the code BEDISK~\citep[][]{sigut07}, and its complementary code BERAY~\citep[][]{sigut11}, also have been employed to investigate the physical properties of Be star disks~\citep[e.g.,][]{halonen10, arcos18, sigut18}.\par 

\begin{table}[t]
\caption{\label{table_radiative_transfer_codes} Summary of basic characteristics of non-LTE and line-blanketed radiative transfer codes for modeling massive hot stars. Adapted from Chap.~5 of~\citet{crivellari2019radiative}}
\centering
\renewcommand{\arraystretch}{3.00}
\begin{adjustbox}{width=1.00\textwidth}
\begin{tabular}{lccccccccc}
\toprule
\toprule
\multicolumn{1}{l}{Code} & \multicolumn{1}{c}{Geometry} & \multicolumn{1}{c}{\makecell{Line- \\ blanketing}} & \multicolumn{1}{c}{\makecell{Radiative \\ transfer}} & \multicolumn{1}{c}{\makecell{Temperature \\ structure}} & \multicolumn{1}{c}{\makecell{Wavelength \\ range}}  & \multicolumn{1}{c}{\makecell{Major \\ application}}  & \multicolumn{1}{c}{\makecell{Execution \\ time}}  & \multicolumn{1}{c}{Comments} &\multicolumn{1}{c}{\makecell{Basic \\ references}} \\
\midrule 

TLUSTY & \makecell{Plane- \\ parallel} & Yes & \makecell{Observer's \\ frame} & \makecell{Radiative \\ equilibrium} & No limitations & \makecell{Hot stars \\ with negligible \\ winds} & Hours &  No winds &~\citet{hubeny95} \\

CMFGEN & Spherical & Yes & CMF & \makecell{Radiative \\ equilibrium} & No limitations & \makecell{OBA-stars, \\ WR, CSPN, \\ SNe} & Hours & \makecell{Start model \\ required} &~\citet{hillier98} \\

PoWR & Spherical & Yes & CMF & \makecell{Radiative \\ equilibrium} & No limitations & \makecell{WR, \\ O-stars} & Hours & --- &~\citet{grafener02} \\

FASTWIND & Spherical & Approx. & \makecell{CMF and \\ Sobolev approx.} & \makecell{$\mathrm{e^{-}}$ thermal \\ balance} & Visible and IR & \makecell{OB-stars, \\ early A-sgs} & Up to 0.5h & \makecell{User-specified \\ atomic models} &~\citet{puls05} \\

WM-basic & Spherical & Yes & Sobolev approx. & \makecell{$\mathrm{e^{-}}$ thermal \\ balance} & UV & \makecell{Hot stars \\ with dense winds, \\SNe} & 1-2h & No clumping &~\citet{pauldrach01} \\

\bottomrule
\end{tabular}
\end{adjustbox}
\end{table}

In advance of discussion, HDUST is not included, in Table \ref{table_radiative_transfer_codes}, since it provides a statistical method for solving the radiative transfer equation, jointly to the radiative and statistical equilibrium equations. In addition, to date, HDUST provides a fully hydrogen composition (for gaseous environments), and then not taking into account line-blanketing effects due to heavy elements (which are discussed in the next section).\par

Table \ref{table_radiative_transfer_codes} provides a global picture of state-of-the-art non-LTE line-blanketed radiative transfer codes, which are used for the analysis of hot massive stars. From Table \ref{table_radiative_transfer_codes}, TLUSTY~\citep[][]{hubeny95, lanz03, lanz07} is the only code with a plane-parallel geometry, and, hence, suited to model lines formed (almost) in the photosphere. This means that TLUSTY is dedicated to the (photospheric) analysis of hot stars with less strong winds, such as B dwarfs (and late O dwarfs, if the primary focus is to determine the photospheric parameters). Also, note that the TLUSTY solution is performed in the observer's frame since it is a purely photospheric model code, that is, taking into account only the hydrostatic region of the atmosphere.\par

Furthermore, the code FASTWIND~\citep[][]{santolaya97, puls05} was originally designed to provide an approximate treatment of line-blanketing. This means that a group of lines of `explicit elements'' (e.g., H, He, C, N, O, and others) is treated in the comoving frame (CMF), while most part of the lines due to ``background elements'' (mostly from the iron group) are treated using the Sobolev approximation. Using this approach, FASTWIND is able to provide a realistic line synthesis for hot stars, comparable to one provided by more robust codes, such as CMFGEN~\citep[see, e.g.,][]{massey13}. Nevertheless, recent efforts have been made for a fully consistent treatment of line-blanketing with FASTWIND, by means of treating all the lines in the CMF approach~\citep[][]{puls17}, and thus allowing a realistic line synthesis in the UV region. Note that only the visible and infrared regions are indicated for FASTWIND in Table \ref{table_radiative_transfer_codes}. Further details on atmosphere model codes of massive hot stars can be found in the review of~\citet{sander17_review}.\par

Finally, note that all the codes for extended atmosphere, listed in Table \ref{table_radiative_transfer_codes}, use a spherically symmetric geometry, being thus 1-D models. This is a very fair assumption for stellar winds of non-rotating single stars. On the other hand, such codes cannot reproduce the break of spherical symmetry that is introduce by fast rotation, such as the formation of equatorial disks around classical Be stars. As it will discussed in Sect.~\ref{sec_radiative_transfer_modeling_hdust}, such geometric limitation is well overcome by the code HDUST due to its alternative method, in comparison with the CMF approach, to solve the radiative transfer problem.\par


\section{The code CMFGEN}
\label{sec_radiative_transfer_modeling_cmfgen}

CMFGEN\footnote{Acronym for comoving frame general.}~\citep{hillier98} is a state-of-the-art non-LTE radiative transfer code, which was originally developed to study the properties of massive hot stars, such as WR and LBV stars~\citep[e.g,][]{hillier99, dessart00, hillier01}. Nevertheless, it also has been widely used for the analysis of central stars of planetary nebula~\citep[e.g.,][]{marcolino07}, OB-type~\citep[e.g.,][]{bouret12}, and, more recently, core-collapse supernovae~\citep[e.g.,][]{dessart16}. CMFGEN has been currently updated by D.~J.~Hillier (University of Pittsburgh, USA) and is publicly available\footnote{CMFGEN website: \url{http://kookaburra.phyast.pitt.edu/hillier/web/CMFGEN.htm}.}

CMFGEN iteratively solves the radiative transfer equation and the statistical and radiative equilibrium equations in the comoving frame, in a simultaneous way. The assumption of radiative equilibrium, used to constrain the temperature structure, means that all the energy in the atmosphere is fully transported by radiation, that is, neglecting conduction and convection. In this case, all the (radiative) energy that is absorbed at a volume of the atmosphere is equal to the total energy that is emitted at the same volume. Formally, holding for each point of the atmosphere, this can be expressed as follows:

\begin{equation}
\int_{0}^{\infty} \chi_{\nu} (J_{\nu} - S_{\nu}) \, d\nu = 0.
\end{equation}

As discussed in Sect.~\ref{sec_radiative_transfer_modeling_elementary_concepts}, holding LTE, the state of the gas is given by the local temperature (and the particle density). In this case, the statistical equilibrium equations are described by the Maxwell-Boltzmann and Saha distributions. In the non-LTE approach, as using CMFGEN, ``in a simultaneous way'' (as stated above) means that the radiative transfer equation and the ones of statistical and radiative equilibrium equation form a coupled problem to be solved. Whereas the statistical equilibrium equations, describing the populations, depends on the radiation field, the radiation field also depends on the solution of the statistical equilibrium equations. Then a simultaneous solution of them is needed. Further details on the expressions of the statistical equilibrium equations in non-LTE can be found in Sect.~5.5 of~\citet{mihalas78}.\par 

From a practical point-of-view, CMFGEN requires a start model (see Table \ref{table_radiative_transfer_codes}), providing an initial guess of the populations. After a number of iterations (for example, 60), from the correction of the populations, a new converged model can be found when the radiation field and the populations satisfied simultaneously the radiative transfer equation and the statistical equilibrium equations (under radiative equilibrium) for the new input parameters. The reader can found details on the solution methods of CMFGEN in Sect.~3 of~\citet{hillier98}.\par 

The input parameters in CMFGEN have been discussed in details in a series of works. In short, as a non-extensive list, these are the main parameters to define the atmosphere model: in addition to the chemical composition, the stellar luminosity, effective temperature, gravity surface acceleration, radius, mass, mass-loss rate, wind clumping factor, terminal velocity, and wind velocity law exponent. The reader can find further details in~\citet{groh11},~\citet{hillier12}, Appendix B of~\citet{martins04_thesis}, and in the reference manual of CMFGEN, which is available in the CMFGEN website (see footnote 3 of this chapter).\par

In the following, we summarize some of the principal characteristics of CMFGEN:

\begin{enumerate}[label=(\roman*)]
\setlength\itemsep{1em}

\item It assumes a stationary spherically symmetric wind. Currently, CMFGEN does not solve the radiative transfer equations in a consistent way with the wind hydrodynamics. This means that the wind density and velocity structures must be provided, they are input parameters of the model. The density is given by Eq.~\ref{eq:mass_continuity}, while Eq.~\ref{eq:beta_law_geral} sets the velocity law in its simplest form. Nevertheless, somewhat more complex velocity laws can be adopted, for example, a two-component $\beta$ velocity law, describing the inner and outer regions of the wind~\citep[see., e.g.,][]{hillier99}. However, more recently, CMFGEN allows a temporal dependent treatment of the radiative transfer and statistical equations (non-stationary outflow), being useful to model the light curves and spectra of supernovae~\citep[e.g., see, again,][]{dessart16}.\par

\begin{figure}[t]
\centerfloat
\centerline{\resizebox{0.75\textwidth}{!}{\includegraphics{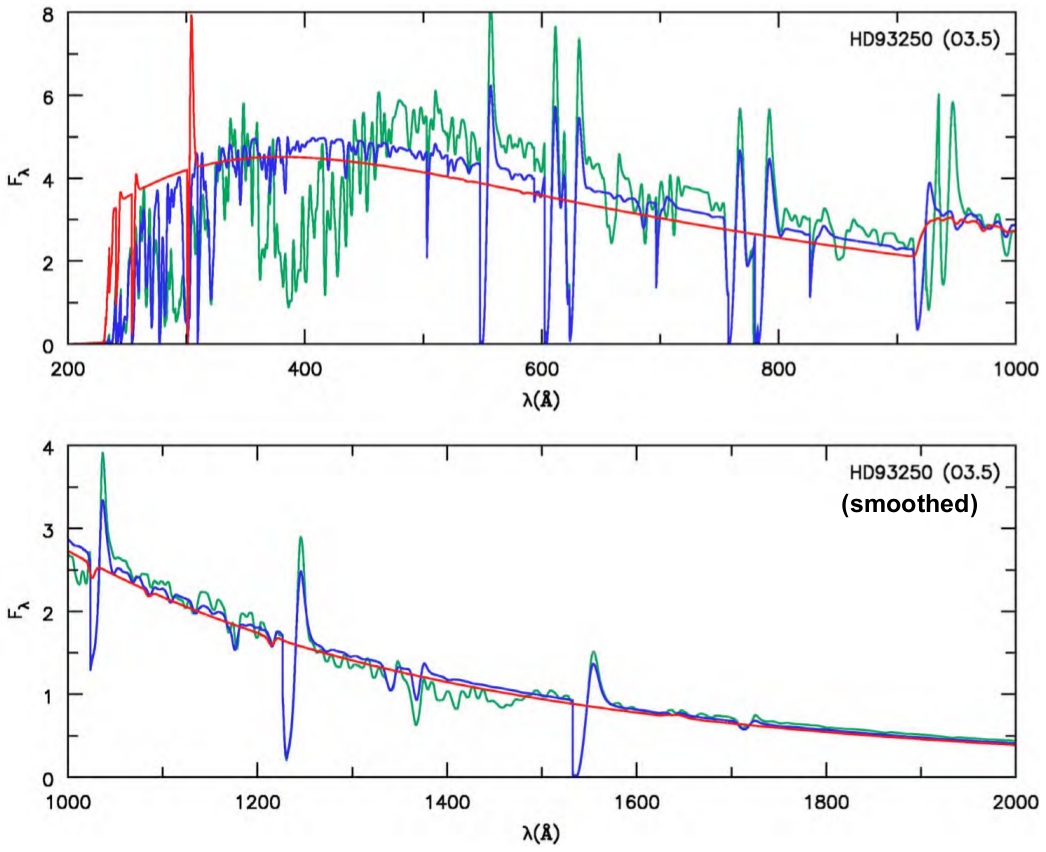}}}
\caption[Line-blanketing effect on the emergent spectrum.]{Line-blanketing effect on the emergent spectrum. Comparison among three CMFGEN models with the following chemical compositions: H + He (red), H + He + CNO (blue), and H + He + CNO + Fe group elements (green). All the models have the following parameters: $\log L_{\star}/L_{\odot}$ = 6.11, $T_{\mathrm{eff}}$ = 45700 K, $\log g$ = 4.0, $R_{\star}$ = 18.3 $\mathrm{R_{\odot}}$, $\dot{M}$ = $5.6\e{-7}$ $\mathrm{M_\odot}$ yr\textsuperscript{-1}, and $v_{\infty}$ = 3000 km s\textsuperscript{-1}. This set of physical parameters fairly describes the spectrum of HD 93250 (spectral type O4). See text for discussion. Reproduced from~\citet{groh11}.}
\label{sec_modeling_radiative_transfer_groh11_fig11}
\end{figure}

\item CMFGEN allows us to include a large quantity of atomic species in the modeling, namely, ions of: H, He, C, N, O, F, Ne, Na, Mg, Al, Si, P, S, Cl, Ar, K, Ca, Ti, Cr, Mn, Fe, Co, and Ni. Fig.~\ref{sec_modeling_radiative_transfer_groh11_fig11} compares the synthetic spectrum in the UV from CMFGEN models with different chemical compositions: H + He, H + He + CNO, and H + He + CNO + Fe group elements. The parameters are set according to a O4 star (HD 93250). One sees that the proper modeling of the UV region for such a hot star demands the inclusion of elements heavier than helium. This is also needed in order to treat the so-called line-blanketing effect that is ``exactly'' treated in CMFGEN. The larger opacity introduced by metals blocks the radiation and then increases the temperature at the inner region of the atmosphere (backwarning effect). This affects not just the lines, but the continuum region itself (as shown in Fig.~\ref{sec_modeling_radiative_transfer_groh11_fig11}). This means that models with the inclusion of metals (as realistically as possible) requires lower values of effective temperature to reproduce the same ionization (as without metals). Thus, it impacts the determination of the effective temperature: line-blanketed models (correctly) downward the inferred $T_{\mathrm{eff}}$ of hot stars~\citep[e.g., see][and references within]{najarro06}.\par

\item As discussed in Sect.~\ref{sec_intro_theory_radiative_line_winds}, intrinsic instabilities in the line force creates non-homogeneous winds (i.e., ``clumps'' in the wind density structure), and thus originating significant emission of X-Rays due to shocks in the wind of massive hot stars~\citep[e.g., see][and references within]{oskinova16}. X-Ray emission must be taken into account in order to reproduce, as realistically as possible, the ionization balance in the wind, and then the spectral lines (in particular the ones formed in the UV, which are important wind diagnostics). CMFGEN enables to take the Auger ionization into account ~\citep{cassinelli79, olson81} by X-Rays~\citep[see Sect.~6 of][]{hillier98}. Lastly, the wind clumping is modelled in CMFGEN in terms of a volume filling factor $f(r)$:

\begin{equation}
f(r) = f_{\infty} + (1 - f_{\infty})\mathrm{e}^{-\frac{v(r)}{v_{\mathrm{initial}}}},
\label{eq:cmfgen_clumping}
\end{equation}
being $f_{\infty}$ and $v_{\mathrm{initial}}$ the (input) free parameters chosen to include the effect of clumping. This simple parameterization assumes that the interclump medium is void and that the wind clumps are smaller in comparison with the photon mean-free path for all the values of wavelength (called microclumping approximation).\par

Physically, $v_{\mathrm{initial}}$ means the location in the wind where the formations of clumps begins to be important, and $f_{\infty}$ gives the value of clumping at larger distances, that is, $f(r\to\infty)$, with lower values of $f_{\infty}$ meaning more structured winds. From Eq.~\ref{eq:cmfgen_clumping}, $f$ is set to unity when $f_{\infty}$ = 1.0 (homogeneous wind, no clumps).\par

Taking into account the volume filling factor $f(r)$, the wind density (Eq.~\ref{eq:mass_continuity}) is then modified as follows:

\begin{equation}
\rho(r) = \frac{\dot{M}}{4\pi r^{2} v(r) f(r)}.
\label{eq:cmfgen_density_modified_clumping}
\end{equation}

Several spectroscopic studies on hot stars showed that the inclusion of clumping is needed to determine realistic values of mass-loss rates for hot stars winds. In short, the inclusion of wind clumping in the models implies downward revisions on their wind mass-loss rates by a factor up to $\sim$10, when compared with the values derived using homogeneous wind models~\citep[e.g.,][]{bouret03, repolust04, bouret12}.\par

\end{enumerate}

\begin{figure}[t]
\centerfloat
\centerline{\resizebox{0.75\textwidth}{!}{\includegraphics{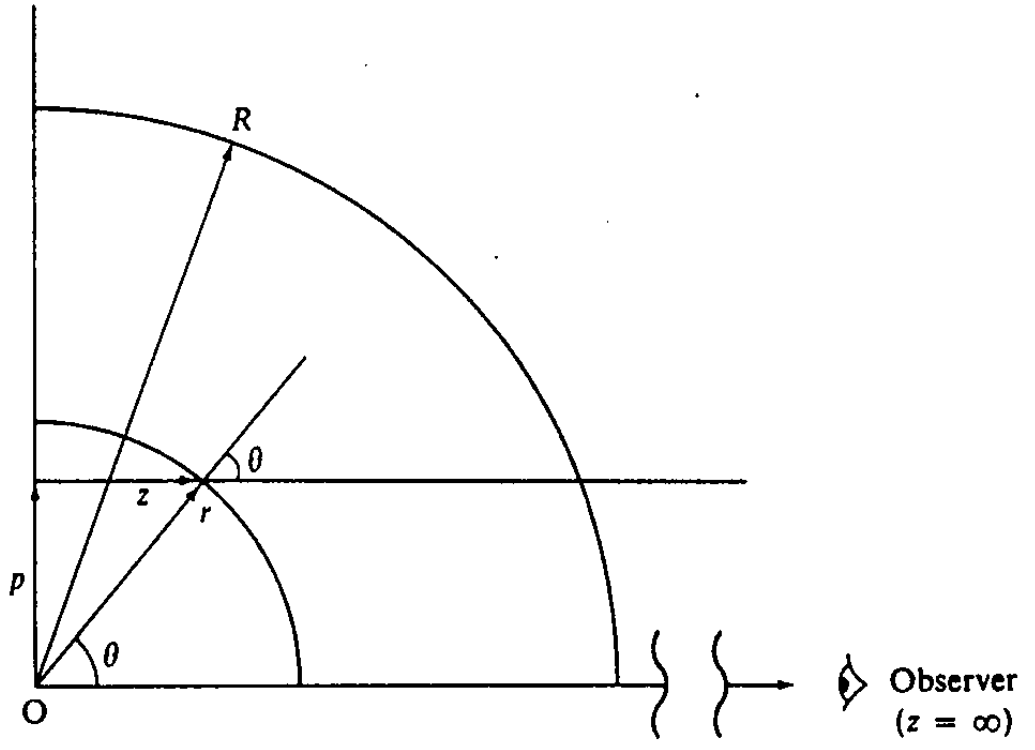}}}
\caption[Schematics of the $(p,z)$ coordinate system, which is used in the code CMFGEN to solve the radiative transfer problem (ray-by-ray solution method).]{Schematics of the $(p,z)$ coordinate system, which is used in the code CMFGEN to solve the radiative transfer problem (ray-by-ray solution method). The coordinate $p$ is called as the impact parameter, measuring the perpendicular distance to the rays given by the direction $z$. The coordinates $(p,z)$ are related to the polar ones $(r, \theta)$ by: $z = r \cos{\theta}$, $p = r \sin{\theta}$, and $r = \sqrt{p^2 + z^2}$. For instance, Fig.~ \ref{sec_modeling_radiative_transfer_pcygni_e8_intensity_vs_lambda_all_p1} shows the theoretical intensity (as a function of wavelength) for each value of $p$ from a model calculated using CMFGEN. Further details on the ray-by-ray solution method can be found in Sect.~5.3.2 of~\citet{crivellari2019radiative}. Reproduced from~\citet{mihalas78}.}
\label{sec_modeling_radiative_transfer_pz_system_mihalas78}
\end{figure}

In order to calculate the predicted visibility, it is needed a description of the intensity as a function of distance from the center of the star (impact parameter $p$, see Fig.~\ref{sec_modeling_radiative_transfer_pz_system_mihalas78}) for a certain value of wavelength. In Fig.~\ref{sec_modeling_radiative_transfer_pcygni_e8_intensity_vs_lambda_all_p1}, we show the intensity profiles, as a function of wavelength (around the H$\alpha$ line), for each value of impact parameter, calculated using CMFGEN. A certain agglomeration of profiles is seen closer to the photosphere since the grid points of the model are not equally spaced in the photosphere and in the wind. In particular, this model was calculated in the context of my work about P Cygni (Sect.~\ref{sec_results_pcygni}).\par 

One sees that H$\alpha$ is forming a strong P Cygni profile, related to the high value of mass-loss rate ($\dot{M}$ = $1.0\e{-5}$ $\mathrm{M_\odot}$ yr\textsuperscript{-1}). In Fig.~\ref{sec_modeling_radiative_transfer_pcygni_e8_intensity_vs_lambda_all_p1}, lower values of intensity in the continuum (close-by H$\alpha$) come from larger distances in the model (wind region). Note how H$\alpha$ is filled in emission toward larger distances. Moreover, weak spectral features around H$\alpha$, such as the \ion{C}{II} $\lambda \lambda$6580,6585 doublet, are formed in pure-absorption in the photosphere (dark intensity profiles), but in emission through the wind (light intensity profiles).\par 

\begin{figure}[t]
\centerfloat
\centerline{\resizebox{1.00\textwidth}{!}{\includegraphics{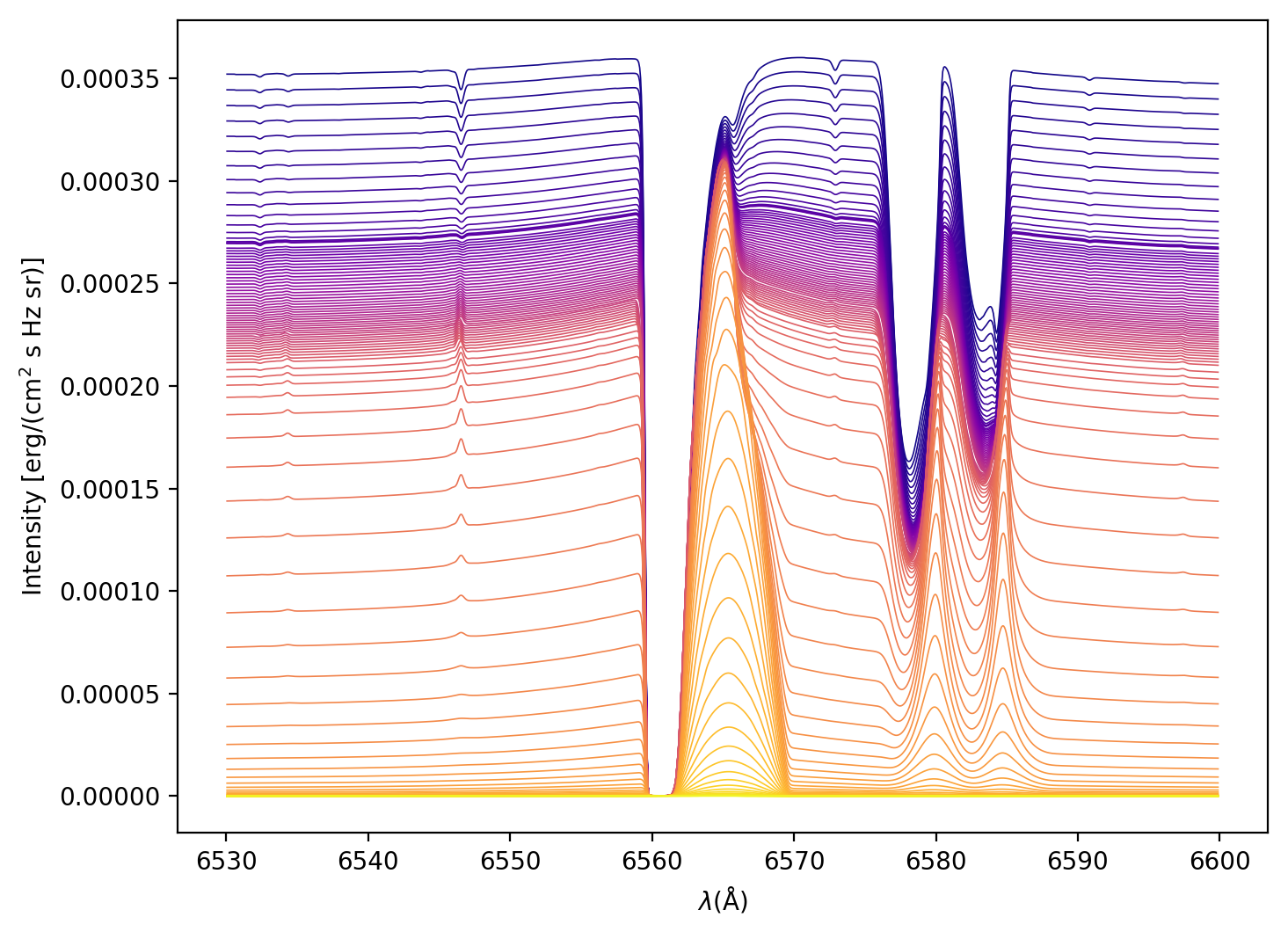}}}
\caption[Intensity profiles, as a function of the wavelength ($\sim$6530--6600 {\AA}, around H$\alpha$), calculated by myself using CMFGEN]{Intensity profiles, as a function of the wavelength ($\sim$6530--6600 {\AA}, around H$\alpha$), calculated by myself using CMFGEN with the following stellar and wind parameters: $\log L_{\star}/L_{\odot}$ = 5.73, $T_{\mathrm{eff}}$ = 18700 K, $\log g$ = 2.3, $R_{\star}$ = 70.6 $\mathrm{R_{\odot}}$, $\dot{M}$ = $1.0\e{-5}$ $\mathrm{M_\odot}$ yr\textsuperscript{-1}, and $v_{\infty}$ = 185 km s\textsuperscript{-1}. Intensity profiles are shown for each value of distance from the center of the star (in total, 115 points). Lighter (toward yellow) $I$ versus $\lambda$ profiles come from larger distances through the wind region (extension of 50 $R_{\star}$). See text for discussion.}
\label{sec_modeling_radiative_transfer_pcygni_e8_intensity_vs_lambda_all_p1}
\end{figure}

From Fig.~\ref{sec_modeling_radiative_transfer_pcygni_e8_intensity_vs_lambda_all_p1}, we are able to create an image cube, describing the intensity map of the object for each value of wavelength, and then predict the model visibility from the Fourier transform of the intensity map (valid for each value of wavelength). Since CMFGEN is essentially a 1-D model, it is needed to take into account the geometry of the model (spherically symmetric wind) in order to interpolate an image (2-D).\par 

Also notice that the resulting spectrum (Fig.~\ref{sec_modeling_radiative_transfer_pcygni_e8_spectrum_6530_6600}), around $\sim$6530-6000 {\AA}, is computed by adding the different intensity curves shown in Fig.~\ref{sec_modeling_radiative_transfer_pcygni_e8_intensity_vs_lambda_all_p1}. For example, the emergent spectrum in \ion{C}{II} $\lambda \lambda$6580,6585 forms two weak P Cygni profiles (note the shift, in wavelength, between the absorption and emission contributions).\par

Of course, the variation of the mass-loss rate in the model will impact the intensity profiles, especially in H$\alpha$. Fixing all the other parameters, increasing $\dot{M}$ implies a stronger P Cygni profile, with a more intense emission component\footnote{Note that, in this case, the absorption component of the P Cygni profile in H$\alpha$ is saturated, that is, null intensity.}.\par 

In short, despite the high computational cost (as will be discussed in Sect.~\ref{sec_radiative_transfer_modeling_computational_cost}), this shows how radiative transfer modeling is a powerful tool to probe the physical conditions of circumstellar environments. From a set of (almost) fundamental physical and chemical parameters, it can be used to model spectroscopic and (spectro-)interferometric data at different transitions.\par

\begin{figure}[t]
\centerfloat
\centerline{\resizebox{1.00\textwidth}{!}{\includegraphics{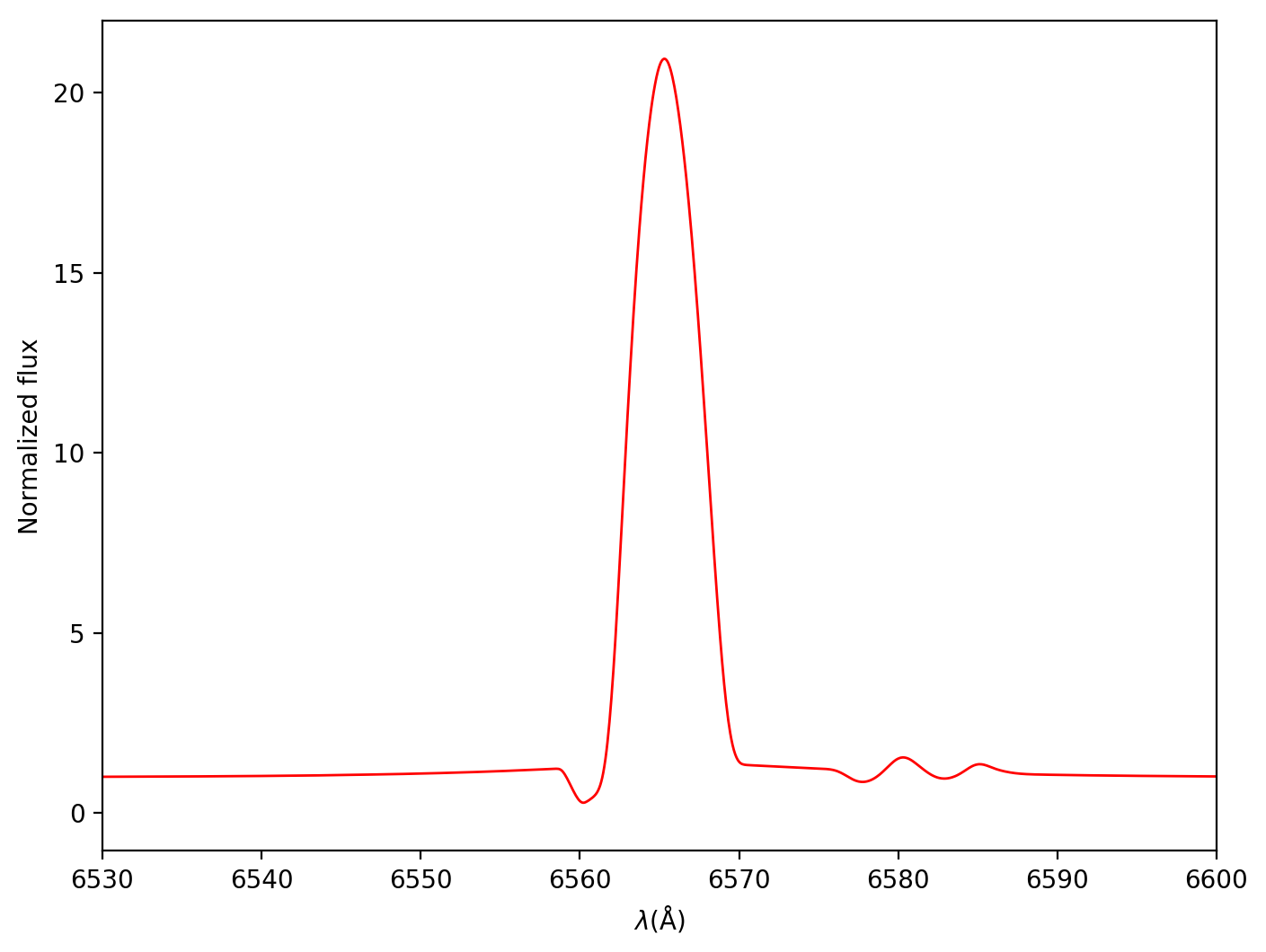}}}
\caption[Emergent flux (normalized to the continuum), computed from the same CMFGEN model in Fig.~\ref{sec_modeling_radiative_transfer_pcygni_e8_intensity_vs_lambda_all_p1}, around the H$\alpha$ line (strong P Cygni profile).]{Emergent flux (normalized to the continuum), computed from the same CMFGEN model in Fig.~\ref{sec_modeling_radiative_transfer_pcygni_e8_intensity_vs_lambda_all_p1}, around the H$\alpha$ line (strong P Cygni profile). Note that \ion{C}{II} $\lambda \lambda$6580,6585 lines also form P Cygni profiles but very weak in comparison with the H$\alpha$ transition.}
\label{sec_modeling_radiative_transfer_pcygni_e8_spectrum_6530_6600}
\end{figure}


\section{The code HDUST and the BeAtlas grid}
\label{sec_radiative_transfer_modeling_hdust}

One alternative to CMF-based codes, which explicitly (numerically) solve the radiative transfer and statistical equilibrium equations in the comoving frame, relies on Monte Carlo method. In this case, the radiation field is reconstructed by classifying the photons packages by position, propagation, direction, and wavelength. This allows us to solve the radiative transfer problem self-consistently in a statistical sense. In particular, the Monte Carlo method is a very useful when dealing with the radiative transfer in more complex geometries than the 1-D assumption as in CMFGEN, and other CMF-based codes, for radiative winds (see, again, Table \ref{table_radiative_transfer_codes}).\par 

HDUST~\citep[][]{carciofi06a, carciofi08} is a state-of-the-art 3-D non-LTE radiative transfer code that solves the radiative transfer and statistical equilibrium equations using the Monte Carlo method. The Monte Carlo method implemented in HDUST largely follows the one developed by~\citet{lucy02, lucy03, lucy05}, which
ensures the condition of radiative equilibrium, differing with respect to the process of atom-photon interaction~\citep[see Sect.~2 of][]{carciofi06a}.\par

In HDUST, the circumstellar environment can be treated under arbitrary 3-D geometry and kinematics. This provides a suitable approach to model both winds and disks around hot stars. For purely-gaseous environment, such as in Be disks, the chemical composition is assumed by atomic hydrogen (and free electrons). As suggested by its name, besides a hydrogen gaseous composition, HDUST also allows us to take into account the opacity of dust grains, following the so-called Mie theory~\citep[see][]{carciofi04}. This is particularly relevant for the analysis of objects such as B[e] stars, which have gaseous and dusty environments. In short, HDUST has been successfully used in a series of studies to interpret different observational properties of Be and B[e] stars. For instance, see~\citet{bjorkman05},~\citet{carciofi10},~\citet{klement15}, and~\citet{domiciano15}.\par

In this thesis, all the HDUST models analysed are part of the BeAtlas project~\citep[][]{faes15, mota19}. This model grid have been developed by D.~M.~Faes, B.~C.~Mota, and A.~C.~Carciofi (Universidade de São Paulo, Brazil) in particular for the analysis of the Be disks, as well as the central star.\par

In the models of BeAtlas, the kinematics of the disk is adopted as following a Keplerian rotation (Eq.~\ref{eq:keplerian}), according to the VDD scenario (Sect.~\ref{sec_intro_disk_dynamics}). BeAtlas is composed of $\sim$14000 models with images (specific intensity maps), SEDs, and spectra calculated in natural and polarized light, over several spectral regions, including the H$\alpha$ and Br$\gamma$ lines, and also the K-band region. Hence, it is very suitable for interpreting spectro-interferometric data, such as from CHARA/VEGA and VLTI/AMBER. Besides the purely theoretical investigation by~\citet{faes15} of differential phase signatures for Be disks, this thesis provides to date the first application of BeAtlas for the spectro-interferometric analysis of a Be star (Sect.~\ref{sec_results_omicron_aquarii}).\par

\begin{table}[t]
\centering
\begin{adjustbox}{width=0.75\textwidth}
\begin{threeparttable}
\caption{\label{table_beatlas} List of HDUST parameter in the BeAtlas grid. The first row indicates the spectral type corresponding to the stellar mass for B dwarfs~\citep{townsend04}. All the model are calculated with the following fixed parameters: scale height at the disk base $H_{0}$ = 0.72, fraction of H in the core $X_{c}$ = 0.30, metallicity $Z$ = 0.014, and disk radius = 50 $R_{\mathrm{eq}}$. Adapted from~\citet{faes15}}
\renewcommand{\arraystretch}{1.50}
\begin{tabular}{ll}
\toprule
\toprule
Parameter &  Value \\
\midrule
Spectral type & \makecell[l]{B0.5, B1, B1.5, B2, B2.5, B3, B4, B5 \\ B6, B7, B8}\\
\midrule
$M_\star$ ($\mathrm{M_\odot}$) & \makecell[l]{14.6, 12.5, 10.8, 9.6, 8.6, 7.7, 6.4, 5.5 \\ 4.8, 4.2, 3.8}\\
$i$ (deg) & \makecell[l]{0.0, 27.3, 38.9, 48.2, 56.3, 63.6, 70.5 \\ 77.2, 83.6, 90.0}\\
Oblateness ($R_{\mathrm{eq}}/R_{\mathrm{p}}$) & 1.1, 1.2, 1.3, 1.4, 1.45\\
$\Sigma_{0}$ (g cm\textsuperscript{-2})\,\tnote{a} & 0.02, 0.05, 0.12, 0.28, 0.68, 1.65, 4.00\\
$m$\,\tnote{b} & 3.0, 3.5, 4.0, 4.5 + non-isothermal\\
\bottomrule
\end{tabular}
\begin{tablenotes}
\item[a] Surface density at the base of the disk.
\item[b] Disk mass density law exponent. ``non-isothermal'' stands for HDUST models computed without the assumption of isothermal disk scale height.
\end{tablenotes}
\end{threeparttable}
\end{adjustbox}
\end{table}

Table \ref{table_beatlas} shows the parameter spaces covered in BeAtlas. We stress that these values correspond to the BeAtlas grid developed by~\citet{faes15}, being this version employed in the present thesis. BeAtlas models with finer parameter steps have been calculated by~\citet{mota19}.\par

Since we are dealing with a pre-calculated model grid, a BeAtlas-based analysis is, at least at a certain level, biased by the limited parameter space and selected parameter values of the grid, which are not homogeneously distributed (for example, see Fig.~\ref{sec_modeling_radiative_transfer_carta_sigma0_massa}). From that, one sees a gap of models with higher values of $\Sigma_{0}$, as the stellar mass decreases since convergence issues with HDUST arises in this case ~\citep[see Sect.~3.4.2 of][]{mota19}. Nevertheless, this particular limitation in the grid is not expected to highly affect the analysis of Be disks since late-type Be stars (i.e., lower stellar mass) are more likely to have less dense disks~\citep[see][]{vieira17, arcos17}.\par

In total, five physical parameters vary in the grid. Three out of five parameters describe the central star: the stellar mass, stellar oblateness, inclination angle. Each value of $R_{\mathrm{eq}}/R_{\mathrm{p}}$ is uniquely related to $W$, assuming that the central star is a rigid rotator under the Roche model. From Table  \ref{table_beatlas}, these values of $R_{\mathrm{eq}}/R_{\mathrm{p}}$ correspond to $W$ = 0.447, 0.663, 0.775, 0.894, and 0.949, respectively. We recall the reader that the rotational $W$ is defined according to Eq.~\ref{eq:rotational_rate_w}. Since rotational effects are relevant for the analysis of Be stars, the gravity darkening effect \footnote{See, again, Sects.~\ref{sec_intro_stellar_rotation_von_zeipel_effect} and Sect.~\ref{sec_intro_fast_rotation}.} is included in the BeAtlas models following the prescription from~\citet{espinosa11}.\par

\begin{figure}[t]
\centerfloat
\centerline{\resizebox{1.00\textwidth}{!}{\includegraphics{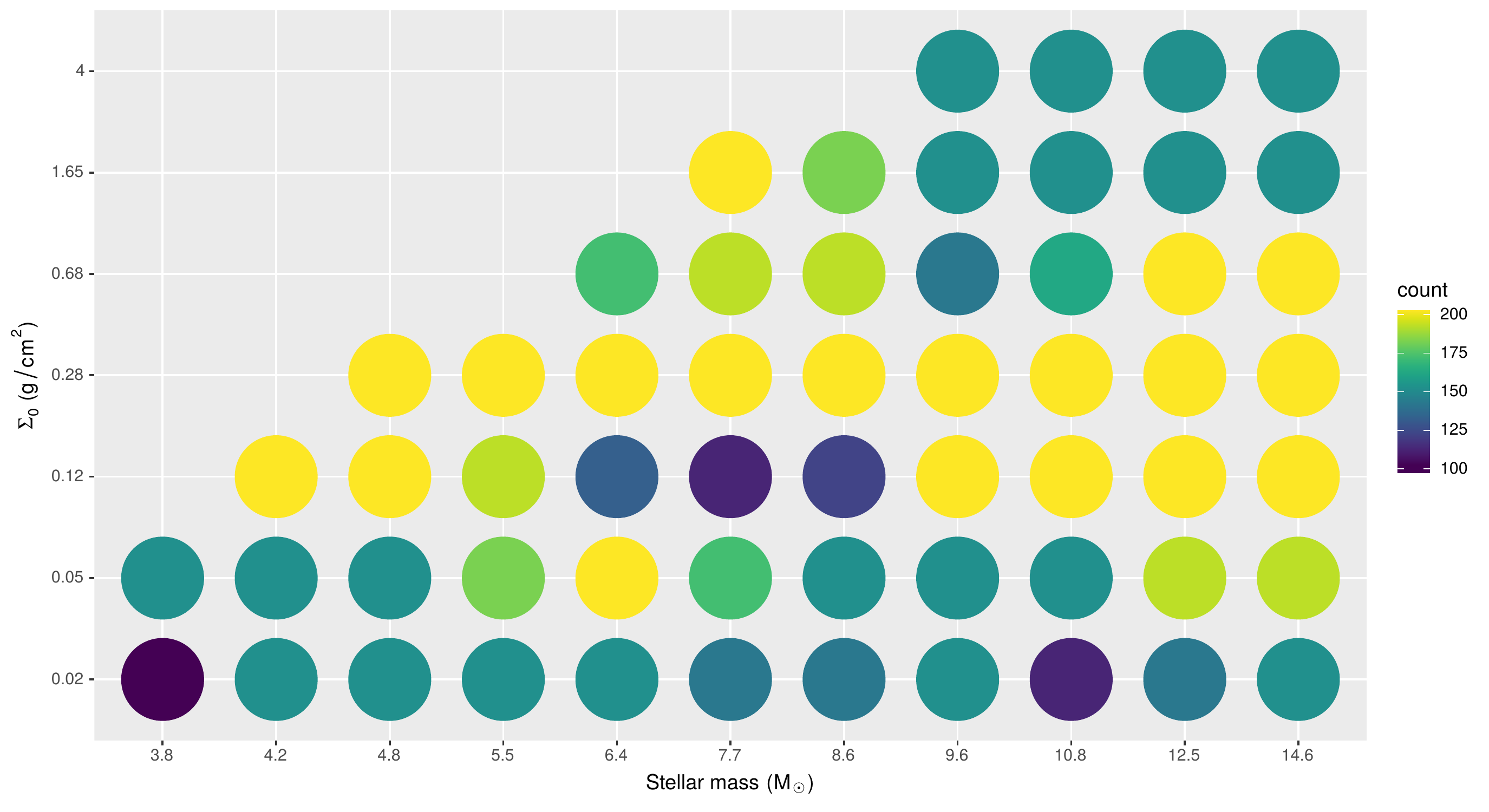}}}
\caption[Distribution of the BeAtlas models with respect to the stellar mass and base disk surface density.]{Distribution of the BeAtlas models with respect to the stellar mass and base disk surface density. Note the absence of models with high values of base disk density for the lower stellar mass. See text for discussion.}
\label{sec_modeling_radiative_transfer_carta_sigma0_massa}
\end{figure}

The polar stellar radius $R_{\mathrm{pol}}$ is obtained from each combination of stellar mass and stellar oblateness. For example, models with $M_{\star}$ = 3.4 $\mathrm{M_{\odot}}$ have $R_{\mathrm{pol}}$ between $\sim$3.36 and 3.37 $\mathrm{R_{\odot}}$ for the different values of rotational rates considered in the grid (between 0.447 and 0.949, respectively). For $M_{\star}$ = 14.6 $\mathrm{M_{\odot}}$, the polar radius is $\sim$ 7.50 and $\sim$ 7.38 $\mathrm{R_{\odot}}$ for these extreme values of $W$. The stellar mass and rotational rate also define the value of the stellar rotational velocity $v_{\mathrm{rot}}$. For $M_{\star}$ = 3.4 $\mathrm{M_{\odot}}$, we have  $v_{\mathrm{rot}}$ $\sim$ 160 and 300 km s\textsuperscript{-1} for $W$ = 0.447 and 0.949. For $M_{\star}$ = 14.6 $\mathrm{M_{\odot}}$, $v_{\mathrm{rot}}$ $\sim$ 210 and 420 km s\textsuperscript{-1} for these same values of rotational rate.\par

\begin{figure}[t]
  \begin{adjustbox}{minipage=\textwidth,scale=1.00}
  \centering
  \includegraphics[width=1.00\columnwidth]{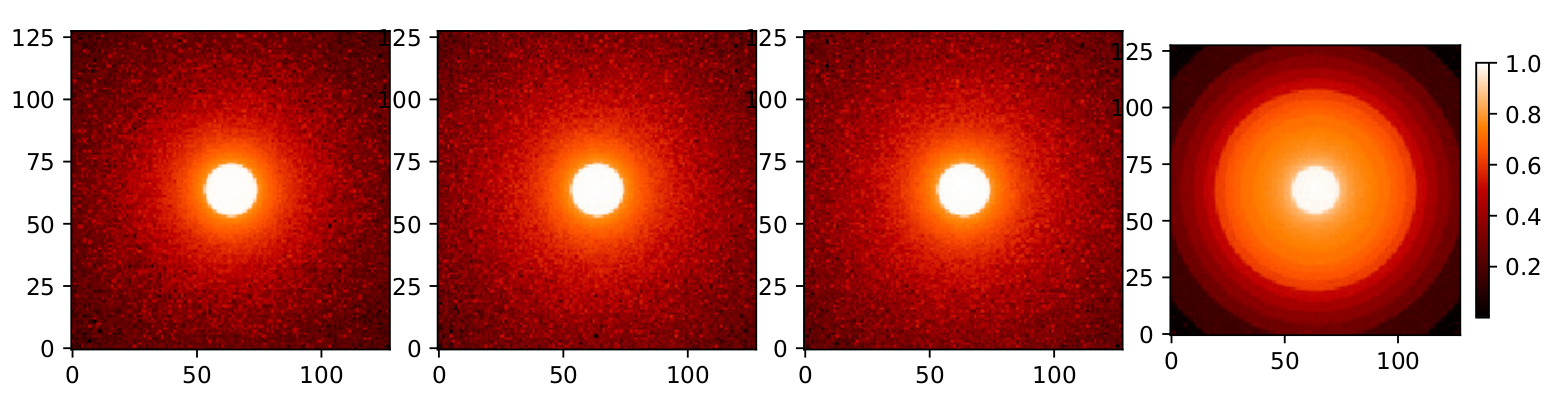}
  \medskip
  \includegraphics[width=1.00\columnwidth]{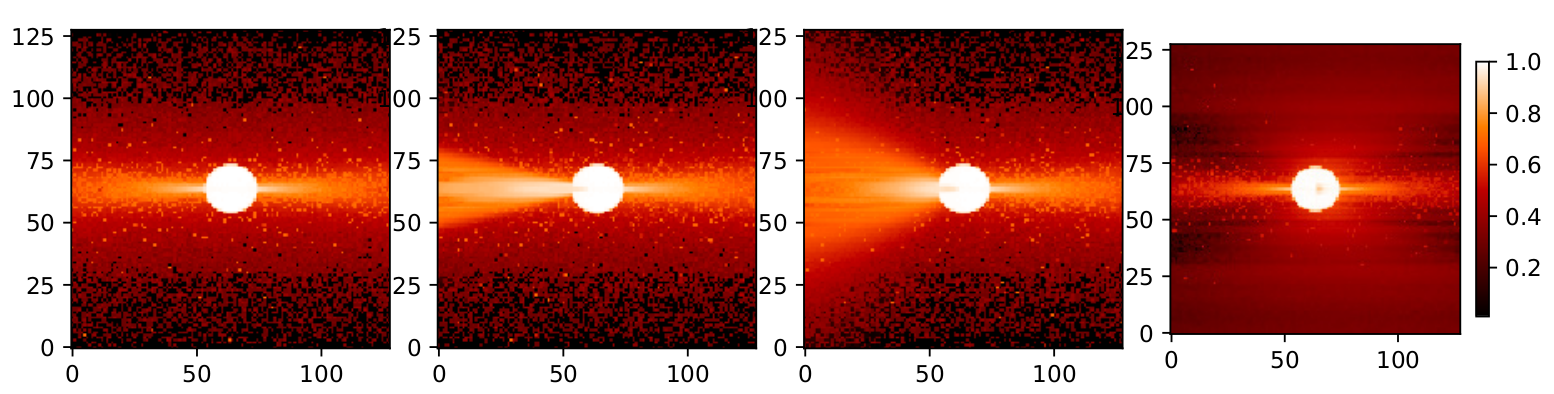}
  \includegraphics[width=0.48\columnwidth]{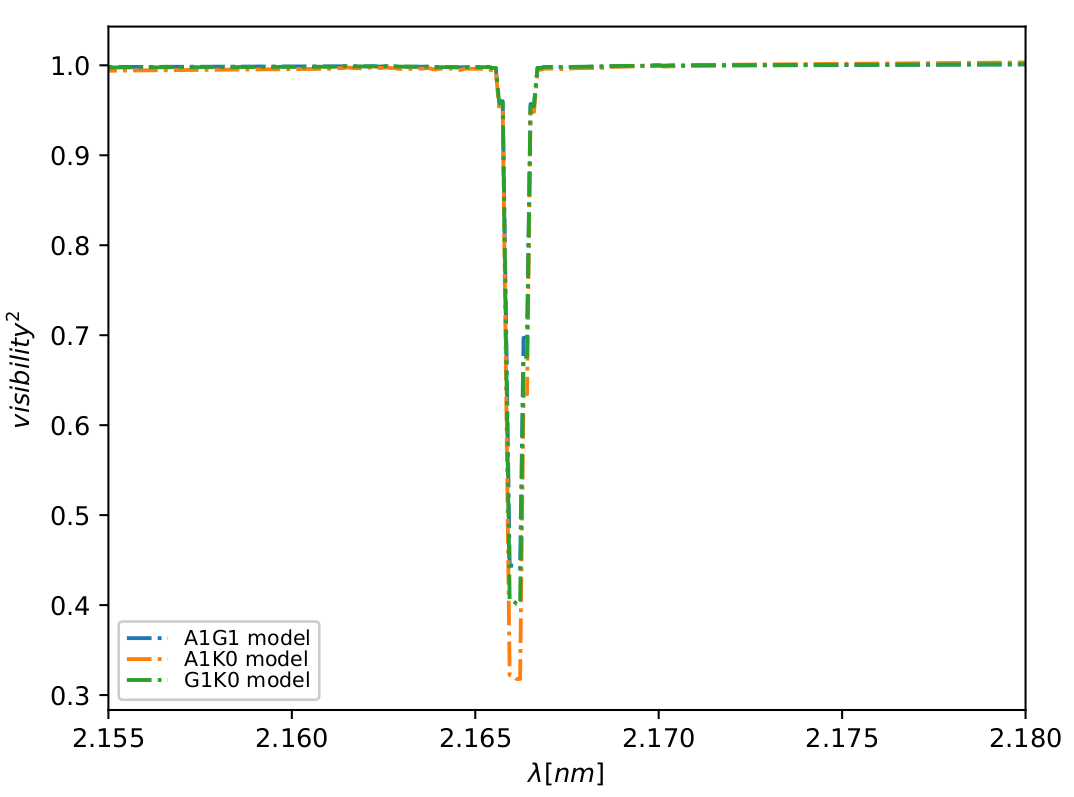}
  \medskip
  \includegraphics[width=0.48\columnwidth]{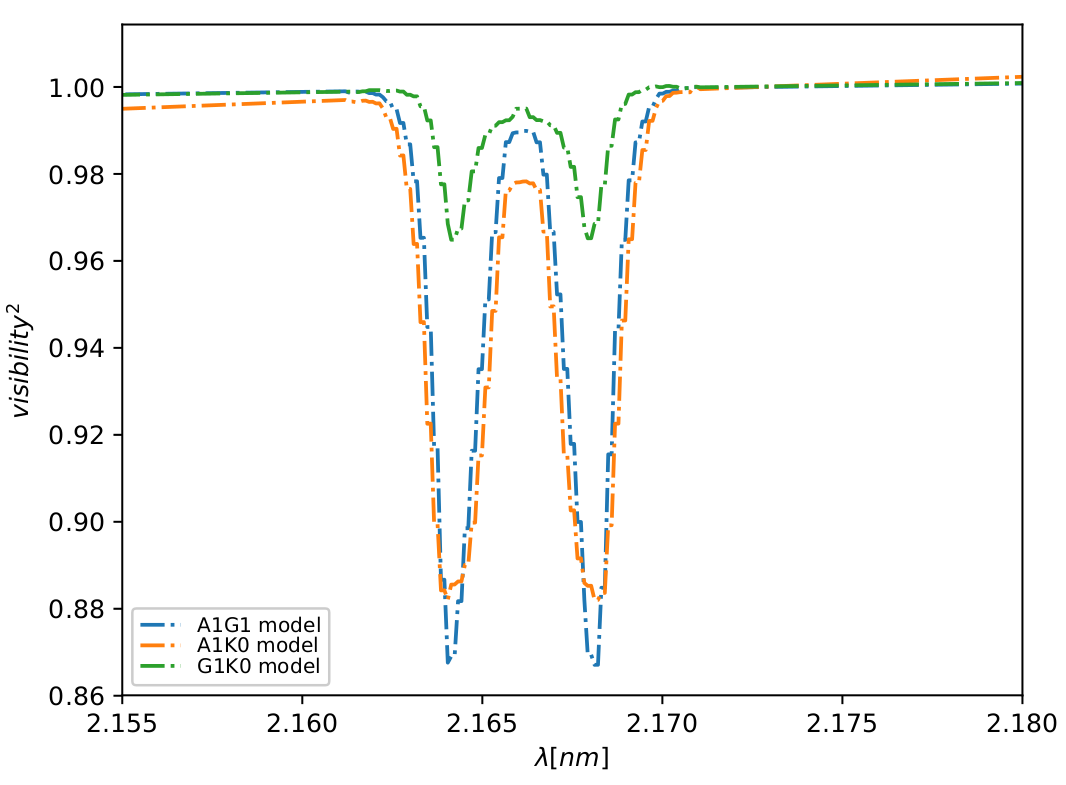}
  \end{adjustbox}
  \caption[First two rows: intensity maps (128 x 128 pixels) at different values of wavelength close to Br$\gamma$ (from left to right: 2.161, 2.164, 2.165, and 2.166 $\mu$) of two HDUST models, from the BeAtlas grid, with different values of inclination angle: $i$ = 0 (first row) and $i$ = \ang{90} (second row).]{First two rows: intensity maps (128 x 128 pixels) at different values of wavelength close to Br$\gamma$ (from left to right: 2.161, 2.164, 2.165, and 2.166 $\mu$) of two HDUST models, from the BeAtlas grid, with different values of inclination angle: $i$ = 0 (first row) and $i$ = \ang{90} (second row). All the other parameters are fixed as follows: $M_{\star}$ = 14.6 $M_{\odot}$ (type B0.5), $R_{\mathrm{eq}}/R_{\mathrm{p}}$ = 1.1, $\Sigma_{0}$ = 0.5 g cm\textsuperscript{-2}, and $m$ = 3.0 (see Table \ref{table_beatlas}). The flux/pixel scale is shown in arbitrary units. Simulated VLTI/AMBER visibilities from these 2 models are shown in the last row: $i$ = 0 (left panel) and $i$ = \ang{90} (right panel). See text for discussion. }
\label{sec_modeling_radiative_hdust_image_mod1}
\end{figure}

Regarding the disk properties, the last two parameters in Table \ref{table_beatlas} are parameterizations for the density structure of the VDD model description: the base surface density ($\Sigma_{0}$) and the radial density exponent ($m$). More generally, two kinds of disk models are found in the BeAtlas grid:

\begin{enumerate}[label=(\roman*)]
\setlength\itemsep{1em}
\item Non-isothermal steady-state models: self-consistent solution of the viscous diffusion equation considering a steady-state disk, that is, a null temporal variation of the disk density:

\begin{equation}
\diffp{\Sigma(r,t)}{t} = 0,
\end{equation}
considering a non-null mass injection in the disk, $\Sigma_{\mathrm{in}} \neq 0$.\par

Then the radial (surface) disk density is described by the mass-loss rate $\dot{M}$ and the Shakura-Sunyaev's viscous parameter $\alpha$, as given by Eq.~\ref{eq:sigma_analytical_steady}.\par

\item Parametric models: it is assumed an isothermal disk scale height (Eq.~\ref{eq:disk_height_scale}). In this case, the radial disk density is parameterized by the base disk density $\Sigma_{0}$, corresponding to different values of $\dot{M}$, and the density law exponent $n$, as given by Eq.~\ref{eq:radial_surface_density}.\par

\end{enumerate}

We recall the reader that each value of $n$ are related to the (volume) density law exponent by Eq.~\ref{sec_intro_relation_n_m}, as well as $\Sigma_{0}$ and $\rho_{0}$ are related to each other by Eq.~\ref{sec_intro_relation_sigma0_rho0}\par

In Sect.~\ref{sec_results_omicron_aquarii}, our analysis is based on the parametric models of BeAtlas, as the non-isothermal (hydrodynamically self-consistent) models currently cover a quite limited range of parameter values, with $\alpha$ fixed at 0.5, composing a small fraction of the grid ($\sim$14\%) from~\citet{faes15}. Further details on self-consistent models using HDUST can be found in~\citet{carciofi08} and~\citet{klement15}.

In conclusion, the range of values for $\Sigma_{0}$ and $m$ in BeAtlas encompass somewhat extreme cases in the literature for the circumstellar disk of Be stars. The listed values of $\Sigma_{0}$ correspond to $\rho_{0}$ from $\sim$ $10^{-12}$ g cm\textsuperscript{-3} to $\sim$ $10^{-10}$  g cm\textsuperscript{-3}. Parametric models with $m$ = 3.5 are equivalent to the steady-state solution of the viscous diffusion equation considering an isothermal disk scale height. Thus, regarding the mass density law exponent $m$, models with $m > 3.5$ would represent a disk passing by an accretion phase, while the ones with $m < 3.5$ a disk in an ongoing process of dissipation~\citep[see, e.g.,][]{haubois12, vieira17}.\par

Finally, Fig.~\ref{sec_modeling_radiative_hdust_image_mod1} compares the intensity map around the Br$\gamma$ of two BeAtlas models with different values of inclination angle: a perfectly pole-on ($i$ = 0) and one seen edge-one ($i$ = \ang{90}). All the other parameters are fixed, as indicated in Fig.~\ref{sec_modeling_radiative_hdust_image_mod1}. From that, note how the disk emitting envelope is much larger in the Br$\gamma$ line than in the close-by continuum (first three columns of the top panels), and then affecting the predicted value of differential visibility (simulated to a AMBER triplet configuration, bottom panels).\par


\section{For what CMFGEN and HDUST are suited?}
\label{sec_radiative_transfer_modeling_comparing_cmfgen_hdust}

From our previous discussion, both CMFGEN and HDUST are state-of-the-art codes, used in a series of works on hot massive stars. Despite being a very limited metric of impact, up to date, the basic reference of CMFGEN~\citep{hillier98} have been cited in $\sim$820 papers, while the one of HDUST~\citep[][]{carciofi06a} have received $\sim$150 citations. For sure, there is a time-delay between these two codes, but such discrepancy can be understood in terms of a larger applicability of CMFGEN, which, for example, fully takes into account the effect of line-blanketing, while HDUST is currently limited to hydrogen composition, when considering a pure gaseous environment. In addition, unlike HDUST\footnote{For access and collaborations with HDUST, contact A.~C.~Carciofi.}, CMFGEN is publicly available.\par

The code CMFGEN allows us to include several chemical elements (at different levels of ionization), and for each one the lines are treated in a ``exact'' way, that is, computed from solving in the comoving frame the radiative transfer, statistical and radiative equilibrium equation simultaneously. For instance, from our discussion of Fig.~\ref{sec_modeling_radiative_transfer_groh11_fig11}, one sees how unsuitable is HDUST in order to model the wind of an early-type O star (physical parameters for a O4 star in Fig.~\ref{sec_modeling_radiative_transfer_groh11_fig11}) due the absence of metals in the evaluation of the opacity, being this issue less severe for B-type stars (showing lower values of effective temperature).\par 

On the other hand, unlike CMFGEN, HDUST allows to take into account the opacity due to dust grains, being this a very important feature, among the massive hot stars, when modeling the environments of B[e] supergiants. Nevertheless, we point out that current efforts have been made to include several chemical elements, beyond the hydrogen (and free electron) gas composition, in the modeling with HDUST~\citep[HDUST3,][]{carciofi17}.\par

In addition, HDUST does not take the task of explicitly solve the radiative transfer equations, providing a statistical approach, by a Monte Carlo method, to find the radiative field and state of the gas. In particular, one caveat of Monte Carlo radiative transfer codes relies on convergence issues when dealing with the calculation of optically thick lines, as the number of interaction particle-photon is increased due to a larger opacity of the medium.\par 

However, the use of a Monte Carlo approself-ach introduces a much higher flexibility on the modeled geometry. Again, all the CMF-based codes, shown in Table \ref{table_radiative_transfer_codes} assume a spherical symmetric outflow. Thus, for example, HDUST is very suited to model the environment of a Be star, formed by an equatorial disk (and possibly, in addition to a polar wind), due to complex geometry in this case. This is a proper task to be performed using a Monte Carlo-based code, such as HDUST. Lastly, unlike CMFGEN, notice that HDUST is able to predict the degree of light polarization as a result of the asymmetric geometry of the envelope. Nevertheless, efforts have been made to implement non-LTE CMF codes for hot stars beyond the 1-D geometry. For example, based on CMFGEN, see the code of~\citet{busche05}.\par


\section{Computational cost of physical models}
\label{sec_radiative_transfer_modeling_computational_cost}

Due to the complexity of solving the radiative transfer problem, under the non-LTE treatment, both CMFGEN and HDUST take several hours to convergence a single model.\par 

Considering the non-parallelized version of CMFGEN\footnote{Currently, it is possible to install a parallelized version of CMFGEN. This leads to faster calculations by up to a factor of 2-3 times when using 4 processors.}, it must be needed $\sim$6-12 hours of computing time in order to convergence a single CMFGEN model, allocating $\sim$2-3 GB of RAM memory. However, these reference values are highly dependent on the chosen solution method, the number of atomic species that are included in the modeling, as well as, the number of iterations and depth points that are chosen by the user to calculate the photosphere and wind\footnote{For instance, models for O stars typically must be computed using a total of $\sim$60-80 depth points for the photosphere plus wind structure. See the description for the control file MODEL\_SPEC in the CMFGEN manual.} .\par

Furthermore, since in practices it is required a start CMFGEN model, in order to estimate the populations, for sure, the convergence time (as the convergence itself) depends on how the parameters are changed in the new model, compared with the start model. For instance, including more heavy elements in the model, and thus increasing the number of equations to be solved, increases the allocation of memory and it can lead to a computing time for up to more than one day. For comparison, the code FASTWIND demands about 1 to 10 minutes to converge a model (Table \ref{table_radiative_transfer_codes}), as it works using the Sobolev approximation to treat a large number of lines (mainly of the iron group).\par 

In short, both Monte Carlo- (as HDUST) and CMF- (as CMFGEN) based codes have a significant computational cost. Hence, it is impossible to implement a ``real-time'' fitting process using these codes. For example, considering the time of 6 hours for converging a CMFGEN model, and employing a Markov Chain Monte Carlo method (as will be latter discussed in Sect.~\ref{sec_analytical_modeling_mcmc}), with 180000 tested samples of parameters, it would be needed $\sim$120 years for searching the best-fit parameters that explain the data!\par

As will be discussed in Chap.~\ref{chapter_analytical_models}, this issue described above can be overcome using  analytical and simple numerical model, having a computational cost of a few seconds, or much less than that.\par

In conclusion, obtaining the ``best-fit'' models, using the CMFGEN and HDUST codes, relies on physical insights from the user in order to change a specific parameter of interest for comparison with the data, and typically requiring to compute a large number of models (for example, $\sim$50) to obtain the best-fit model. Alternatively, one can use a pre-calculated model grid (as the BeAtlas grid, discussed in Sect.~\ref{sec_radiative_transfer_modeling_hdust}). However, this approach is also biased with respect to exploration of the parameter space in order to better explain the data. Regarding CMFGEN, current efforts have been made to calculate large public model grids, and then requiring the use of CPU clusters~\citep[e.g.,][]{zsargo13, fierro15, zsargo17}.\par
\chapter{Model fitting of interferometric data}
\label{chapter_analytical_models}
\minitoc

\section{Geometric models}
\label{sec_analytical_modeling_geometric_models}

\subsection{Applying the Zernike-van Cittert theorem to analytical models}
\label{sec_analytical_modeling_geometric_models_overview}

Despite not being able to constrain the physical conditions in the photosphere and the envelope, such as the density and temperature structures, as well its kinematics, geometric models are useful modeling tools to study the geometry of the central star and its environment~\citep[e.g., see][]{berger03}.\par 

This is the simplest possible approach to interpret interferometric data. As discussed in Sect.~\ref{sec_interf_quantities_zernike}, the so-called Zernike-van Cittert theorem (Eq.~\ref{eq:van_cittert_theorem}) relates the brightness of the source projected on the sky to the degree of the light coherence measured by the interferometer at given wavelength and spatial frequencies. For comparison with observational interferometric quantities, the intensity distribution of the star is modeled by a simple analytical function and then the complex visibility calculated by applying the Zernike-van Cittert theorem, that is, a Fourier transform applied to the stellar intensity map on the sky.\par

In the following, we discuss just some simplified cases of geometric modeling, but before setting the problem itself, it is useful to remind the reader about some basic properties of Fourier transform (FT), liking the brightness distribution and the interferometric quantities, as follows:

\begin{enumerate}
    \item Addition property: $\mathrm{FT}\,\{I_{1}(\alpha,\beta) + I_{2}(\alpha,\beta)\}=\widetilde{V_{1}}(u,v) + \widetilde{V_{2}}(u,v)$. 
    
    \item Similarity property: $\mathrm{FT}\, \{ I_{1}(a \alpha, b \beta)   \}  = \frac{1}{|ab|} \widetilde{V_{1}}(u/a,v/b) $.
    
    \item Translation property: $\mathrm{FT}\, \{  I_{1}(\alpha-\alpha_{0}, \beta-\beta_{0} )  \} =  \widetilde{V_{1}}(u,v) \exp{[2\pi i (u\alpha_{0} + v\beta_{0})]}  $.
    
    \item Convolution property: $\mathrm{FT}\,\{I_{1}(\alpha,\beta) \circledast I_{2}(\alpha,\beta)\} = \widetilde{V_{1}}(u,v) \, \widetilde{V_{2}}(u,v)$.
    
\end{enumerate}

In the equations above, $\widetilde{V}(u,v)$ denotes the complex visibility without taking into account the scaling factor due to the total intensity of the source, that is, it is not normalized. On the other hand, we remind the reader that $\widetilde{V}(u,v)$, as expressed by Eq.~\ref{eq:van_cittert_theorem}, is the complex visibility normalized by the total intensity of the source. The functions $I_{1}$ and $I_{2}$ denote two different intensity maps and the notation of the variables stands as used in Sect.~\ref{sec_interf_quantities}, $a$, $b$, $\alpha_{0}$, and $\beta_{0}$ are constants, and the convolution operator is denoted by ``$\circledast$''.\par

\subsection{The simplest case: an example of one-component model}
\label{sec_analytical_modeling_geometric_models_one_component_model}

The simplest possible geometric modeling scenario consistent of assuming that the stellar intensity is described by a uniform disk. In this case, the (squared) visibility of a uniform disk with angular diameter $\theta$ is given by a first-order Bessel function of the first kind, $J_{1}$:

\begin{equation}
V_{\mathrm{UD}}^{2} = \left(\frac{2J_{1}(\pi \theta \sqrt{u^2 + v^2})}{\pi \theta \sqrt{u^2 + v^2}}\right)^2,
\label{eq:visibility_uniform_disk}
\end{equation}
where the variables $u$ and $v$ are the spatial frequency coordinates (Eq.~\ref{eq:spatial_frequencies}). In the case of a uniform disk, since the intensity map of the object is center-symmetric the phase's component of the complex visibility is null, simplifying it to a real number.\par

\begin{figure}[t]
\centerfloat
\centerline{\resizebox{0.75\textwidth}{!}{\includegraphics{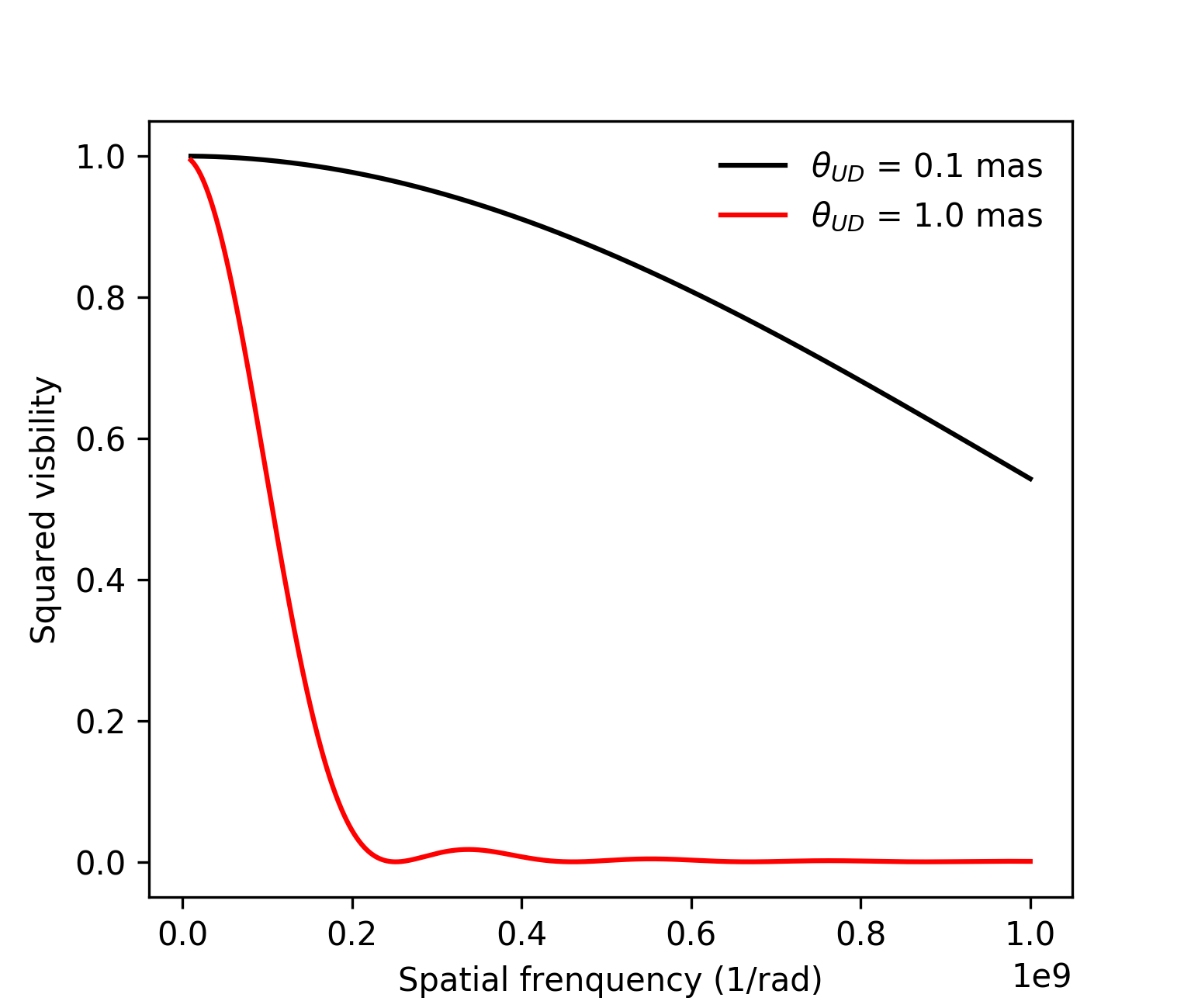}}}
\caption{Comparison between the visibility curves (squared visibility, $V^2$) of two uniform disks with different values of angular diameter: $\theta_{\mathrm{UD}}$ = 0.1 mas (black) and 1.0 mas (red).}
\label{sec_analytical_modeling_plot_visibility_uniform_disks}
\end{figure}

For instance, Fig. \ref{sec_analytical_modeling_plot_visibility_uniform_disks} shows $V_{\mathrm{UD}}^2$ for uniform disks of different angular diameters. From that, notice the visibility lobes from the Bessel function for the larger uniform disk, while the smaller one is barely resolved at lower spatial frequencies ($V_{\mathrm{UD}}^2$ close to 1).\par

\subsection{Adding other components: an example of two-component model}
\label{sec_analytical_modeling_geometric_models_two_component_model}

Somewhat more complex model configurations, that is, with higher number of free parameters, can be employed. For instance, Fig. \ref{sec_analytical_modeling_fig8_sigut15} shows the geometric modeling by~\citet{sigut15} to NPOI visibilities of the Be star $\omicron$ Aquarii.\par 

Their best-fit model (reduced $\chi^2$ of 1.097) is a two-component geometric model: a 0.22 mas uniform disk contributing to 86\% of the total flux, representing the photosphere, plus an elliptical Gaussian, representing the flux contribution from the disk, with parameters indicated in Fig. \ref{sec_analytical_modeling_fig8_sigut15}.\par

From Fig.~\ref{sec_analytical_modeling_fig8_sigut15}, their second component contributes to about 14\% of the total flux~\citep[see Table 3 of][]{sigut15}. They found a disk major-axis Full width at half maximum (FWHM) of $\sim$2.58 mas, corresponding to $\sim$11.6 $D_{\star}$, under their assumption of 0.22 mas for the stellar diameter, with position angle of \ang{110}.\par

Following the addition property of Fourier transform, in such a two-component geometric approach, in this case above, the total model visibility, $V_{\mathrm{total}}$, is then expressed as~\citep[see Eq.~6 of][]{sigut15}: 

\begin{equation}
V_{\mathrm{total}} = F_{\star}V_{\mathrm{UD}}(0.22 \, \mathrm{mas}) + (1 - F_{\star})V_{\mathrm{GB}}.
\label{eq:visility_total_model_visibility}
\end{equation}

\begin{figure}[t]
  \begin{adjustbox}{minipage=\textwidth,scale=1.00}
  \centering
  \includegraphics[width=0.495\columnwidth]{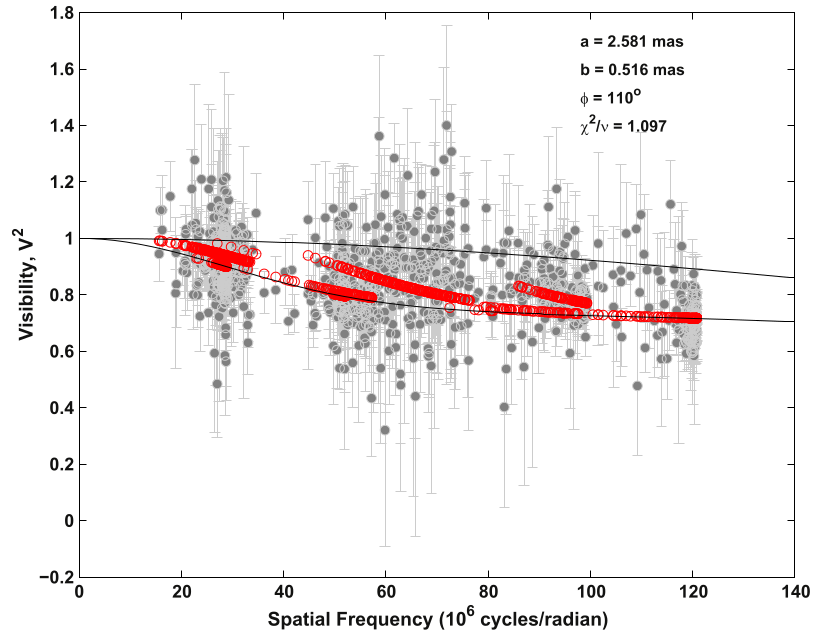}
  \includegraphics[width=0.495\columnwidth]{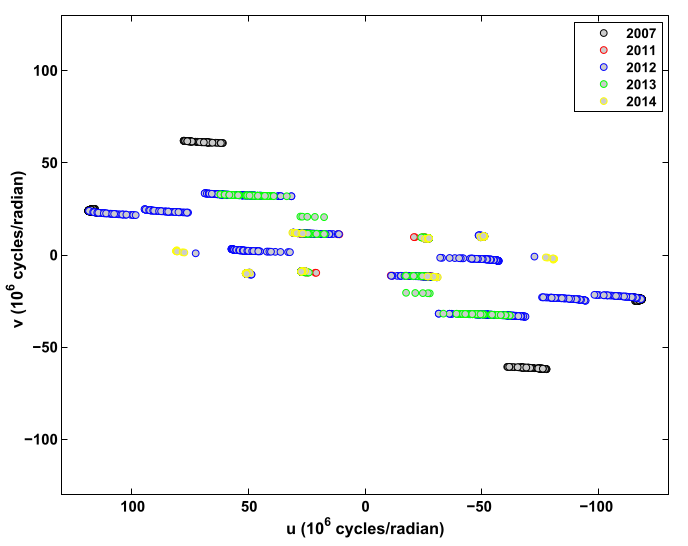}
  \end{adjustbox}
  \caption[Left panel: modeling to NPOI observations (interferometric data centered on H$\alpha$) of the Be star $\omicron$ Aquarii using geometric models.]{Left panel: modeling to NPOI observations (interferometric data centered on H$\alpha$) of the Be star $\omicron$ Aquarii using geometric models. The best-fit model to $V^2$ is shown in red points (simulated visibilities for each value of spatial frequency): a two-component geometric model composed by a uniform disk and an elliptical Gaussian distribution (axial ratio $r$ = 0.2). The visibilities of the major and minor axes of the fitted elliptical Gaussian (values provided in the figure, a and b, respectively) are shown in black solid lines. $a$ and $b$ correspond to the major and minor axes (Full width at half maximum) of the Gaussian distribution. $\phi$ = \ang{110} is the major-axis position angle of the disk on the sky. The corresponding $uv$ plan coverage of the observations is shown in the right panel. Reproduced from~\citet{sigut15}.}
\label{sec_analytical_modeling_fig8_sigut15}
\end{figure}

In Eq.~\ref{eq:visility_total_model_visibility}, $F_{\star}$ is the (normalized) flux contribution from the central star, and then $1 - F_{\star}$ the one from the disk. $V_{\mathrm{UD}}(0.22 \, \mathrm{mas})$ is given by Eq.~\ref{eq:visibility_uniform_disk} for $\theta$ = 0.22 mas, while the visibility from the elliptical Gaussian distribution is described by~\citep[e.g., see][]{tycner06}:

\begin{equation}
V_{\mathrm{GB}}(s) = \exp{\left[ -\frac{(\pi a s)^2}{4 \ln{2}} \right]},
\end{equation}
where $a$ is the major-axis Full width at half maximum (FWHM) of the Gaussian, and the variable $s$ is defined as follows:

\begin{equation}
s = \sqrt{r^2(u\cos{\phi} - v\sin{\phi})^2 + (u\sin{\phi} + v\cos{\phi})^2},
\end{equation}
where $\phi$ and $r$ are respectively the major-axis position angle and axial ratio ($a$/$b$) of the disk, represented by the Gaussian, following the same notation as in Fig.~\ref{sec_analytical_modeling_fig8_sigut15}. Note that for a circular Gaussian distribution ($r$ = 1, non-flattened), $s$ is simply replaced by the term $\sqrt{u^2 + v^2}$, as in Eq.~\ref{eq:visibility_uniform_disk} for the uniform disk.


\section{Analytical model fitting: the software LITpro}
\label{sec_analytical_modeling_litpro}

\subsection{Overview: a tool dedicated to interferometric modeling}
\label{sec_analytical_modeling_litpro_overview}

LITpro\footnote{Acronym for Lyon Interferometric Tool prototype.}~\citep{tallonbosc08} is a model-fitting software for visible and infrared interferometric data developed by the Jean-Marie Mariotti Center (JMMC). It allows us to combine multiple geometric model components and to derive the model parameters that minimizes the reduced $\chi^2$ to data. In order to search for the local $\chi^2$ minimum, this modeling software is based on a Levenberg-Marquardt algorithm in combination with a Trust Region method. Details on these numeric methods can be found in~\citet{tallonbosc08} and references within.\par

The data is read by LITpro in the OI Exchange Format~\citep[OI-FITS][]{pauls05}, allowing to store, following the OIFITS notation: squared visibilities (VIS2), amplitude and phase of complex visibility (VISAMP and VISPHI), amplitude and phase of bispectrum (T3AMP and T3PHI). It allows to include data from several OIFITS files, reorganizing the arrays in the memory in order to make the data analysis more efficient. LITpro has been updated and currently improved by JMMC and is publicly available\footnote{LITpro website: \url{http://www.jmmc.fr/litpro}}, in addition to many other softwares dedicated to optical interferometry that also are supported by JMMC\footnote{See \url{http://www.jmmc.fr/english/the-jmmc/who-are-we/}.}.\par

\begin{figure}[t]
\centerfloat
\centerline{\resizebox{1.00\textwidth}{!}{\includegraphics{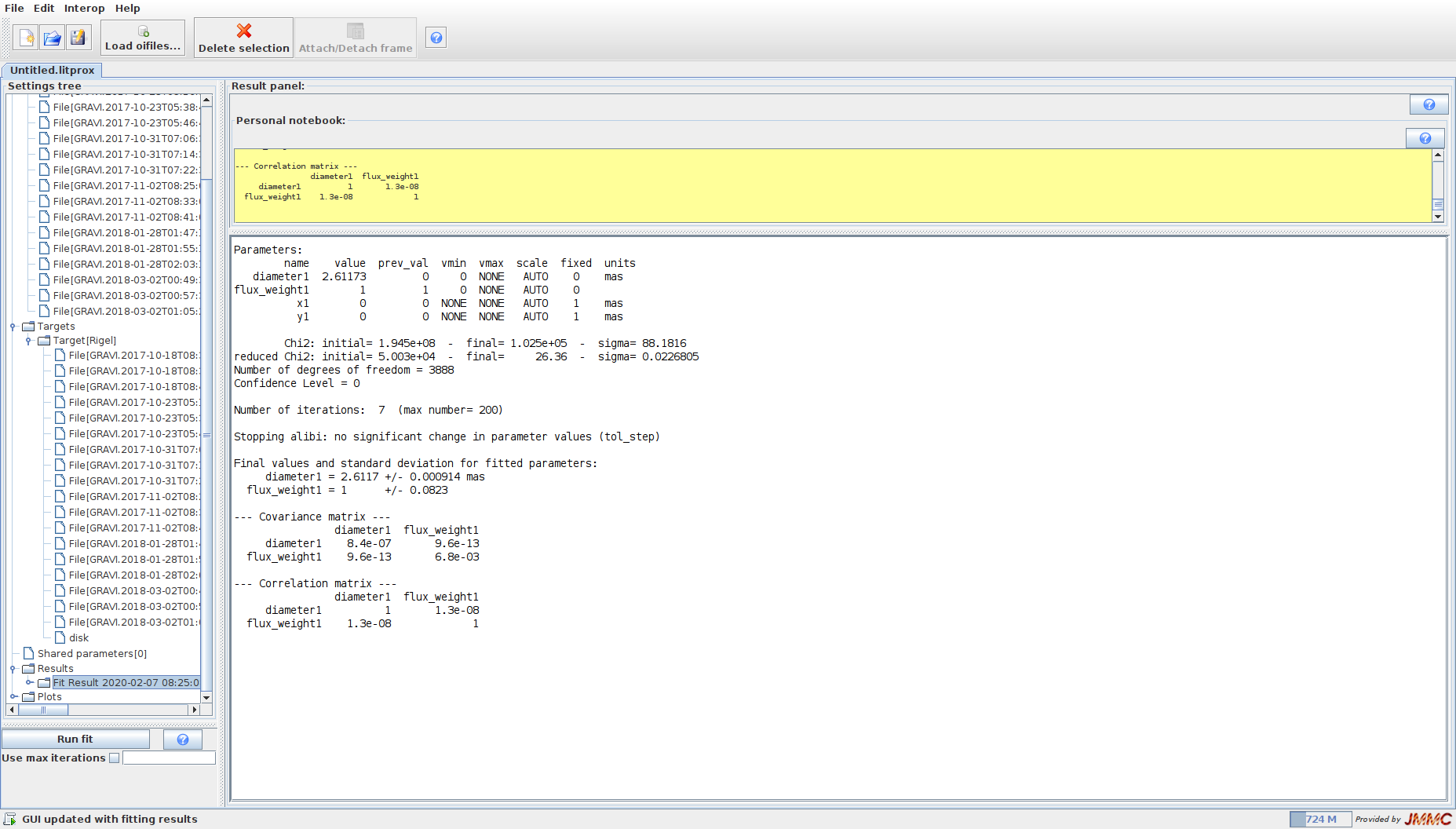}}}
\caption{Example of the graphical interface of LITpro, used in this case to model VLTI/GRAVITY data of Rigel using a uniform disk model: best-fit model with angular diameter of $\sim$2.61 mas. See text for discussion.}
\label{sec_analytical_modeling_plots_litpro1}
\end{figure}

LITpro has several built-in functions to be chosen by the user to the data analysis, such as uniform and elliptical disk and uniform and elliptical Gaussian distributions, which are of interest to interpret interferometric data of Be stars. The complete list of available models in LITpro can be found in the LITpro Reference Guide (Table 1) in the LITpro website (see footnote 2 of this chapter). For instance, somewhat more complex models are allowed, such as limb-darkened disk model, which are of interest to evaluate the limb-darkening effect using interferometric data~\citep[e.g., see][and references within]{nardetto16}. It is important to stress that LITpro works directly in the Fourier space, using the analytical solutions of the Fourier transform for the chosen built-in functions.\par

\subsection{Example of modeling: VLTI/GRAVITY data of Rigel}
\label{sec_analytical_modeling_litpro_example_rigel}

For instance, Fig. \ref{sec_analytical_modeling_plots_litpro1} shows the graphic interface of LITpro (version 1.0.16). It works using Graphic User Interface in JAVA language. In this case, we consider a simple uniform disk to model our VLTI/GRAVITY squared visibility (absolute visibility) of the B supergiant Rigel (see Sect.~\ref{sec_ongoing_studies_rigel_spectro_interferometry}). The fitted-visibility curve to our data is shown in Fig. \ref{sec_analytical_modeling_plots_litpro2}. This is the visibility at the close-by continuum region, $\sim$2.145-2.155 $\mu$m, to the Br$\gamma$ line.\par

Considering a uniform disk, the best-fit model, $\theta$ $\sim$ 2.61 mas, is very efficiently found by LITpro, as just 7 iterations are needed to be performed from a total of 200 maximum iterations that were chosen in this case. The large value of minimum reduced $\chi^2$ ($\sim$26) comes from the unrealistic very low error bars considered in the fitting. However, one sees how the minimum $\chi^2$ is well constrained around $\theta$ = 2.61 mas (see the left panel of Fig. \ref{sec_analytical_modeling_plots_litpro2}).\par

In addition to the example above, I extensively used the software LITpro in my spectro-interferometric study on the Be star $\omicron$ Aquarii, as will be discussed in Sect.~\ref{sec_results_omicron_aquarii}. In advance of discussion, the best-fit geometric models to VEGA calibrated data of $\omicron$ Aquarii are presented in Fig.~2 of~\citet{paperIII}, as well as Table~1 of the same article shows the summary of the results for the (angular) extension of its photosphere and circumstellar disk, based on that data and LITpro modeling.\par

\begin{figure}[t]
  \begin{center}
  \begin{adjustbox}{minipage=\textwidth,scale=1.00}
  \includegraphics[width=0.444\columnwidth]{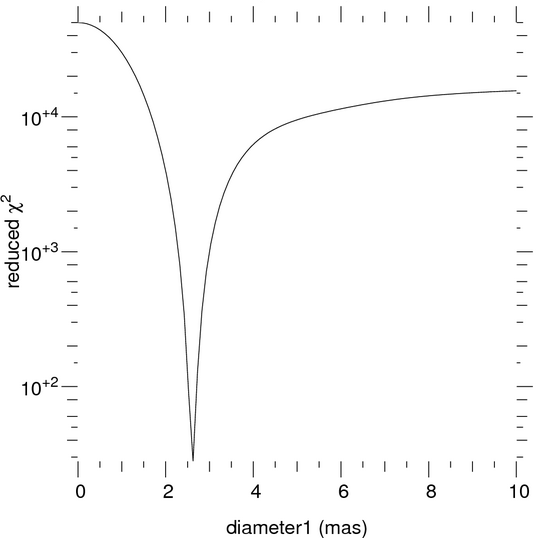}
  \includegraphics[width=0.495\columnwidth]{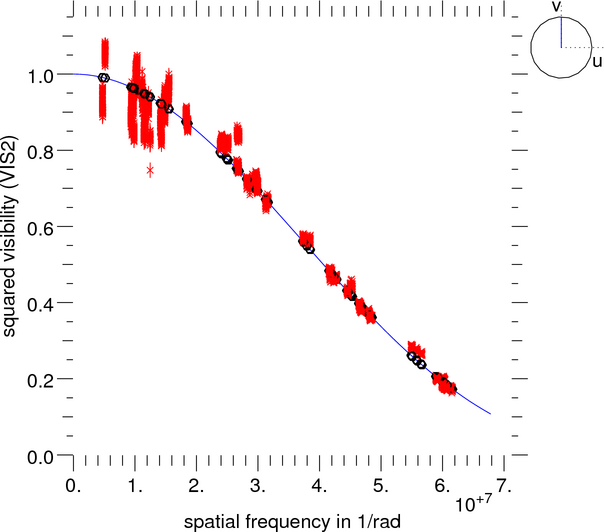}
  \end{adjustbox}
  \caption[Geometric modeling, using the software LITpro (Fig.~\ref{sec_analytical_modeling_plots_litpro1}), of our VLTI/GRAVITY data of Rigel: squared visibilities at the close-by continuum region ($\sim$2.145-2.155 $\mu$m) to Br$\gamma$.]{Geometric modeling, using the software LITpro (Fig.~\ref{sec_analytical_modeling_plots_litpro1}), of our VLTI/GRAVITY data of Rigel: squared visibilities at the close-by continuum region ($\sim$2.145-2.155 $\mu$m) to Br$\gamma$. The $\chi^2$-map delivered by LITpro is shown in the left panel. The GRAVITY data (red points) is compared with the fitted geometric model (open points, blue line) in the right panel. See text for discussion.}
  \label{sec_analytical_modeling_plots_litpro2}
  \end{center}
\end{figure}

\subsection{Limitations: what is the global $\chi^2$ minimum?}
\label{sec_analytical_modeling_litpro_limitations}

Since LITpro is a robust fitting-tool that uses a Levenberg-Marquardt algorithm, it allows us to find local $\chi^{2}$ minimum, or, minima, to explain the data, based on the partial derivatives of the selected functions to model the data.\par

From Fig.~\ref{sec_analytical_modeling_plots_litpro2}, one sees that the global minimum using a uniform disk model is easily found (around $\theta$ = 2.61 mas) in this case. However, discriminate the global minimum among several local minima can be a much harder task to be performed, especially considering more complex models, and ultimately the own data. More generally, this is one major difficulty of non-linear model fitting methods, as the Levenberg-Marquardt method.\par

Lastly, as a practical point-of-view, we point out that the results found by LITpro are somewhat sensitive to the initially adopted parameter values. Some tools have been developed to guide the user to find the global minimum (as the $\chi^{2}$-map shown Fig.~\ref{sec_analytical_modeling_plots_litpro2}). To verify if LITpro is not indeed converging to a particular local minimum solution, the user should test different adoptions for the initial values of the model parameters.\par


\section{The kinematic code}
\label{sec_analytical_modeling_kinematic_models}

\subsection{Overview: the model parameters}
\label{sec_analytical_modeling_kinematic_models_parameters}

The kinematic code was developed by A.~Meilland (Observatoire de la Côte d'Azur) with the particular purpose of constraining the geometry and kinematics of circumstellar disks using spectro-interferometric data (Sect.~\ref{sec_interf_spectro_interferometry}), especially of Be stars.\par

It was firstly used by~\citet{delaa11} to model VEGA observations of the Be stars 48 Persei and $\phi$ Persei, and since then it has been extensively used in other interferometric studies of Be stars~\citep[e.g.,][]{meilland11, meilland12, jamialahmadi15, cochetti19}. The interested reader can find a broader discussion on this code in these references above.\par

More recently, the kinematic code was also employed to interpret the VEGA and AMBER spectro-interferometric data of $\omicron$ Aquarii, presented in this thesis (Sect.~\ref{sec_results_omicron_aquarii}). In advance of discussion, a summary of our best-fit kinematic models (to VEGA and AMBER data) is found in Table~2 of~\citet{paperIII}.\par

Before discussing on how the kinematic code works, a full list of parameters of the kinematic model is given as follows:

\begin{enumerate}[label=(\roman*)]
\setlength\itemsep{1em}

\item The global simulation parameters: size of the simulation in pixels ($n_{x} = n_{y}$), field of view in stellar diameter ($fov$), number of wavelength ($n_{\lambda}$), central wavelength of the emission line ($\lambda_{0}$), step in wavelength ($\delta\lambda$), and spectral resolution (R).

\item The global geometric parameters: stellar radius ($R_{\star}$), distance ($d$), inclination angle ($i$), and disk major-axis position angle ($PA$).

\item The disk continuum parameters: disk major-axis FWHM in the continuum ($a_{c}$), disk continuum flux normalized by the total continuum flux ($F_{c}$).

\item The disk emission line parameters: disk major-axis FWHM in the line ($a_{l}$) and line equivalent width ($EW$).

\item The global kinematic parameters: stellar rotational velocity ($v_\mathrm{rot}$), expansion velocity at the photosphere ($v_{0}$), terminal velocity ($v_{\infty}$), exponent of the expansion velocity law ($\gamma$), and exponent of the rotational velocity law ($\beta$).

\end{enumerate}

Thus, in the total, 13 parameters describe the geometry and kinematics of the central stars and the disk.\par

\subsection{Model description: the central star and the circumstellar environment}
\label{sec_analytical_modeling_kinematic_models_description}

This codes uses simple geometric models to describe the emissions from the star and its environment, but taking into account the Doppler effect on the considered spectral channel (line), being thus suited to model spectro-interferometric data (see, again, Fig. \ref{sec_inter_meilland14_fig8_9}).\par 

The flux contribution from the star is modelled as a uniform disk, while the circumstellar disk is modelled as two (elliptical) Gaussian distributions: one for the disk emission in the continuum and the other for the disk emission in the line.\par

The disk intensity map in the line is computed taking into account the Doppler effect in the considered spectral channel due to the disk velocity of expansion and rotation. The expansion velocity is parameterized as follows:

\begin{equation}
v_{\mathrm{radial}}(r) = v_{0} + (v_{\infty} - v_{0})\left(1 - \frac{R_{\star}}{r}\right)^{\gamma}.
\label{eq:kinematic_model_rdisk}
\end{equation}

We aware the reader that Eq.~\ref{eq:kinematic_model_rdisk} is indeed Eq.~\ref{eq:beta_law_geral} (Sect.~\ref{sec_intro_stellar_winds}), and thus following the CAK-theory, but changing the variable $\beta$ in Eq.~\ref{eq:beta_law_geral} to $\gamma$ (as in the parameter list in Sect.~\ref{sec_analytical_modeling_kinematic_models_parameters}). This notation for the wind velocity-law in Sect.~\ref{sec_intro_stellar_winds} follows the practices in the literature of radiative line-driven winds~\citep[e.g.,][]{puls09}. As described below, this change is done because the variable $\beta$ is used here~\citep[and in][]{paperIII} to describe the rotational velocity law.\par

The rotational velocity of the disk in the kinematic model is given by a simple power law with exponent $\beta$:

\begin{equation}
v_{\mathrm{azimuthal}}(r) = v_{\mathrm{rot}} {\left(\frac{r}{R_\star}\right)}^{\beta}.
\label{eq:kinematic_model_vdisk}
\end{equation}

Note that, as Be disks show very slow expansion velocity, the parameters $v_{0}$ and $v_{\infty}$ can be set zero ($v_{\mathrm{azimuthal}} >> v_{\mathrm{radial}}$).\par

Taking into account Eqs. \ref{eq:kinematic_model_rdisk} and \ref{eq:kinematic_model_vdisk}, the resulting velocity field, $v_{\mathrm{proj}}$ , along the observer's line of sight ($\phi$) is then computed as follows:

\begin{equation}
v_{\mathrm{proj}} = ( v_{\mathrm{azimuthal}}\sin{\phi} - v_{\mathrm{radial}}\cos{\phi} )  \sin{i}.
\label{eq:kinematic_model_vproj}
\end{equation}

We remind the reader that $i$ is defined as the stellar inclination angle (Sect.~\ref{sec_analytical_modeling_kinematic_models_parameters}). From $v_{\mathrm{proj}}$, iso-velocity maps ($R(x, y, \lambda,\delta \lambda)$), projected along the line of sight, are calculated for each considered spectral channel:

\begin{figure}[t]
\centerfloat
\centerline{\resizebox{0.75\textwidth}{!}{\includegraphics{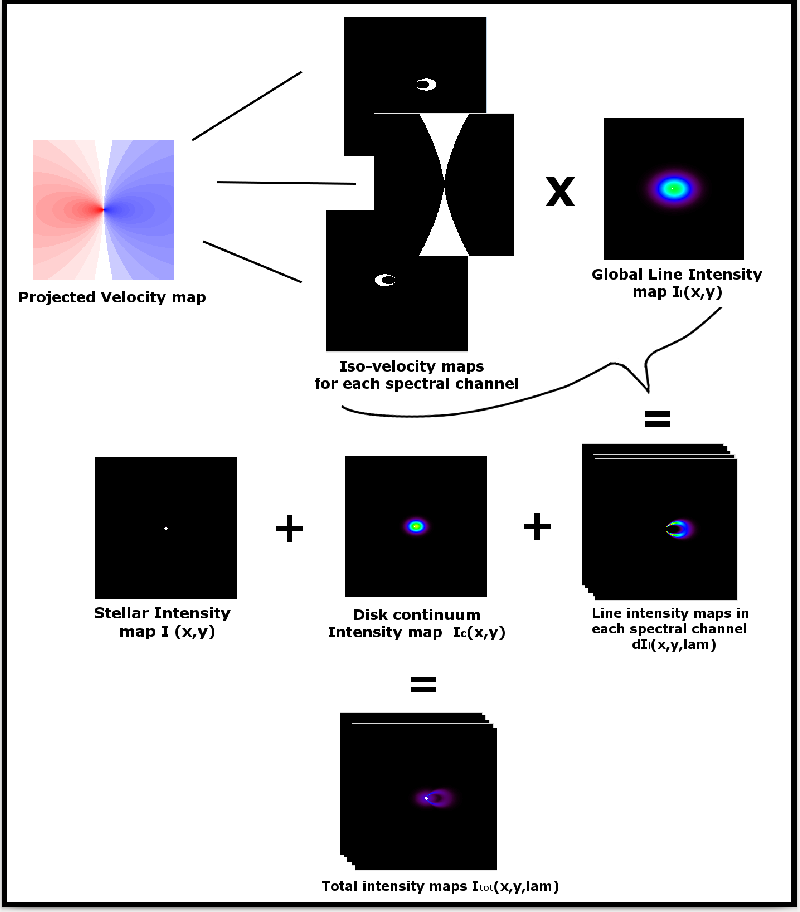}}}
\caption{Chart showing the calculation of the intensity map of the central star plus disk system (total intensity map shown in the bottom) with the kinematic code. See text for discussion.}
\label{sec_analytical_modeling_schematic_kinematic_model}
\end{figure}

\begin{equation}
R(x, y, \lambda,\delta \lambda) = \frac{1}{\sigma \sqrt{2\pi}} \exp{ \left[  \left(-\frac{v_{\mathrm{proj}} (x,y) - v_{\mathrm{doppler}}(\lambda)}{\sqrt{2}\,\sigma}\right)^{2}  \right]  },
\label{eq:iso_velocity_map}
\end{equation}
where $x$ and $y$ are the coordinates of the object (projected on the sky), and $v_{\mathrm{doppler}}$ is the Doppler displacement in the line profile, with $\sigma$ given by:

\begin{equation}
\sigma = \frac{\delta \lambda c}{2 \lambda \sqrt{2 \ln{2}}}.
\end{equation}

Finally, the total intensity map ($I_{\mathrm{total}}$) can be calculated as follows:

\begin{equation}
I_{\mathrm{total}} = I_{\star}F_{\star} + I_{c}^{env}F_{c} + R(x, y, \lambda,\delta \lambda) I_{\mathrm{line}} EW,
\label{eq:total_intensity_map}
\end{equation}
where $I_{\star}$ is the star intensity (constant brightness, a uniform disk), $I_{c}^{env}$ is the circumstellar intensity in the continuum. The parameters $F_{\star}$ and $F_{c}$ are, respectively, the stellar and disk continuum fluxes normalized by the total continuum flux, and the other variables were described so far (see the list of parameters shown at the beginning of this section). The last term of \ref{eq:total_intensity_map} corresponds to the circumstellar disk intensity in the spectral line. Further details on Eqs.~\ref{eq:kinematic_model_vproj}, \ref{eq:iso_velocity_map}, and \ref{eq:total_intensity_map} can be found in~\citet{delaa11}.\par 

A schematic of the calculation with the kinematic mode is presented in Fig. \ref{sec_analytical_modeling_schematic_kinematic_model}. In the end, an image cube (intensity map for each value of wavelength) of the modeled star plus disk system is calculated and then the interferometric quantities, as a function of wavelength, are extracted by applying the discrete Fourier transform using a fast Fourier transform algorithm\footnote{
Currently, the kinematic code is written in IDL language. Details about the IDL implementation of the fast Fourier transform algorithm can be found at \url{https://www.harrisgeospatial.com/docs/FFT.html}.} to the intensity maps at each value of wavelength.\par

\begin{figure}[t]
\centerfloat
\centerline{\resizebox{1.00\textwidth}{!}{\includegraphics{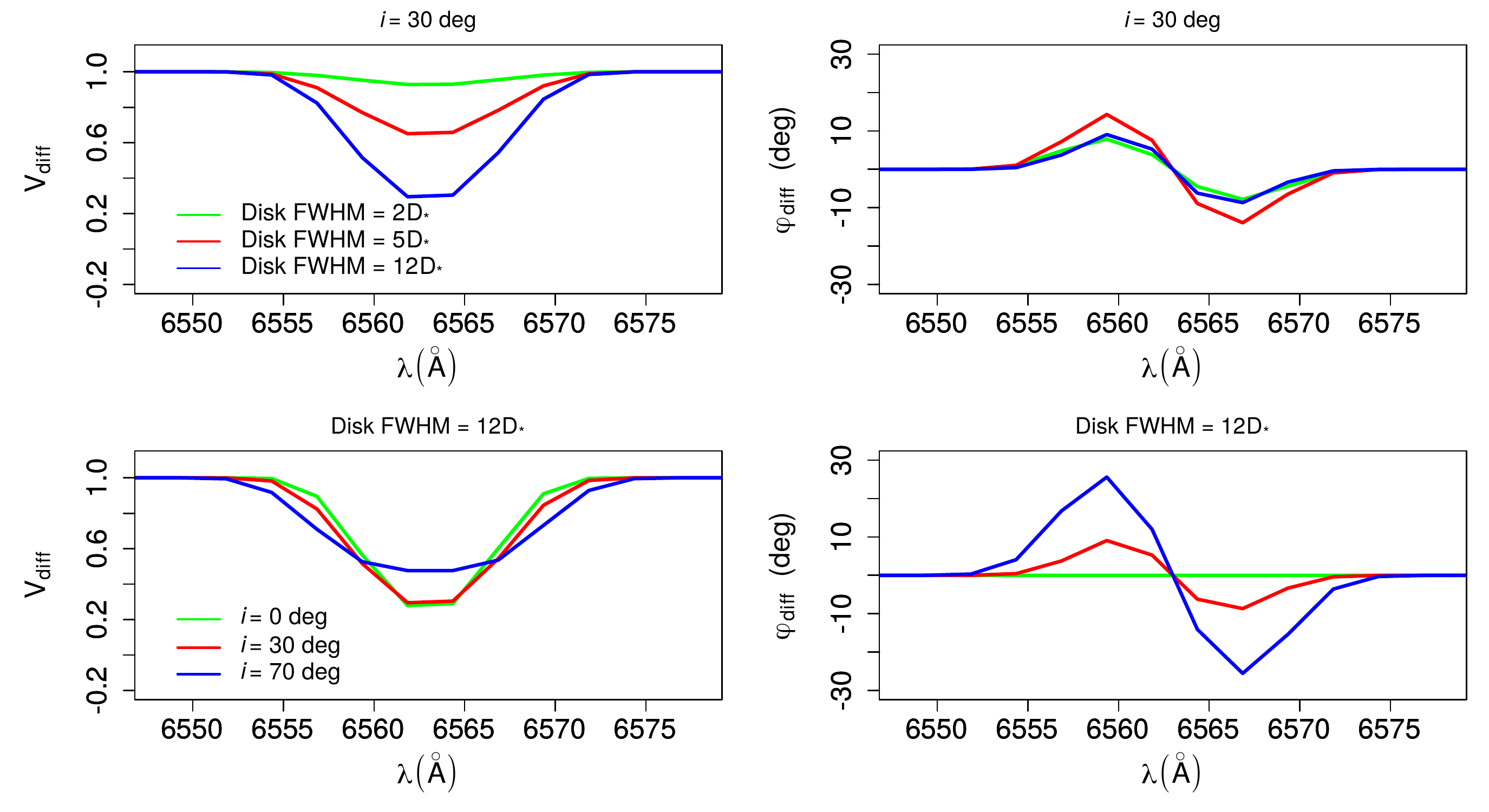}}}
\caption[Comparison between kinematic models, calculated around the H$\alpha$ line, by varying just one selected parameter:]{Comparison between kinematic models, calculated around the H$\alpha$ line, by varying just one selected parameter: the disk size in the line (top panels, FWHM from 2 to 12 $D_{\star}$ with $i$ fixed at \ang{30}) and the stellar inclination angle (bottom panels, $i$ from 0 to \ang{70} with FWHM fixed at 12 $D_{\star}$). Simulated differential visibility and phase for one of our CHARA/VEGA measurements of $\omicron$ Aquarii. Note how the variation of these parameters changes the visibility and phase in a different way. See text for discussion.}
\label{sec_analytical_kinematic_model_varying_disk_size_incl}
\end{figure}

\subsection{Parameters effects on the interferometric quantities}
\label{sec_analytical_modeling_kinematic_models_effect_varying_parameters}

An example of observables (differential interferometric quantities and the spectrum in H$\alpha$) using the kinematic code was presented in Fig. \ref{sec_inter_meilland14_fig8_9} (Sect.~\ref{sec_interf_spectro_interferometry}).\par 

In Fig. \ref{sec_analytical_kinematic_model_varying_disk_size_incl}, we compare kinematic models, calculated in the H$\alpha$ line, with different values of disk size in the H$\alpha$ line ($a_{l}$) and stellar inclination angle ($i$). All the other parameters are fixed, then the effects of these two parameters can be easily seen in the predicted differential visibility and phase in H$\alpha$. From that, in general, one sees how the drop in visibility is more affected by varying the extension of the disk in the kinematic model, while the differential phase is more affected by changing the inclination.\par 

As expected, a disk with a larger extension shows a larger drop in visibility. In particular, when considering $i$ = 0, and thus no Doppler effect in the line, with $v_{\mathrm{proj}}$ = 0 (see Eq.~\ref{eq:kinematic_model_vproj}), the differential phase is equal to zero as the photocenter of the system (star plus disk) is symmetric in the line (and in the continuum). Nevertheless, notice that varying $i$ will also affect the visibility width and drop in the line. A more comprehensive discussion on the effects of the kinematic model parameters on the visibility and phase can be found in Sect.~5.3.2 of~\citet{meilland12} and Sect.~3.1 of~\citet{cochetti19}.

\subsection{An example of model fitting: AMBER data of the nova T Pyxidis}
\label{sec_analytical_modeling_kinematic_models_example_chesneau11}

Despite being widely used in studies about Be stars, the kinematic code can also be applied to the interferometric analysis of other stellar sources.\par 

\begin{figure}[t]
\centerfloat
\centerline{\resizebox{0.85\textwidth}{!}{\includegraphics{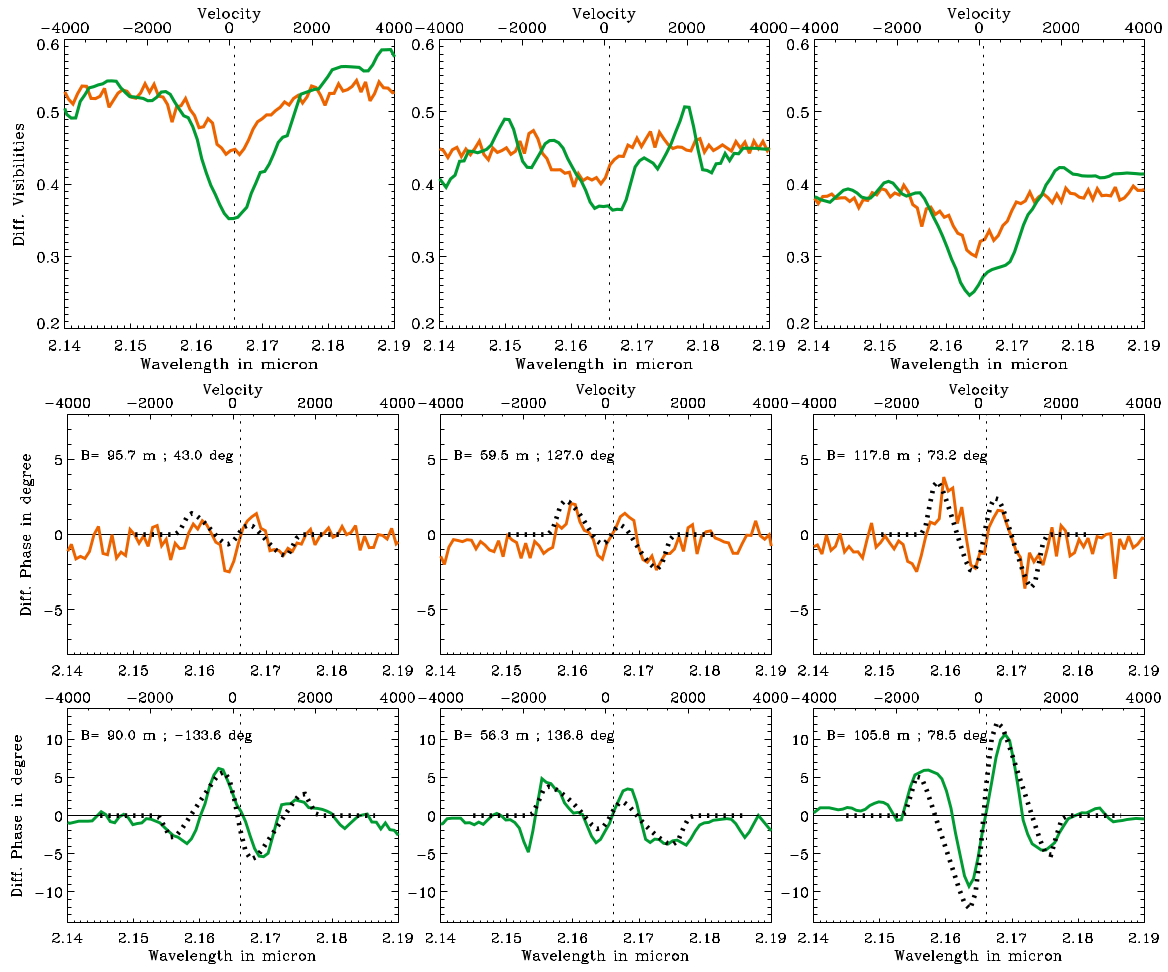}}}
\caption[VLTI/AMBER observations in Br$\gamma$ (color lines) of the nova T Pyxidis at different days since the outburst onset (28.76 days in orange and 35.77 days in green).]{VLTI/AMBER observations in Br$\gamma$ (color lines) of the nova T Pyxidis at different days since the outburst onset (28.76 days in orange and 35.77 days in green). The best-fit kinematic models of~\citet{chesneau11} to one each of these observations is shown dashed-line. See text for discussion. Adapted from~\citet{chesneau11}.}
\label{sec_analytical_modeling_chesneau11_letter_fig5}
\end{figure}

For instance, Fig.~\ref{sec_analytical_modeling_chesneau11_letter_fig5} shows the results of~\citet{chesneau11} to modeling AMBER differential data (in the Br$\gamma$ line) of the nova object T Pyxidis\footnote{Novae are outbursts ignited by thermonuclear fusion on the surface of white dwarfs due to material transferred from an evolved close binary companion, a red giant star~\citep[e.g., see][]{bode08}.} observed at different days from the outburst (about one week of difference). In this case,~\citet{chesneau11} used a modified version of the kinematic code, as presented in Sect.~\ref{sec_analytical_modeling_kinematic_models_description}, just considering radial expansion of the novae shell (with two components, polar and equatorial velocities).\par 

From Fig.~\ref{sec_analytical_modeling_chesneau11_letter_fig5}, one sees how the AMBER visibility drops in Br$\gamma$ as the circumstellar material evolves since the outburst onset. Despite being dependent on the specific baseline configuration, an inverse S-shaped phase is observed for this object, being successfully modelled using the kinematic code with different values of polar and equatorial velocities for the two observations at different epochs~\citep[see][]{chesneau11}.\par 

In short, based on the kinematic code, the study~\citet{chesneau11} provided the first direct access to the morphology and kinematics of novae at an early-stage from the first days of the explosion. This is one of the outstanding questions in astrophysics. One remarkable result found by these authors is that novae shell expand in a bipolar-shape since the very early-stage, instead of a spherically symmetric outflow, in line with some previous studies of novae using interferometry~\citep[e.g.,][]{chesneau07}.\par


\section{Kinematic model fitting using MCMC} 
\label{sec_analytical_modeling_mcmc}

\subsection{Why use a MCMC fitting-method to the kinematic code?}
\label{sec_analytical_modeling_mcmc_for_what}

When dealing with just one free parameter, such as in the geometric modeling with a uniform disk (Eq.~\ref{eq:visibility_uniform_disk}), the best estimate parameter to the data can be easily found, considering that the parameter space was sufficiently well explored for finding the global minimum (if it exists) to the data. However, this task becomes harder as the number of free parameters of our analytical model increases. As discussed in Sect.~\ref{sec_analytical_modeling_kinematic_models}, the kinematic model has a total of 13 free parameters.\par

\begin{figure}[t]
\centerfloat
\centerline{\resizebox{1.00\textwidth}{!}{\includegraphics{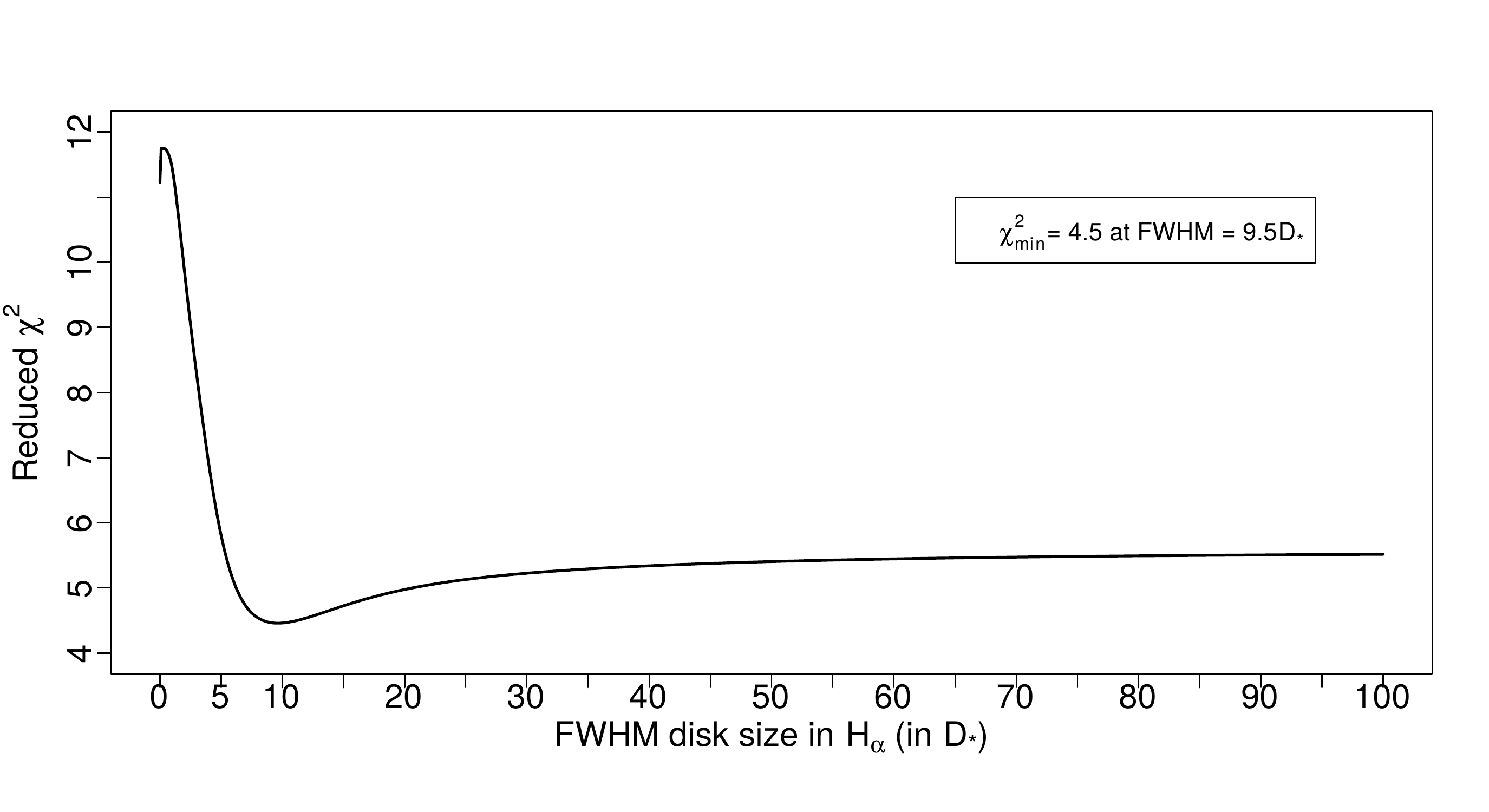}}}
\caption[$\chi^{2}$ map to VEGA data of $\omicron$ Aquarii for kinematic models (calculated around H$\alpha$) with different values of disk size (major-axis FWHM of an elliptical Gaussian distribution, in stellar diameter).]{$\chi^{2}$ map to VEGA data of $\omicron$ Aquarii for kinematic models (calculated around H$\alpha$) with different values of disk size (major-axis FWHM of an elliptical Gaussian distribution, in stellar diameter). All the other parameters are fixed, as follows: $R_{\star}$ = 3.7 $\mathrm{R_{\odot}}$, $d$ = 134 pc, $i$ = \ang{60}, $PA$ = \ang{110}, $a_{c}$ = 0, $F_{c}$ = 0, $EW$ = 19 {\AA}, $v_\mathrm{rot}$ = 450 km s\textsuperscript{-1}, and $\beta$ = -0.5 (see Sect.~\ref{sec_analytical_modeling_kinematic_models_parameters}).}
\label{sec_analytical_modeling_kinematic_model_1d_chi2}
\end{figure}

For example, we show, in Fig. \ref{sec_analytical_modeling_kinematic_model_1d_chi2}, the $\chi^{2}$ map of kinematic models, calculated around H$\alpha$, to model our VEGA dataset of $\omicron$ Aquarii, varying just one parameter: the disk FWHM major-axis in H$\alpha$ ($a_{l}$, in stellar diameters). In Fig \ref{sec_analytical_modeling_kinematic_model_1d_chi2}, all the other model parameters are fixed. Here, $v_{0}$, $v_{\infty}$, and $\gamma$ are fixed at zero. One sees that the reduced $\chi^2$ is minimized at $a_{l}$ $\sim$ 9.5 $D_{\star}$, with $\chi^2$ = 4.5. However, is this indeed the real minimum $\chi^2$ to explain our dataset and then $a_{l}$ = 9.5 $D_{\star}$ the best estimate? Of course, this result depends on the assumptions for the other parameters, which are fixed here.

Answering this question above demands to explore the parameter spaces of our kinematic code, this can be a hard task when dealing with more complex analysis than the one presented in Fig. \ref{sec_analytical_modeling_kinematic_model_1d_chi2} (one free parameter). For instance, varying the parameter values also in 1000 regular steps, as in Fig. \ref{sec_analytical_modeling_kinematic_model_1d_chi2} for the disk size, yields to the calculation of $10^{15}$ kinematic models, being this an unpractical task to performed.\par

In conclusion, it is conspicuous the necessity of a wiser approach to explore the model parameter spaces and finding the set of best-fit parameters that explain the data. All the previous studies on Be stars cited at the beginning of Sect.~\ref{sec_analytical_modeling_kinematic_models_parameters}, employ the kinematic code by varying manually their parameters in order to find the best-fit parameters and their associated uncertainties, and then being biased by the very limited number of model tests to their datasets.\par

We could approach this question above in terms of the same fitting-method of LITpro (Sect.~\ref{sec_analytical_modeling_litpro}), that is, employing a Levenberg-Marquardt method to constrain the kinematic model parameters. However, as discussed in Sect.~\ref{sec_analytical_modeling_litpro_limitations}, one major problem of this fitting-method relies on possibly getting stuck at local minima that are far-away from the global minimum to explain the data. Such common issue of frequentist approaches, as the Levenberg-Marquardt method, can be overcome by employing a Bayesian statistical inference technique, as the so-called Markov-Chain Monte Carlo (MCMC) method that is discussed below.\par

\subsection{The MCMC method}
\label{sec_analytical_modeling_mcmc_python}

This issue above can be overcome using a MCMC fitting procedure. The MCMC method is a way to estimate the probability density function of the model parameters by sampling the parameter spaces. Then MCMC is inherently a Bayesian statistical inference method, that is, relying on the so-called Bayes' theorem~\citep[e.g., see Sect.~3.8 of][]{feigelson12}.\par 

One of its main advantages is to allow us to infer the best-fit parameters and their associate uncertainties by directly calculating statistics, such as the mean and standard deviation, as well as the median and quartiles, from the probability density function. Also, as being a Bayesian inference method, it allows us to find the best-fit parameters based on prior information about the object of science, both from an empirical and theoretical point-of-views.\par 

The basic idea behind MCMC is perform a random walk covering the model parameter spaces, in a way that the probability of a ``walker'' to arrive at a certain set of parameter values is proportional to the probability density function. The random walk is performed by following a Markov Chain. This means that the selection of the next successive point to be covered depends only on the immediate previous state of the chain.\par

\subsection{The code EMCEE: a MCMC implementation}

In this implemented an automatic-fitting procedure to the kinematic model using the code EMCEE\footnote{Publicly available package. See the EMCEE website: \url{https://emcee.readthedocs.io/en/stable/}.}~\citep[][]{foreman13}. It is a MCMC implementation in Python, based on the MCMC method of~\citet{goodman10}. EMCEE is more efficient and simpler to use than more traditional MCMC methods, such as the ones based on the Metropolis-Hastings algorithm~\citep[see, e.g.,][]{foreman13}. For example, considering a N-dimensional parameter space, these latter MCMC codes require the evaluation of $\sim$ $N^2$ tuning parameters, while this problem is limited to just one or two parameters in EMCEE, such as by setting a number of walkers. Then EMCEE has been more widely used in a series of astrophysical studies\footnote{A list of astrophysical studies using EMCEE can be found at \url{https://EMCEE.readthedocs.io/en/stable/testimonials/}.}. In particular, it has been used in some recent works on stellar spectroscopy and interferometry~\citep[see, e.g.,][]{monnier12, domiciano14, sanchez17, mossoux18, bouchaud20}.\par

Denoting $y$ as the predicted model value and $x$ the observed one, the likelihood function $p_{\mathrm{like}}$ of the model is given by the following quantity~\citep[see Eq.~3.67 of][]{feigelson12}: 

\begin{equation}
p_{\mathrm{like}} = \frac{1}{\sqrt{2\pi \sigma}} \exp{ \left[ \frac{-(y_{i} - x_{i})^{2}}{2\sigma^2}  \right] },
\end{equation}
where $\sigma$ is the standard deviation of the data. Thus, following the definition of model $\chi^2$, and apart from a constant, the (logarithm) of the likelihood function is given by:

\begin{equation}
\ln{p_{\mathrm{like}}} = -\frac{\chi^2}{2}.
\label{eq:likelihood_function}
\end{equation}

From Eq.~\ref{eq:likelihood_function}, one sees that the maximization of the likelihood function means to minimize the model $\chi^2$.

\subsection{Using prior information with EMCEE}

Instead of working on the search of the maximum likelihood function, a MCMC method works with the posterior probability function (denoted here by $p_{\mathrm{post}}$) from Bayesian inference. This means that the tested sample by MCMC is sensitive to preset information given by a prior probability function $p_{\mathrm{prior}}$:

\begin{equation}
\ln{p_{\mathrm{post}}} = \ln{p_{\mathrm{like}}} +  \ln{p_{\mathrm{prior}}}.
\label{eq:posterior_probability_function}
\end{equation}

Such a prior probability function typically can be set by physical conditions known to our problem. For example, based on the projected rotational velocity, $v \sin{i}$, we can define $p_{\mathrm{prior}}$ as follows:

\begin{equation}
p_{\mathrm{prior}} = -w\left(  \frac{[ (v \sin{i})_{\mathrm{model}} - (v \sin{i})_{\mathrm{obs}}) ]^{2} }{2\sigma^2}   \right),
\label{eq:prior_probability_function}
\end{equation}
where the $(v \sin{i})_{\mathrm{model}}$ is simply calculated from the sampled MCMC values for $v_{\mathrm{rot}}$ and $i$ of the kinematic model, and $(v \sin{i})_{\mathrm{obs}}$ is the prior information. Here, the variable $w$ denotes an arbitrary weight on the prior probability function, and $\sigma$ the error bar on the measured value of $v \sin{i}$.\par 

Note that $w$ = 0 means to set Eq.~\ref{eq:posterior_probability_function} equal to \ref{eq:likelihood_function}, this being the simplest case. As will be discussed in Sect.~\ref{sec_results_omicron_aquarii}, this particular prior function is included in our modeling of $\omicron$ Aquarii data using the EMCEE code to find the best-fit kinematic model parameters.\par

\subsection{Example of modeling: VLTI/AMBER data of $\omicron$ Aquarii}

\begin{figure}[t]
\centerfloat
\centerline{\resizebox{0.95\textwidth}{!}{\includegraphics{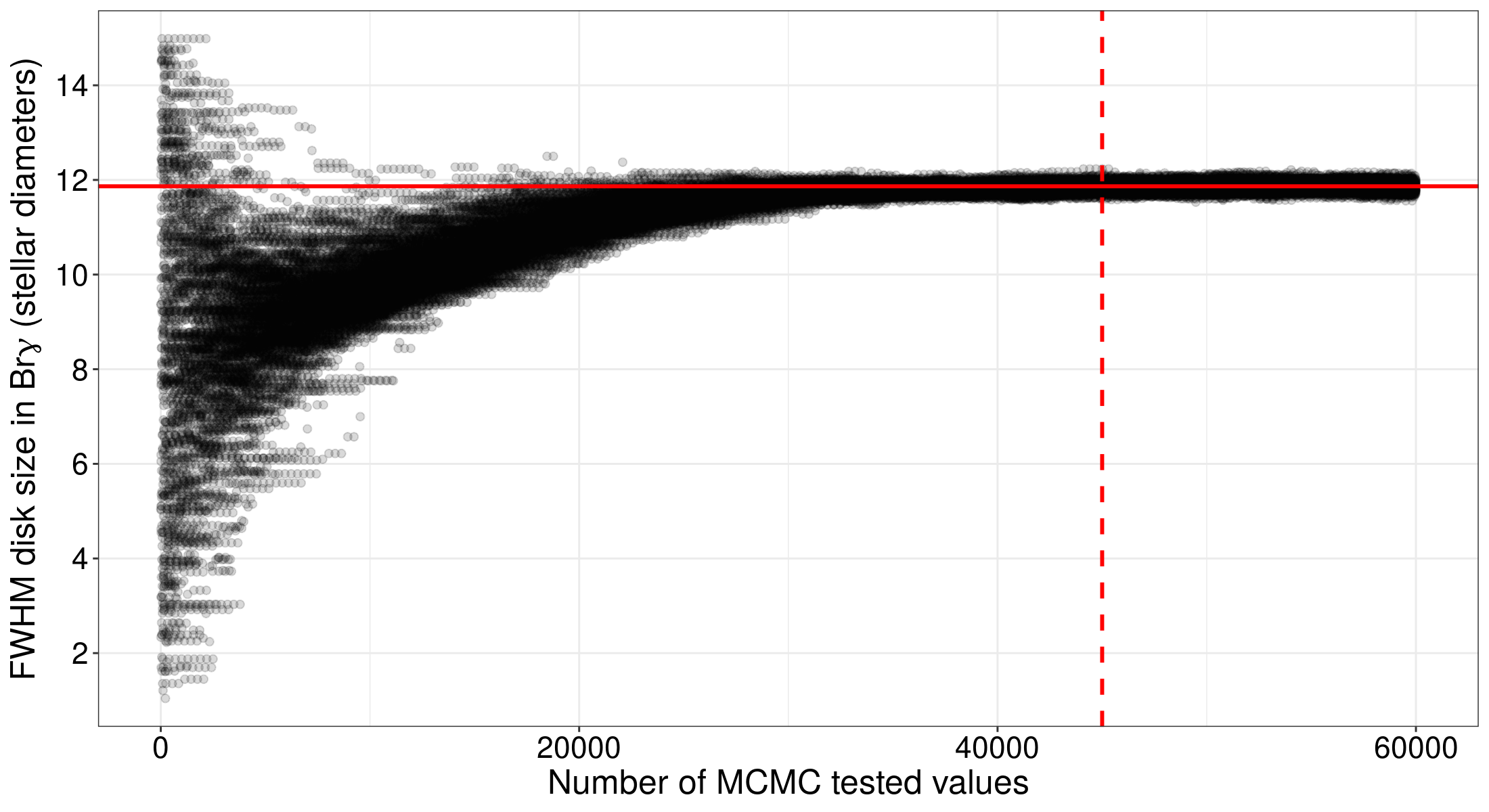}}}
\caption[Kinematic model disk size (in Br$\gamma$, fit to AMBER data of $\omicron$ Aquarii) sampled in a MCMC run using EMCEE with 300 walkers and 200 iterations in total.]{Kinematic model disk size (in Br$\gamma$, fit to AMBER data of $\omicron$ Aquarii) sampled in a MCMC run using EMCEE with 300 walkers and 200 iterations in total. The vertical dashed red line marks the end of the burn-in phase of the MCMC run. The horizontal solid red line marks the value of disk size equal to 11.8 $D_{\star}$. See text for discussion.}
\label{sec_analytical_modeling_mcmc_amber_model3_prior_vsini_hipparcos}
\end{figure}

Fig. \ref{sec_analytical_modeling_mcmc_amber_model3_prior_vsini_hipparcos} shows the results for the disk emitting size in Br$\gamma$ from one of our MCMC tests to fit the AMBER data of $\omicron$ Aquarii (Sect.~\ref{sec_results_omicron_aquarii}) using the kinematic model. Notice that this MCMC run quickly converges to the value of $\sim$11.8 stellar diameters for the disk extension in Br$\gamma$ of the Be star $\omicron$ Aquarii.\par 
More generally, the quantity of parameter samples in the EMCEE run is controlled by the total number of walkers and iterations performed in each walker: \textnumero\, of walkers multiplied by \textnumero\, of iterations. Since we chosen 300 walkers and 200 iterations in the MCMC run shown in Fig. \ref{sec_analytical_modeling_mcmc_amber_model3_prior_vsini_hipparcos}, the number of tested values for $a_{\mathrm{line}}$ (Br$\gamma$) is equal to 60000.\par 

The calculation of the statistics to find the best estimate is performed after the burn-in phase of the run (150 iterations in this case), that is, all the values tested before that are discarded in the determination of the best-fit parameters. The acceptance fraction ($a_{f}$), in each walker, is an important parameter to evaluate if the MCMC sampling represents the posterior distribution of the model parameters well. In each walker, the acceptance fraction expresses the ratio between the accepted steps and the total number of steps. For instance, if $a_{f}$ is very close to one, this means that almost all the proposed steps are accepted in the MCMC run and then the tested values are highly independent among them, producing chains that are not representative to the target's posterior distribution. As a rule of thumb, the acceptance fraction, at least the averaged value among all the walkers, should be $\sim$0.2-0.5~\citep[see][and references within]{foreman13}. The mean $a_{f}$ is $\sim$0.56 in our MCMC run showed in Fig. \ref{sec_analytical_modeling_mcmc_amber_model3_prior_vsini_hipparcos}.

\subsection{Further improvements}

Lastly, some coding improvements are needed for a better computational performance of our MCMC fitting procedure.\par 

Since the kinematic code is written in the IDL language, the current MCMC fitting code employs a Python library (pIDLy\footnote{Available at \url{https://pypi.org/project/pIDLy/}}) to run the kinematic code. One con of this procedure comes from the execution time that is needed to calculate a single kinematic mode. For instance, it can take $\sim$6 and 9s to simulate the interferometric quantities for a large sample of VEGA and AMBER observations, as presented in this thesis\footnote{These computing times stand for calculations were performed using CPU clusters from the CentOS 7.}. Although still a feasible approach in terms of time, in this case, a robust MCMC run using a large number of walkers and iterations, such as 300 and 200, requires a computation time of about 4 and 6 days.\par 

As our MCMC code was written in Python, it is needed to employ specific Python libraries to import the outputs from the kinematic code (IDL), and then wasting computing time to perform this process. A natural solution for a better performance on the computing time is then to re-write the kinematic in Python language. Currently, we have been working on MCMC tests with the kinematic code written in Python. This improvement in terms of coding will be useful for our future project on a large VEGA and AMBER\footnote{As discussed in Sect.~\ref{sec_interf_amber}, despite the decommissioning of the VLTI/AMBER instrument in 2017, several studies on hot stars have been performed since then.} survey of Be stars (Sect.~\ref{sec_ongoing_studies_classical_be_stars}).\par 
\pagestyle{empty}
\cleardoublepage
\pagestyle{fancy}

\chapter{Published studies}
\label{chapter_results}
\minitoc

This chapter presents the results that I obtained during my PhD on three different studies about massive hot stars. They are presented here in chronological order of publication.\par 

Sect.~\ref{sec_results_ostars} discusses my first-authored A\&A paper~\citep[][hereafter Paper I]{paperI} on spectroscopic analysis of O-type stars. An early summary about this topic is published in the proceedings of the XV Latin American Regional IAU Meeting~\citep{paperI_proceedings} and is also presented in Sect.~\ref{sec_results_ostars}. Sect.~\ref{sec_results_pcygni} discusses my co-authored MNRAS paper~\citep[][hereafter Paper II]{paperII} on intensity interferometry of the LBV star P Cygni. Lastly, Sect.~\ref{sec_results_omicron_aquarii} discusses my first-authored A\&A paper~\citep[][hereafter Paper III]{paperIII} on spectro-interferometry of the Be star $\omicron$ Aquarii.\par 

This study about $\omicron$ Aquarii is the principal paper of my thesis, for which I devoted most part of my work time during the PhD from learning about interferometric observations to their interpretation using different modeling methods (Chaps.~\ref{chapter_radiative_transfer_modeling} and \ref{chapter_analytical_models}).\par

\section{Mini-survey of O stars}
\label{sec_results_ostars}

\subsection{Problems with the theory of line-driven winds: weak winds}

From the discussion in Chap.~\ref{chapter_introduction}, O stars lose a significant fraction of mass through radiative line-driven winds during their short lifetimes. This affects their physical properties across the HR diagram such the stellar rotation and chemical surface abundances. However, although the theoretical bases of line-driven winds were set around half a century ago, it is still uncertain what are the real mass-loss rates among different luminosity classes of late- and early-type O stars. Based on spectroscopic analysis performed at different wavelength regions, the values of mass-loss rate for these stars seem to (at least partially) diverge, as discussed below.\par

This paper investigates the so-called weak wind phenomenon, also called as the weak wind problem~\citep[e.g.,][]{martins05_weakwinds, marcolino09, muijres12}. As pointed out by~\citet{puls09}, it is one of the main open issues in the massive star literature. As discussed in Sect.~\ref{sec_intro_radiative_winds_hr_diagram}, this problem is found for late O dwarfs (O8-9V stars) being characterized by a large disagreement between the mass-loss rates derived from UV spectroscopic studies and the values that are predicted based on hydrodynamical simulations~\citep{vink00, vink01}, the latter ones are larger by up to two orders of magnitude.\par

It is still not completely clear if the weak wind problem only arises based on the analysis of UV spectra of O stars.~\citet{marcolino09} verified that the UV mass-loss rates, that is, determined from fitting UV spectra, for their sample of five late O dwarfs match the observed H$\alpha$ profiles well in overall. On the other hand, both their final models (with UV mass-loss rates) and the theoretical mass-loss rates using the recipe from~\citet{vink00} are able to reproduce fairly well the observed H$\alpha$ profiles for two out of their five objects. This means that with respect to the visible region, both models with lower (from spectroscopy) and higher (from theory) mass-loss rates are somewhat indistinct for these stars.\par

Furthermore, some studies claim that only X-ray observations could provide reliable estimates of the mass-loss rate for late O dwarfs~\citep[see][and references within]{oskinova16}. In this case, for a low-density wind, as in late O dwarfs, the bulk of the wind would be in a hot gas state, filling a much larger volume than the cool gas. This would prevent us from deriving the real mass-loss rate based purely on the modeling of UV and visible spectra. Currently, this is a debatable question regarding weak winds due to the low statistics of analysed objects with X-ray spectra: it is partially supported by some studies~\citep[e.g.,][]{huenemoerder12}, while other studies do not support this idea. For instance,~\citet{cohen14} found a weak wind for the late O dwarf $\zeta$ Ophiuchi (O9.7V), deriving a mass-loss rate of $\sim$$1.5\e{-9}$ $\mathrm{M_\odot}$ yr\textsuperscript{-1} from modeling X-ray line profiles. This result is in line with the mass-loss rate determined by~\citet{marcolino09} for this star, namely, $\sim$$1.6\e{-9}$ $\mathrm{M_\odot}$ yr\textsuperscript{-1} (mass-loss rate derived from fitting UV lines). On the other hand, the mass loss recipe of~\citet{vink00} predicts a mass-loss rate of the order of $10^{-7}$ $\mathrm{M_\odot}$ yr\textsuperscript{-1} for $\zeta$ Ophiuchi~\citep{marcolino09}.\par

Lastly, state-of-the-art stellar evolution models~\citep[e.g.,][]{ekstrom12,meynet15} use the mass-loss recipe of~\citet{vink00, vink01} for the wind of massive stars during the H-burning phase, and thus being likely overestimated for the less luminous massive stars. In short, these questions above show the current importance of investigating the phenomenon of weak winds on O stars, demanding a multi-wavelength spectroscopic analysis of their wind properties.\par

\subsection{Master thesis work}

I started to work on this project during my Master thesis~\citep{masterthesis} under the supervision of W.~L.~F.~Marcolino (Observatório do Valongo, Brazil). We aimed to investigate weak winds among Galactic late-type O giant stars (O8-9.5III). These stars are somewhat more evolved objects than dwarfs and have luminosities around the critical value proposed in the literature of $\log(L_\star/L_\odot) \approx 5.2$, for which the weak wind problem seems to begin (onset of weak winds). So far, this problem has only been reported in late O and early B dwarfs~\citep[see][and references within]{hillier20}. We performed a detailed UV and visible spectroscopic analysis using sophisticated non-LTE atmosphere models calculated with the code CMFGEN of nine O8-O9.5III stars, namely: HD 156292, HD 24431, HD 105627, HD 116852, HD 153426, HD 218195, HD 36861, HD 115455, and HD 135591.\par 

For this purpose, we used high resolution UV and visible spectra to derive the stellar and wind properties of our sample. The UV spectra were observed with the International Ultraviolet Explorer (IUE) telescope, using the Short Wavelength Prime (SWP) instrument\footnote{Public data available in the Barbara A.~Mikulski Archive for Space Telescopes (MAST): https://archive.stsci.edu/iue/}. The spectra cover the wavelength interval of $\sim$1200-1975 {\AA}. As discussed in Sect.~\ref{sec_spectro_line_diagnostics}, this spectral region is important for the determination of the physical properties of O-type stars, which display lines formed both in the photosphere (the so-called UV iron forest) and in the wind region (such as \ion{N}{V} $\lambda$1240, \ion{Si}{IV} $\lambda$$\lambda$1394,1403, \ion{C}{IV} $\lambda$$\lambda$1548,1551 and \ion{N}{IV} $\lambda$1718). For the visible data, covering the H$\alpha$ line, we observed six stars of our sample using the FEROS spectrograph mounted at the ESO/MPG 2.2\,m telescope in La Silla (Chile) from March 17 to 22, 2016. For the other three stars of our sample, we used public visible data obtained with the CFHT/ESPADONS and 2-m Télescope Bernard Lyot/NARVAL instruments.\par

This allowed us to probe the weak wind phenomenon on massive stars others than O dwarfs, based on a multi-wavelength quantitative spectroscopic approach, that is, using radiative transfer models (CMFGEN) to fit both the UV and visible (H$\alpha$) observed spectra. Furthermore, combined UV and visible spectroscopic studies focusing on stellar and wind properties of late O giants are still somewhat scarce in the literature. To date, the study of~\citet{mahy15} performed one of the most comprehensive UV and visible spectroscopic analysis of late O giants in the Galaxy, also using the code CMFGEN: five objects in total, but with UV spectra available for only two out of these five stars.\par

\subsection{Improvements during my PhD}

I did the bulk of the work on this topic during my Master thesis, but I substantively improved the analysis during my PhD thesis, in particular, based on suggestions provided by J.-C.~Bouret (Observatoire de Marseille, France). These improvements are summarized in the following:

\begin{enumerate}[label=(\roman*)]
\setlength\itemsep{1em}

\item Determination of the stellar and wind parameters from the visible region for the complete star sample (Sects.~2 and 4 of Paper I). Previously, in~\citet{masterthesis}, this was performed for the most part of the sample (seven out of nine objects), leaving to be performed for two stars (HD 24431 and HD 218195). Regarding the wind properties, this was particularly important for checking our mass-loss rates (derived using UV lines) to the observed H$\alpha$ line profile for all the star sample.\par

\item Inclusion of more atomic species to evaluate their effects on the spectroscopic modeling, especially regarding the UV region (Sect.~4.1 of Paper I). Namely, at the end, the following ions were added to our basic CMFGEN models (Table 2 of Paper I): \ion{C}{II}, \ion{N}{II}, \ion{O}{II}, \ion{Ne}{II}, \ion{Ne}{III}, \ion{Ne}{IV}, \ion{Ne}{V}, \ion{P}{IV}, \ion{P}{V}, \ion{S}{III}, \ion{S}{IV}, \ion{Ar}{III}, \ion{Ar}{IV}, \ion{Ar}{V}, \ion{Ar}{VI}, \ion{Cr}{III}, \ion{Cr}{IV}, \ion{Cr}{V}, \ion{Cr}{VI}, \ion{Ni}{III}, \ion{Ni}{IV}, \ion{Ni}{V}, \ion{Ni}{VI}. Due to the high computational cost of including so much more elements, this was performed based on our best-fit CMFGEN model for one star of our sample (HD 116852).\par

\item Analysis of the spectral energy distribution (SED) for all the stars of our sample, using Gaia DR2 parallaxes~\citep{gaia18}, and then exploring possible effects due to interstellar extinction (Sect.~4.2.1 of Paper I). Previously, we have adopted the luminosities values for each star based on sophisticated calibration for O stars provided by~\citet{martins05_calibration}, without checking the predicted SED from our best-fit CMFGEN models in comparison with observations.\par

\item Comparison between the mass-loss rates determined in Paper I with the theoretical values of~\citet{vink00} and~\citet{lucy10, lucy10_grid} (Sect.~4.3.2 of Paper I). Previously, all the comparison between our results on the wind mass-loss rate and the predicted values was relied on the study of~\citet{vink00}.\par

\item Evaluation of binary effects on the H$\alpha$ line profiles for three stars (spectroscopic double-lined binary systems) of our sample: HD 156292, HD 153426, and HD 156292 (Sect.~5.2 of Paper I). For that, multi-epoch public FEROS data of these stars were collected. We recall the reader that CMFGEN provides a spherically symmetric wind modeling. In short, as discussed in Sect.~5.2 of Paper I, the H$\alpha$ wings seems to be indeed affected by the binary nature of these stars. Nevertheless, our determined mass-loss rates are unlikely biased by binary effects, as the H$\alpha$ line core (sensitive to the wind mass-loss rate) is not substantively affected.\par

\item Analysis of luminosity effects on the line profiles, in particular, \ion{Si}{IV} $\lambda \lambda$ 1394,1403, \ion{C}{IV} $\lambda \lambda$ 1548,1551, in order to evaluate possible changes on the determination of the mass-loss rates of our star sample (Sect.~5.3.3 of Paper I). As discussed in Sect.~\ref{sec_spectro_stellar_classification}, the \ion{Si}{IV} $\lambda \lambda$ 1394,1403 line profiles are well-known to be sensitive to the luminosity class of O stars.\par

\end{enumerate}

\subsection{Results and conclusions}

Further details on the observational data used in this study, as well as on the modeling assumptions using CMFGEN, are found in Sects.~2 and 3.1 of Paper I, respectively. The UV and visible line-diagnostics used to derive the fundamental stellar (e.g., effective temperature and surface gravity acceleration) and wind parameters (mass-loss rate and terminal velocity) are discussed in details in Sect.~3.2. The determined parameters for all the nine stars of our sample of late-type O giants are shown in Table 3 of Paper I.\par

Figs.~12 and 13 of Paper I compare our derived mass-loss rates for late O giants with the values predicted from the theoretical works discussed above. In addition to our results, spectroscopic mass-loss rates found in the literature for others O stars with different luminosity (dwarfs, early giants, and supergiants\footnote{Here, early-type B supergiants are also included in the analysis.}) are also compared. First, we point out that the spectroscopic mass-loss rates shown in these figures are unclumped mass-loss rates, they are equal to the clumped mass-loss rates (i.e., derived using clumped models) multiplied by a factor of $1/\sqrt{f_\infty}$ (clumping filling factor, Eq.~\ref{eq:cmfgen_clumping}).\par 
For our results, this conversion factor is $\sim$3.14 since the $f_\infty$ was fixed at 0.1 in the modeling of all the star sample. Despite being not realistic to consider unclumped (i.e., a homogeneous density structure) winds for O stars, we need to consider unclumped mass-loss rates (Table 3 of Paper I), when comparing our results with the theoretical mass-loss rates of~\citet{vink00} and~\citet{lucy10_grid}, since these theoretical studies do not take wind clumping into account. Following the notation used in Paper I, the mass-loss rates predicted using the recipes from~\citet{vink00} and~\citet{lucy10_grid} are denoted here by $\dot{M}_\mathrm{Vink}$ and $\dot{M}_\mathrm{Lucy}$, respectively.\par

Comparing the spectroscopic results from the literature with $\dot{M}_\mathrm{Vink}$ (Fig.~12 of Paper I), weak winds are found in late O dwarfs with $\log(L_\star/L_\odot) \lesssim 5.2$. These stars show spectroscopic mass-loss rate $\sim10^{-9}$ $\mathrm{M_\odot}$ yr\textsuperscript{-1}, while the values for $\dot{M}_\mathrm{Vink}$ reach up to $10^{-7}$ $\mathrm{M_\odot}$ yr\textsuperscript{-1}. On the other hand, the spectroscopic mass-loss rates for the more luminous OB star (supergiants, luminosity class I) are in line with the values predicted using the mass-loss recipe of~\citet{vink00}.\par

As discussed above, our results for the mass-loss rate are also compared with the theoretical values using~\citet{lucy10_grid} (Fig.~13 of Paper I). We found that the predictions from~\citet{lucy10_grid} provide a better match to our determined mass-loss rates of late O giants, when compared with $\dot{M}_\mathrm{Vink}$. This result is in line with previous studies on late O dwarfs, showing a better agreement between the $\dot{M}_\mathrm{Lucy}$ and the mass-loss rates derived from spectroscopy~\citep{lucy10_grid}. In this case, the weak wind problem is substantially attenuated, with discrepancies up to $\sim$1.0 dex between $\dot{M}_\mathrm{Lucy}$ and the derived $\dot{M}$ for our star sample.\par 

However, one sees that study of~\citet{lucy10_grid} underestimates the mass-loss rates for OB supergiants, for which Vink's predictions are in line with the spectroscopic results. The results found in Figs.~12 and 13 are summarized in Fig.~14 of Paper I, where the discrepancies between the spectroscopic and theoretical mass-loss rates are compared among the different classes of OB stars. It is hard to compare the predictions of~\citet{vink00} with the ones from~\citet{lucy10_grid} because they employ different approaches. While~\citet{vink00} find $\dot{M}$ that is globally (through the wind) consistent with the conservation of energy,~\citet{lucy10_grid} predicts indeed the mass-loss rate from first principles (solving the wind momentum equation). Nevertheless, the values of $\dot{M}_\mathrm{Vink}$ for OB supergiants are supported by more recent hydrodynamical simulations from~\citet{muijres12}.\par

The UV synthetic lines from our final models are compared with the ones using $\dot{M}_\mathrm{Vink}$ for all the star sample (Fig.~15 of Paper I). We limited this spectral analysis to mass-loss rates using~\citet{vink00} since these theoretical values are widely used in the literature in comparison with~\citet{lucy10_grid}. Again, we remind the reader that our final models considered in this comparison use unclumped mass-loss rates. All the other parameters in the models are fixed according to Table 3 of Paper I. In short, when using $\dot{M}_\mathrm{Vink}$ in the CMFGEN models, the synthetic \ion{Si}{IV} $\lambda \lambda$ 1394,1403 line profiles are systematically much more intense than the observed spectra. With respect to \ion{C}{IV} $\lambda$$\lambda$1548,1551, the models with $\dot{M}_\mathrm{Vink}$ display saturated line profiles, in contrast with the observations for the HD 24431, HD 105627, and HD 153426 (unsaturated profiles in \ion{C}{IV} $\lambda$$\lambda$1548,1551). This means that the predictions using~\citet{vink00} are ruled out based on the modeling of the UV spectra of our star sample.\par

Fig.~16 of Paper I provides an analogous comparison to the one performed in Fig.~15, but for the H$\alpha$ line. As discussed in details in Sect.~5.1 of Paper I, the situation when modeling the H$\alpha$ line is somewhat more complex than in the analysis of UV region. Considering our UV mass-loss rates (within the derived error bars), we are able to reproduce the observed H$\alpha$ profile for four objects of our sample, while the models using  $\dot{M}_\mathrm{Vink}$ highly misfit the observed data in H$\alpha$ for 5 objects of our sample.\par 

Nevertheless, these discrepancies found between the models with $\dot{M}_\mathrm{Vink}$ and observations are clearly enhanced in comparison to the results found from~\citet{marcolino09} for O8-9.5V stars. While their sample of late O dwarfs has a mean value of $\sim$ $9.0\e{-8}$ $\mathrm{M_\odot}$ yr\textsuperscript{-1} for $\dot{M}_\mathrm{Vink}$, our sample of late O giants has mean $\dot{M}_\mathrm{Vink}$ six times higher than that ($\sim$$5.0\e{-7}$ $\mathrm{M_\odot}$ yr\textsuperscript{-1}). Since the H$\alpha$ line profile is more sensitive on the mass-loss rate regime of $\dot{M} \gtrsim 10^{-7}$ $\mathrm{M_\odot}$ (Sect.~\ref{sec_spectro_line_diagnostics}), we were able able to provide a better check for the Vink's predictions for our sample with respect to the fit to the H$\alpha$ line.\par

Regarding the simultaneous fit to the UV and visible (H$\alpha$) spectra, weak winds are favoured by our study when compared with the predicted mass-loss rates for late O giants in overall.\par 

From Sect.~6 of Paper I: \say{in conclusion, our results indicate the weak wind phenomenon in O8-9.5III stars. It is the first time that weak winds are found for spectral types other than O8-9.5V. Despite our efforts, we are not able to model at the same time both the UV wind diagnostic lines and the H$\alpha$ profile for all the stars of our sample. This issue could be solved by investigations regarding macroclumping implementation in the modeling with CMFGEN and potential H$\alpha$ variability (as observed in late OB supergiants) among late O giants. Apart from this problem, low $\dot{M}$ (weak winds) are favored to model the spectra (UV + optical regions) of late O dwarfs and giants in comparison with values provided by theory. In other words, the measured mass-loss rates of these stars are systematically lower than the predictions of~\citet{vink00}. This is important as they are low luminosity O stars (latter spectral types), implying that the majority of the O-type stars must undergo a weak wind phase. Therefore, we suggest that the mass-loss recipe in the majority of modern stellar evolution codes must severely overestimate $\dot{M}$ during the H-burning phase. Further investigations are needed to evaluate the consequences of this in terms of physical parameters for massive stars (e.g., angular momentum and CNO surface abundances).}.\par

We precisely identified the luminosity region of $\log(L_\star/L_\odot) \approx 5.2$ as critical for weak winds, but we were not able to address the physical causes for that in this paper. In short,~\citet{muijres12} found that their simulations for O stars fail to drive a wind for the late-type O dwarfs due to the lack of \ion{Fe}{V} lines at the base of wind for these stars, when compared with the earlier O stars. More recently,~\citet{vilhu_kallman19} suggested that weak winds can be explained in terms of a velocity-porosity (vorocity) effect of the wind stratification (clumps) for the later O dwarfs. Indeed, to date it is still an open issue in the literature and further investigations are needed on the physical causes of weak winds.\par

Another interesting issue raised from weak wind regards the so-called modified wind momentum-luminosity relation for massive hot stars. The modified wind momentum ($D_{mom}$) was firstly introduced by~\citet{kudritzki95}:

\begin{equation}
D_{mom} = \dot{M} v_{\infty} \sqrt{R_{\star}},
\label{eq:mod_wind_momentum}
\end{equation}
being the term ``modified'' used for this quantity due to the stellar radius $\sqrt{R_{\star}}$, apart from the wind momentum ($\dot{M}v_{\infty}$).\par

As shown by the~\citet{kudritzki95}, this quantity for radiative-line driven winds is expected to scale with the stellar luminosity:

\begin{equation}
\log(D_{mom}) = a \log(L_\star/L_\odot) + b,
\label{eq:mod_wind_momentum_luminosity_relation}
\end{equation}
where the linear and angular coefficients above are specifically dependent on the spectral type and metallicity.\par 

This relation has been supported by subsequent spectroscopic studies~\citep[e.g.,][]{puls96, kudritzki99, mokiem07a, mokiem07b}, and it simply states that more luminous massive stars are likely to have more intense wind (larger wind momentum), as expected from the CAK-theory.\par

Thus, based on prior information about $D_{mom}$, for instance, from theoretical investigations~\citep[e.g.,][]{vink00,vink01}, one could be able to estimate the stellar luminosity. This is particularly important as independent tool to estimate distances, up to Mega-parsec scales, as will also be discussed in Sect.~\ref{sec_results_pcygni} (Paper II). As supported by Paper I, this relation should be taken with caution since it breakdowns at $\log(L_\star/L_\odot) \approx 5.2$, when compared with the theoretical relation predicted from~\citet{vink00} (see Fig.~11 of Paper I).\par

Despite focusing our discussion here on the weak wind problem, other interesting issues were found by Paper I, also deserving further investigations. For instance, our best-fit models overestimate the emission component of \ion{C}{IV} $\lambda \lambda$ 1548,1551 for all the star sample. As discussed in Sect.~3.1 of Paper I, the wind velocity law exponent $\beta$ was initially fixed at 1.0 in our analysis based on previous results found for O-type stars (Sect.~\ref{sec_intro_radiative_line_driven_winds}). However, we needed to reduce $\beta$ from 1.0 to very low values, namely, $\sim$0.3. Such low values of $\beta$ are unreliable since they are not acceptable at all in the framework of line-driven winds (standard CAK-theory).

Therefore, we were just able to reduce the discrepancies between our final models ($\beta$ = 1.0) and data when taking into account a fuller ion composition in the models, but in addition to a high value of photospheric microturbulence (up to 20-30 km s\textsuperscript{-1}). On the other hand, such a high value of microturbulence is not expected at all for late O giants stars (Sect.~4.1 of Paper I). Moreover, this latter adoption on the microturbulence prevents us from determining $T_{\mathrm{eff}}$ by the UV and visible lines in a self-consistent way. Fixing the photospheric microturbulence at 10 km s\textsuperscript{-1}, the fit to the emission component of \ion{C}{IV} $\lambda \lambda$ 1548,1551 of our star sample seems to be improved just when setting $\beta$ $\sim$ 0.3 (see Fig.~10 of Paper I). One possibility to explain this issue could rely on the quite simple $\beta$ parameterization used in Paper I (following Eq.~\ref{eq:beta_law}).\par

\includepdf[pages=-]{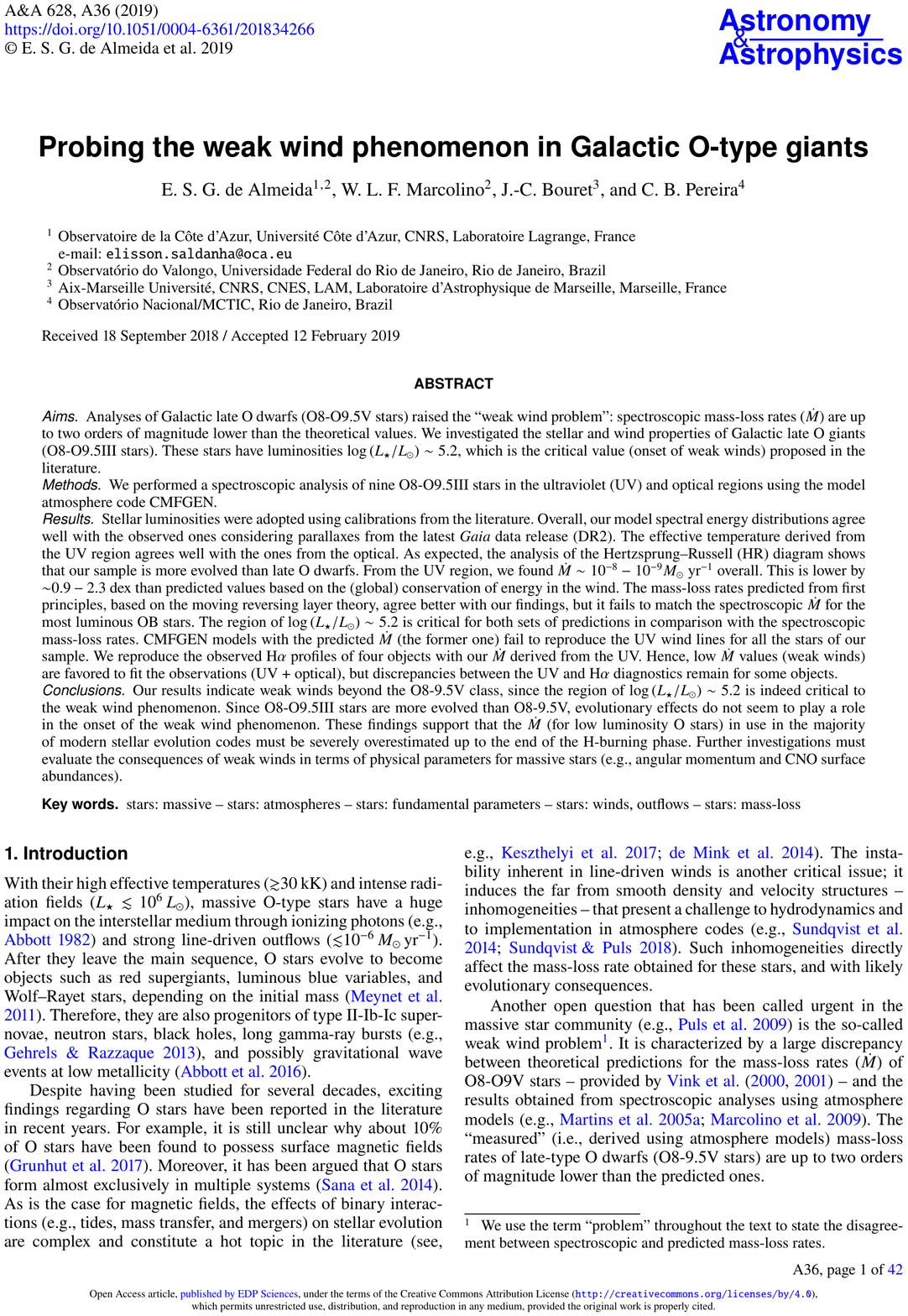}

\includepdf[pages=-]{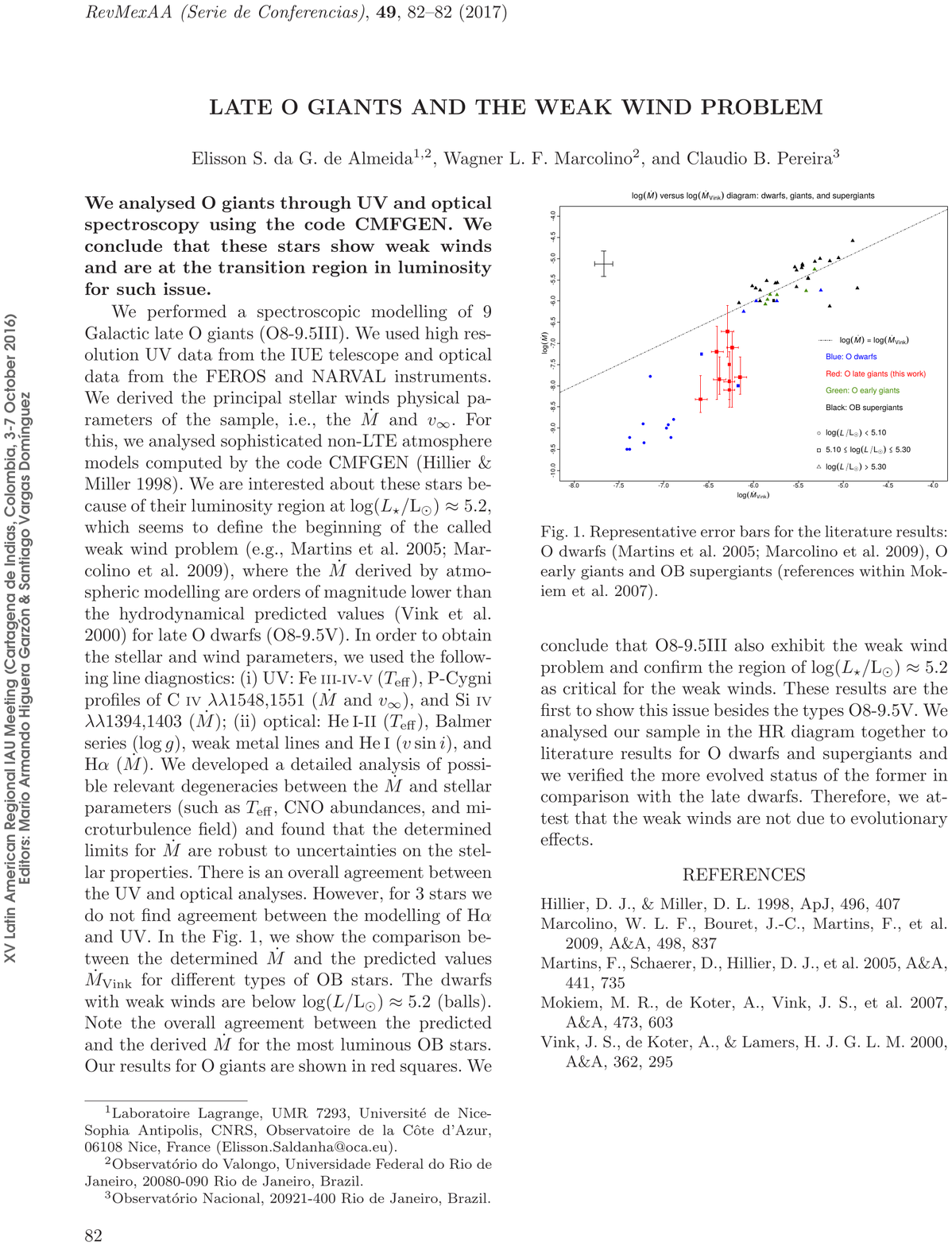}


\section{The LBV star P Cygni}
\label{sec_results_pcygni}

\subsection{Winds and episodic outbursts of LBVs}

P Cygni (HD 193237, B1Iape\footnote{The suffix ``ep'' is used to designate emission lines with peculiarity.}) is one of the most observed LBV stars, for which extensive models have been developed~\citep[e.g.,][]{lamers83, najarro97, najarro01, richardson13}. P Cygni is a prototype of the LBV stars, also called as S Doradus variables.\par 

These stars evolved from the more massive O-type stars ($\mathrm{M_{ZAMS}}$ $\gtrsim$ 40 $\mathrm{M_\odot}$; Table \ref{sec_intro_ekstrom13_table1_conti_scenario}) and display photometric and spectroscopic variability covering a wide range of timescales from weeks to years and over centuries~\citep[e.g.,][]{lamers95, markova01a, richardson11}. LBVs possess the strongest radiative line-driven winds among the hot massive stars, with mass-loss rates typically ranging from $\sim$ $10^{-5}$ up to $10^{-3}$ $\mathrm{M_\odot}$ yr\textsuperscript{-1} (Fig.~\ref{sec_intro_smith14_fig3})\footnote{For example, P Cygni has a wind mass-loss rate in the order of $10^{-5}$ $\mathrm{M_\odot}$ yr\textsuperscript{-1}~\citep[e.g.,][]{najarro01}.}.\par 

In addition, they undergo violent episodic mass loss, ejecting a much larger amount of material than the mass loss process by stellar winds. It is understood that $\eta$ Carinae (another LBV star, more luminous than P Cygni) ejected an astonishing quantity of material, $\sim$15-20 $\mathrm{M_{\odot}}$, due to a giant eruption that lasted about a whole decade during the 19th century~\citep{smith06}. We point out that the mechanisms linked to LBV eruptions are not well known yet. Nevertheless, one possibility relies on a continuum-driven outburst arising when deep photospheric layers occasionally excess the Eddington limit (see Eq.~\ref{eq:eddington_parameter}), and then the radial velocity in the photosphere suddenly increases to high values up to about 200 km s\textsuperscript{-1}~\citep[e.g.,][]{guzik97, cox98}.\par

As discussed in Sect \ref{sec_intro_stellar_winds}, P Cygni also passed by two eruptions during the 17th century, reaching a maximum of brightness of V $\sim$ 3 and a subsequent drop up to V $\sim$ 6. The condensation of dust grains in its circumstellar environment could explain such a drop in the visual magnitude~\citep[see][and references within]{israelian99}. Since then the visual magnitude is more or less constant around the mean value of 4.82, but showing a gradual decrease in the visual magnitude of $\sim$0.15 mag/century~\citep[][]{degroot92}. Indeed, the analysis of historical observations of P Cygni by these authors provided the first direct evidence of evolution in the HR diagram.~\citet{markova01b} showed that P Cygni's effective temperature is slowly decreasing by a rate of $\sim$12 K yr\textsuperscript{-1}. This is consistent with the picture that P Cygni evolves toward the red part of the HR diagram, lying between the shell hydrogen burning phase and the core helium phase~\citep[e.g.,][]{el_eid93}.\par 

Moreover, the wind mass-loss rate of P Cygni is known to vary by $\sim$19\% over a period of about 7 years of short S Doradus phase. This affects its apparent stellar radius~\citep{markova01b}, thus being relevant for interpreting interferometric observations of this star. Therefore, our discussion above shows that quasi-contemporaneous photometric and spectroscopic observations are needed to correctly interpret interferometric observations of LBV stars.\par

\subsection{Intensity interferometry in a nutshell}

The technique of intensity interferometry is distinct from the OLBI technique, which is used in Paper III, as it is based on the so-called Hanbury Brown and Twiss effect~\citep{hanbury56, hanbury_twiss56}. Unlike OLBI, in the technique of intensity interferometry, the light received by a separate pair of telescopes is not brought together using optical systems, measuring the degree of correlation of the photon fluxes, instead of electric fields, as in the case of OLBI.\par

These pioneer works of Robert Hanbury Brown and Richard Twiss resulted in the development of intensity interferometers working from the visible up to longer wavelengths in the sub-millimetric and radio regions. For instance, the Narrabri 200-m baseline interferometer~\citep[][]{hanbury67a} provided the first interferometric catalogue of angular diameters for 32 early-type stars with spectral types from O7 to F8, observed in the visible~\citep[][]{habury74}. This instrument had a lasting impact in the field of stellar high angular resolution observations by means of systematic measurements of stellar diameters, in particular for hot and massive stars~\citep[e.g.,][]{hanbury67b, hanbury70, davis70, code76}.\par

\subsection{My collaboration with the I2C team}

Using the technique of intensity interferometry (see Sect.~\ref{sec_interf_historical_overview}),
the visibility of P Cygni was measured using a two-telescope array (1-m class telescopes) mounted at the Centre Pédagogique Planète et Univers (C2PU/Observatoire de la Côte d’Azur) facility at the Plateau de Calern, France. This project on stellar intensity interferometry is called as Intensity Interferometry at Calern~\citep[I2C,][]{lai18, rivet18} and is composed by a multidisciplinary group of researchers, comprising both astrophysicists and cold-atom physicists from Observatoire de la Côte d'Azur and Institute de Physique de Nice (Université Côte d'Azur). In addition to P Cygni, a few bright stars have been recently observed by the I2C team~\citep[see][]{guerin17, guerin18}: Rigel (B8Iae), Vega (A0Va), Capella (G8III + G0III), Arcturus (K0III), Gomeisa (F5IV-V), and Pollux (K0III).\par

I have collaborated to this project by calculating a small set of radiative transfer models using the code CMFGEN, and then analysing these intensity interferometric observations of P Cygni centered on the H$\alpha$ emission line.\par

In particular, we were interested in determining the distance of P Cygni from modeling its visibility curve. Despite being well-studied for a long time, the exact distance of P Cygni is a current issue. As summarized in Table 1 of~\citet{turner01}, different distance determination methods provide distances varying between $\sim$1.2 kpc and 2.3 kpc for this star. More recently, Gaia global astrometry mission (Data Release 2) finds a distance of 1.36 $\pm$ 0.24 kpc~\citep{gaia18}. However, this result is unlikely very accurate as P Cygni is too much bright for the normal scanning operation with Gaia, as it was optimally designed to observe fainter sources with a visual magnitude higher than $\sim$11.\par

\subsection{Results and conclusions}

For this purpose, we compared our H$\alpha$ visibility data with the predicted visibility from sophisticated non-LTE radiative transfer models calculated using the code CMFGEN. We point out that a spherically symmetric wind is a good assumption for P Cygni~\citep[e.g.,][]{richardson11}, and CMFGEN was used on some previous spectroscopic and interferometric studies about P Cygni~\citep[e.g.,][]{richardson13, najarro01}. Following the approach of~\citet{richardson13}, we adopted the basic stellar and wind parameters of our CMFGEN model from the results of~\citet{najarro01}. This latter study provided a detailed spectroscopic modeling of P Cygni from the ultraviolet to the mid-infrared region. The values of the fundamental parameters in our CMFGEN model are shown in Table 4 of Paper II. In particular, I computed approximately 30 different models for fixing the set of parameters for this study, as well as performing some additional modeling tests.\par

As an independent cross-check validation of our physical model, the synthetic spectrum of our model is compared with quasi-simultaneous observed spectrum of P Cygni in the visible region\footnote{Public data retrieved from the Astronomical Ring for Access to Spectroscopy (ARAS) database: \url{http://www.astrosurf.com/aras/}.}. In short, we are able to match the visible spectrum of P Cygni fairly well (see Fig.~5 of Paper II). Despite being an interferometric study focused on H$\alpha$ observations, we used a complex chemical composition for the atmosphere of P Cygni, as in~\citet{najarro01}. The basic atomic our CMFGEN model is shown in Table 3 of Paper II. From that, in addition to the strong H and He P Cygni profiles, our CMFGEN is also able to reproduce weak metal lines in the visible spectrum of P Cygni, such as the \ion{C}{II} $\lambda$6580 and $\lambda$6585 lines (close to H$\alpha$). Unlike the atomic data used in the CMFGEN models of Paper I (Sect.~\ref{sec_results_ostars}), we needed to include low-ionisation species, as of C, N, O, and Fe in the models for P Cygni due to its lower effective temperature ($T_{\mathrm{eff}}$ $\sim$ 18700 K) in comparison with late O giants ($T_{\mathrm{eff}}$ $\sim$ 30000-35000 K).\par

From matching the observed visibility curve with the predicted one using CMFGEN, we estimated the distance of P Cygni as $d$ = 1.56 $\pm$ 0.25 kpc (Fig.~7 of Paper II). Despite being a fit based on just two visibilities measurements, note that we are able to find an error bar on P Cygni's distance that is comparable (compatible within the error bars) to the one from the Gaia DR2 parallaxes, namely, 1.36 $\pm$ 0.24 kpc. In short, the results found by Paper II indicate a smaller value of distance for P Cygni in comparison with the most adopted value for this star in the literature of $\sim$1.8 kpc.\par

In conclusion, despite the current technical limitation on the technique of intensity interferometry, Paper II shows that intensity interferometry in the H$\alpha$-band allows us to retrieve useful information on the fundamental parameter of the central star and its environments.\par

Furthermore, this study relies on a quite unusual interpretation of interferometric quantities. For instance, the interferometric data of the Be star $\omicron$ Aquarii (Sect.~\ref{sec_results_omicron_aquarii}) was interpreted considering a fixed value of distance from the Gaia DR2 parallax. As will be discussed in Sect.~\ref{sec_ongoing_studies_rigel_intensity_interferometry}, we aim to apply the modeling method of Paper II to analyse intensity interferometric observations of other bright massive stars such as Rigel. Thus, Paper II sets an independent calibration method for estimating distances, being particularly interesting for evolved (bright) stars.\par 

Lastly, this method also can be used as an independent check to the modified wind momentum-luminosity relation (Sect.~\ref{sec_results_ostars}), another distance determination method, but based on the wind fundamental parameters of massive stars. This is currently important since, as discussed in Sect.~\ref{sec_results_ostars}, Paper II evidences problems with respect to the theoretical modified wind momentum-luminosity relation for low luminous O stars.\par

\includepdf[pages=-]{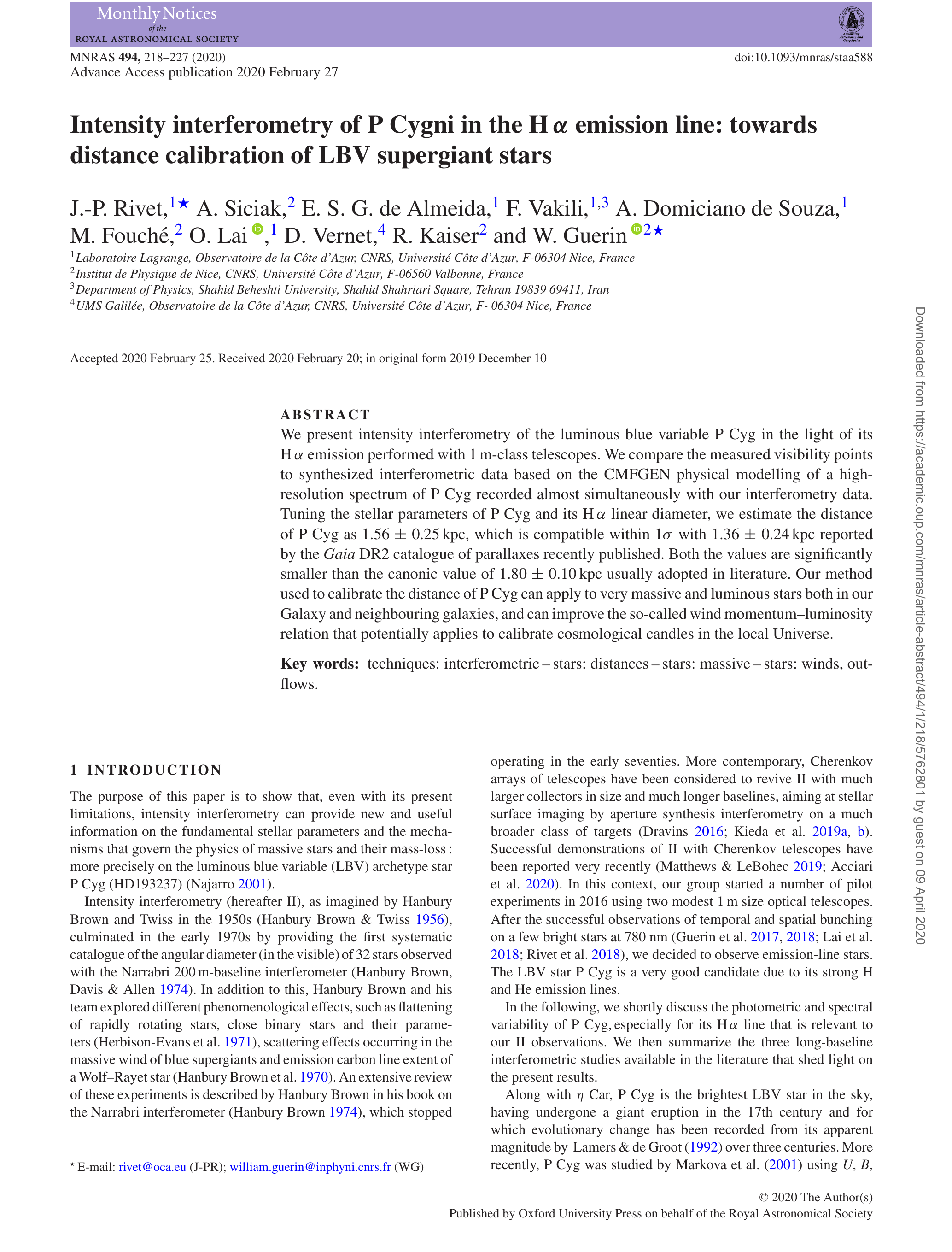}


\section{The Be-shell star $\omicron$ Aquarii}
\label{sec_results_omicron_aquarii}

\subsection{Probing the Be phenomenon: why study $\omicron$ Aquarii with interferometry?}

Understanding the origin of the decretion disks around Be stars is one the current challenges in astrophysics. In particular, a central puzzle on the Be phenomenon regards the physical mechanisms underlying the formation of their circumstellar disks.\par 

Unlike the more massive objects, such as O-type and LBVs stars, which have rotational velocities far-below their critical values, fast rotation plays a key role in the process of mass loss in Be stars. Moreover, their Keplerian decretion disks are efficient sources of angular momentum transport, when compared with the momentum loss due to their stellar winds~\citep[e.g.,][]{krticka11}. However, and discussed in Chap.~\ref{chapter_introduction}, how close Be stars are indeed to the break-up limit is still an open question in the literature, and it is an important question regarding the Be phenomenon itself.\par

Most Be stars appear to rotate below the critical value, showing rotational rates as low as $\sim$30-40\% in some cases~\citep[e.g.,][]{cranmer05, zorec16}. This means that other channels in addition to rotation are needed to remove material from the stellar surface, as provided by pulsations, magnetic fields, and the own radiative force for the earlier Be stars. In addition, more recently, binary effects are understood to be important in the process of spin-up of Be stars, which can show rotational rates very close to the critical value, requiring the search for binary companions among these stars~\citep[e.g.,][]{klement19, hastings20}.\par

$\omicron$ Aquarii (HD 209409, type B7IVe) is a good target to probe the Be phenomenon, especially through interferometry, since is quite near~\citep[$d$ = 144 pc from][]{gaia18} and bright (V = 4.7). Interestingly,~\citet{rivinius06} identified the presence of a central quasi-emission (CQE) feature in the \ion{Mg}{II} $\lambda$4481 line of $\omicron$ Aquarii. As discussed in Sect.~\ref{sec_intro_variability_be_stars}, this is a purely-geometric effect that can arise due to circumstellar disks seen close to the edge-on orientation, this being the case of $\omicron$ Aquarii, as discussed below. This means that the kinematics and physical conditions of its disk can be better probed by interferometry since such a edge-on view results in clear specific signatures in the interferometric quantities. For instance, see, again, in Fig.~\ref{sec_analytical_kinematic_model_varying_disk_size_incl}, the S-shape differential phase that is found for a disk seen edge-on.\par

The first spectro-interferometric analysis of $\omicron$ Aquarii was performed by~\citet{meilland12}, using the kinematic code (Sect.~\ref{sec_analytical_modeling_kinematic_models}) to model VLTI/AMBER (K-band) observations of eight bright Be stars. Despite the very limited number of observations for $\omicron$ Aquarii (just one measurement), and the also low quality data, these authors were able to constrain its inclination angle, as $i$ = 70 $\pm$ \ang{20}. In addition, they determined a disk extension of FWHM = 14 $\pm$ 1 $D_{\star}$ in the Br$\gamma$ line and a rotational rate (see Sect.~\ref{sec_intro_stellar_rot_oblateness}) of $\Omega/\Omega_{\mathrm{crit}}$ = $0.93^{+0.06}_{-0.17}$.\par 

This study was followed-up by~\citet{cochetti19}, also analysing $\omicron$ Aquarii with the kinematic code, in their large AMBER survey of Be stars (26 objects). Due to the larger and good quality data (8 measurements) for this star,~\citet{cochetti19} better constrained the disk extension in Br$\gamma$ and the inclination angle, FWHM = 8 $\pm$ 0.5 $D_{\star}$ and $i$ = 70 $\pm$ \ang{5}, and then improving the precision on the determination of $\omicron$ Aquarii's rotation rate as $\Omega/\Omega_{\mathrm{crit}}$ = 0.93 $\pm$ 0.07. From that, the results found by these authors indicate that $\omicron$ Aquarii must be a Be star rotating nearly close to the break-up limit.\par

To date,~\citet{sigut15} provided one of the most comprehensive picture of $\omicron$ Aquarii's disk, employing radiative transfer modeling with the code BEDISK (Sect.~\ref{sec_radiative_atmosphere_codes_hot_star}). Unlike in~\citet{meilland12}, these authors did not use spectro-interferometric observations, but they combined large band (15 nm) interferometric data (period of 2007-2014) centered on H$\alpha$, obtained from the Navy Precision Optical Interferometer, with H$\alpha$ spectroscopy from the Lowell Observatory Solar/Stellar Spectrograph (2005-2014).\par

\citet{sigut15} managed to reproduce different types of observables from their unified physical model for the disk: absolute visibility, H$\alpha$ line profile, spectral energy distribution, and also provided a qualitative explanation for the CQE feature observed in the \ion{Mg}{II} $\lambda$4481 line. They found that the H$\alpha$ $EW$ is quite stable (within about 5\%) up to about ten years. Besides that, they also could not detect disk variability from the analysis of the interferometric data considering different time spans within ten years. Another interesting result noted by~\citet{sigut15} is that the disk extension in H$\alpha$ (Gaussian FWHM $\sim$ 13 $D_{\star}$) found by their analysis is comparable to the one determined by~\citet{meilland12} in Br$\gamma$ ($\sim$ 14 $D_{\star}$). This result deserves to be better investigated because it is commonly thought that Be disks show a larger extension in the H$\alpha$ line than in Br$\gamma$, as discussed in Sect.~\ref{sec_intro_disk_geometry_extension}.\par

\subsection{Observing $\omicron$ Aquarii in the AMBER and VEGA Be surveys}

In addition to the VLTI/AMBER data already analysed in the surveys of~\citet{meilland12} and~\citet{cochetti19}, we performed new observations of $\omicron$ Aquarii using the CHARA/VEGA beam-combiner, centered at H$\alpha$, in order to obtain a detailed interferometric picture of this star, both in the visible and in the infrared.\par

Taking advantage of the longer baselines provided by the CHARA Array, compared with the ones from VLTI, we were able to archive a higher level of spatial resolution of $\omicron$ Aquarii and its environment, allowing us to partially resolve them in different spectral channel using the VEGA instrument (see Fig.~2 of Paper III).\par

During my PhD, I was engaged in the remote observations of the VEGA team, performing observations of 20 nights in total. In particular, with respect to $\omicron$ Aquarii data, I observed in remote mode the 2016 VEGA data analysed in this thesis. In total, we observed $\omicron$ Aquarii 50 times with the VEGA instrument between 2012 and 2016, as part of four different programs of the VEGA team at Laboratoire Lagrange/Observatoire de la Côte d'Azur (PI: D.~Mourard), as follows:

\begin{enumerate}[label=(\roman*)]
\setlength\itemsep{1em}
\item V51: 2$^{nd}$ Be stars small survey (2012 - one measurement).\par

\item V56: Connecting stellar surface and circumstellar environment of Be stars (2013, 8 measurements).\par

\item V62: Critical rotation and mass-loss: new insights from the study of edge-on Be stars (2014, 38 measurements).\par

\item V66: Be stars large survey  (2016, 3 measurements).\par
\end{enumerate}

\subsection{A multi-technique modeling approach: from analytical to numerical models}

As pointed out above, these VEGA observations of $\omicron$ Aquarii are part of a larger program to study the photosphere and circumstellar environment of Be stars (to date program V66, PI: A.~Meilland).\par 

In addition to our interest to achieve a better characterization of physical properties of $\omicron$ Aquarii and its disks, this study also aimed to set a more systematic modeling method using both the called kinematic model and the BeAtlas grid of HDUST radiative transfer models, as discussed in the following.\par

To date, our study presents the largest spectro-interferometric dataset analysed for one Be star, encompassing different spectral channels. For this purpose, we used modeling methods of increasing complexity in order to explain our data, and then allowing us to probe different properties of $\omicron$ Aquarii: 

\begin{enumerate}[label=(\roman*)]
\setlength\itemsep{1em}
\item Using the LITpro fitting-tool (Sect.~\ref{sec_analytical_modeling_litpro}), simple geometric models for fitting the calibrated visibility measurements: morphology of the central star and the disk.\par

\item Kinematic models for fitting the differential visibility and phase measurements: morphology of the central star and the disk, as well as their kinematics (as the stellar rotation rate and disk velocity law).\par

\item Radiative transfer models calculated with the HDUST code (BeAtlas grid) for fitting all the observables:  morphology of the central star and the disk, as well as their kinematics, in addition to the physics of the disk (as the disk mass density distribution).\par
\end{enumerate}

For the analysis using the numerical models, that is, calculated using the kinematic and HDUST codes, I wrote several scripts in Python, IDL, and R language. For example, using the library PYHDUST\footnote{Publicly available at \url{https://pyhdust.readthedocs.io/en/latest/}.}, I wrote scripts in Python for extracting in a automatic way the image cubes in .FITS files from the BeAtlas grid. Details on the PYHDUST library can be found in Sect.~2.1.4 of~\citet{faes15}. Fig.~\ref{sec_modeling_radiative_hdust_image_mod1} illustrated image cubes around the Br$\gamma$ line of BeAtlas models, as the predicted visibility (simulated to AMBER observations). The predicted interferometric quantities from each BeAtlas image is then calculated using the Discrete Fourier transform (scripts written in IDL) in order to obtain the reduced $\chi^2$ between the model and the VEGA and AMBER datasets. Given the automatic computing of model $\chi^2$ for all the BeAtlas grid, I wrote script mainly in R language in order to visualize the results and calculation statistics of interest. For instance, Fig.~9 of Paper III shows the lowest $\chi^2$ of BeAtlas from fitting our VEGA and AMBER data as a function of model parameters, and statistics from the HDUST parameters are shown in Table 4 of Paper III.\par 

Furthermore, for first time, an automatic modeling fitting procedure was implemented to fit spectro-interferometric with the kinematic code using a MCMC approach (Sect.~\ref{sec_analytical_modeling_mcmc_python}). From that, I was able to precisely constrain the following parameters for $\omicron$ Aquarii: stellar inclination angle, stellar rotation velocity, disk position angle, disk emitting extension, and the disk rotational velocity law exponent.\par

In short, despite being written in different computing languages, these scripts conjointly provide a somewhat automatic interferometric analysis of any VEGA or AMBER data for Be stars using the BeAtlas grid, ultimately finding the group of BeAtlas models that better explain the datasets in terms of reduced $\chi^2$, and also in terms of the kinematic code using a MCMC fitting-model procedure. This successful attempt of modeling the VEGA and AMBER data in our study, providing robust model parameters and associated uncertainties for $\omicron$ Aquarii, paved the way for a systematic analysis of Be stars using spectro-interferometric data from multiple instruments, such as AMBER, VEGA, MATISSE (Sect.~\ref{sec_perspetives_be_survey}).\par

\subsection{Results and conclusions}

From the simplest modeling approach using geometric modeling (LITpro) to fit the VEGA calibrated squared visibility, we constrained the stellar radius of $\omicron$ Aquarii with a 8$\%$ accuracy, that is, R$_\star$ = 4.0 $\pm$ 0.3 R$_\odot$, from our best-fit two-component uniform disk model to the VEGA data in the H$\alpha$ band (Table 1 of Paper III).\par 

This value of stellar radius is significantly larger than the one adopted in the study~\citet{sigut15} of 3.2 R$_\odot$~\citep[polar radius for a B7 dwarf from][]{townsend04}. We point out that this could be expected as $\omicron$ Aquarii is likely flattened due to its fast rotation, and then our measurement here for its radius represents a mean value between its polar and equatorial radii. Indeed, we were not able to detect any effect of oblateness from this simple geometric analysis, but the situation is quite different when looking the differential data modeled with HDUST (see Tables 4 and 5 of Paper III).\par

Table 2 of Paper III summarizes our results found using the kinematic code to fit both the differential VEGA and AMBER datasets in separate way, and Fig.~4 shows the MCMC histogram distributions and correlation plots of our best-fit kinematic models. Interestingly, we found that the H$\alpha$ (from fitting VEGA) and Br$\gamma$ (from fitting AMBER) emitting disk sizes presents a quite similar extension of $\sim$11-12 $D_{\star}$. This provided a more robust constraint on the previous results found by~\citet{meilland12} and~\citet{sigut15}, that performed independent analysis of $\omicron$ Aquarii in the near-infrared (AMBER) and H$\alpha$ (NPOI), as discussed above. As discussed in Sect.~8.1 of Paper III, this result could be explained in terms of the high inclination angle of $\omicron$ Aquarii, but it is one of the open issues that should be addressed by analysing a large sample of Be stars (Sect.~\ref{sec_perspetives_be_survey}).\par

Furthermore, the inclination angle derived from fitting our VEGA data with the kinematic code is quite lower than the one derived from AMBER, in about \ang{15}. Such a lower value of $i$ is also supported by our analysis with the BeAtlas grid (again, see Fig.~9 and Table 4 of Paper III). This result deserves further investigations, but one possibility to explain that relies on non-isothermal effect arising from $\omicron$ Aquarii's disk, as discussed in Sect.~8.2 of Paper III.\par 

In Fig.~\ref{sec_publised_studies_plot_hdust_gauss_hit_m3.0_non_iso}, we provide a very simple and preliminary analysis on this question above. We show the major-axis FWHM Gaussian size fitted from each HDUST model, similarly to the analysis provided in Fig.~13 of Paper III, but here we also consider non-parametric HDUST models of the BeAtlas grid (with viscosity parameter $\alpha$ fixed at 0.5). These models are no longer parameterized by $\Sigma_{0}$ and $m$ (see Eq.~\ref{eq:radial_surface_density}), as discussed in Sect.~\ref{sec_radiative_transfer_modeling_hdust}. In addition, we also show our results based on our best-fit HDUST model (Table 5 of Paper III) from the BeAtlas, that is, the disk density exponent law is fixed at $m$ = 3.0.\par

\begin{figure}[t]
\centerfloat
\centerline{\resizebox{0.75\textwidth}{!}{\includegraphics{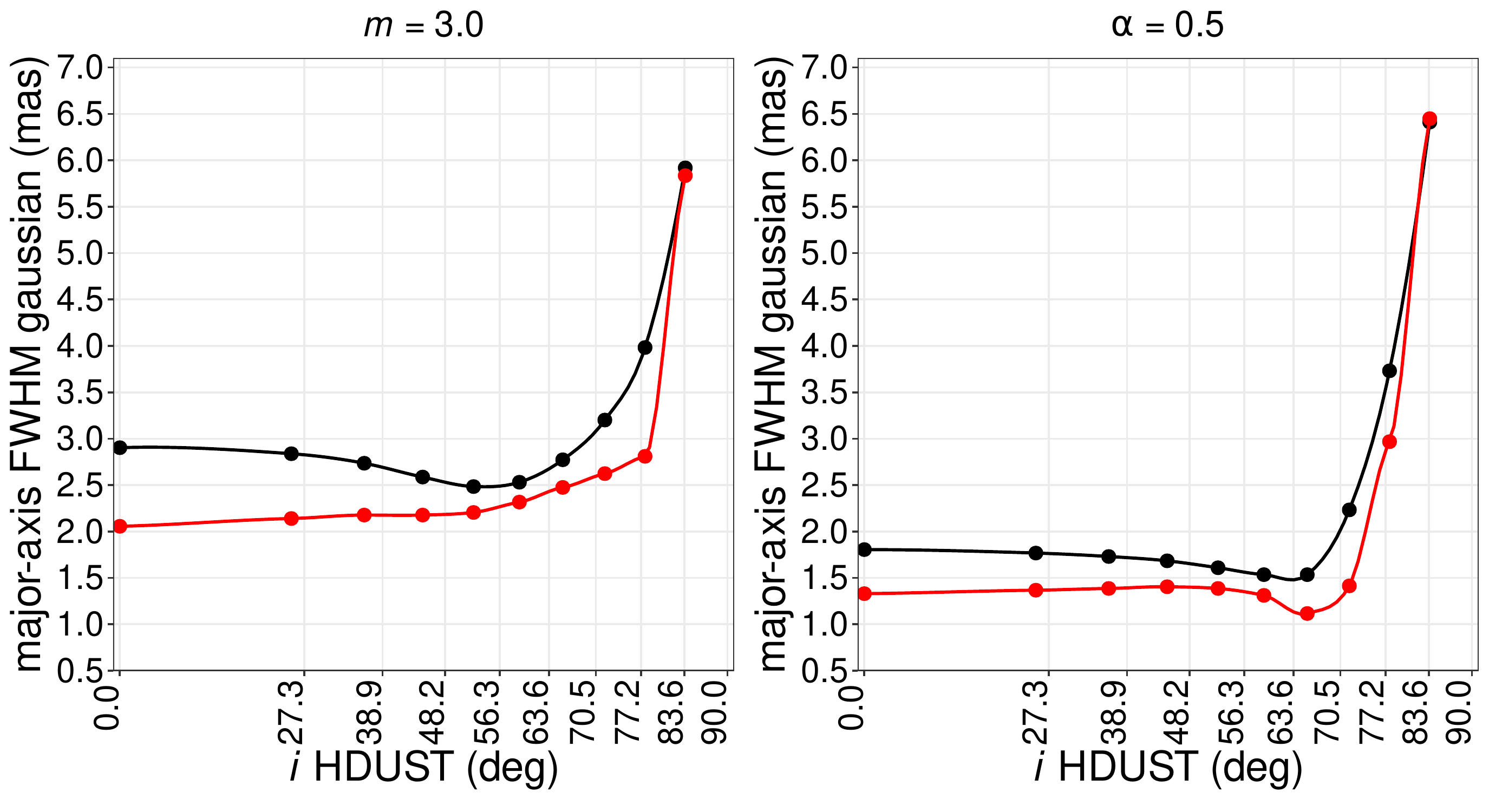}}}
\caption[Comparison between parametric HDUST models ($m$ = 3.0, left panels) and steady-state non-isothermal HDUST models (non-parametric models, right panels).]{Comparison between parametric HDUST models ($m$ = 3.0, left panels) and steady-state non-isothermal HDUST models (non-parametric models, right panels). The major-axis FWHM from a Gaussian fit to each HDUST image is shown as a function of HDUST inclination angle. Notice that, unlike the parametric case with $m=3.0$, the FWHM extension is very similar in both H$\alpha$ (black) and Br$\gamma$ (red) when considering the non-isothermal HDUST models in the grid. See text for discussion.}
\label{sec_publised_studies_plot_hdust_gauss_hit_m3.0_non_iso}
\end{figure}

From that, one sees that the circumstellar disk extension (major-axis FWHM Gaussian) presents a very similar response to the variation in $i$ for the group of non-isothermal models, and then it could be expected to derive closer value for $i$ from VEGA and AMBER, when compared with the parametric case with $m$ = 3.0. Therefore, compared with our best-fit model ($m$ = 3.0), we would be able to obtain a better agreement between $i$ derived from the fit to VEGA and AMBER, when considering the group of non-isothermal models in the BeAtlas. On the other hand, in this case of non-parametric models, the fit to AMBER is largely affected ($\chi^2_\mathrm{r} \sim$ 15). Notice, from Table 4, that the $\chi^2$ of the best-fit HDUST to AMBER is quite lower than that ($\chi^2_\mathrm{r} \sim$ 4.7), being this a parametric model. Finally, we remind the reader the BeAtlas grid is currently quite limited in terms of these self-consistent models ( Sect.~\ref{sec_radiative_transfer_modeling_hdust}), and additional calculations, exploring different values of disk viscosity, are needed in order to verify if the different values of $i$ found by our study from H$\alpha$ and Br$\gamma$ arise due to non-isothermal effects on the vertical structure of $\omicron$ Aquarii's disk.\par

In conclusion, our study shows that $\omicron$ Aquarii seems to fit in the global picture of Be stars, when looking at the continuum and Br$\gamma$ line, that is, a Keplerian rotation disk well-described by the so-called VDD model (Sect.~\ref{sec_intro_Be_disks}). However, the H$\alpha$ data cannot be fully understood in the framework of the VDD model: it seems needed to consider a super-Keplerian disk, that is, $\beta$ higher (more positive) than -0.5, to explain the H$\alpha$ interferometric data, the same being indicated considering the H$\alpha$ line profile. This shows that more complex physical processes in Be disks are needed to be taken into account when modeling quantities measured in H$\alpha$ in comparison to Br$\gamma$.\par 

Apart from these biases regarding H$\alpha$, we were able to successively model using the BeAtlas grid both the VEGA and AMBER data with the same physical description for the circumstellar disk, finding a base disk surface density $\Sigma_{0}$ = 0.12 g cm\textsuperscript{-2} and a radial density law exponent $m$ = 3.0. Our best-fit HDUST model, purely found from adjusting interferometric data, is able to reproduce other different types of observables of $\omicron$ Aquarii well, the spectral energy distribution, the H$\alpha$ and Br$\alpha$ line profile (Sect.~7 of Paper III), and also to predict fairly well the level of intrinsic polarization in the V-band, namely, 0.41\% from our physical model in comparison with 0.49 $\pm$ 0.03 \% that is derived from our polarimetric data (Sect.~8.4.3 of Paper III).\par

Furthermore, based on multi-technique observations (spectroscopy, interferometry, and polarimetry), our results supports that $\omicron$ Aquarii presents a globally stable disk structure for several years (up to two decades). Interestingly, from the modeling both of our visible and near-infrared interferometric data, our best-fit physical model (BeAtlas grid) presents a volume disk density exponent $m$ lower than the canonical value of 3.5 for the steady-sate regime in the VDD model, namely, $m$ = 3.0. As discussed in details in Sect.~8.4.4. of Paper III, others stable Be stars with stable disks also present values of $m$ lower than 3.5, such as $\beta$ Canis Minoris (B8Ve), $\alpha$ Arae (B2Vne), and $\alpha$ Columbae (B9Ve). Since $m$ = 3.5 stands for the simplified scenario of the single Be stars with a isothermal vertical height scale disk, this could indicate binary or non-isothermal effects on Be disks.\par

One possibility to explain such a long-term stability of $\omicron$ Aquarii's disk relies on its own high stellar rotational rate, very close ($\sim$ 96\%) to the critical value (391 $\pm$ 27 km s\textsuperscript{-1}). Thus, apart from other possible known mechanisms of mass loss in Be stars (Sect.~\ref{sec_intro_fast_rotation}), this could provide a nearly constant rate of mass-injection in the disk of $\omicron$ Aquarii.\par 

Lastly, the high temporal disk stability and rotational rate found for $\omicron$ Aquarii (B7IV) are consistent with the more recent scenario proposed in the literature that late-type Be stars are more likely to have stable disks and rotate near to the break-up limit, when compared with early-type Be stars~\citep[e.g.,][]{cranmer05, vieira17}.\par

\includepdf[pages=-]{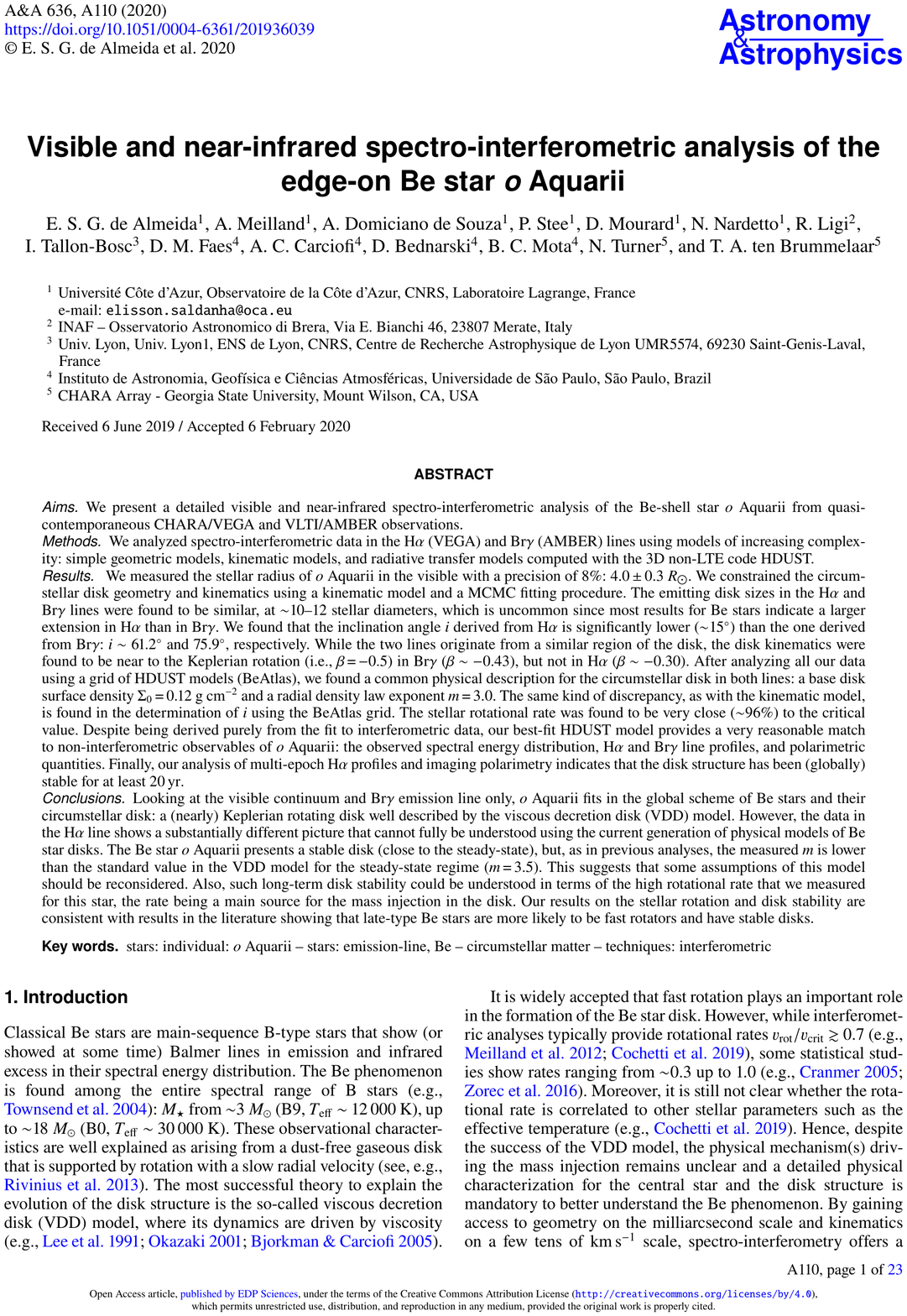}


\pagestyle{empty}
\cleardoublepage
\pagestyle{fancy}

\chapter{Ongoing studies}
\label{chapter_perspectives} 
\minitoc

My ongoing and near-future studies on the two main lines of research of this thesis are discussed here: radiative line-driven winds and disks of massive hot stars.\par 

Sect.~\ref{sec_ongoing_studies_rigel} discusses in details my work on the fundamental wind parameters of the blue supergiant star Rigel through a multi-wavelength interferometric approach (visible, near-infrared, and mid-infrared), while Sect.~\ref{sec_ongoing_studies_classical_be_stars} is devoted to the study of Be disks also by means of interferometry at multi-spectral channels.\par

\section{The radiative line-driven wind of Rigel}
\label{sec_ongoing_studies_rigel}


\subsection{Rigel: an evolved massive star}
\label{sec_ongoing_studies_rigel_overview}

Rigel ($\beta$ Orionis, HD 34085) is the MK-standard of spectral type B8Ia. This B supergiant star belongs to class of $\alpha$ Cygni stars, presenting variability at different time-scales, ranging from $\sim$1.2 to $\sim$74.7 days, due to non-radial pulsations~\citep[e.g.,][]{kaufer96, kaufer97, moravveji12a, moravveji12b}. Rigel is thought to have an initial mass of $\sim$25 $\mathrm{M_{\odot}}$ (O-type star) and to have already passed by a red supergiant phase before the blue-loop in the HR diagram~\citep{saio13}.\par

The circumstellar environment of Rigel was analysed through spectro-interferometry observations in the visible (H$\alpha$) with CHARA/VEGA~\citep{chesneau10} and in the near-infrared (Br$\gamma$) with VLTI/AMBER~\citep{chesneau14}. This latter study performed one of the largest interferometric campaigns of Rigel with observation collected in the periods of 2006-2007 and 2009-2010.\par 

One of the main findings of~\citet{chesneau14} is the wind variability of Rigel with the mass-loss rate changing in $\sim$20\% in the time-scale of one year (periods of 2006-2007 and 2009-2010). These authors found a mean mass-loss rate of $\sim$$9.4\e{-7}$ $\mathrm{M_\odot}$ yr\textsuperscript{-1} in the period of 2006-2007, while it is $\sim$$7.6\e{-7}$ $\mathrm{M_\odot}$ yr\textsuperscript{-1} in the latter one from modeling AMBER visibilities (measured in Br$\gamma$). In addition, they verified a substantial discrepancy between the mass-loss rate derived from the H$\alpha$ transition ($\sim$$1.5\e{-7}$ $\mathrm{M_\odot}$ $\mathrm{M_\odot}$ yr\textsuperscript{-1}), when compared with one found from Br$\gamma$ (up to about one order of magnitude).\par

Thus, further investigations on the wind parameters of Rigel are still necessary, allowing us to have a better picture of the winds around OBA supergiants. As discussed above, these stars show interesting photospheric phenomena, such as pulsations, which are undoubtedly connected to their wind properties~\citep[e.g.,][and references within]{kraus15, haucke18}. They also impose a current challenge to the standard CAK-theory of line-driven winds (see again our discussion on the wind acceleration of hot supergiants in Sect.~\ref{sec_intro_theory_radiative_line_winds}).


\subsection{Multi-band spectro-interferometry: CHARA and VLTI}
\label{sec_ongoing_studies_rigel_spectro_interferometry}

Rigel is one the brightest (V $\sim$ 0.13) massive stars, and thus an ideal target to further tests on the capability of the recent operating GRAVITY instrument (K-band region, VLTI second generation, Sect.~\ref{sec_interf_vlti}) to probe the structure and intensity of radiative line-driven winds around massive stars. Commissioned in 2016, GRAVITY has been successfully used to probe the circumstellar environments of massive stars, such as the study of wind-wind collision regions of $\eta$ Carinae~\citep{gravity_collaboration18} and HD 93206 A~\citep{sanchez17}.\par

\begin{figure}[t]
\centerfloat
\centerline{\resizebox{0.75\textwidth}{!}{\includegraphics{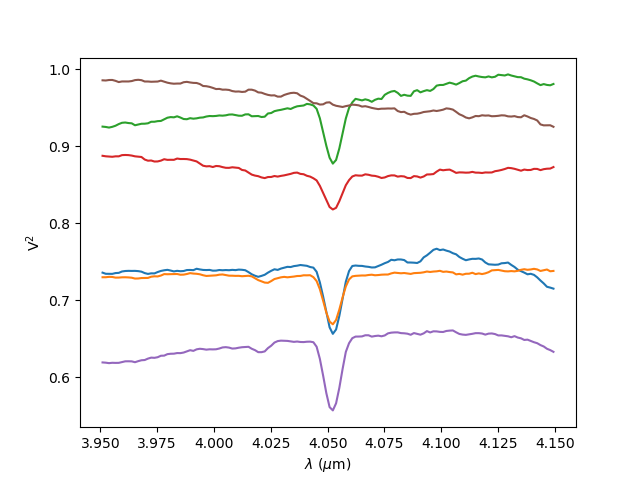}}}
\caption[VLTI/MATISSE observation of Rigel performed on September 25, 2019.]{VLTI/MATISSE observation of Rigel performed on September 25, 2019. The squared visibility for each baseline is shown as a function of wavelength around the Br$\alpha$ line. Note the large visibility drop in the Br$\alpha$ line due to the larger flux contribution from the wind in this line, when compared with the continuum.}
\label{sec_ongoing_studies_rigel_20190925_matisse_bralpha}
\end{figure}

The main objective of this project is investigate the mass-loss rate of Rigel through interferometric data obtained with VLTI/GRAVITY, in the K-band, conjointly with data from CHARA/VEGA (visible), VLTI/AMBER (near-infrared), and VLTI/MATISSE (mid-infrared). Part of our AMBER data was already analysed by~\citep{chesneau14}, as discussed above. In addition, we have new unpublished AMBER data obtained from our team at Laboratoire Lagrange, covering the period of 2006-2016 in total. At the end of 2019, Rigel was also observed by our team using the new VLTI mid-infrared beam-combiner MATISSE\footnote{Acronym for ``Multi AperTure mid-Infrared SpectroScopic Experiment''.}~\citep{lopez14}, during its Commissioning/Science Verification phase. Fig.~\ref{sec_ongoing_studies_rigel_20190925_matisse_bralpha} illustrate our MATISSE observations of Rigel (calibrated visibility around the Br$\alpha$ line) using different VLTI baseline configurations. Then we aim to address the wind variability of Rigel in terms of more recent multi-wavelength spectro-interferometric data covering the period of 2016-2019.\par

Together with my PhD advisors and other collaborators from Laboratoire Lagrange, I proposed to observe Rigel with the VLTI/GRAVITY instrument, as the PI of the ESO program 0100.D-0332\footnote{co-PIs: A.~Meilland, A.~Domiciano de Souza, M.~Carbillet, S.~Kannan, E.~Lagadec, F.~Millour, and P.~Stee.}. Our proposal was approved (ESO period 100, 6h of observation in the total) and all the observations were performed in service mode, between October 2017 and March 2018, using the high spectral resolution mode of GRAVITY ($R$ = 4000). To obtain different levels of spatial resolution of Rigel, covering both small and high frequencies in the Fourier plan, our observations were performed with all the possible auxiliary telescope quadruplets at VLTI, A0-B2-C1-D0, D0-G2-J3-K0, and A0-G1-J2-J3, allowing us to reach larger baseline lengths, respectively.\par

\begin{figure}[t]
\centerfloat
\centerline{\resizebox{0.85\textwidth}{!}{\includegraphics{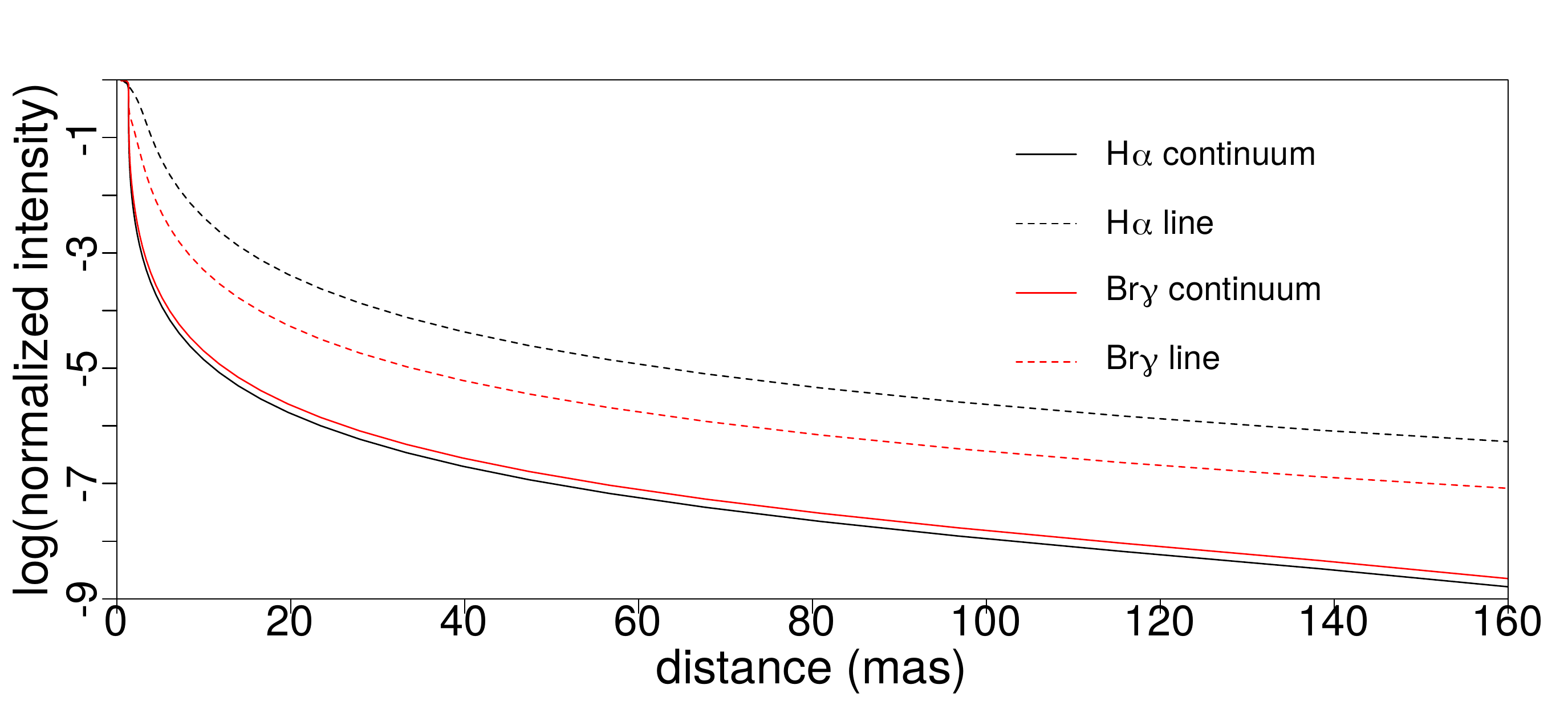}}}
\caption[Intensity profiles from one of our adopted CMFGEN models for Rigel, used as a reference model in our VLTI/GRAVITY proposal.]{Intensity profiles from one of our adopted CMFGEN models for Rigel, used as a reference model in our VLTI/GRAVITY proposal. These profiles are calculated at different spectral regions, in the H$\alpha$ and Br$\gamma$ lines, and in their close-by continuum regions. The intensity is shown as a function of distance from the center of the star (in milliarcsecond).}
\label{sec_ongoing_studies_rigel_cmfgen_e15_intensity_distance_halpha_brgamma}
\end{figure}

For preparing this observational proposal, I computed a small grid of atmosphere models using the code CMFGEN (1-D non-ELT radiative transfer) for different values of mass-loss rate, taking as reference the value of $\dot{M}$ tested by~\citet{chesneau14}. Also (partially) based on Chesneau' study about Rigel, the other parameters in our model are fixed as follows: $\log L_{\star}/L_{\odot}$ = 5.45, $T_{\mathrm{eff}}$ = 20000 K, $\log g$ = 2.6, $R_{\star}$ = 44.4 $\mathrm{R_{\odot}}$, and $v_{\infty}$ = 300 km s\textsuperscript{-1}.\par 

We see a notable difference between the CMFGEN model intensity in both the H$\alpha$ (VEGA) and Br$\gamma$ (AMBER and GRAVITY) lines, when compared with their close-by continuum regions, up to large distances from the photosphere (Fig.~\ref{sec_ongoing_studies_rigel_cmfgen_e15_intensity_distance_halpha_brgamma}). Such difference is highly dependent on the mass-loss rate used in the models due to the larger opacity of the wind in these lines than in the continuum. Fixing the other model parameters allows us to predict the drop in visibility in these lines for a certain value of $\dot{M}$, as shown in Fig.~\ref{sec_modeling_radiative_esaldanhaP100_fig1}. As expected, higher values of mass-loss rates yield a larger drop in the Br$\gamma$ visibility, and also a more intense Br$\gamma$ line profile.\par

\begin{figure}[t]
\centerfloat
\centerline{\resizebox{1.25\textwidth}{!}{\includegraphics{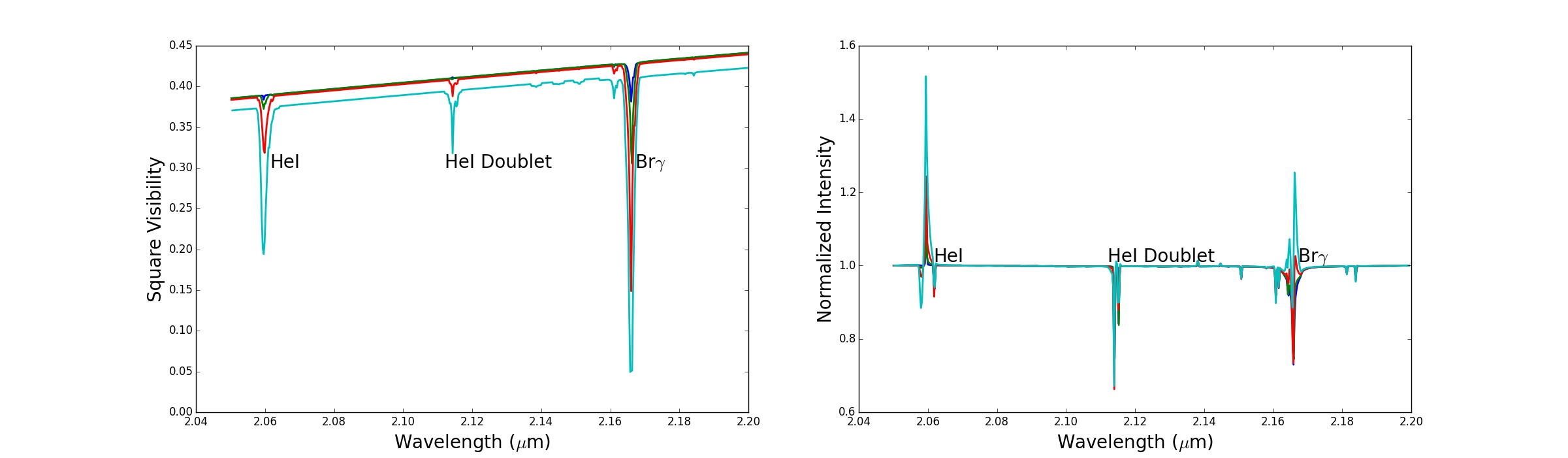}}}
\caption[Simulated visibilities (left panel) and spectra (right panel) in the K-band (GRAVITY wavelength region) from CMFGEN models with different values of mass-loss rate]{Simulated visibilities (left panel) and spectra (right panel) in the K-band (GRAVITY wavelength region) from CMFGEN models with different values of mass-loss rate: $\dot{M}$ = $1\e{-7}$ (blue), $2\e{-7}$ (green), $4\e{-7}$ (red), and $8\e{-7}$ (cyan) $\mathrm{M_\odot}$ yr\textsuperscript{-1}. All the other parameters are fixed as discussed in the text.}
\label{sec_modeling_radiative_esaldanhaP100_fig1}
\end{figure}

Fig.~\ref{sec_ongoing_studies_rigel_gravity_data_cmfgen_model} shows an example of our GRAVITY data (calibrated $V^2$) for one observational night, taken using the K0-G2-D0-J3 quadruplet (six different baselines). Unfortunately, information on the visibilities measured in the \ion{He}{I} lines is lost due to atmospheric effects. Nevertheless, the data quality is good with a clear signal of drop in Br$\gamma$ visibility, being this transition our principal interest in this project (well-known wind diagnostic, as discussed in Sect.~\ref{sec_spectro_line_diagnostics}).\par 

As a preliminary analysis, also shown in Fig.~\ref{sec_ongoing_studies_rigel_gravity_data_cmfgen_model}, we compare one of our CMFGEN model ($\dot{M}$ = $4\e{-7}$ $\mathrm{M_\odot}$ yr\textsuperscript{-1}) to calibrated $V^2$ data of Rigel from all the observations with GRAVITY (shown as a function of spatial frequency, instead of wavelength). In this case, our predicted visibility of the model is calculated considering a stellar diameter of 2.72 mas from~\citet{bourges17}.\par 

First, from just looking the calibrated GRAVITY visibility, we are able to almost resolve the stellar diameter with the longer baselines. Comparing our adopted CMFGEN model to data, we see that the assumed value of stellar diameter overestimates the overall drop in the continuum visibility, indicating a somewhat smaller value for the photosphere size of Rigel. In addition, this model also seems to overestimate the observed visibility drop in the Br$\gamma$, and then indicating a value of mass-loss rate lower than $4\e{-7}$ $\mathrm{M_\odot}$ yr\textsuperscript{-1}.\par

\begin{figure}[t]
  \begin{center}
  \begin{adjustbox}{minipage=\textwidth,scale=0.90}
  \includegraphics[width=0.95\textwidth]{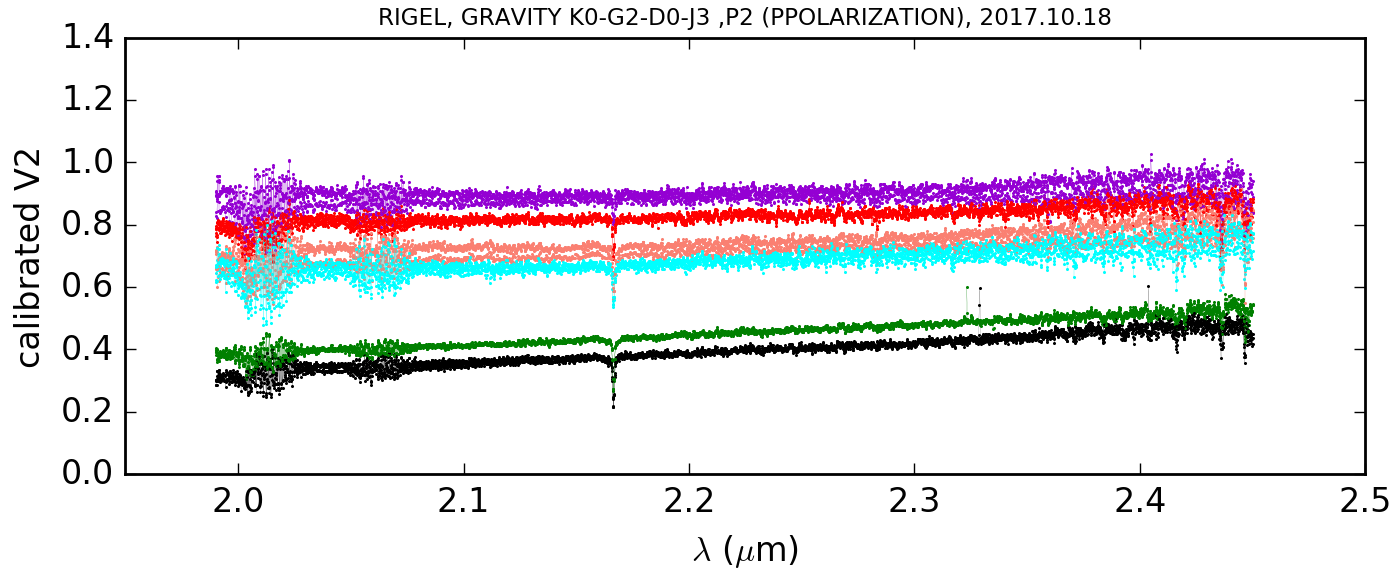}
  \includegraphics[width=1.00\textwidth]{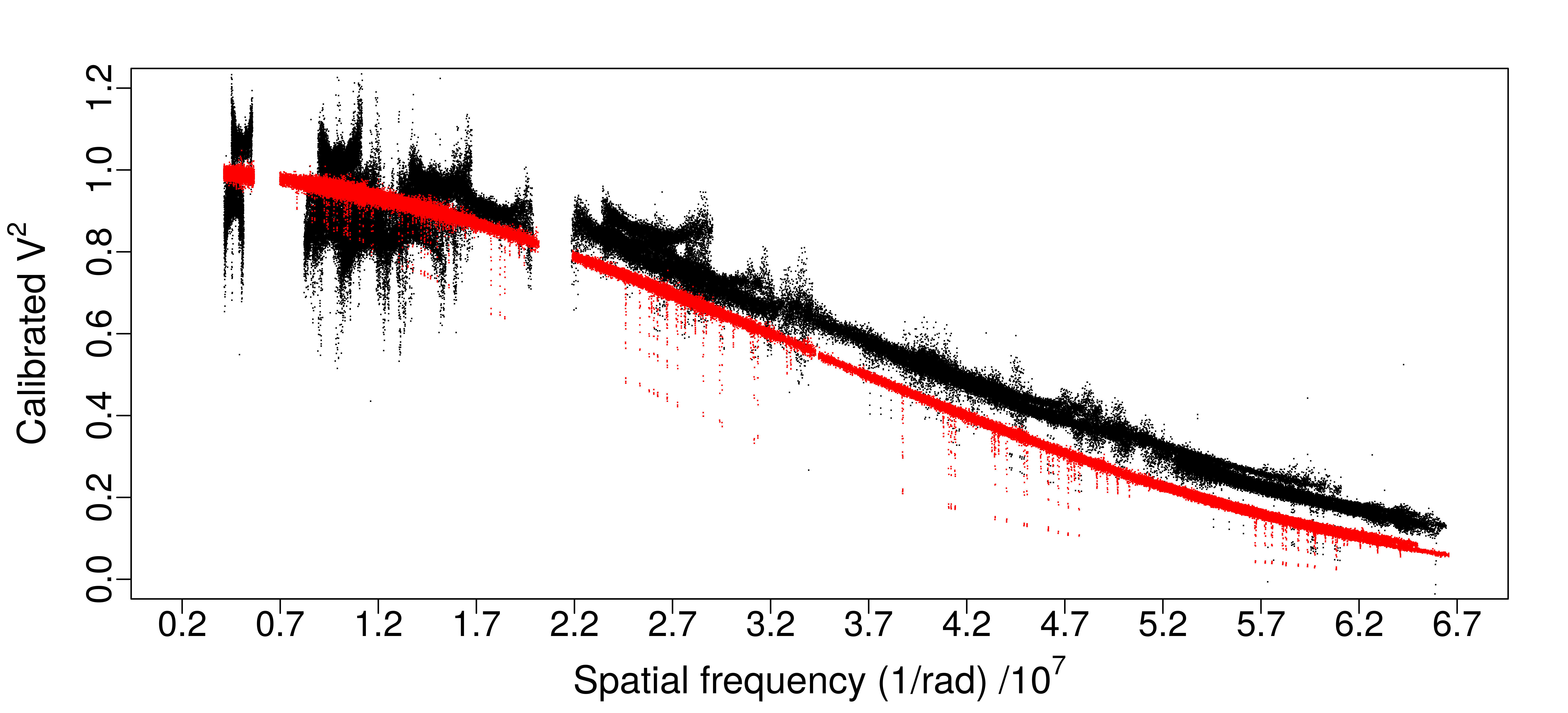}
  \end{adjustbox}
  \caption[]{Top panel: an example of reduced GRAVITY data for Rigel, observed in one night (18/10/2017) using the VLTI quadruplet K0-G2-D0-J3, as a function of wavelength (K-band). The baselines configurations are shown in different colors. Bottom panel: our complete GRAVITY dataset (8 nights, 6 h of observations in total) as function of spatial frequency (black), compared with the same CMFGEN shown in Fig.~\ref{sec_ongoing_studies_rigel_cmfgen_e15_intensity_distance_halpha_brgamma} (red). See text for discussion.}
  \label{sec_ongoing_studies_rigel_gravity_data_cmfgen_model}
  \end{center}
\end{figure}

Using the LITpro tool, I almost finished the work regarding the geometric modeling of calibrated GRAVITY visibility. These results are summarized in Fig.~\ref{sec_ongoing_studies_rigel_gravity_continuum_brgamma_litpro_modeling}. The GRAVITY visibility of Rigel is shown in the continuum region close to Br$\gamma$ (2.165-2.168 $\mu$m), and in the core of this line. As suggested by the analysis shown in Fig.~\ref{sec_ongoing_studies_rigel_gravity_data_cmfgen_model}, it is needed a smaller extension for the stellar diameter, a uniform disk with a diameter of $\sim$2.6 mas. However, the best-fit uniform disk found from fitting the $V^2$ measured in the core of Br$\gamma$ is significantly larger than this value from the continuum, with a diameter of $\sim$ 3.1 mas, due to the larger flux contribution from the wind in the Br$\gamma$ line. Therefore, this preliminary analysis from our GRAVITY data allows us to detect the wind of Rigel.\par

Further work is in progress on this study about Rigel. In particular, I finished the reduction of all the GRAVITY observations, using the standard GRAVITY data reduction pipeline\footnote{Publicly available at \url{https://www.eso.org/sci/facilities/paranal/instruments/gravity/tools.html}} provided by ESO~\citep{lapeyrere14}, and the analysis of calibrated visibility using geometric models is well-advanced, as discussed above.\par

Nevertheless, I still need to improve the grid of CMFGEN models for a proper analysis of these data. Our reference model has a substantially higher effective temperature ($T_{\mathrm{eff}}$ = 20000 K) than a reliable value for Rigel with $T_{\mathrm{eff}}$ $\sim$ 12000~\citep[e.g.,][]{przybilla06, chesneau14}. This difference comes from the fact that I started to work on models for Rigel based on my CMFGEN grid for O-type stars (Paper I), combined with the deadline for submitting our proposal to ESO. To have a fixed value of luminosity from~\citet{chesneau14}, our reference model has a quite smaller stellar radius of $\sim$44 $\mathrm{R_\odot}$, instead of the value adopted by these authors (115 $\mathrm{R_\odot}$). In addition, the model for Rigel presented here has the same chemical composition of the basic CMFGEN models of Paper I, and I have been working to improve this later point for the case of a late B supergiant as Rigel (B8Ia). In short, it is still a ``dirty'' CMFGEN model for Rigel.\par

\begin{figure}[t]
\centerfloat
\centerline{\resizebox{1.00\textwidth}{!}{\includegraphics{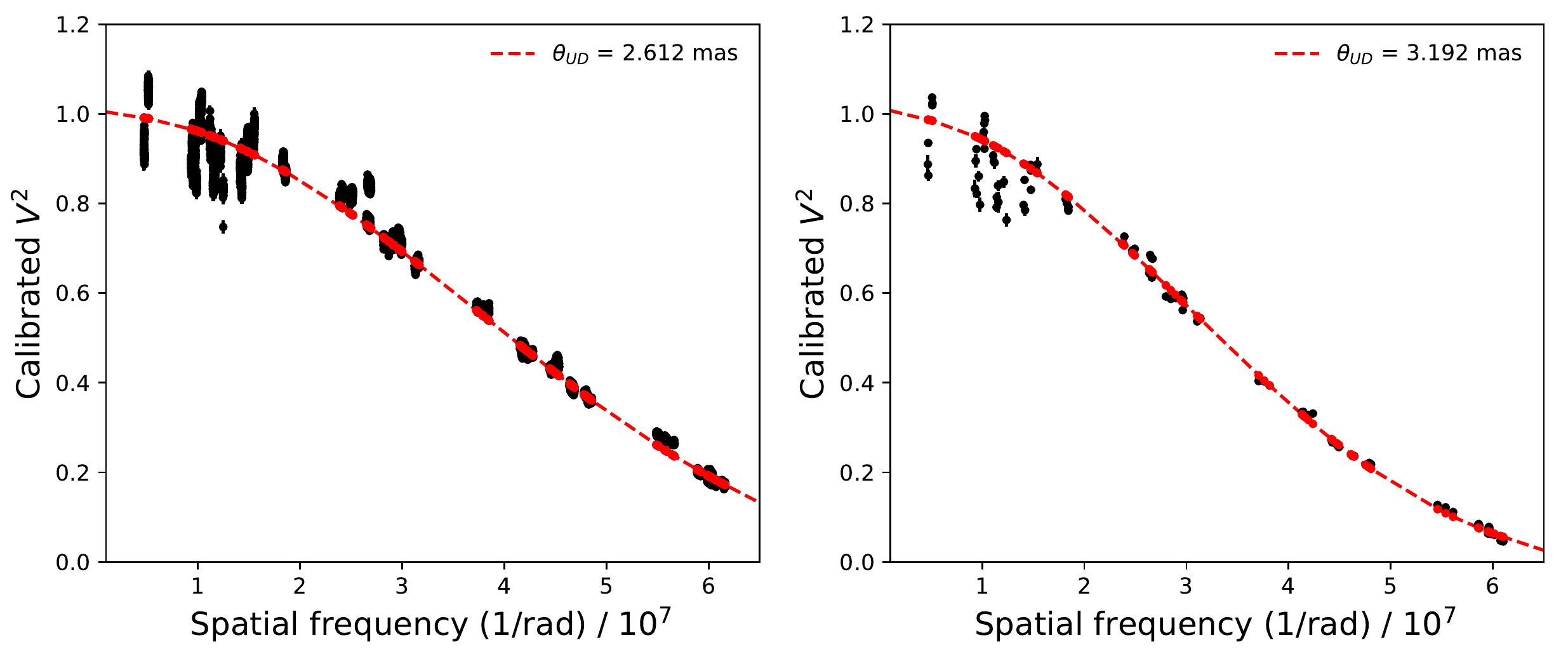}}}
\caption[LITpro best-fit uniform disk models found from fitting our GRAVITY data for Rigel, in the continuum region close to Br$\gamma$ (left panel) and in the core of Br$\gamma$ (right panel).]{LITpro best-fit uniform disk models found from fitting our GRAVITY data for Rigel, in the continuum region close to Br$\gamma$ (left panel) and in the core of Br$\gamma$ (right panel). Note the larger size of the fitted uniform disk when looking at the Br$\gamma$ line, compared with its close-by continuum.}
\label{sec_ongoing_studies_rigel_gravity_continuum_brgamma_litpro_modeling}
\end{figure}


\subsection{H$\alpha$ intensity interferometry: I2C team}
\label{sec_ongoing_studies_rigel_intensity_interferometry}

Furthermore, the I2C team observed Rigel (total of $\sim$50 h of observations) in a campaign performed from January 27 to February 14, 2020. Also as a short-term perspective, we aim to use our best-fit CMFGEN model for Rigel also to interpret these new H$\alpha$-band intensity interferometric data, applying on Rigel the same method used in Paper II for deriving the distance of P Cygni.\par 

In conclusion, we plan to write two papers based on these interferometric observations of Rigel, in both of them using the code CMFGEN, but with different aims, that is, investigating its wind parameters from spectro-interferometric data (VEGA, AMBER, GRAVITY, and MATISSE) and estimating its distance from H$\alpha$ intensity interferometry. We point out that Rigel's distance is estimated to be $\sim$265 pc with $\sim$10\% of uncertainty based on Hipparcos parallax measurements~\citep{vanleeuwen07}. Then, such an interferometric study of Rigel allows us to constrain its distance based on a different distance determination method. In addition, depending on the quality of these new H$\alpha$ intensity interferometric observations of Rigel, we could try to test our results for the wind mass-loss rate found from the best-fit model to the spectro-interferometric datasets.\par

Also using CMFGEN models, we intend to analyse a large sample of hot supergiants in the framework of this distance determination method. In addition to the importance of more accurate and precise distance determinations on many branches of astrophysics, such a large survey of OBA supergiants allows us to provide an independent check on the modified-wind momentum luminosity relation for massive stars, and then directly touching on our understanding about their wind properties.\par


\subsection{Direct imaging: VLT/SPHERE}
\label{sec_ongoing_studies_rigel_imaging}

Besides the power provided by interferometry to probe the structure of massive stars winds, another very interesting observational technique relies on direct imagining. For instance, based on near-diffraction limited observations, using adaptive optical system,~\citet{chesneau00} were able to resolve the H$\alpha$-emitting region of P Cygni and to detect clear signatures of clumps in the wind of P Cygni from the inner ($\sim$20 $R_{\star}$) up to the outer wind region ($\sim$1000 $R_{\star}$) (see Fig.~\ref{sec_analytical_chesneau2000_fig9}).\par

\begin{figure}[t]
  \begin{center}
  \begin{adjustbox}{minipage=\textwidth,scale=0.90}
  \includegraphics[width=0.475\columnwidth]{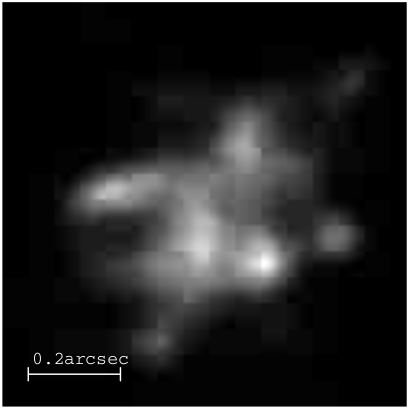}
  \includegraphics[width=0.475\columnwidth]{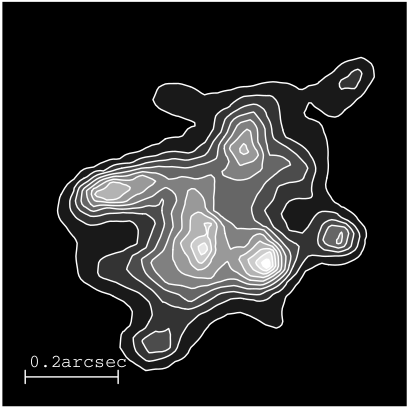}
  \end{adjustbox}
  \caption[Image reconstruction by~\citet{chesneau00} of P Cygni large-scale circumstellar environment (up to $\sim$1000 $R\star$) using Observatoire de Haute-Provence (OHP) observations with adaptive optics.]{Image reconstruction by~\citet{chesneau00} of P Cygni large-scale circumstellar environment (up to $\sim$1000 $R\star$) using Observatoire de Haute-Provence (OHP) observations with adaptive optics. The central star has an angular diameter of about 0.2 mas. Reproduced from~\citet{chesneau00}.}
  \label{sec_analytical_chesneau2000_fig9}
  \end{center}
\end{figure}

The VLT/SPHERE instrument is the current state-of-the-art of the technique of high dynamical coronographic imaging, giving us access on how the circumstellar environment is distributed up to larger distance from the stellar surface. Despite being more widely used for studies on exoplanet and evolved cool stars, observation of massive stars with SPHERE are for sure of great interest, for example, when looking for departures of spherical symmetry in their radiative line-driven winds.\par 

Fig.~\ref{sec_ongoing_studies_esaldanhaP102_fig3} shows our simulations of SPHERE observations (in Pa$\beta$ and Br$\gamma$) based on our reference CMFGEN for Rigel. I calculated this model in the framework of my small set of CMFGEN models for preparing an observational proposal of Rigel with GRAVITY and SPHERE. Thus, using SPHERE, we would be able to detect the wind of Rigel both in Pa$\beta$ and Br$\gamma$ and constrain its geometry at a larger scale than using long-baseline interferometry, as with GRAVITY, allowing us to determine the mass-loss rate of Rigel's wind at different scales.\par

\begin{figure}[t]
\centerfloat
\centerline{\resizebox{1.00\textwidth}{!}{\includegraphics{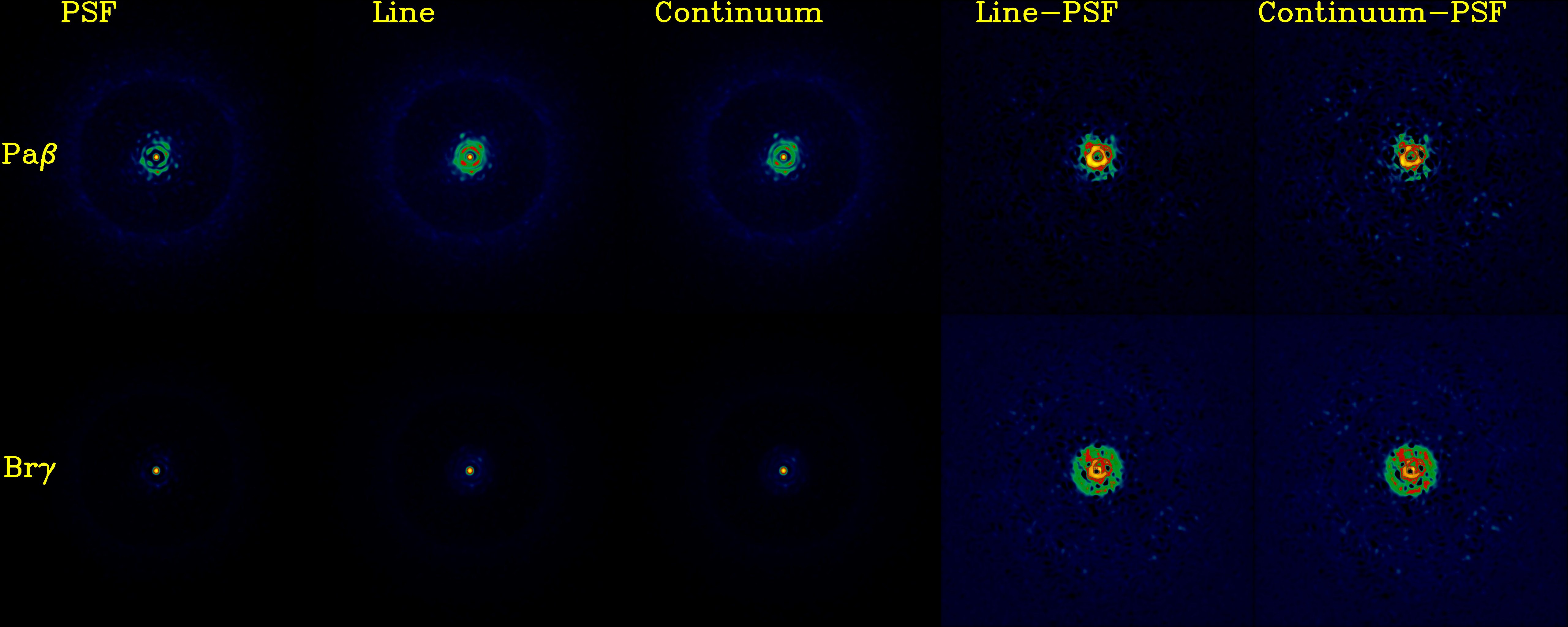}}}
\caption[SPHERE simulated observations (raw images) from our reference CMFGEN model for Rigel (with mass-loss rate of $4\e{-7}$ M$_\odot$yr$^{-1}$).]{SPHERE simulated observations (raw images) from our reference CMFGEN model for Rigel (with mass-loss rate of $4\e{-7}$ M$_\odot$yr$^{-1}$). Note the difference between the PSF and the line/continuum image in both lines (Pa$\beta$ and Br$\gamma$).}
\label{sec_ongoing_studies_esaldanhaP102_fig3}
\end{figure}

Unlike our request with GRAVITY (discussed in Sect.~\ref{sec_ongoing_studies_rigel_spectro_interferometry}), unfortunately our last request for observing Rigel with SPHERE was denied. Nevertheless, this science case about Rigel shows the current large potential for studying the morphology and physical conditions on massive stars winds using high angular resolution techniques other than interferometry. Of course, these simulations for the wind of Rigel encourage us to request again time to observe this star with VLT/SPHERE in the near-future.\par


\section{Classical Be stars}
\label{sec_ongoing_studies_classical_be_stars}

\subsection{Drawing a big picture of Be disks}

Our detailed analysis of the Be star $\omicron$ Aquarii (Sect.~\ref{sec_results_omicron_aquarii}; Paper III) helped us to constrain its parameters, as the equatorial rotation velocity and the kinematical and physical condition of its disk, and draw a unified picture of this star and its environment in the visible and near-infrared region.\par 

My work on $\omicron$ Aquarii paved the way for a more systematic analysis of Be stars, based on the same methodology used in Paper III. This means to use the kinematic model with a MCMC mode fitting procedure and also a large grid of pre-calculated HDUST models (BeAtlas grid), together with my routines for data analysis, to model spectro-interferometric data measured at different transitions.\par

We aim to analyse a large sample of Be stars, based both on published and new spectro-interferometric data taken by our group at OCA. Such a multi-wavelength interferometric Be survey allows us to find a more complete description on the morphology, kinematics, and physical properties of both the cooler and hotter Be stars with disks seen under different directions. For instance, this should help us to test more firmly, in a statistical way, the picture proposed in Paper III that the quite similar disk extension of $\omicron$ Aquarii in both H$\alpha$ and Br$\gamma$ arises due to its nearly edge-on view.\par

In Sect.~\ref{sec_perspetives_be_survey}, we present a very brief discussion on our VEGA and AMBER Be stars datasets, and Sect.~\ref{sec_ongoing_studies_be_stars_matisse} discusses our very preliminary observations of Be stars using the new VLTI mid-infrared beam combiner MATISSE. Lastly, Sect.~\ref{sec_perspetives_individual_stars} summarizes some of our on-going studies about individual objects.\par

\subsection{The VEGA and AMBER large surveys}
\label{sec_perspetives_be_survey}

\begin{figure}[t]
\centerfloat
\centerline{\resizebox{1.00\textwidth}{!}{\includegraphics{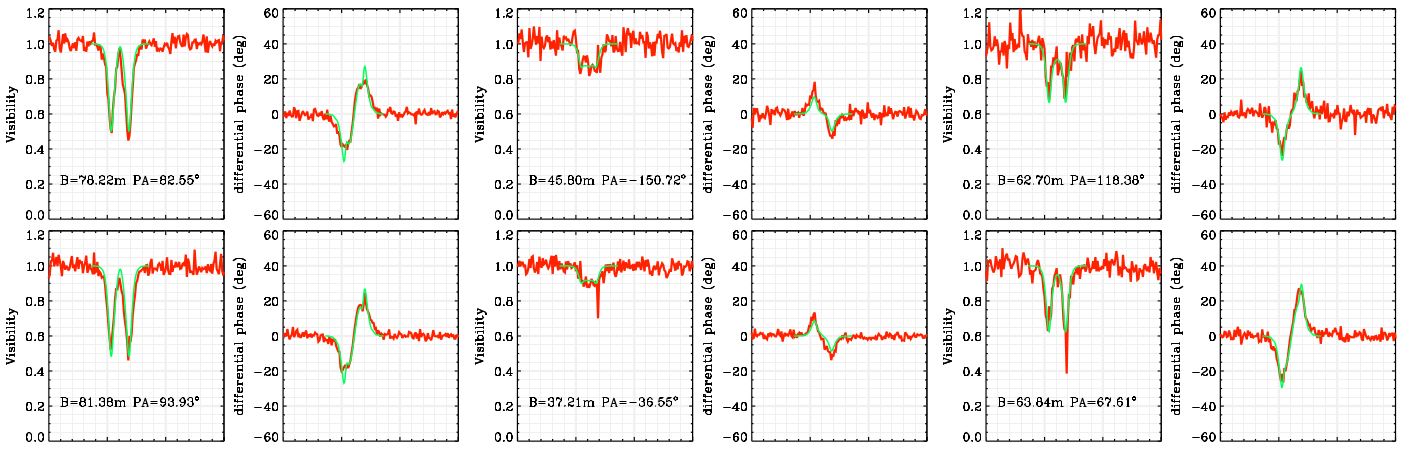}}}
\caption[Example of AMBER differential data (red line) of $\alpha$ Arae observed in the Be survey of~\citet{meilland12}.]{Example of AMBER differential data (red line) of $\alpha$ Arae observed in the Be survey of~\citet{meilland12}. The best-fit kinematic model found by these authors is shown in green line. Adapted from~\citet{meilland12}.}
\label{sec_ongoing_studies_fig9_meilland12_adapted}
\end{figure}

\begin{figure}[!h]
  \begin{center}
  \begin{adjustbox}{minipage=\textwidth,scale=1.05}
  \includegraphics[width=0.49\columnwidth]{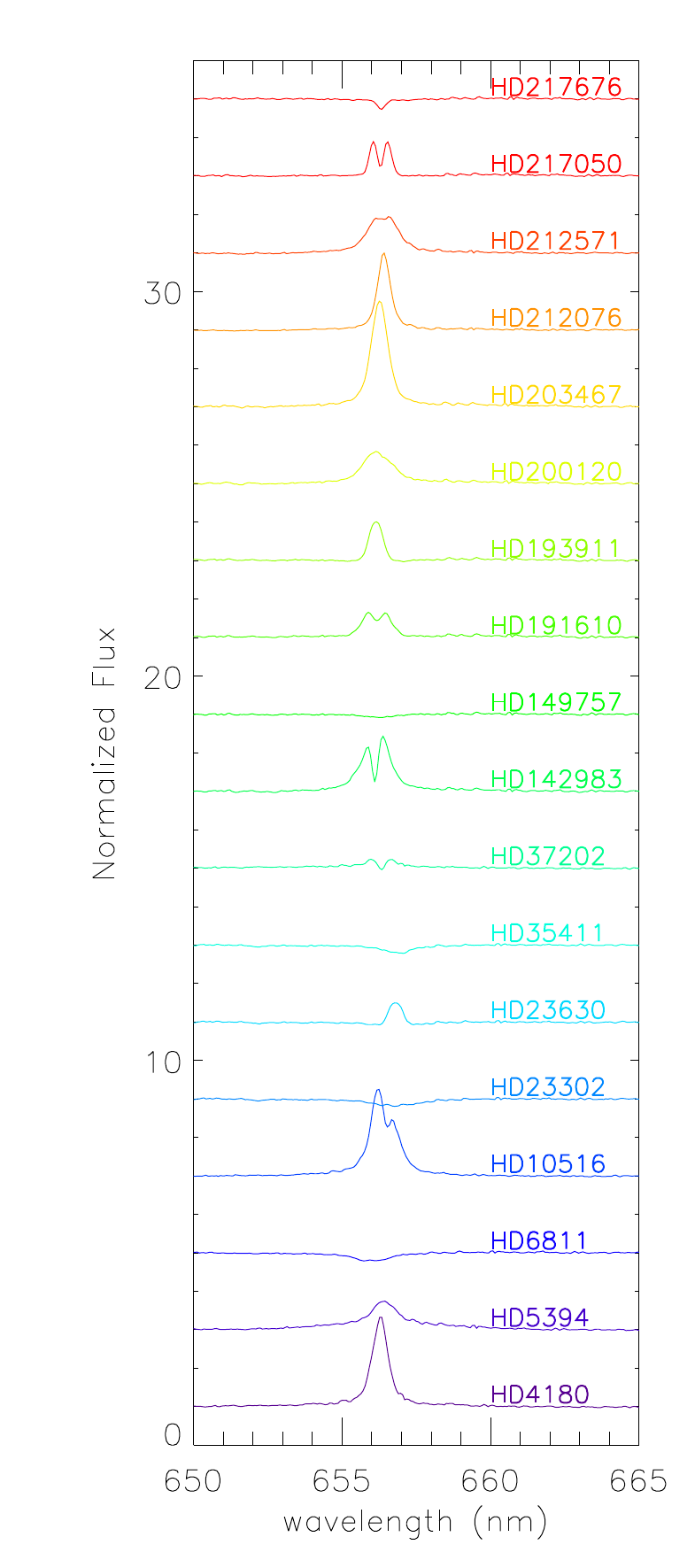}
  \medskip
  \includegraphics[width=0.46\columnwidth]{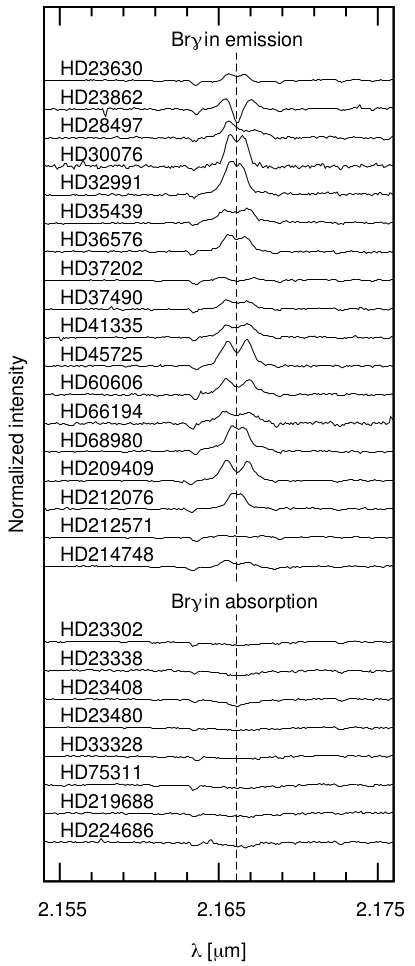}
  \end{adjustbox}
  \caption[Left: example of H$\alpha$ line profiles observed with the VEGA instrument for 17 Be stars of our VEGA survey (34 objects observed in total).]{Left: example of H$\alpha$ line profiles observed with the VEGA instrument for 17 Be stars of our VEGA survey (34 objects observed in total). Right: Br$\gamma$ line profiles of the Be star sample (26 objects) analysed by~\citet{cochetti19}. The right panel is reproduced from~\citet{cochetti19}.}
  \label{sec_ongoing_studies_be_survey_vega_spectrum}
  \end{center}
\end{figure}

Regarding the K-band region, all the AMBER data for this project was already published in the Be surveys of~\citet{meilland12} (8 objects) and~\citet{cochetti19} (26 objects). Details on their star samples can be found in Tables 1 and 2 of~\citet{meilland12} and~\citet{cochetti19}, respectively.\par 

For instance, the Br$\gamma$ profiles of the sample analysed by~\citet{cochetti19} are shown in Fig.~\ref{sec_ongoing_studies_be_survey_vega_spectrum}. In addition, Fig.~\ref{sec_ongoing_studies_fig9_meilland12_adapted} shows part of the AMBER dataset of $\alpha$ Arae (HD 158427, B2Vne) from~\citet{meilland12} together with the best-fit kinematic model found by these authors to fit their AMBER data to this star. Interestingly, as discussed in Sect.~\ref{sec_intro_disk_dynamics}, the AMBER study of~\citet{meilland07a} on this star provided the first direct detection of a Keplerian rotating disk in Be stars, in addition to find evidences of an enhanced polar wind in $\alpha$ Arae, that is, contributing to the flux distribution of its environment along the rotational axis.\par

We remind the reader that these studies are based purely on the kinematic code, but without the implementation of an automatic fitting procedure as performed in this thesis using the code EMCEE. Furthermore, as these authors do not employ radiative transfer models on their analysis, we have the opportunity to model their AMBER datasets using our BeAtlas grid of HDUST models to derive the disk density parameters of the objects in their samples.\par

In complement to the analysis in the Br$\gamma$ line, A.~Meilland has been leading a large observational program of Be stars with the VEGA instrument since about 2015 (observations centered at H$\alpha$).\par 

To date, we have observed 34 Be stars using VEGA, in particular, almost choosing the CHARA two-telescope configurations with different lengths and orientations, S1S2 ($\sim$34 m), E1E2 ($\sim$66 m), W1W2 ($\sim$108m), allowing us to resolve our targets with different levels of angular resolution. This program is almost finished, only missing observations of more 7 stars (for which any data were taken). 

Fig.~\ref{sec_ongoing_studies_be_survey_vega_spectrum} shows the observed H$\alpha$ line profiles for 17 objects of our VEGA sample. One sees that most part of them present prominent emission-lines in H$\alpha$ due to the flux contribution arising form their circumstellar disks. In addition, our VEGA sample encompasses both Be stars with low and high inclination angles, as one can see by the shape of H$\alpha$ in Fig.~\ref{sec_ongoing_studies_be_survey_vega_spectrum}. Based on the H$\alpha$ spectroscopic study of~\citet{silaj10}, HD 212076 (single-peak H$\alpha$ profile) has $i$ $\sim$ \ang{20}, while HD 217050 (double-peak H$\alpha$ profile) is seen close to edge-on showing $i$ $\sim$ \ang{70}.\par


\subsection{First observations with MATISSE}
\label{sec_ongoing_studies_be_stars_matisse}

\begin{figure}[t]
\centerfloat
\centerline{\resizebox{1.25\textwidth}{!}{\includegraphics{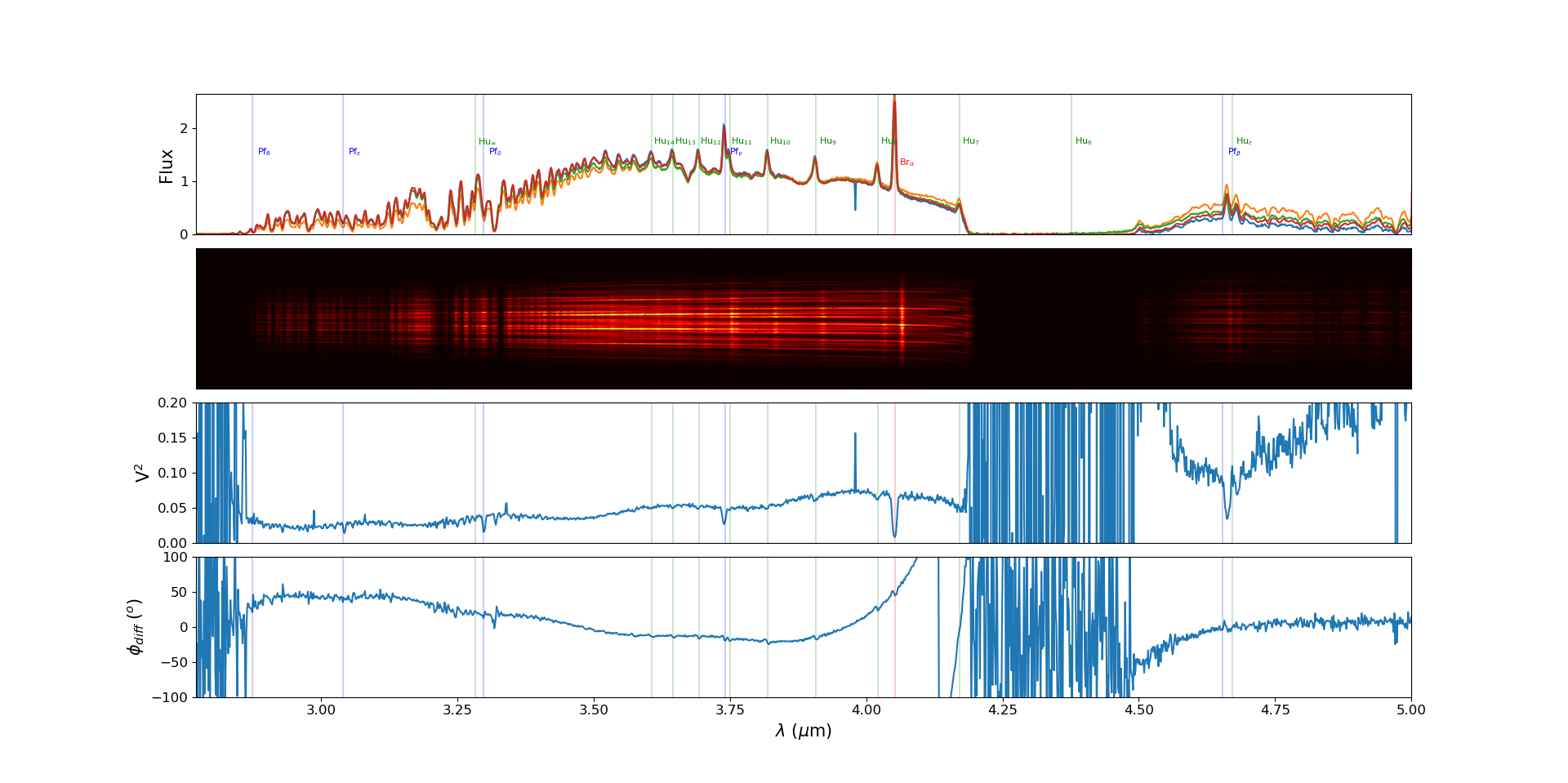}}}
\caption[Observation of the Be star $\alpha$ Arae with VLTI/MATISSE from the L-band to the M-band (covering $\sim$3.0-5.0 $\mu$m).]{Observation of the Be star $\alpha$ Arae with VLTI/MATISSE from the L-band to the M-band (covering $\sim$3.0-5.0 $\mu$m). The first row shows the observed spectrum of $\alpha$ Arae, and the principal spectral lines, mainly due to hydrogen (Brackett, Pfund, and Humphreys series), are indicated. The second row shows the measured MATISSE fringes and the measured interferometric quantities (the squared visibility and differential phase) are shown in the last two rows.}
\label{sec_ongoing_studies_example_matisse_observations_delta_centauri}
\end{figure}

Our team at OCA has been involved in the Commissioning/Science Verification phase of the VLTI/MATISSE instrument, and a small sample of bright Be stars (seven objects) were observed during that between 2018 and 2020, mostly of earlier spectral types: $\alpha$ Arae (B2V), $\delta$ Scorpii (B0.3IV), $\delta$ Centauri (B2V), $\eta$ Centauri (B2V), 48 Librae (B5III), Achernar (B6V), and $\mu$ Centauri (B2V). Then, besides the visible and near-infrared analyses provided by VEGA and AMBER, these new data will allow us to constrain the circumstellar disk parameters also in the mid-infrared region.\par

Fig.~\ref{sec_ongoing_studies_example_matisse_observations_delta_centauri} shows observations of $\delta$ Centauri with MATISSE in the L- and M-bands (from $\sim$3.0 to 5.0 $\mu$m). The observed spectrum is shown here in addition to the measured interferometric fringes and the computed quantities (visibility and differential phase). This illustrates how rich is the mid-infrared spectral region of Be stars, showing several emission-lines being the Br$\alpha$ and Pf$\gamma$ the strongest ones. As discussed in Sect.~\ref{sec_spectro_line_diagnostics}, Br$\alpha$ and Pf$\gamma$ are usually the best lines to trace the density structure of the environments in massive stars~\citep[e.g,][]{lenorzer04, najarro11}. Indeed, from Fig.~\ref{sec_ongoing_studies_example_matisse_observations_delta_centauri} the largest visibility drops (when comparing to the nearby continuum) are found in these transitions.\par

\begin{figure}[t]
\centerfloat
\centerline{\resizebox{1.00\textwidth}{!}{\includegraphics{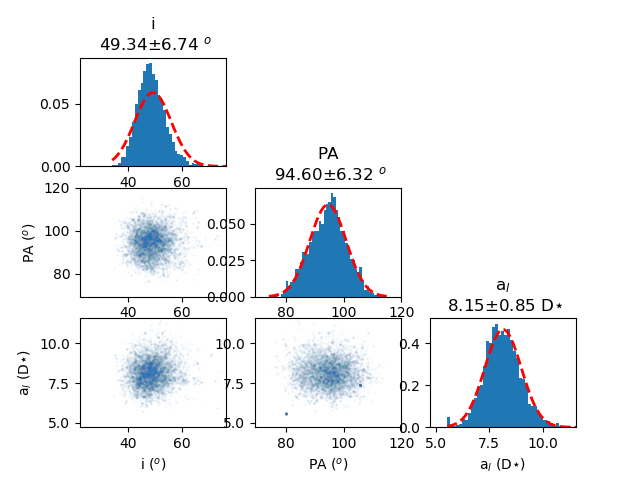}}}
\caption[Preliminary MCMC analysis of VLTI/MATISSE differential data (in the Br$\alpha$ line) of the Be star $\alpha$ Arae]{Preliminary MCMC analysis of VLTI/MATISSE differential data (in the Br$\alpha$ line) of the Be star $\alpha$ Arae, fitted using the kinematic code with three free parameters: stellar inclination angle, disk major-axis position angle, and the disk major-axis FWHM in Br$\alpha$. See text for discussion.}
\label{sec_ongoing_studies_example_matisse_mcmc_fit_alpha_arae}
\end{figure}

As a very preliminary analysis of these MATISSE data, Fig.~\ref{sec_ongoing_studies_example_matisse_mcmc_fit_alpha_arae} shows a simple evaluation of our MCMC kinematic model fitting code to fit differential MATISSE data (in the Br$\alpha$ line) of $\alpha$ Arae. Despite being a simple initial test, we were able to constrain the three kinematic model parameters that were free in the fitting fairly-well: inclination angle of $\sim$\ang{49}, disk major-axis position angle of $\sim$\ang{95}, and the disk size in Br$\alpha$ of $\sim$8 $D_{\star}$.\par 

We point out that this star was also analysed in the AMBER Be surveys of~\citet{meilland12} and~\citet{cochetti19}, and they determined these parameters as follows: $i$ = 45 $\pm$ \ang{5}, disk $PA$ = 88 $\pm$ \ang{2}, and $a_{\mathrm{line}}$ (Br$\gamma$) = 5.8 $\pm$ 0.5 $D_{\star}$. Then our results presented here for the inclination angle and the disk position angle seems to be in line with the values derived these authors.\par 

From Fig.~\ref{sec_ongoing_studies_example_matisse_mcmc_fit_alpha_arae}, we note that the disk extension seems to be substantially larger about 1.4 time) in Br$\alpha$ than in Br$\gamma$. A deeper analysis is needed due to the preliminary status of our results, but this could result from a possible larger flux contribution of $\alpha$ Arae's disk in the mid-infrared, when compared with shorter wavelengths, and it shows the current importance of investigating Be disks at different spectral bands.\par


\subsection{Toward a detailed view on Be stars (other than $\omicron$ Aquarii)}
\label{sec_perspetives_individual_stars}

\begin{figure}[t]
\centerfloat
\centerline{\resizebox{0.75\textwidth}{!}{\includegraphics{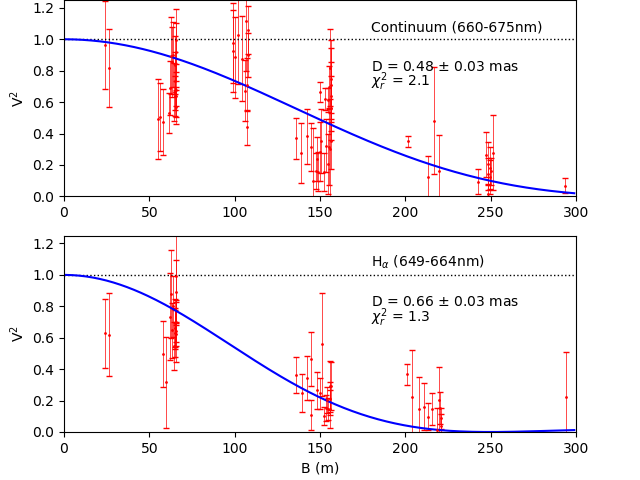}}}
\caption[Geometric modeling of VEGA squared visibilities of the late-type Be star $\kappa$ Draconis (B6IIIe).]{Geometric modeling of VEGA squared visibilities of the late-type Be star $\kappa$ Draconis (B6IIIe). Note that the VEGA visibility is quite lower in the H$\alpha$-band, in comparison with the close-by continuum band, and then a larger uniform disk is found in the H$\alpha$-band.}
\label{sec_ongoing_studies_kappa_draconis_geometric_modeling_vega}
\end{figure}

Lastly, besides the large spectro-interferometric programs of Be stars described above, our modeling approach in Paper III is very useful to provide a more robust analysis of specific objects, as for the case of $\omicron$ Aquarii.\par 

In the following, we present a non-extensive list of old and new data of Be stars that have been currently investigated by our group and collaborators:

\begin{enumerate}[label=(\roman*)]
\setlength\itemsep{1em}
\item Published VEGA and MIRC data of $\phi$ Persei by~\citet{mourard15}. In spite of using image reconstruction models and the kinematic code, it still lacks an interpretation of their data with physical models, and this can be done using the BeAtlas grid.\par

\item New 2016 VEGA and FRIEND data of $\gamma$ Cassiopeiae observed in four-telescope configuration.\par

\item VEGA imaging program on $\kappa$ Draconis (PI: A.~Meilland; 2015-2019). This program was finished and the analysis is going-on. Fig.~\ref{sec_ongoing_studies_kappa_draconis_geometric_modeling_vega} shows the geometric modeling of the calibrated VEGA data in two different bands: continuum band (close to H$\alpha$) and H$\alpha$-band.\par

\item VEGA imaging program on $\beta$ Canis Majoris (PI: R.~Klement; 2017-present). This program is close to be finished. $\beta$ Canis Majoris is known to have a stable disk and was recently investigated in details by~\citet{klement15} with multi-instrument spectroscopic and interferometric data, interpreted using the code HDUST, and it shows evidences of a truncated disk by an unseen binary companion.\par

\end{enumerate}


\pagestyle{empty}
\cleardoublepage
\pagestyle{fancy}

\chapter{Conclusions and perspectives}
\label{chapter_conclusions}

The objective of this thesis was to study the mechanisms of mass loss in different types of massive hot stars. In addition to the own physical properties of their environments (e.g., the mass-loss rate and the density law profile), this task demands a detailed analysis of their photospheric parameters, such as the effective temperature and rotational rates, as performed in this thesis.\par 

This means that we investigated both the less massive hot stars showing relevant extended atmospheres (as the Be stars), for which the stellar rotation is understood to be (at least partially) a key mechanism to form their circumstellar disks, as well as the more massive hot stars (as the O and LBV stars), which show larger lost material from their surfaces through strong radiative line-driven winds arising from their very high luminosities and effective temperatures.\par

For this purpose, we combined different observational methods, mainly focusing on interferometry and spectroscopy at different wavelength regions (UV, visible, and infrared). In Paper I, spectroscopic data taken with multiple instruments was used to constrain the photospheric and wind parameters of a small sample of late-type O giants in the Galaxy. In Paper II, the LBV star P Cygni was studied using H$\alpha$ intensity interferometry. Finally, Paper III studied quasi-simultaneous spectro-interferometric observations of the Be star $\omicron$ Aquarii in the visible (CHARA/VEGA) and in the near-infrared (VLTI/AMBER).\par

Besides the particular modeling methods used in each one of these studies, such as the multi-tool approach by Paper III, non-LTE radiative transfer models are key among all of them: CMFGEN (Papers I and II), one of the state-of-the-art codes for hot stars with intense radiative line-driven winds, and HDUST that is well-suited for hot stars with highly asymmetrical environments as in Be stars (Paper III). Our studies evidence the success of these forefront codes to model different types of observables in hot stars: photometry, polarimetry, spectroscopy, and interferometry.\par 

Based on different observational methods and spectral bands, we were then able to draw a unified picture of the physical conditions on the surface and environments of massive stars using these codes. Apart of constraining their fundamental parameters, radiative transfer modeling of interferometry can also be explored to constrain other quantities of interest, as shown by Paper II, allowing us to estimate distances to these stars in an independent way to other methods.\par

We tried hard to completely conciliate the UV and H$\alpha$ mass-loss rates of O giant stars, but it was not possible for all the stars of our sample. Nevertheless, we clearly showed that the theoretical mass-loss rates for the most part of our sample fail to match simultaneously the observed UV and visible spectra, and this departure is much more explicit than in the case of late O dwarfs.\par 

We showed for first time that more evolved massive stars, late O giants, present the weak wind phenomenon (originally found in late O dwarfs). Our derived mass-loss rates are much lower (by up to $10^2$) than the values found by theoretical studies. The luminosity region around $\log (L_\star/\mathrm{L_\odot}) \sim 5.2$ is indeed critical for the onset of weak winds in O stars.\par

In short, our study suggests that weak winds are unlikely linked to evolutionary effects in O stars, and weak winds should occur along the H-burning phase of less luminous O stars ($\log (L_\star/\mathrm{L_\odot}) \lesssim 5.2$). This means that the majority of state-of-the-art stellar evolution models are taking severely overestimated mass-loss rates into account for these stars. Further investigations should evaluate the impact of weak winds on their physical properties, probably affecting their rotation and surface chemistry during their evolutionary stages.\par

To date, our interferometric study on $\omicron$ Aquarii allowed us to draw one of the most complete pictures of a Be star and its disk both in the visible and the near-infrared. From that, we were able to precisely constrain both the central star and disk parameters of $\omicron$ Aquarii. We showed that its disk extension is quite similar in both H$\alpha$ and Br$\gamma$. Such a quite uncommon feature for Be disks could be explained in terms of an opacity effect in these lines for Be stars seen under high inclination angles, and it deserves further investigations. We showed that the disk density is well-described by a common physical model, that is, the same values for the disk density parameters in the H$\alpha$ and Br$\gamma$ lines, and this result is linked to the similar disk extension found at these lines. 

We found that $\omicron$ Aquarii has a stable disk over a very long period: combined to older literature studies, our results suggest a global disk stability for up to at least 40 years. Such a long-term disk stability can be interpreted in terms of our results showing that $\omicron$ Aquarii has a rotational rate very close to the break-up limit. This means that the rotational rate could be a main source of mass injection in the disk of $\omicron$ Aquarii.\par 

Despite being a result for an individual star, it matches well the more recently suggested picture of the Be phenomenon that late-type Be stars, such as $\omicron$ Aquarii, are more likely to have a quasi-critical rotation and long-term stable disks, when compared with the earlier Be stars. In comparison with the later Be stars, early-type Be stars present a more relevant contribution from the radiative force to break the hydrostatic equilibrium at the photosphere, and then the rotational rate should be less important for the process of mass loss in more massive Be stars.\par

In short, $\omicron$ Aquarii fits in the global scheme of Be stars and their circumstellar disk, that is, a Keplerian rotating disk well described by the VDD model when looking at the Br$\gamma$ emission line. However, the determination of the disk kinematics seems to be significantly biased when looking at H$\alpha$, apart from our efforts to conciliate our picture of its kinematics at these lines simultaneously. As indicated by some previous studies on Be stars, this bias could arise due to a higher effect of non-coherent scattering on the H$\alpha$ line formation than in Br$\gamma$.\par

The findings of this thesis clarified some issues regarding the environments of massive hot stars, but, of course, future investigations are needed to constrain their fundamental parameters. The results found for the mass-loss rates of line-driven winds in massive stars diverge, when looking to their winds at several spectral bands (X-ray, UV, visible, and infrared regions). Furthermore, it is still not firmly understood how the stellar properties of Be stars, such as the effective temperature and the rotational rate, correlate among themselves, and consequently are linked to the formation of the envelopes in these stars. Chap.~\ref{chapter_perspectives} presented my ongoing and near-future studies on some of these open issues: it clearly shows the current need to continue to study both winds and disks of massive hot stars at different spectral regions.\par 

Finally, the infrared region should be more explored to constrain the fundamental parameters of the environments of these stars: the spectral lines in this region are less investigated when compared with more traditional wind diagnostics, as H$\alpha$, in the visible~\citep[e.g.,][]{lenorzer04, najarro11, marcolino17}. A quantitative spectroscopic analysis of a significant sample of O stars simultaneously in the UV, visible, near- and mid-infrared regions is still missing. As pointed out by~\citet{najarro11}, together with other lines in the L-band, as Pf$\gamma$ (3.731 $\mu$m), the Br$\alpha$ (4.051 $\mu$m) line is very sensitive to the wind mass-loss rate of massive stars. In particular, Br$\alpha$ seems to be a very promising mass loss diagnostic for low-density winds. Then Br$\alpha$ is an interesting diagnostic to probe the weak wind phenomenon in the infrared region. Despite being more suited for studying dusty environments, the VLTI/MATISSE spectro-interferometric instrument (operating at the L-, M-, and N-bands) just have started to investigate at these lines the morphology and physical conditions of the environments around Be, O, and B supergiant stars. Moreover, in the following years, the instruments at the James Webb Space Telescope will allow us to investigate their physical properties by means of spectroscopy at even longer wavelengths up to $\sim$30 $\mu$m.\par

}
\bibliographystyle{aa}

{\setstretch{1.25}
\bibliography{references}
}


\end{document}